%% file: Thesis.tex
\documentclass[12pt,a4paper]{report}

\include{I-packages}
\include{I-config}
\include{I-packages-2} 

\begin{document}

\pagestyle{empty}
\input{Preliminaries/1-titlepage}

\clearpage
\pagestyle{preliminary}
\input{Preliminaries/2-quote}
\clearpage
\input{Preliminaries/3-abstract_EN}
\clearpage
\input{Preliminaries/4-abstract_FR}
\clearpage
\input{Preliminaries/5-publications}
\clearpage
\input{Preliminaries/6-acknowledgments}
\clearpage
\input{Preliminaries/7-expression}
{
    \setstretch{1}
    \hypersetup{linkcolor=black}
    \tableofcontents
}

\clearpage

\pagestyle{chapter}

\subfile{Chapters/Chapter1-Introduction}

\subfile{Chapters/Chapter2-Quantum}

\subfile{Chapters/Chapter3-Metrology}

\subfile{Chapters/Chapter4-GraphStates}

\subfile{Chapters/Chapter5-QECC}

\subfile{Chapters/Chapter6-Crypto}

\subfile{Chapters/Chapter7-Remarks}

\appendix

\subfile{Appendices/AppendixA}

\subfile{Appendices/AppendixB}

\subfile{Appendices/AppendixC}

\printbibliography[title={Bibliography}, heading=bibintoc]

\end{document}

%% file: I-packages.tex
\usepackage{graphicx} 
\usepackage{caption}
\usepackage{subcaption}
\usepackage{fancyhdr} 
\usepackage[utf8]{inputenc}
\usepackage[T1]{fontenc} 
\usepackage{setspace} 
\usepackage[left=3cm,right=3cm,top=4cm,bottom=4cm]{geometry}
\usepackage{slantsc}
\usepackage{titlesec}
\usepackage{fix-cm}
\usepackage{xcolor}
\usepackage{mfirstuc} 
\usepackage{calc}
\usepackage{bm}
\usepackage{amsmath}
\usepackage{amssymb}
\usepackage{pifont}
\usepackage{physics}
\usepackage{mathrsfs}
\usepackage{array}
\usepackage[backend=biber,style=ieee-alphabetic,citestyle=alphabetic]{biblatex}
\usepackage{hyperref} 
\usepackage{afterpage}
\usepackage{float}
\usepackage{CJKutf8}
\usepackage{standalone}
\usepackage{tikz}
\usepackage{pgfplots}
\pgfplotsset{compat=1.17}
\usetikzlibrary{shapes.geometric,positioning,arrows,calc,math,angles,quotes,arrows.meta, bending, decorations.markings}
\tikzset{
    tri/.style={
        draw,
        shape border rotate=90,
        isosceles triangle,
        isosceles triangle apex angle=60,
        fill=red,
        node distance=0.5ex
    }
}
\newcommand{\xmark}{\ding{55}}
\newcommand{\tikzcircle}[2][red,fill=red]{\tikz[baseline=-0.7ex]\draw[#1,radius=#2] (0,0) circle ;}
\usepackage{blochsphere}


%% file: I-config.tex
\definecolor{gray75}{gray}{0.75}
\newcommand{\hsp}{\hspace{0pt}}

\fancypagestyle{preliminary}{
    
    \fancyhead[RCL]{}

    \pagenumbering{roman}
}
\fancypagestyle{chapter}{
    
    \fancyhead[L]{\normalfont\itshape\fontsize{10pt}{12pt}\selectfont\nouppercase{\leftmark}}
    \fancyhead[R]{}

    \fancyfoot[C]{\thepage}

    \pagenumbering{arabic}
}

\newcommand{\headingBaseline}{12}
\newcommand{\headingBaselineDiv}{10}
\newlength{\chapterFontSize}
\newlength{\sectionFontSize}
\newlength{\subsectionFontSize}
\newlength{\chapterBaseline}
\newlength{\sectionBaseline}
\newlength{\subsectionBaseline}

\titleformat{\chapter}[hang]{\flushright
\fontseries{b}\fontsize{35}{35}\selectfont}{\fontseries{b}\fontsize{100}{130}\selectfont \textcolor{gray75}\thechapter\hsp}{0pt}{\\ \vspace{-30pt}\textcolor{gray75}}[]
\titlespacing*{\chapter}{0pt}{-15pt}{35pt}

\titleformat{name=\chapter,numberless}[hang]
  {\bfseries\LARGE}{}{1pt}{\center}[]
\titlespacing*{name=\chapter,numberless}{0pt}{-45pt}{10pt}

\titleformat{\section}[hang]
    {\normalfont\bfseries\fontsize{\sectionFontSize}{\sectionBaseline}\selectfont}{\thesection}{5pt}{\capitalisewords}
\titlespacing*{\section}{0pt}{25pt}{15pt}

\titleformat{\subsection}[hang]
    {\normalfont\bfseries\fontsize{\subsectionFontSize}{\subsectionBaseline}\selectfont}{\thesubsection}{5pt}{\capitalisewords}
\titlespacing*{\subsection}{0pt}{20pt}{10pt}

\usepackage{hyperref}
\hypersetup{
    colorlinks = true,
    linkcolor = blue, 
    citecolor = blue, 
    urlcolor = blue, 
    filecolor = black 
}

%% file: I-packages-2.tex
\usepackage{subfiles}

%% file: Preliminaries/1-titlepage.tex
\begin{center}

\setstretch{1}

\begin{spacing}{1.7}

\rule{\textwidth}{0.8mm}

{\fontfamily{bch}\selectfont {\LARGE\bfseries Quantum Information Techniques for Quantum Metrology}}

\vspace{-0.4\baselineskip}

\rule{\textwidth}{0.8mm}

\vspace{0.7\baselineskip}

{\sffamily \LARGE{Nathan Shettell}}
\end{spacing}

\vspace{25pt}

\includegraphics[height=90pt]{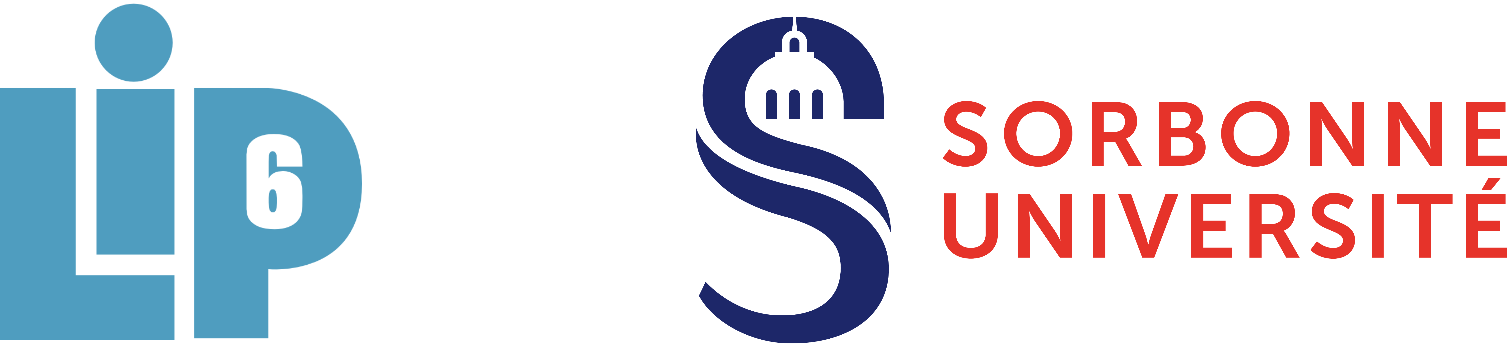}

\vspace{25pt}

{\large LIP6 \\ \vspace{8pt} SORBONNE UNIVERSITÉ}

\vspace{70pt}

{\footnotesize
\begin{tabular}{l r}
    \textbf{Jacob Dunningham}, \textit{Professeur, University of Sussex, Angleterre} & Rapporteur \\
    \textbf{Pieter Kok}, \textit{Professeur, University of Sheffield, Angleterre} & Rapporteur \\
     \textbf{Lorenzo Maccone}, \textit{Professeur, Università di Pavia, Italie} & Examinateur \\
     \textbf{Nicolas Treps}, \textit{Professeur, Sorbonne Université, France} & Examinateur \\
     \textbf{Pérola Milman}, \textit{Directrice de recherche, CNRS, France} & Examinateur \\
     \textbf{Damian Markham}, \textit{Chargé de recherche, CNRS, France} & Directeur de Thèse
\end{tabular}

}

\vspace{75pt}

{\scshape\setstretch{1.25}\Large Submitted for the degree of \\ Doctor of Philosophy\\
}

\vspace{30pt}

DECEMBER 2021

\vfill

\end{center}

%% file: Preliminaries/2-quote.tex
\vspace*{\fill}
\begin{quote}
\textit{Try and be nice to people, avoid eating fat, read a good book every now and then, get some walking in, and try and live together in peace and harmony with people of all creeds and nations.} \begin{flushright} -Monty Python's The Meaning of Life \end{flushright}
\end{quote}
\vspace*{\fill}

%% file: Preliminaries/3-abstract_EN.tex
\begin{center}
\LARGE\textbf {Abstract (English)}
\end{center}
\vspace{5pt}

\noindent
\textit{Quantum metrology} is an auspicious discipline of quantum information which is currently witnessing a surge of experimental breakthroughs and theoretical developments. The main goal of quantum metrology is to estimate unknown parameters as accurately as possible. By using quantum resources as probes, it is possible to attain a measurement precision that would be otherwise impossible using the best classical strategies. For example, with respect to the task of phase estimation, the maximum precision (the Heisenberg limit) is a quadratic gain in precision with respect to the best classical strategies. Of course, quantum metrology is not the sole quantum technology currently undergoing advances. The theme of this thesis is exploring how quantum metrology can be enhanced with other quantum techniques when appropriate, namely: graph states, error correction and cryptography.

Graph states are an incredibly useful and versatile resource in quantum information. We aid in determining the full extent of the applicability of graph states by quantifying their practicality for the quantum metrology task of phase estimation. In particular, the utility of a graph state can be characterised in terms of the shape of the corresponding graph. From this, we devise a method to transform any graph state into a larger graph state (named a bundled graph state) which approximately saturates the Heisenberg limit. Additionally, we show that graph states are a robust resource against the effects of noise, namely dephasing and a small number of erasures, and that the quantum Cramér-Rao bound can be saturated with a simple measurement strategy.

Noise is one of the biggest obstacles for quantum metrology that limits its achievable precision and sensitivity. It has been showed that if the environmental noise is distinguishable from the dynamics of the quantum metrology task, then frequent applications of error correction can be used to combat the effects of noise. In practise however, the required frequency of error correction to maintain Heisenberg-like precision is unobtainable for current quantum technologies. We explore the limitations of error correction enhanced quantum metrology by taking into consideration technological constraints and impediments, from which, we establish the regime in which the Heisenberg limit can be maintained in the presence of noise.

Fully implementing a quantum metrology problem is technologically demanding: entangled quantum states must be generated and measured with high fidelity. One solution, in the instance where one lacks all of the necessary quantum hardware, is to delegate a task to a third party. In doing so, several security issues naturally arise because of the possibility of interference of a malicious adversary. We address these issues by developing the notion of a cryptographic framework for quantum metrology. We show that the precision of the quantum metrology problem can be directly related to the soundness of an employed cryptographic protocol. Additionally, we develop cryptographic protocols for a variety of cryptographically motivated settings, namely: quantum metrology over an unsecured quantum channel and quantum metrology with a task delegated to an untrusted party.

Quantum sensing networks have been gaining interest in the quantum metrology community over the past few years. They are a natural choice for spatially distributed problems and multiparameter problems. The three proposed techniques, graph states, error correction and cryptography, are a natural fit to be immersed in quantum sensing network. Graph states are an well-known candidate for the description of a quantum network, error correction can be used to mitigate the effects of a noisy quantum channel, and the cryptographic framework of quantum metrology can be used to add a sense of security. Combining these works formally is a future perspective.

%% file: Preliminaries/4-abstract_FR.tex
\begin{center}
\LARGE\textbf {Résumé (Français)}
\end{center}
\vspace{5pt}

\noindent
\textit{La métrologie quantique} est une discipline prometteuse de l'information quantique qui connaît actuellement une vague de percées expérimentales et de développements théoriques. L'objectif principal de la métrologie quantique est d'estimer des paramètres inconnus aussi précisément que possible. En utilisant des ressources quantiques comme sondes, il est possible d'atteindre une précision de mesure qui serait autrement impossible en utilisant les meilleures stratégies classiques. Par exemple, en ce qui concerne la tâche d'estimation de la phase, la précision maximale (la limite d'Heisenberg) est un gain de précision quadratique par rapport aux meilleures stratégies classiques. Bien entendu, la métrologie quantique n'est pas la seule technologie quantique qui connaît actuellement des avancées. Le thème de cette thèse est l'exploration de la manière dont la métrologie quantique peut être améliorée par d'autres techniques quantiques lorsque cela est approprié, à savoir : les états graphiques, la correction d'erreurs et la cryptographie.

Les états de graphes sont une ressource incroyablement utile et polyvalente dans l'information quantique. Nous aidons à déterminer l'étendue de l'applicabilité des états de graphes en quantifiant leur utilité pour la tâche de métrologie quantique de l'estimation de phase. En particulier, l'utilité d'un état de graphe peut être caractérisée en fonction de la forme du graphe correspondant. À partir de là, nous concevons une méthode pour transformer tout état de graphe en un état de graphe plus grand (appelé "bundled graph states") qui sature approximativement la limite de Heisenberg. En outre, nous montrons que les états de graphe constituent une ressource robuste contre les effets du bruit (le déphasage et un petit nombre d'effacements) et que la limite quantique de Cramér-Rao peut être saturée par une simple stratégie de mesure.

Le bruit issu de l’environnement est l'un des principaux obstacles à la métrologie quantique, qui limite la précision et la sensibilité qu'elle peut atteindre. Il a été démontré que si le bruit environnemental peut être distingué de la dynamique de la tâche de métrologie quantique, des applications fréquentes de correction d'erreurs peuvent être utilisées pour combattre les effets du bruit. En pratique, cependant, la fréquence de correction d'erreurs requise pour maintenir une précision de type Heisenberg est impossible à atteindre pour les technologies quantiques actuelles. Nous explorons les limites de la métrologie quantique améliorée par la correction d'erreurs en prenant en compte les contraintes et les obstacles technologiques, à partir desquels nous établissons le régime dans lequel la limite d'Heisenberg peut être maintenue en présence de bruit.

La mise en œuvre complète d'un problème de métrologie quantique est technologiquement exigeante : des états quantiques intriqués doivent être générés et mesurés avec une grande fidélité. Une solution, dans le cas où l'on ne dispose pas de tout le matériel quantique nécessaire, consiste à déléguer une tâche à un tiers. Ce faisant, plusieurs problèmes de sécurité se posent naturellement en raison de la possibilité d'interférence d'un adversaire malveillant. Nous abordons ces questions en développant la notion de cadre cryptographique pour la métrologie quantique. Nous montrons que la précision du problème de la métrologie quantique peut être directement liée à la solidité d'un protocole cryptographique employé. En outre, nous développons des protocoles cryptographiques pour une variété de paramètres motivés par la cryptographie, à savoir : la métrologie quantique sur un canal quantique non sécurisé et la métrologie quantique avec une tâche déléguée à une partie non fiable.

Les réseaux de détection quantique ont suscité un intérêt croissant dans la communauté de la métrologie quantique au cours des dernières années. Ils constituent un choix naturel pour les problèmes distribués dans l'espace et les problèmes multiparamètres. Les trois techniques proposées, les états de graphes, la correction d'erreurs et la cryptographie, s'intègrent naturellement dans les réseaux de détection quantique. Les états de graphes sont un candidat bien connu pour la description d'un réseau quantique, la correction d'erreurs peut être utilisée pour atténuer les effets d'un canal quantique bruyant et le cadre cryptographique de la métrologie quantique peut être utilisé pour ajouter un sentiment de sécurité. La combinaison formelle de ces travaux est une perspective future. 

%% file: Preliminaries/5-publications.tex
\begin{center}
\LARGE\textbf {List Of Publications}
\end{center}
\vspace{5pt}

\noindent \textbf{In Publication}

\begin{itemize}
    \item N. Shettell and D. Markham, “Graph states as a resource for quantum metrology”, \textit{Physical Review Letters}, vol. 124, no. 11, p. 110 502 (2020).
    \item N. Shettell, W. J. Munro, D. Markham and K. Nemoto, “Practical limits of error correction for quantum metrology”, \textit{New Journal of Physics}, vol. 23, no. 4, p. 043 038 (2021).
    \item Y. Ouyang, N. Shettell and D. Markham, “Robust quantum metrology with explicit symmetric states”, \textit{IEEE Transactions on Information Theory} (2021).
    \item N. Shettell, E. Kashefi and D. Markham “A cryptographic approach to quantum metrology”, \textit{Physical Review A}, vol. 105, p. L010401 (2022).
\end{itemize}

\noindent \textbf{Pre-Print}

\begin{itemize}
    \item N. Shettell and D. Markham, “Quantum Metrology with Delegated Tasks”, \textit{arXiv preprint arXiv:2112.09199} (2021).
\end{itemize}

%% file: Preliminaries/6-acknowledgments.tex
\begin{center}
\LARGE\textbf {Acknowledgements}
\end{center}
\vspace{5pt}

\noindent The beauty of physics lies within its impossibly large scope. To paraphrase french philosopher Maupertuis: `the movement of animals and the vegetative growth of plants are consequences of the laws of nature'. These laws nature which govern a physical system take into account innate properties, but also interactions with other systems. The liminal space that is a PhD is not subjected to the aforementioned laws of nature, yet, there is an analogous statement to be made about my PhD journey, in that it was molded by properties of my own self, like passion and persistence, but also, the interactions I had with others. Before delving into the contents of this thesis, I would like to extend my gratitude to those who helped shape the last three years of my life into such a rich and fulfilling experience.

First and foremost, I would like to thank my supervisor Damian for his support and flexibility. The environment fostered by Damian is a perfect balance of guidance and freedom. I am extremely grateful for his insights and perspectives. For the opportunity to work in Tokyo for six months. For instigating Friday night beers. For being an exemplary researcher.

Thanks to my friends and colleagues of LIP6. The colourful cast of characters made for an entertaining work environment, which, as far as I'm concerned, is unparalleled. Thank you for the coffee breaks. Thank you for the post-work trips to the bar. Thank you for the insightful academic discussions and the delightful balderdash. Good luck to all for great beginnings.

Thanks to those at NII and to those at Sakura house in Tabata. My time spent in Japan was unforgettable: the remarkable food, overnight karaoke and the magical countryside. \begin{CJK}{UTF8}{min} ありがとうございました。\end{CJK} I would like to thank Prof. Kae Nemoto and Dr. Bill Munro, for taking on an unofficial role of co-supervisors while I was present. Thank you for the mentorship and guidance, and for integrating me in your group. I thoroughly enjoyed my time at NII.

Thank you to my friends at the Maison des étudiants canadiens. Thank you for reminding me of home, the shenanigans on the terrace, Wednesday night karaoke at Fleurus, and the Sunday brunches. Specifically, for those present during restrictions brought on by the pandemic of Covid-19: la confinement, couvrefeu, fermeture des bars et restaurants, ou peu importe. Thank you for the sense of community.

Last, but by no means the least, thank you to my family: Mom, Dad, Brit and Jake. I am grateful for the unconditional love and support from across the Atlantic. Without you, I would not be the person I am today.

%% file: Preliminaries/7-expression.tex
\thispagestyle{empty}
\vspace*{\fill}
\begin{quote}
\begin{center}
\textit{Per aspera ad astra.}
\end{center}
\end{quote}
\vspace*{\fill}

%% file: Chapters/Chapter1-Introduction.tex
\chapter{Introduction}

\section{Quantum Technologies}

The advent of quantum theory has completely revolutionized modern physics. The underlying dynamics are perplexing and counter intuitive - e.g. depending on the circumstance, electrons exhibit wave-like or particle-like behaviour \cite{davisson1928} - and has since changed our perspective of the universe at the microscopic level.

\begin{quote}
    \textit{Those who are not shocked when they first come across quantum theory cannot possibly have understood it.} \begin{flushright} -Niels Bohr \end{flushright}
\end{quote}

Erwin Schrödinger received a Nobel prize in 1933 for his work establishing the basis of quantum mechanics and atomic theory. Be that as it may, nearly twenty years later in 1953, he begins a lecture in Dublin with a humorous forewarning that the contents of the lecture may seem `lunatic' \cite{bitbol1996}. Clearly said in jest, there is inherent truth in this statement. Quantum theory allows for dynamics which are not observed at the macroscopic level, and as a result are difficult to envisage. The most prominent of which are: entanglement and superposition. \textit{Quantum entanglement} is a term coined to indicate non-classical correlations between quantum systems. When a single constituent of an entangled quantum system is measured, the effects propagate amongst the complete system. \textit{Quantum superposition} is the principle that any configuration of superposed quantum states is also an allowable quantum state.

The first theoretical prototypes of quantum computers were pioneered in the 1980's \cite{benioff1980, feynman1982, deutsch1985}; this was the beginning of the quantum information zeitgeist. Such a computer would be compromised of microscopic objects subjected to the realm of quantum mechanics. In particular, a two level quantum system, such as the spin of an electron, is characterized as a quantum version of the traditional binary bit - usually abbreviated to \textit{qubit}. By virtue of quantum mechanical effects, such as entanglement and superposition, a quantum computer can greatly outperform the abilities of a classical (i.e. inherently \textit{not}-quantum) computer \cite{preskill2012}. For example, Shor's algorithm (an algorithm designed to be carried out on quantum computers) can find the prime factorization of large numbers in a small amount of time \cite{shor1994}; a task which is extremely difficult for the world's most state of the art supercomputer. In 2019, Google demonstrated that their 53 qubit quantum computer could execute a sampling task in 200 seconds \cite{arute2019}. Even though IBM showed that this task could be executed by a classical computer in two and a half days \cite{pednault2019}, it quickly converges to an impossible problem for a classical computer as the number of qubits increase incrementally. Practically, we are entering the era where classical computers cannot compete.

Quantum computing is not the unique technology proposed as an advantageous version of its classical analogue. For the past few decades, academic and government institutions, and even some companies such as Google and IBM, have increased their investment and support in the quest of designing quantum technologies \cite{dowling2003}. In China, satellites are being used for long distance quantum key distribution \cite{liao2017}. In Europe, a rudimentary version of a quantum internet is in development \cite{kimble2008, wehner2018}. Quantum technologies are often divided into four categories depending on their scope: quantum computation, quantum simulation, quantum communication, and quantum metrology and sensing \cite{acin2018}. The focal point of this thesis is quantum metrology and sensing technologies enhanced by other quantum information techniques, namely graph states (computation and communication), quantum error correction (computation) and quantum cryptography (communication).

\textit{Quantum metrology and sensing} is a relatively new and auspicious type of quantum technology \cite{paris2009, toth2014, degen2017}, in which quantum phenomena are exploited to accurately estimate physical parameters with a precision which cannot matched with the best classical strategies \cite{caves1981}. Since the publication of \textit{Quantum-enhanced measurements: beating the standard quantum limit} \cite{giovannetti2004} by Giovannetti, Lloyd and Maccone, there has been a surge of interest in the field. Current research is flourishing at a theoretical and experimental level.

\section{Metrology: From Classical to Quantum}

\textit{Metrology}, the science of measurement and precision, is often not discussed and regularly misunderstood as meteorology (the science of weather). Be that as it may, metrology plays a critical role in the advancement of science. Scientific theories are tested by observing a physical processes predicted by said theory; in physics and chemistry this step is often carried out by performing a measurement. As the accuracy of technology improves, more theories are put to the test. In 2016, LIGO (in collaboration with VIRGO) announced successful observations of gravitational waves\footnote{The experiment is currently being upgraded to use squeezed light which will allow for an even more accurate measurement  \cite{LIGOSqueezed}.}\footnote{A Michelson interferometer with arms which spanned four kilometers in length was used in the experiment, and the achieved precision was comparable to measuring the distance from Earth to the nearest star (besides the sun) with an uncertainty smaller than the width of a human hair \cite{LIGOfacts}.} \cite{LIGO}, a phenomenon predicted by Einstein's theory of general relativity. In 2021, the standard model for particle physics was put under scrutiny after Fermilab released their measurement results of the anomalous magnetic dipole moment of the muon \cite{fermilab2021}, in which the measured value was different than the predicted value by the current theory. In a similar vein to its importance to science, metrology is an unsung hero of engineering, architecture and design. A chair/table/house/bridge in which the lengths are measured up to the nearest tenth of millimeter is more reliable and safe than a counterpart in which the lengths are measured up to the nearest centimeter.

Alas, most physical parameters of interest cannot be associated with a direct measurement process. A more accurate description is to say such a parameter is \textit{estimated}. The underlying tool of constructing an estimate is still a measurement of a related (measurable) quantity. In the LIGO experiment, the gravitational wave introduced a relative phase in the light source. A relative phase is not a directly measurable quantity, instead the phase was estimated from the observed interference pattern. Formally, \textit{estimation theory} is the branch of statistics which establishes techniques and the mathematical formalism pertaining to estimating unknown parameters from measured empirical data \cite{kay1993, cox2006}. It is the principal mathematics of metrology.

There are two major philosophies of estimation theory: the Bayesian approach and the frequentist approach. The Bayesian approach is used for stochastic parameters and the frequentist approach is used for deterministic parameters. This thesis focuses uniquely on the frequentist approach to quantum metrology\footnote{A summary of Bayesian estimation theory is provided in \textbf{Chapter 3} for completeness.}. With sufficient measurement data, the frequency of observations will begin to mimic the true probability distribution, hence the name `frequentist'. In principle, a deterministic parameter can be estimated to any degree of precision with a sufficiently large set of empirical data. The precision of an estimate is denoted by the mean-squared error. Within the frequentist framework, this is ultimately bounded by the reciprocal of the Fisher information - a measure of how much information the measurable data contains about the unknown data \cite{fisher1925, kullback1997}. This bound is called the Cramér-Rao bound \cite{cramer1946, radhakrishna1945}.

Estimation theory was formally adapted to realm of quantum information in the latter half of the 20th century by Helstrom \cite{helstrom1967, helstrom1968, helstrom1969} and Holevo \cite{holevo1973, holevo1982}. The established terminology to describe quantum parameter estimation is difficult to misconstrue; as the rhetorical tradition dictates, existing terminology is preceded by the word \textit{quantum}, for example \textit{quantum Cramér-Rao bound}, \textit{quantum Fisher information}, et cetera \cite{braunstein1994, hayashi2005}. In quantum parameter estimation problems, an unknown parameter is encoded into a quantum probe by a physical interaction. As a result of quantum phenomena, quantum parameter estimation problems can attain a precision impossible to a purely classical system \cite{caves1981, bondurant1984}. An experimental quantum advantage has been reported using optical systems \cite{okamoto2008, kacprowicz2010, xiang2011}, atomic systems \cite{meyer2001, taylor2008, facon2016, chalopin2018, dietsche2019} and superconducting circuits \cite{wang2019}.

\textit{Phase estimation} is the canonical problem of quantum metrology \cite{holland1993, giovannetti2004, toth2014}. An unknown phase is encoded in an $n$ qubit highly entangled GHZ state, and a simple measurement strategy can be implemented to estimate the unknown phase such that the mean squared error scales as $1/n^2$. This notion of precision (where the quantum Cramér-Rao bound is saturated) is referred to as the \textit{Heisenberg limit}: the ultimate limit of precision enabled by quantum mechanics \cite{giovannetti2006}. With respect to phase estimation, the Heisenberg limit is a quadratic advantage over the analogous scenario sans non-classical correlations (i.e. the $n$ qubits are not entangled). Here the mean-squared error is dictated by the central limit theorem and scales as $1/n$, this notion of precision is commonly referred to as the \textit{standard quantum limit}, classical limit or the shot-noise limit.

The applicability of quantum metrology spans a number of domains. These include, but are not limited to, magnetometry \cite{taylor2008, wasilewski2010, sewell2012, brask2015, razzoli2019}, thermometry \cite{neumann2013, toyli2013, correa2015}, spectroscopy \cite{meyer2001, leibfried2004, kira2011, dorfman2016,shaniv2018}, imaging \cite{lugiato2002, barzanjeh2015, genovese2016}, gravimetry \cite{qvarfort2018, kritsotakis2018} and clock synchronization \cite{giovannetti2001, appel2009, ludlow2015, schioppo2017}. Quantum metrology is particularly appealing for biology and medicine \cite{pena2012, schirhagl2014, taylor2016, mejia2018}, where probing a sample is often destructive in nature, and so the non-classical correlations of quantum systems may lead to a reduction in the number of probes required whilst still attaining a required precision.

\section{Motivation}

The overarching theme of this thesis is the incorporation of other quantum information techniques within the usual quantum metrology framework. Specifically, we explore the immersion of graph states \cite{SM20}, quantum error correction \cite{SMMN21}, and quantum cryptography \cite{SMK21, SM21}. All of these technologies offer a unique functionality to the standard quantum metrology problem with respect to different circumstances.

Firstly, in the case of graph states, having an multi-purposeful resource is very desirable for the realm of quantum technologies, as focusing on a specific class of quantum states will greatly facilitate the design and implementation of quantum hardware. Graph states \cite{hein2004} come to mind as a potential `super resource', as they are used for many tasks in quantum computation \cite{schlingemann2001a, raussendorf2003} and quantum communication \cite{markham2008, meignant2019, hahn2019}. In this context then, it is a natural question to ask which graph states are an efficient resource for quantum metrology \cite{SM20}.

Secondly, we consider the utility of error correction. One of the biggest obstacles for early generations quantum hardware will be its susceptibility to quantum noise. It is known that said noise imposes many challenge for quantum metrology \cite{escher2011a, escher2011b, demkowicz2012, kolodynski2013}. It has been shown that quantum error correction can be used to completely mitigate the effects of noise \cite{demkowicz2017, zhou2018}. Unfortunately, the necessary frequency of error correction is impossible for current quantum hardware \cite{cramer2016, ofek2016}. Thus, it is important to determine the utility of quantum error correction in a real world scenario \cite{SMMN21}. 

Finally, we consider a cryptographic framework. Another obstacle for the early generations of quantum hardware is the lack of `all-in-one' devices. Because quantum metrology is technologically demanding, one solution is to delegate some of the difficult tasks to a third party with more computational power. In this event, quantum information will have to be transmitted through a quantum channel. This raises several security issues, as quantum channels can be intercepted by malicious adversaries. It is critical to properly adapt the parameter estimation problem in such a cryptographic setting as many of the standard assumptions, namely having an unbiased estimator, may not necessarily be true \cite{SMK21}. An equally important task is to create cryptographic protocols which do not interfere with the underlying quantum metrology problem, but provide a sense of privacy and security \cite{SMK21, SM21}.

Formally, multiple parties communicating through a quantum channel is known as a quantum network \cite{chiribella2009}. Quantum networks have been proposed as a resource for spatially separated quantum metrology and multiparameter quantum metrology \cite{komar2014, komar2016, eldredge2018, ge2018, proctor2018, zhuang2018, qian2019, rubio2020a, guo2020}. The quantum technologies discussed in this thesis (graph states, error correction and cryptography) all fit in naturally within the framework of quantum networks. A future perspective is to combine these works in interesting and useful ways. Currently, we are combining the cryptographically themed results to establish a notion of a secure quantum sensing network.

\section{Thesis Outline}

The subsequent chapters of this thesis are partitioned into two preliminary chapters, three research chapters and a discussion chapter. The research chapters provide insight on the projects I worked on during my PhD in a pedagogical fashion. Following the main chapters are three appendices, which contain proofs omitted from the main text due to length or complexity.

The preliminary chapters equip the reader with the necessary definitions and mathematical tools to comprehend the subsequent research chapters. \textbf{Chapter 2} acts a crash course on the mathematics of quantum mechanics specific to quantum information. Key concepts such as quantum states, entanglement and quantum measurements are explained. \textbf{Chapter 3} overviews the foundations of the parameter estimation problem and its adaptation to the realm of quantum information. The canonical example of a highly entangled quantum state used for phase estimation is explored in this chapter and it is regularly used as a comparison in the research chapters.

\textbf{Chapter 4} is based on the work \textit{Graph states as a resource for quantum metrology} \cite{SM20}. We characterize the use of graph states for quantum metrology by linking the quantum Fisher information to the shape of the corresponding graph. We construct a class of graph states which approximately achieve the Heisenberg limit for phase estimation and are thus a practical resource for quantum metrology. We name this class of graph states bundled graph states, as many vertices in the corresponding graph are in bundles which are permutation invariant. We also show that the Heisenberg limit can maintain a quantum advantage in the presence of noise and that the Cramér-Rao bound can be saturated with a simple measurement strategy.

\textbf{Chapter 5} is based on the work \textit{Practical limits of error correction for quantum metrology} \cite{SMMN21}. We analyze the effectiveness of a realistic quantum error correction scheme to mitigate the impact of noise for quantum metrology. This is accomplished by incorporating impediments an implementation of an error correction code may face, such as a delay in any error correction operations, noisy ancillary qubits and imperfect operations. We outline the circumstances in which the Heisenberg limit may be recovered. Even though this work focuses on a specific error correction code (the parity check code), we hypothesize that other error correction strategies encounter the same limitations.

\textbf{Chapter 6} is based on the work \textit{A Cryptographic approach to Quantum Metrology} \cite{SMK21} as well as \textit{Quantum Metrology with Delegated Tasks} \cite{SM21}. We provide a rigorous framework of the functionality of quantum metrology problems in a cryptographically motivated setting. By integrating an appropriate cryptographic protocol, the functionality of the parameter estimation scheme is mostly unchanged. We show that the added bias and additional uncertainty in the cryptographic framework can be bounded in terms of the soundness of the protocol. We establish protocols for a variety of possible settings, such as exchanging information over an unsecured quantum channel \cite{SMK21}, and delegating a portion of the quantum metrology scheme to an untrusted party \cite{SM21}.

\textbf{Chapter 7} is a discussion chapter; the key ideas from the main research chapters are summarized and future perspectives are listed. Insight on a current project is given, where the core concept is an amalgamation of quantum networks and the cryptographic framework for quantum metrology to devise a notion of a secure quantum sensing network.

%% file: Chapters/Chapter2-Quantum.tex
\chapter{Mathematical Foundations of Quantum Information}

Quantum theory is an extensive area of physics with a rich mathematical history. The majority of its subtleties are beyond the scope of this thesis. This chapter is intended to familiarize the reader with the underlying mathematics of the subsequent chapters. See \cite{griffiths2018} for a broader overview of quantum mechanics, and \cite{nielsen2002} for a more detailed analysis of quantum information.

\begin{quote}
    \textit{As Deepak Chopra taught us, quantum physics means anything can happen at any time for no reason!}
    \begin{flushright} -Professor Farnsworth \end{flushright}
\end{quote}

\section{Quantum States}

\subsection{Qubits}

The bit is the primitive building block of information theory. It can be thought of as a physical switch, or any object subjected to a binary state: 0 or 1, yes or no, on or off, et cetera. The quantum bit, commonly referred to as a qubit, is the analogous primitive building block of \textit{quantum information}. Just as a bit can be in the states 0 and 1, a qubit can be in the states $\ket{0}$ and $\ket{1}$\footnote{The notation $\ket{\square}$, known as Dirac notation or bra-ket notation, is ubiquitously used in quantum mechanics to describe quantum states.}. Unlike a classical bit, the state of a qubit can be any linear combination of $\ket{0}$ and $\ket{1}$:
\begin{equation}
    \alpha \ket{0} + \beta \ket{1}
\end{equation}
with $\alpha$ and $\beta$ being complex numbers subjected to $|\alpha|^2+|\beta|^2=1$. This is the \textit{superposition principle}, which asserts that any linear combination of valid quantum states is also a valid quantum state.

\begin{figure}
    \centering
    \input{Figures/Chapter2/blochsphere}
    \caption{The Bloch sphere is a geometric representation of single qubit quantum states. A point on the surface of the sphere with polar angle $\theta$ and azimuthal angle $\phi$ represents the quantum state $\ket{\psi}= \cos ( \theta/2) \ket{0} + e^{i \phi} \sin ( \theta /2) \ket{1}$.}
    \label{fig:blochsphere}
\end{figure}
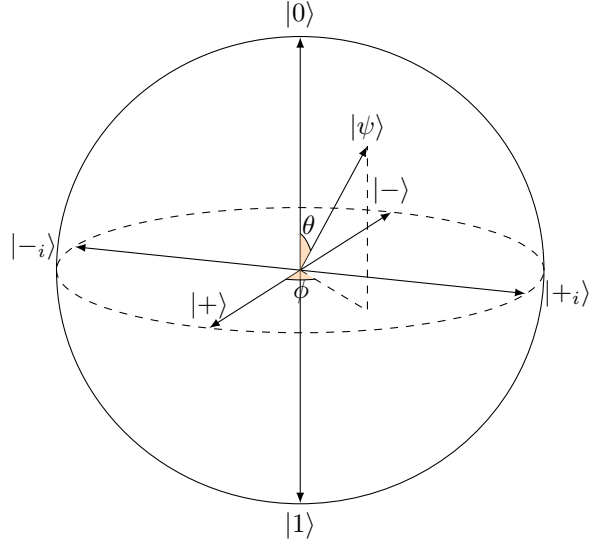

Just as a bit can be thought of as a physical object, so can a qubit. There exists a variety of physical implementations to realize a qubit, for example, the spin of an electron \cite{childress2006,dutt2007}, the direction of current in a superconducting circuit \cite{wendin2017} or the polarization of a photon \cite{strauf2007}. Having said that, in this thesis (unless it is otherwise stated) a quantum state should be thought of as a mathematical element of a Hilbert space $\mathcal{H}$, formally one writes $\ket{\psi} \in \mathcal{H}$. For a qubit, $\mathcal{H}$ is a two dimensional space, hence $\{ \ket{0}, \ket{1} \}$ corresponds to an orthonormal basis. This is not the sole basis representation for qubits; other commonly used bases are $\{ \ket{+}, \ket{-} \}$, where $\ket{\pm} = \frac{1}{\sqrt{2}} \big( \ket{0} \pm \ket{1} \big)$ and $\{ \ket{+_i}, \ket{-_i} \}$, where $\ket{\pm_i} = \frac{1}{\sqrt{2}} \big( \ket{0} \pm i\ket{1} \big)$. For any orthonormal basis $\{ \ket{x}, \ket{y} \}$, the inner product of quantum states $\ket{\psi}= \alpha \ket{x} + \beta \ket{y}$ and $\ket{\phi}= \gamma \ket{x} + \delta \ket{y}$ is defined to be
\begin{equation}
    \braket{\phi}{\psi} = \braket{\psi}{\phi}^* = \alpha \gamma^* + \beta \delta^*,
\end{equation}
where an asterisk is used to signify the complex conjugate.

The 2-dimensional Hilbert space of a qubit naturally generalizes to $d$-dimensional spaces. These quantum states are commonly known as \textit{qudits} and can be represented via
\begin{equation}
    \sum_{k=0}^{d-1} \alpha_{k} \ket{k},
\end{equation}
where $\sum_{k=0}^{d-1} |\alpha_k|^2=1$ and $\{\ket{0},\ldots,\ket{d-1} \}$ forms a basis for said Hilbert space. Even though the results presented throughout this thesis are derived with respect to systems composed of qubits, the techniques presented (quantum metrology, graph states, error correction and cryptography) have higher dimensional forms, and thus the results presented can be generalized to systems composed of qudits.

\subsection{Multiple Qubits and Quantum Entanglement}

A bipartite quantum system composed of $\ket{\psi_A} \in \mathcal{H}_A$ and $\ket{\psi_B} \in \mathcal{H}_B$ is represented via
\begin{equation}
    \ket{\psi_{AB}} = \ket{\psi_A} \ket{\psi_B} \in \mathcal{H}_{AB}=\mathcal{H}_A \otimes \mathcal{H}_B.
\end{equation}
The above quantum states are called \textit{separable}, as the composite system is (by construction) a product of quantum states each belonging to a separate Hilbert space. By the superposition principle, the composite Hilbert space $\mathcal{H}_{AB}=\mathcal{H}_A \otimes \mathcal{H}_B$ also contains superpositions of separable quantum states. The two-qubit quantum state
\begin{equation}
    \frac{1}{\sqrt{2}} \big( \ket{0}_A \ket{0}_B + \ket{1}_A \ket{1}_B \big)
\end{equation}
cannot be written as a product of two one-qubit quantum states. In other words, each qubit in the composite system cannot be described independently from one another. This property is better known as \textit{entanglement} and is a peculiarity unique to quantum mechanics. Quantum entanglement is the root of the well-known (and frequently misinterpreted in popular media\footnote{If someone has forgotten whether or not they have food in their fridge, their fridge is not in a macroscopic superposition of `empty' and `full'. Instead, they are a simply a forgetful person.}) Schrödinger's thought experiment \cite{schrodinger1935}. In the thought experiment, a hypothetical cat is placed in a box with a radioactive source and a flask of poison. The poison is released upon detecting that the radioactive source has decayed: killing the cat. The premise is that the nature of the cat is entangled with the radioactive source. When the state of the source evolves to a superposition of `not-decayed' and `decayed', the cat would ultimately evolve to be in a macroscopic superposition of `alive' and `dead'.

\begin{figure}[!ht]
    \centering
    \includegraphics[width=0.5\textwidth]{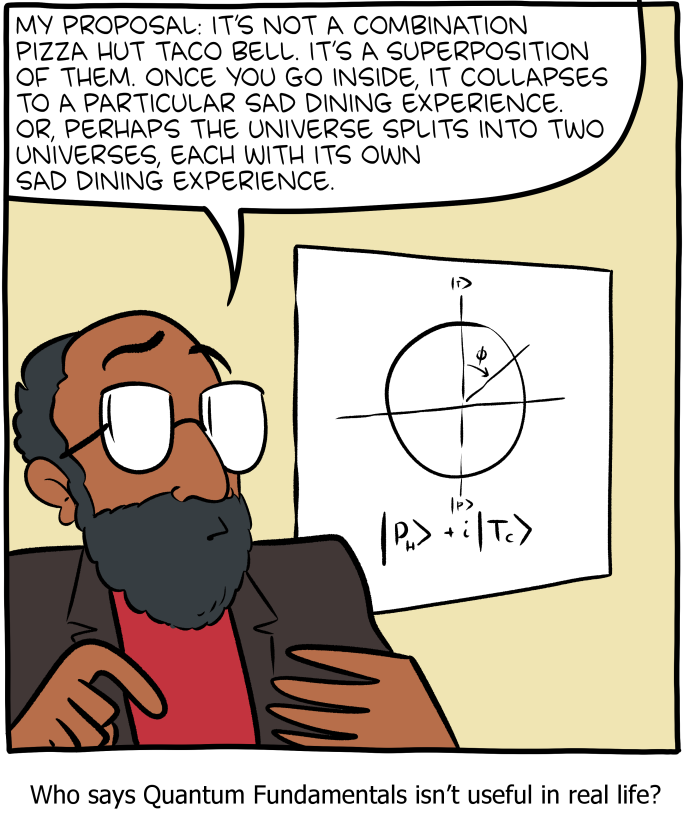}
    \caption{It is worth stressing that quantum properties such as \textit{superposition} and \textit{entanglement} are theoretically possible at a macroscopic level, but are not observed \cite{zurek2006}. Ergo, quantum effects are difficult to visualize. Illustration by Zach Weinersmith, \textit{Saturday Morning Breakfast Cereal: Quantum-2} (2019), see \cite{SMBC} in the bibliography for the source details.}
\end{figure}

In general, a quantum state $\ket{\psi}$ in the composite Hilbert space $\bigotimes_{k=1}^n \mathcal{H}_{A_k}$ is called separable if and only if there exists $\ket{\psi_{A_k}} \in \mathcal{H}_{A_k}$ for all $k$ such that
\begin{equation}
    \ket{\psi} = \bigotimes_{k=1}^n \ket{\psi_{A_k}},
\end{equation}
otherwise it is entangled. For example, the $n$ qubit Greenberger–Horne–Zeilinger (GHZ) state
\begin{equation}
    \label{eq:GHZstate}
    \ket{\psi_\text{GHZ}}=\frac{1}{\sqrt{2}} \big( \ket{0}^{\otimes n}+\ket{1}^{\otimes n} \big),
\end{equation}
is a highly entangled state with many practical applications, including quantum metrology. GHZ states are the canonical resource for the quantum metrology problem of phase estimation \cite{giovannetti2004, toth2014}. The utility of a GHZ state is frequently referenced in this thesis and used as a benchmark in \textbf{Chapter 4} and \textbf{Chapter 5}.

Although quantum entanglement was originally coined as \textit{spooky} by Einstein \cite{einstein1935}, it has since been shown to be a valuable resource for the field of quantum information. Numerous quantum-based protocols (e.g. superdense coding \cite{bennett1992communication}, teleportation \cite{bennett1993teleporting}) are contingent on the non-classical correlations of entangled quantum states. Quantum metrology is no different: entanglement\footnote{In continuous variable systems, non-classical correlations can also be achieved through a process called squeezing \cite{lvovsky2015}. Squeezing is not the same as entanglement, but also leads to a quantum advantage for metrology problems \cite{caves1981, demkowicz2015, schnabel2017}.} allows for estimation strategies to surpass the limits of classical statistics \cite{giovannetti2004, giovannetti2006, giovannetti2011, toth2014}.

\subsection{Mixed States}

It is often practical to consider statistical ensembles of quantum states $\{(p_i, \ket{\psi_i}) \}$, where $p_i$ is the probability of the system being in the quantum state $\ket{\psi_i}$. This abstraction is useful to incorporate stochastic processes and classical randomness into the description of a quantum system. Mathematically this is represented as a linear and positive semi-definite\footnote{$\rho$ is positive semi-definite if $\mel{\phi}{\rho}{\phi} \geq 0 \; \forall \ket{\phi} \in \mathcal{H}$.} operator

\begin{equation}
    \rho = \sum_i p_i \dyad{\psi_i},
\end{equation}
which is often referred to as a density operator, density matrix or (most commonly) a \textit{mixed state}. Because the set $\{ p_i \}$ represents a set of classical probabilities, we must have that $\sum_i p_i=1$, from which it follows that all mixed states have unit trace $\Tr \rho =1$. The purity of a mixed state is a measure on the classical randomness present in a quantum system and defined by $\Tr \rho^2$. For a general mixed state $0 \leq \Tr \rho^2 \leq 1$, and the upper-bound is saturated if and only if there is no inherent classical randomness present, i.e. the system is in a definite quantum state - more commonly referred to as a \textit{pure state}. Density operator formalism is predominantly used in this thesis and, depending on the context, may signify a general mixed state or specifically a pure state.

When dealing with composite systems, $\rho_{AB} \in \mathcal{H}_A \otimes \mathcal{H}_B$, it can be beneficial to describe a subsystem when one does not have access to the other systems, e.g. $A$ does not have access to $B$. This is better known as a reduced density operator and can be computed via the partial-trace
\begin{equation}
    \rho_A = \Tr_B \rho_{AB} = \sum_{k} \mel{b_k}{\rho_{AB}}{b_k},
\end{equation}
where $\{ \ket{b_k} \}$ is any orthonormal basis of $\mathcal{H}_B$. If a composite system is an entangled pure quantum state, then the reduced density operator is guaranteed to be a mixed state with purity less than one. Therefore, by discarding a portion of a composite quantum system, one introduces classical randomness into the non-discarded systems.

The opposite is similarly true, in that it can be beneficial to extend the Hilbert space of a mixed state to a composite system in which it is a pure state. This is known as a purification process. If $\rho_A=\sum_i p_i \dyad{\psi_i} \in \mathcal{H}_A$ and $\mathcal{H}_B$ is an auxiliary Hilbert space with orthonormal basis $\{ \ket{\phi_i}\}$, then the pure state
\begin{equation}
    \label{eq:purification}
    \ket{\Psi_{AB}}= \sum_i \sqrt{p_i} \ket{\psi_i} \ket{\phi_i}
\end{equation}
is a purification of $\rho_A$ because $\Tr_B \dyad{\Psi_{AB}}=\rho_A$. The purification is not unique.

\subsection{Vector and Matrix Representation}

Up until now, pure states have been represented as an abstract mathematical element of a Hilbert space, and general mixed states as a linear and non-negative operator acting on said Hilbert space. For the most part of this thesis, this abstract representation is sufficient. However, some of the mathematical derivations in \textbf{Appendix~B} make use of an alternative representation using vectors and matrices.

The vector representation of a pure qubit state is a two dimensional\footnote{This representation extends to qudits, where the vectors are $d$ dimensional objects.} column vector
\begin{equation}
    \alpha \ket{0} + \beta \ket{1} \longleftrightarrow
    \begin{pmatrix}
        \alpha \\
        \beta
    \end{pmatrix},
\end{equation}
and the representation of the corresponding dual is a two-dimensional row vector
\begin{equation}
    \alpha^* \bra{0} + \beta^* \bra{1} \longleftrightarrow
    \begin{pmatrix}
        \alpha^* & \beta^*
    \end{pmatrix}.
\end{equation}
Combining the above, the mixed state $\{(p_1, \alpha\ket{0}+\beta \ket{1}), (p_2,\gamma\ket{0}+\delta \ket{1}) \}$ is represented with the matrix
\begin{equation}
    p_1 \begin{pmatrix} \alpha \\ \beta \end{pmatrix} \begin{pmatrix} \alpha^* & \beta^* \end{pmatrix} + p_2 \begin{pmatrix} \gamma \\ \delta \end{pmatrix} \begin{pmatrix} \gamma^* & \delta^* \end{pmatrix} =
    \begin{pmatrix}
        p_1|\alpha|^2 + p_2 |\gamma|^2 & p_1 \alpha \beta^* + p_2 \gamma \delta^* \\
        p_1 \alpha^* \beta + p_2 \gamma^* \delta & p_1|\beta|^2 + p_2 |\delta|^2
    \end{pmatrix}.
\end{equation}

\section{Quantum Operations}

Operator formalism in quantum mechanics is used to describe transformations to quantum states\footnote{In this thesis, quantum states are viewed as the \textit{variables}, this is known as the Schrödinger picture. There is another formulation in which the operators act as the \textit{variables}, better known as the Heisenberg picture \cite{griffiths2018}.}. At the most general level, a \textit{quantum operator} $\Gamma$ is a linear map from an input Hilbert space $\mathcal{H}_1$ to an output Hilbert space $\mathcal{H}_2$
\begin{equation}
\begin{split}
    \Gamma : \mathcal{H}_1 &\rightarrow \mathcal{H}_2 \\
    \rho &\rightarrow \Gamma(\rho).
\end{split}
\end{equation}
It is demanded that $\Gamma$ has two properties. The first is for $\Gamma ( \rho )$ to have unit-trace (for it to qualify as a quantum state); this is known as being \textit{trace-preserving}. The second is for $\Gamma ( \rho )$ to be positive semi-definite, and more so, if a partial trace is taken, then the remaining subsystem is also positive semi-definite; this is known as being \textit{completely positive}. If $\Gamma$ satisfies both properties, it is called a completely positive trace-preserving (CPTP) map. A CPTP map can be written in the form
\begin{equation}
    \label{eq:CPTPKraus}
    \Gamma ( \rho ) = \sum_{j} A_j \rho A_j^\dagger,
\end{equation}
where $\{ A_j \}$ are known as Kraus operators \cite{hellwig1969} which satisfy $\sum_j A_j A_j^\dagger = \mathbb{I}$.

\subsection{Pauli and Clifford Operators}

The three Pauli operators, $X$, $Y$ and $Z$, are conceivably the most widely used operators in the field quantum information. The Pauli operators are Hermitian and involutory operators which act on single qubit quantum states, and along with the identity map, form a group. Listed are the bra-ket and matrix representations of the Pauli operators:
\begin{align}
    X &= \dyad{0}{1}+\dyad{1}{0} \rightarrow \begin{pmatrix} 0 & 1 \\ 1 & 0 \end{pmatrix} \\
    Y &= -i \dyad{0}{1} + i \dyad{1}{0} \rightarrow \begin{pmatrix} 0 & -i \\ i & 0 \end{pmatrix} \\
    Z &= \dyad{0}{0}-\dyad{1}{1} \rightarrow \begin{pmatrix} 1 & 0 \\ 0 & -1 \end{pmatrix}.
\end{align}

The Pauli group $\{ \mathbb{I}, X, Y, Z \}$ is a basis for all $2 \times 2$ complex matrices, and thus a single qubit quantum state can be expressed as
\begin{equation}
    \rho = \frac{1}{2} \Big(\mathbb{I} + \Tr (X \rho) X + \Tr (Y \rho) Y + \Tr (Z \rho) Z \Big).
\end{equation}
In general, defining the $m$th degree Pauli group to be $\mathcal{P}_m= \{\mathbb{I}, X, Y, Z \}^{\otimes m}$, a quantum system composed of $m$ qubits can be expressed as
\begin{equation}
    \rho = \frac{1}{2^m} \sum_{P \in \mathcal{P}_m} \Tr(P \rho) P.
\end{equation}

Another class of operators which are well known is the Clifford group. The Clifford group is an important set of unitary operators in the realm of quantum computing and quantum algorithms, as they were shown to be efficiently simulated with a classical computer \cite{gottesman1998}. Mathematically, the Clifford group of degree $m$, denoted $\mathcal{C}_m$, is the set of unitary operators which normalize $\mathcal{P}_m$ (up to a phase of $\pm 1$), thus $\forall C \in \mathcal{C}_m$ and $\forall P \in \mathcal{P}_m$
\begin{equation}
    C P C^\dagger \in \pm \mathcal{P}_m.
\end{equation}
The set of local Clifford operations $\mathcal{C}_1$ can be decomposed as a sequence of a Pauli operations or a $\pi/4$ phase shift $e^{\pm i\frac{\pi}{4}P}$ (with $P \in \{X,Y,Z \}$). Evidently, $C_1$ is much simpler to implement than an arbitrary local unitary \cite{niemann2014}. For this reason, all but one of the cryptographic protocols we devise in $\textbf{Chapter 6}$ consist solely of local Clifford operations.

\subsection{Dynamics}

\begin{figure}[!ht]
    \centering
    \begin{subfigure}{.495\textwidth}
        \centering
        \scalebox{1.35}{\input{Figures/Chapter2/dynamics1}}
        \caption{Unitary evolution.}
        \label{fig:rotationdynamics}
    \end{subfigure}
        \begin{subfigure}{.495\textwidth}
        \centering
        \scalebox{1.35}{\input{Figures/Chapter2/dynamics2}}
        \caption{Environmental decoherence.}
        \label{fig:decoherencedynamics}
    \end{subfigure}
    \caption{Visual representation of the dynamics of a single qubit (initialized in the $\ket{+}$ state) is governed by the master equation $\frac{d \rho (t)}{dt}=-\frac{i}{\hbar}[\frac{\omega}{2}Z,\rho(t)]+\gamma \big((X \rho(t) X-\rho(t) \big)$. The evolution of the qubit traces a path in the $X-Y$ plane of the Bloch sphere. In (a) the environmental term is ignored ($\gamma = 0$) and the qubit forever oscillates between $\ket{+}$ and $\ket{-}$ with frequency $\omega$. In (b) the decoherence term ($\gamma \neq 0$) causes the qubit to eventually decohere to the maximally mixed state (the center of the Bloch sphere).}
    \label{fig:dynamics}
\end{figure}
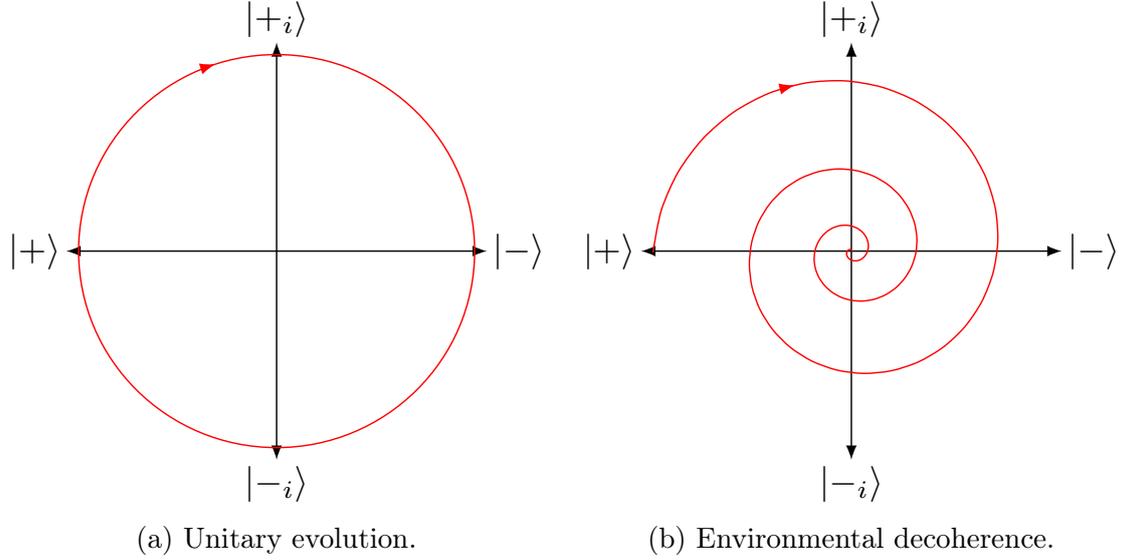

The parameter encoding mechanism is the predominant element in a quantum metrology problem. Formally, this is a physical process which influences the evolution of a quantum state. In a closed and isolated system with a Hamiltonian $H$, the evolution of a quantum state $\rho$ is governed by the Schrödinger equation \cite{schrodinger1926, griffiths2018}
\begin{equation}
    \frac{d \rho(t)}{dt} = -\frac{i}{\hbar} [H, \rho(t)].
\end{equation}
As a result, the evolution is described as a unitary transformation
\begin{equation}
    \rho(t)=U_{t-t_0} \rho(t_0) U_{t-t_0}^\dagger,
\end{equation}
where $U_\tau = e^{-\frac{i}{\hbar}H \tau}$.

It is worth noting that closed and isolated systems do not emulate reality and are effectively a fantasy for experimentalists and engineers. Real world quantum technologies are plagued with noise (the subject of \textbf{Chapter 5}) due to interactions with the environment \cite{gardiner1991, breuer2002}. As a result, information is lost to the surroundings, causing decoherence, dephasing, losses and fluctuations. There is no explicit equation which governs the evolution of a quantum system for a general environmental interaction. However, with some assumptions (namely that the system and environment are weakly-coupled and the interaction is time-independent) then one can model evolution by modifying the Schrödinger equation \cite{breuer2002}
\begin{equation}
    \dot{\rho}(t) = -\frac{i}{\hbar} [H, \rho(t)] + \mathcal{L} \big( \rho(t) \big),
\end{equation}
where the super-operator $\mathcal{L}$ is better known as the Liouvillian. It was demonstrated that for the evolution to yield a valid transformation (CPTP), the Liouvillian will take on the form \cite{lindblad1976}
\begin{equation}
    \mathcal{L} \big( \rho(t) \big)= \sum_{j=1}^{d^2-1} \gamma_j \big[ L_k \rho(t) L_k^\dagger - \frac{1}{2} \big\{ \rho(t),L_k L_k^\dagger \big\} \big],
\end{equation}
where $d$ is the dimension of the Hilbert space, $\gamma_j$ are non-negative decay rates, and $L_1,\ldots,L_{d^2-1}$ are Lindblad operators. This equation is often referred to as the Lindblad master equation.

The contrast between the Schrödinger equation and the Linblad master equation is depicted in Fig.~(\ref{fig:dynamics}). When a single qubit pure state is governed solely by unitary dynamics, it perpetually oscillates between pure states. But, when the system is coupled to the environment, the qubit spirals towards the maximally mixed state.

\section{Quantum Measurements}

\begin{figure}[!ht]
    \centering
    \includegraphics[width=0.9\textwidth]{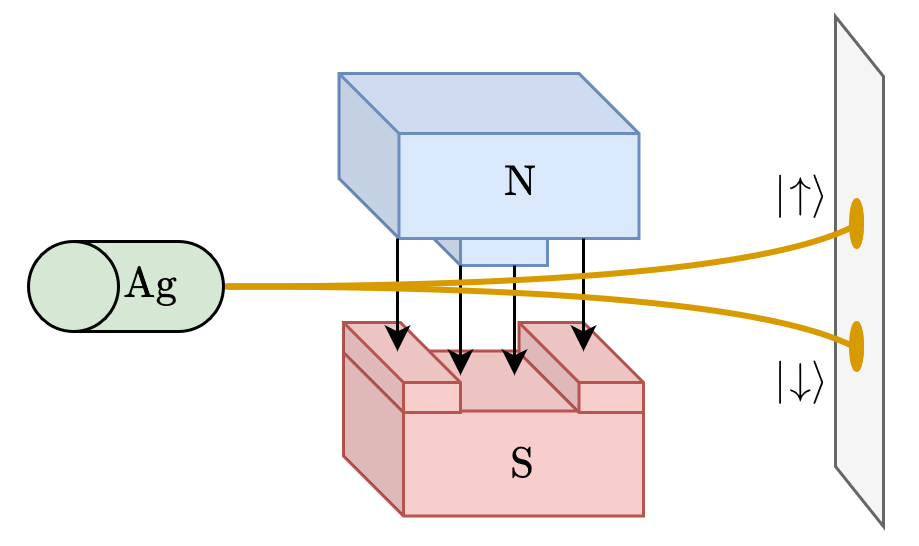}
    \caption{The Stern-Gerlach experiment \cite{gerlach1922} is an early prototype for a quantum measurement. A beam of silver (Ag) atoms is sent through an inhomogeneous magnetic field towards a detector screen. Initially, the spin of the silver atoms are in arbitrary superpositions of $\ket{\uparrow}$ and $\ket{\downarrow}$. Classical physics predicts that the silver atoms would be detected along the length of the detector screen. Instead, the silver atoms were detected in two bunches, one bunch of spin $\ket{\uparrow}$ atoms and one bunch of spin $\ket{\downarrow}$ atoms.}
    \label{fig:sterngerlach}
\end{figure}

The principal goal of quantum metrology is to use a quantum system to estimate the value a physical unknown parameter. With this in mind, it is crucial to extract physical information from a quantum system; in the language of quantum mechanics, this is done by measuring an \textit{observable} \cite{von2018}. Formally, a (finite) observable $O$ is a linear and Hermitian ($O=O^\dagger$) operator. By the spectral value theorem, $O$ can be decomposed into a set of projectors $\{ P_i \}$ satisfying $P_i P_j = P_i \delta_{i,j}$ and $\sum_i P_i = \mathbb{I}$ along with a corresponding set of real-values eigenvalues $\{ o_i \}$ such that $O=\sum_i o_i P_i$. Here, the index $i$ signifies different measurement outcomes. If the quantum state $\rho$ is measured, then outcome $i$ is observed with probability $\Tr (P_i \rho)$ and the expectation value of $O$ is $\expval{O}=\sum_i o_i \Tr (P_i \rho) = \Tr (O \rho )$. This is the simplest description of a quantum measurement, and is called a projection-valued measurement (PVM).

A quantum measurement can be further generalized by abandoning the notion that measurement outcomes are orthogonal. This abstraction is called a positive-operator-valued measure (POVM) \cite{nielsen2002,jacobs2014}. A POVM is designed to accompany any allowable measurement statistics, bearing in mind that the post-measurement state is ambiguous (see the next subsection). A POVM can be described by a set of positive semi-definite operators $\{ M_m \}$ which satisfy the completeness relationship $\sum_m M_m = \mathbb{I}$. The outcome $m$ is observed with probability $\Tr (M_m \rho)$. Comparable to the purification of mixed states, Eq.~\eqref{eq:purification}, it has been shown that a POVM can always be obtained from a PVM acting on a higher dimensional space \cite{nielsen2002}.

In this thesis we focus on single parameter quantum metrology problems. Although many of the results naturally generalize to multiparameter problems, it is important to be cognisant of the incompatibility of simultaneous measurements in the multiparameter setting. Specifically, if two observables, $A$ and $B$, do not commute
\begin{equation}
    [A,B] \neq 0,
\end{equation}
then measuring $A$ and then $B$ is different than measuring $B$ then $A$. In fact, this is one of the major reasons why the cryptographic protocols outlined in \textbf{Chapter~6} can be deemed secure. The incompatibility of simultaneous measurements gives rise to the famous Heisenberg uncertainty principle \cite{robertson1929}
\begin{equation}
    \label{eq:Uncertainty}
    \Delta^2 A \Delta^2 B \geq \frac{1}{4} \big| \expval*{[A,B]} \big|^2,
\end{equation}
where $\Delta^2 A=\expval{A^2}-\expval{A}^2$ is the variance of an observable.

\subsection{Collapse of the wave function}

After a measurement is performed the quantum state undergoes a non-unitary transformation, more commonly referred to as the `collapse of the wave function'\footnote{The collapse of the wave function, a postulate of the Copenhagen interpretation, is arguably the most widely used model for quantum measurements. It is important to note though, to date, the dynamics of quantum measurements are still debated \cite{zeh1970, schlosshauer2005}.}. If a PVM is performed on the state $\rho$ and outcome $i$ is observed, then
\begin{equation}
    \label{eq:collapsePVM}
    \rho \rightarrow \frac{P_i \rho P_i}{\Tr (P_i \rho)}.
\end{equation}
The post-measurement state is drastically more complex when considering a general POVM. As mentioned, the post-measurement state is ambiguous, this is in consequence to the POVM elements $\{ M_m \}$ not having a unique Kraus decomposition \cite{hellwig1969}, as a multitude of measurement schemes may result in the same measurement statistics \cite{jacobs2014}. A Kraus decomposition of $M_m$ is a product of an (not necessarily self-adjoint) operator with its conjugate transpose, i.e for each $M_m$ there exists an $A_m$ such that $M_m = A_m A_m^\dagger$. The set $\{ A_m\}$ are the measurement operators which define a physical process which corresponds with the POVM. For a specific set of measurement operators, if outcome $m$ is observed, then
\begin{equation}
    \label{eq:collapsePOVM}
    \rho \rightarrow \frac{A_m \rho A_m^\dagger}{\Tr (M_m \rho)}.
\end{equation}
By comparing Eq.~\eqref{eq:collapsePVM} and Eq.~\eqref{eq:collapsePOVM}, one can interpret a PVM as a special case of a POVM when the set of measurement operators are all projectors.

\section{Distance Measures}

Quantum states are elements of a Hilbert space, so it is natural to consider the proximity of quantum states. Distance measures can be useful, as quantum states which are \textit{close} to one another can be expected to behave similarly under appropriate transformations. Distance measures, namely the trace-distance and fidelity, play a crucial role in \textbf{Chapter~6}, where the quantum states in question are bounded with respect to the above measures, from which, their utility for quantum metrology can be gauged.

\subsection{Trace Distance}

The trace distance, denoted by $\mathscr{D}$, between quantum states $\rho$ and $\sigma$ can be calculated using
\begin{equation}
    \mathscr{D} \big( \rho, \sigma \big) = \frac{1}{2} \Tr |\rho-\sigma|,
\end{equation}
where $|A|=\sqrt{A^\dagger A}$. An alternative definition of the trace distance can be expressed in terms of POVMs. Let $\{ M_m \}$ be a POVM, in which outcome $m$ is witnessed with probabilities $p_m=\Tr ( M_m \rho )$ and $q_m=\Tr ( M_m \sigma )$. The trace distance is equivalently defined via
\begin{equation}
    \mathscr{D} \big( \rho, \sigma \big) = \max_{ \{ M_m\} } \Big( \frac{1}{2}\sum_m |p_m - q_m |  \Big),
\end{equation}
where the maximization is taken over all POVMs. The contents of the brackets on the right-hand side of the above equation is in fact the definition of the trace distance between probability distributions $\{p_m \}$ and $\{q_m \}$ \cite{nielsen2002}. The second expression listed to compute the trace distance between quantum states is certainly impractical to calculate, however it does provide an insightful inequality: for any POVM $\{ M_m \}$, it follows that
\begin{equation}
    \label{eq:TraceDistanceAlt}
    \frac{1}{2} \sum_m |\Tr \big(M_m (\rho-\sigma) \big)| \leq \mathscr{D} \big( \rho, \sigma \big).
\end{equation}
The trace distance is contractive under a CPTP map $\mathcal{E}$, that is
\begin{equation}
    \label{eq:TDcontractive}
    \mathscr{D} \big( \mathcal{E} (\rho), \mathcal{E}( \sigma) \big) \leq \mathscr{D} \big( \rho, \sigma \big).
\end{equation}

\subsection{Fidelity}

The fidelity between quantum states is perhaps the most renowned measure of closeness in quantum information, even though it is not a metric in the mathematical sense. The fidelity, denoted with $\mathscr{F}$, between quantum states $\rho$ and $\sigma$ can be computed using
\begin{equation}
    \label{eq:fidelity}
    \mathscr{F} \big( \rho,\sigma \big) = \Big( \Tr \sqrt{\sqrt{\rho} \sigma \sqrt{\rho}} \Big)^2,
\end{equation}
which greatly simplifies to $\mathscr{F} \big( \rho,\sigma \big) = \Tr ( \rho \sigma)$ when either $\rho$ or $\sigma$ is a pure state. Note that this version of the fidelity is the square of what is defined in \cite{nielsen2002}. The fidelity and trace distance are related by the Fuchs–van de Graaf inequalities \cite{fuchs1999}
\begin{equation}
    \label{eq:fuchs}
    1-\sqrt{\mathscr{F} \big( \rho,\sigma \big)} \leq \mathscr{D} \big( \rho,\sigma \big) \leq \sqrt{1-\mathscr{F} \big( \rho,\sigma \big)}.
\end{equation}

%% file: Figures/Chapter2/blochsphere.tex
\def\rotationSphere{-112}
\def\radiusSphere{3.2cm}
\def\psiLat{45}
\def\psiLon{45}
\begin{blochsphere}[radius=\radiusSphere,opacity=0,rotation=\rotationSphere]
  \drawLongitudeCircle[]{\rotationSphere} 
  \drawLatitudeCircle[style={dashed}]{0}
  \labelLatLon{ket0}{90}{0};
  \labelLatLon{ket1}{-90}{0};
  \labelLatLon{ketminus}{0}{180};
  \labelLatLon{ketplus}{00}{0};
  \labelLatLon{ketpluspi2}{0}{-90};  
  \labelLatLon{ketplus3pi2}{0}{-270};
  \labelLatLon{psi}{\psiLat}{-\psiLon};
  \draw[-latex] (0,0) -- (ket0) node[above,inner sep=.5mm] at (ket0) {\footnotesize $\ket{0}$};
  \draw[-latex] (0,0) -- (ket1) node[below,inner sep=.5mm] at (ket1) {\footnotesize $\ket{1}$};
  \draw[-latex] (0,0) -- (ketplus) node[above, inner sep=0.6mm] at (ketplus) {\footnotesize$\ket{+}$};
  \draw[-latex] (0,0) -- (ketminus) node[above,inner sep=0.5mm] at (ketminus) {\footnotesize$\ket{-}$};
  \draw[-latex] (0,0) -- (ketpluspi2) node[right,inner sep=1.7mm] at (ketpluspi2) {\footnotesize $\ket{+_i}$};
  \draw[-latex] (0,0) -- (ketplus3pi2) node[left,inner sep=1.6mm] at (ketplus3pi2) {\footnotesize $\ket{-_i}$};
  \draw[-latex] (0,0) -- (psi) node[above]{\footnotesize $\ket{\psi}$};

  \coordinate (origin) at (0,0);
  {
    \setDrawingPlane{0}{0}
    \draw[current plane,dashed] (0,0) -- (-90+\psiLon:{cos(\psiLat)*\radiusSphere}) coordinate (psiProjectedEquat) -- (psi);
    \pic[current plane, draw,fill=orange!50,fill opacity=.5, text opacity=1,"\footnotesize $\phi$", angle eccentricity=2.2]{angle=ketplus--origin--psiProjectedEquat};
  }
  { \setLongitudinalDrawingPlane{\psiLon}
    \pic[current plane, draw,fill=orange!50,fill opacity=.5, text opacity=1,"\footnotesize $\theta$", angle eccentricity=1.5]{angle=psi--origin--ket0};
  }
\end{blochsphere}

%% file: Figures/Chapter2/dynamics1.tex
    \begin{tikzpicture}[
decoration = {markings,mark=at position .3 with
             {\arrowreversed[red]{Latex[length=1.5mm]}}}
                        ]
    \draw[-latex] (0,0) -- (0,2.05) node[above,inner sep=.5mm] at (0,2.05) {\footnotesize $\ket{+_i}$};
    \draw[-latex] (0,0) -- (2.05,0) node[right,inner sep=.5mm] at (2.05,0) {\footnotesize $\ket{-}$};
    \draw[-latex] (0,0) -- (0,-2.05) node[below,inner sep=.5mm] at (0,-2.05) {\footnotesize $\ket{-_i}$};
    \draw[-latex] (0,0) -- (-2.05,0) node[left,inner sep=.5mm] at (-2.05,0) {\footnotesize $\ket{+}$};
\draw[postaction={decorate},red] circle (1.93 cm);
\end{tikzpicture}

%% file: Figures/Chapter2/dynamics2.tex
    \begin{tikzpicture}[
decoration = {markings,mark=at position .84 with
             {\arrowreversed[red]{Latex[length=1.5mm]}}}
                        ]
    \draw[-latex] (0,0) -- (0,2.05) node[above,inner sep=.5mm] at (0,2.05) {\footnotesize $\ket{+_i}$};
    \draw[-latex] (0,0) -- (2.05,0) node[right,inner sep=.5mm] at (2.05,0) {\footnotesize $\ket{-}$};
    \draw[-latex] (0,0) -- (0,-2.05) node[below,inner sep=.5mm] at (0,-2.05) {\footnotesize $\ket{-_i}$};
    \draw[-latex] (0,0) -- (-2.05,0) node[left,inner sep=.5mm] at (-2.05,0) {\footnotesize $\ket{+}$};
\draw[postaction={decorate},red]
    plot[domain=0:22,variable=\t,smooth,samples=101,
           {Latex[length=1mm]}-]
        ({\t r}: {0.004*\t*\t});
\end{tikzpicture}

%% file: Chapters/Chapter3-Metrology.tex
\chapter{Estimation Theory}

Estimation theory is the mathematical language of metrology. Statistical error in classical estimation theory is ultimately constrained by the central limit theorem. \textit{Quantum metrology} overcomes this limitation thanks to quantum entanglement. With the vast number of applications and straightforward proof of principle, it is unsurprising that quantum metrology is witnessing a boon of theoretical and experimental developments \cite{giovannetti2011, degen2017, pirandola2018}.

This chapter is divided into three sections. The first section summarizes important concepts from classical estimation theory \cite{kay1993, cox2006, sheskin2003, rice2006, poor2013}. The second section is devoted to the analogous concepts of quantum estimation theory formalized by Helstrom \cite{helstrom1967, helstrom1968, helstrom1969} and Holevo \cite{holevo1973, holevo1982}. The final section examines example applications of quantum metrology (phase estimation and amplitude estimation) to put into perspective the mathematical tools and concepts introduced throughout the first two sections. For a quantum information perspective on quantum metrology see \cite{toth2014}. For a more mathematical rigorous review of quantum metrology and quantum estimation theory see \cite{simon2017}. For more information on estimation theory and statistical inference see \cite{kay1993, cox2006}.

\begin{quote}
    \textit{An experiment is a question which science poses to Nature and a measurement is the recording of Nature's answer.}
    \begin{flushright} -Max Planck \end{flushright}
\end{quote}

\section{Classical Estimation Theory}

In an abstract sense, the scientific and mathematical knowledge of humankind is reflected in the mathematical models used to describe the contents of the universe: planetary orbits, bacterial growth in a petri dish, even social constructs like financial trends. These models, are not fabricated haphazardly, instead they are a manifestation of a multitude of observations and tested by making predictions. As our efficiency of gathering and interpreting data increases, so do the mathematical models, and in turn our understanding of the universe. For example, the theory of gravity has evolved along with the capabilities of telescopes; from Galilean and Newtonian gravity to Einstein's theory of general relativity to the (currently unconfirmed) theory of dark matter and dark energy.

\textit{Estimation theory} is a branch of statistics at the heart of mathematical modelling. It addresses the question: `What is the most efficient way of extracting information from a set of data?'. This seemingly simple question is difficult to answer. Typically, the variables used to describe a mathematical model can be partitioned in two categories
\begin{enumerate}
    \item observables - an attribute which can be inherently measured (e.g. position and speed).
    \item latent parameters - an attribute which cannot be inherently measured, (e.g. strength of an electromagnetic field).
\end{enumerate}
The \textit{parameter estimation problem} is concerned with the extent at which collected data (observables) can be used to estimate the unknown latent parameters \cite{sheskin2003}. With respect to the listed examples, one could observe the dynamics of a charged particle to estimate the strength of an electromagnetic field.

Formally, observed data $\mathbf{x}=\{x_1, \ldots, x_N \}$ is treated as a realisation of $N$ independent and identically distributed (iid) random variables $X$. A probably density function $p(X | \theta )$ dictates the distribution of observed data, where $\theta$ is a latent parameter. The goal of the parameter estimation problem is to construct an \textit{estimator} $\hat{\theta}(\mathbf{x})$, which should be interpreted as a function whose input is the collected data $\mathbf{x}$ and outputs an estimate of $\theta$. The explicit dependence on $\mathbf{x}$ is sometimes dropped for clarity, $\hat{\theta}(\mathbf{x}) \rightarrow \hat{\theta}$. Estimators are subjected to two conditions. The first condition is that the expected estimate is the true value of the parameter, this is known as having an unbiased estimator
\begin{equation}
    \label{eq:unbiased}
    \expval*{\hat{\theta}} = \int  p(\mathbf{x}|\theta) \hat{\theta}(\mathbf{x}) \mathrm{d} \mathbf{x} = \theta.
\end{equation}
The integral equation is used for observed data which can take on a continuum of values, it is interchangeable with a sum in the discrete case. The second condition is that an estimator tends towards the correct value as the amount of data increases, this is known as being consistent
\begin{equation}
    \lim_{N \rightarrow \infty} \hat{\theta} = \theta.
\end{equation}
An estimator is a manifestation of random variables, and is thus also a random variable, hence, statistical moments such as mean and variance are well-defined.

The statistical inference process adopted to the parameter estimation problem is dependent on the nature of the latent variable: deterministic or stochastic. Usually, a frequentist inference approach is taken for deterministic parameters and a Bayesian inference approach is taken for stochastic parameters \cite{li2018}. Mathematically, these two approaches vary greatly, the primary differences are listed in Tab.~(\ref{tab:estimation_summary}), but they are not mutually exclusive. The subsequent chapters of this thesis employ the frequentist approach, and therefore the frequentist approach is summarized in greater detail in this chapter. That being said, the Bayesian approach has been adapted to the realm of quantum information \cite{holevo1982, tsang2011}, and has been gaining traction in the community \cite{berry2009, gammelmark2013, jarzyna2015, wiebe2016, rubio2020b}. Specifically, to circumvent problems of the frequentist approach: i) lack of a priori knowledge \cite{kolodynski2010, demkowicz2011} and ii) inaccuracies with limited resources \cite{rubio2020b}. Even though it is not applied to the research presented in this thesis, for the sake of completeness, a brief summary of the Bayesian approach used in classical parameter estimation problems and its adaptation to quantum parameter estimation problems is included in this chapter.

\begin{table}[h]
    \centering
    \renewcommand{\arraystretch}{1.5}
    \begin{tabular}{c|c|c}
         & Frequentist Approach & Bayesian Approach  \\
         \hline
        Parameter(s) & Deterministic & Stochastic  \\
        Figure of Merit & Mean squared error & Cost function \\
        Optimization & Local & Global
    \end{tabular}
    \caption{The main differences between the frequentist approach and Bayesian approach for statistical inference. This is a broad perspective and the statistical inference approaches are not restricted by this table.}
    \label{tab:estimation_summary}
\end{table}

\subsection{The Frequentist Approach}

The \textit{frequentist approach} is typically used when $\mathbf{\theta}$ is deterministic (sometimes called static). As $N \rightarrow \infty$ the frequency of collected data tends to reflect the probability density function, hence the etymology. Therefore with a sufficient amount of collected data, the unknown parameter can be estimated to any desired precision. The figure of merit used by the frequentist approach is the mean-squared error (MSE)
\begin{equation}
    \label{eq:MSE}
    \Delta^2 \hat{\theta} = \expval*{ (\hat{\theta} - \theta)^2}=\int  p(\mathbf{x}|\theta) \big(\hat{\theta}(\mathbf{x}) - \theta \big)^2 \mathrm{d} \mathbf{x},
\end{equation}
in which the aim is to find an estimator which minimizes the above equation. Because the estimator is assumed to be unbiased, the MSE is equal to the variance, which is often a more significant statistical quantity.

The first controversy of the the frequentist approach arises due to the fact that an optimal estimator (one where Eq.~\eqref{eq:MSE} is minimized) is potentially dependent on $\theta$. Some estimators may be optimal for specific values of $\theta$ (local), whereas an estimator which is optimal for all values of $\theta$ (global) can only be worse than ones which are locally optimized. At first glance, this appears counter intuitive because a locally optimized estimator requires exact knowledge of $\theta$, which defeats the purpose of parameter estimation. However, it is reasonable to assume that a priori approximate knowledge $\theta \approx \theta_0$ is often known because of theory or previous estimates. In the absence of a priori knowledge, one can construct a locally efficient estimator by increasing $N$. To do so, a fraction of the results are first used to obtain a local approximation $\theta_0$, and the remaining are used within the locally optimized estimator. Unfortunately, the frequentist approach does not provide a method on bounding $N$ such that the local regime can be assured; thus the saturation of an optimal estimator may not be possible without the ability to infinitely increase $N$.

\subsection{Cramér-Rao Bound and Fisher Information}

The \textit{Cramér-Rao Bound} (CRB) is an inequality which assigns a lower bound to the MSE of unbiased estimators \cite{cramer1946}, the derivation of which is straightforward. The unbiased condition, Eq.~\eqref{eq:unbiased}, can be re-written as
\begin{equation}
    \int  p(\mathbf{x}|\theta) \big( \hat{\theta}(\mathbf{x}) - \theta \big)\mathrm{d} \mathbf{x}=0,
\end{equation}
from which it follows that
\begin{equation}
\label{eq:crbproof}
\begin{split}
    0&=\frac{\partial}{\partial \theta} \int  p(\mathbf{x}|\theta) \big( \hat{\theta}(\mathbf{x}) - \theta \big)  \mathrm{d} \mathbf{x} \\
    &=\int  \frac{\partial p(\mathbf{x}|\theta)}{\partial \theta} \big( \hat{\theta}(\mathbf{x}) - \theta \big)\mathrm{d} \mathbf{x}-\int  p(\mathbf{x}|\theta) \mathrm{d} \mathbf{x}  \\
    &=\int p(\mathbf{x}|\theta)  \frac{\partial  \ln p(\mathbf{x}|\theta)}{\partial \theta} \big( \hat{\theta}(\mathbf{x}) - \theta \big)\mathrm{d} \mathbf{x}-1.
\end{split}
\end{equation}
Using the Cauchy–Schwarz inequality
\begin{equation}
\label{eq:CauchySchwarz}
    \bigg| \int  f(x)g(x)\mathrm{d}x \bigg|^2 \leq \bigg( \int  f(x)^2\mathrm{d}x \bigg) \cdot \bigg( \int  g(x)^2\mathrm{d}x \bigg),
\end{equation}
with $x \rightarrow \mathbf{x}$, $f(x) \rightarrow \sqrt{p(\mathbf{x}|\theta)} \frac{\partial \ln p(\mathbf{x}|\theta) }{\partial \theta} $ and $g(x) \rightarrow \sqrt{p(\mathbf{x}|\theta)} \big( \hat{\theta}(\mathbf{x}) - \theta \big) $, Eq.~\eqref{eq:crbproof} is transformed into the inequality
\begin{equation}
    1 \leq \bigg( \int  p(\mathbf{x}|\theta) \big(\hat{\theta}(\mathbf{x}) - \theta \big)^2 \mathrm{d} \mathbf{x} \bigg) \cdot \bigg( \int  p(\mathbf{x}|\theta) \Big( \frac{\partial \ln p(\mathbf{x}|\theta) }{\partial \theta} \Big)^2 \mathrm{d} \mathbf{x} \bigg).
\end{equation}
The above can be manipulated to obtain the CRB
\begin{equation}
\label{eq:CRB}
    \Delta^2 \hat{\theta} \geq \frac{1}{\mathcal{I}\big( p( \mathbf{x} | \theta ) \big)},
\end{equation}
where
\begin{equation}
    \label{eq:FisherInfo}
    \begin{split}
    \mathcal{I}\big( p( \mathbf{x} | \theta ) \big) &=  \int  p(\mathbf{x}|\theta) \Big( \frac{\partial \ln p(\mathbf{x}|\theta) }{\partial \theta} \Big)^2 \mathrm{d} \mathbf{x} \\
    &= \int  \frac{1}{p(\mathbf{x}|\theta)} \Big( \frac{\partial p(\mathbf{x}|\theta) }{\partial \theta} \Big)^2 \mathrm{d} \mathbf{x} \\
    &= - \int  p(\mathbf{x}|\theta)  \frac{\partial^2 \ln p(\mathbf{x}|\theta) }{\partial \theta^2}  \mathrm{d} \mathbf{x}
    \end{split}
\end{equation}
is the \textit{Fisher Information} (FI), where three equivalent (assuming that $p$ is twice differentiable) expressions given. The FI is a non-negative and additive quantity. Because $\mathbf{x}$ is $N$ independent realisations of the random variable $X$, the CRB can be equivalently expressed as
\begin{equation}
    \Delta^2 \hat{\theta} \geq \frac{1}{N \mathcal{I}\big( p( X | \theta ) \big)}.
\end{equation}
The above form of the CRB reflects the limitations of central limit theorem: as $N \rightarrow \infty$ the sample average will take on a normal distribution with a variance of $\mathcal{O} \big( N^{-1} \big)$.

The FI is often interpreted as a measure of how much information about an unknown parameter can be extracted from a probability density function \cite{fisher1925}. In particular, $\theta$ can be learned perfectly when $\mathcal{I} \rightarrow \infty$, and conversely no information can be learned about $\theta$ when $\mathcal{I} = 0$. In fact, when viewing probability density functions as points on a manifold (parameterized by $\theta$), the FI is a Riemannian metric between neighbouring probability density functions $p(X|\theta)$ and $p(X|\theta + \delta \theta)$ \cite{nielsen2013}. Similarly, the statistical angle\footnote{This is the classical version of the Bures angle \cite{wootters1981}.} between probability density functions 
\begin{equation}
    D\big(p_1(x),p_2(x) \big)= \arccos \int \sqrt{p_1(x) p_2(x)} \mathrm{d}x,
\end{equation}
can be expressed as \cite{barndorff1986}
\begin{equation}
    \label{eq:AngularDistance}
    D \big( p(X|\theta),p(X|\theta + \delta \theta) \big) = \frac{1}{2} \sqrt{\mathcal{I}\big( p(X|\theta) \big)} \delta \theta +\mathcal{O} \big( \delta \theta ^2 \big).
\end{equation}
Hence, a probability density function with a high FI will deviate more upon small perturbations $\delta \theta$ than the opposing case of a probability density function with a small FI.

\begin{figure}[ht]
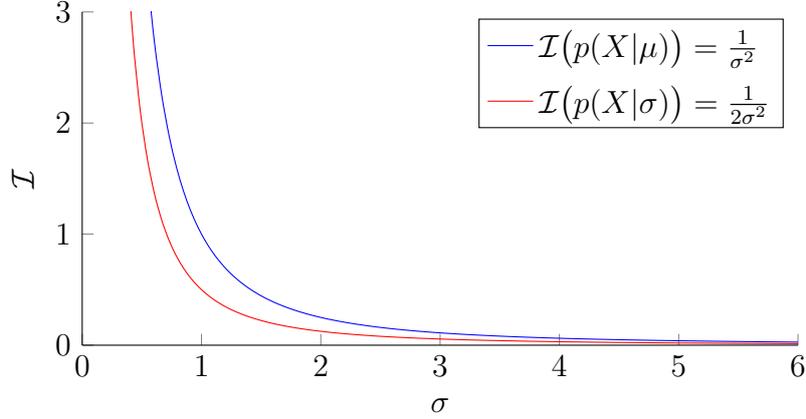

    \centering
    \include{Figures/Chapter3/FIplots}
    \caption{The FI for a normal distribution $p(X|\mu,\sigma)=\frac{1}{\sqrt{2\pi}\sigma}e^{-\frac{1}{2}(\frac{x-\mu}{\sigma})^2}$. Regardless of whether $\mu$ or $\sigma$ is the latent parameter, the FI is exclusively dependent on the standard deviation of the normal distribution. This is logical because it is more difficult to interpret data with a larger standard deviation; which is in accordance with the interpretation of the FI being a measure of extractable information. One could equally consider the scenario in which both $\mu$ and $\sigma$ are unknown parameters. Here multiparameter parameter estimation techniques are needed - which are discussed in a latter part of this chapter.}
    \label{fig:FINormal Distribution}
\end{figure}

The Cauchy-Schwarz inequality, Eq.~\eqref{eq:CauchySchwarz}, is saturated if
\begin{equation}
    \frac{|f(x)|}{|g(x)|}=\frac{\int f(x)^2 \mathrm{d}x}{\int g(x)^2 \mathrm{d}x}
\end{equation}
Therefore an estimator which saturates the CRB for all $\theta$ (global) satisfies
\begin{equation}
    \label{eq:CRBcond}
    \frac{\partial \ln p( \mathbf{x} | \theta )}{\partial \theta}= \mathcal{I} \big( \mathbf{x} | \theta \big) \big( \hat{\theta}(\mathbf{x}) - \theta \big).
\end{equation}
An estimator which saturated the CRB is said to be \textit{efficient}. The above expression can be equivalently written as
\begin{equation}
\label{eq:efficientestimator}
\begin{split}
    p( \mathbf{x} | \theta ) &= \exp \bigg( \int \mathcal{I} \big( \mathbf{x} | \theta \big) \big( \hat{\theta}(\mathbf{x}) - \theta \big) \mathrm{d} \theta \bigg)\\
    &=\exp \bigg( \frac{\partial \mathcal{J}\big( \mathbf{x}|\theta)}{\partial \theta} \big( \hat{\theta}(\mathbf{x}) - \theta \big)+ \mathcal{J}\big(\mathbf{x}|\theta) + c(\mathbf{x}) \bigg),
\end{split}
\end{equation}
where $\mathcal{J}\big( \mathbf{x} |\theta \big)$ is a function which satisfies $\frac{\partial^2 \mathcal{J}\big(\mathbf{x}|\theta \big)}{\partial \theta^2} = \mathcal{I}\big(\mathbf{x}|\theta \big)$ and $c(\mathbf{x})$ is an arbitrary function independent of $\theta$, both of which are chosen such that the unbiased condition, Eq.~\eqref{eq:unbiased}, is satisfied. This general expression for a  probability density function can correspond to a multitude of well-known distributions in statistics with exponential tendencies: Gaussian, Bernoulli, Poisson, et cetera. It should be stressed that an efficient global estimator does not necessarily exist, further it may encounter the earlier stated problem of having a dependence on $\theta$. A locally (approximately) efficient estimator can be constructed with prior knowledge that $\theta \approx \theta_0$ by re-arranging Eq.~\eqref{eq:CRBcond}
\begin{equation}
    \label{eq:localestimator}
    \hat{\theta}_{\text{Local}}=\theta_0+\frac{1}{\mathcal{I} \big( \mathbf{x} |\theta_0 \big) } \frac{\partial \ln p( \mathbf{x} | \theta )}{\partial \theta} \bigg|_{\theta \rightarrow \theta_0}.
\end{equation}
Unfortunately, the locally approximate estimator is ultimately constrained by ones prior knowledge, as shifting $\theta_0 \rightarrow \theta_0 + \delta \theta_0$ will similarly shift Eq.~\eqref{eq:localestimator} by $\mathcal{O}(\delta \theta_0)$. Furthermore, the locally approximate estimator may be ill defined on certain domains, for example one of circular symmetry (such as the problem of phase estimation which is discussed in a later section of this chapter).

\subsection{Maximum Likelihood Estimation}

The \textit{likelihood function} $L(\theta | \mathbf{x})=p(\mathbf{x} | \theta )$ is a goodness of fit between a model and the sampled data. It should be understood that the likelihood function is not a probability density function; the observed data $\mathbf{x}$ is held fixed and the latent parameter $\theta$ is considered a variable. The intuition is simplistic: if $L(\theta_1 | \mathbf{x}) > L(\theta_2 | \mathbf{x} )$, then it is more likely that the true value of $\theta$ is $\theta_1$ rather than $\theta_2$. This is the principal idea of maximum likelihood estimation \cite{berger1988}. The maximum likelihood estimator, $\hat{\theta}_{\text{ML}}$, outputs the value of $\theta$ which maximizes $L(\theta | \mathbf{x} )$
\begin{equation}
    \hat{\theta}_{\text{ML}}(\mathbf{x}) = \underset{\theta}{\text{argmax}} L(\theta | \mathbf{x}) = \underset{\theta}{\text{argmax}} \ln L(\theta| \mathbf{x}).
\end{equation}
Because $p(\mathbf{x} | \theta )$ is a joint probability density function of $N$ independent probability density functions,  $p(\mathbf{x} | \theta ) = \prod_{j=1}^N p(x_j | \theta )$, it is often simpler to maximize the log of the likelihood function, $\ln L (\theta | \mathbf{x} )$, sometimes shortened to the log-likelihood.

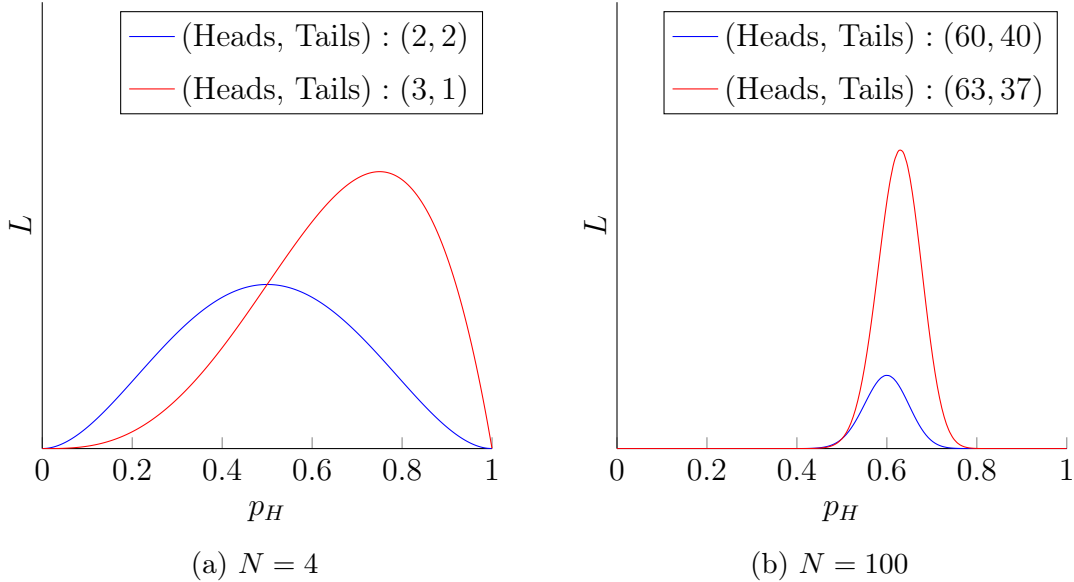
\begin{figure}
    \centering
    \begin{subfigure}{.495\textwidth}
        \centering
        \input{Figures/Chapter3/Likelihood1}
        \caption{$N=4$}
    \end{subfigure}
        \begin{subfigure}{.495\textwidth}
        \centering
        \input{Figures/Chapter3/Likelihood2}
        \caption{$N=100$}
    \end{subfigure}
    \caption{Likelihood function for a biased coin where $p_H$ is an unknown probability of the coin toss resulting in heads. The maximum likelihood estimation strategy outputs $\hat{p}_H=\frac{\text{\# Heads}}{N}$. The estimate is sensitive to small fluctuations in the observed data for small $N$ (a), but becomes more robust to fluctuations as $N$ increases (b).}
    \label{fig:likelihoodplots}
\end{figure}

One controversy with maximum likelihood estimation is that $\hat{\theta}_\text{ML}$ does not generally satisfy the unbiased condition, Eq.~\eqref{eq:unbiased}. Specifically, for small $N$, where the estimator is much more susceptive to statistical outliers within the collected data. However, as $N \rightarrow \infty$ the estimator becomes more unbiased, $\expval*{\hat{\theta}_{\text{ML}}} \rightarrow \theta$. The sensitivity of the maximum likelihood estimator to small fluctuations in $\mathbf{x}$ is illustrated in Fig.~(\ref{fig:likelihoodplots}). Additionally, the MSE of the maximum likelihood estimator tends to saturate the CRB as it becomes more unbiased \cite{kay1993, van2000}. It is important to remark that there is no general formula to determine an appropriate value of $N$. However, within the framework of quantum metrology, unknown parameters are encoded into quantum resources; because of the abundance of these resources the issue of small $N$ is often ignored.

\subsection{Example: Biased Coin}

Consider a biased coin, which when flipped results in heads with an unknown probability $p_H$ and tails with probability $1-p_H$. For the sake of creating a locally optimized estimator, previous coin tosses suggest that the bias is $p_H \approx p_{H,0}$. The FI of a single flip is easy to compute
\begin{equation}
    \mathcal{I}_\text{coin} =  \frac{1}{p_H}  \Big( \frac{\partial p_H}{\partial p_H} \Big)^2 + \frac{1}{1-p_H}  \Big( \frac{\partial (1-p_H)}{\partial p_H} \Big)^2 = \frac{1}{p_H(1-p_H)},
\end{equation}
thus the CRB imposes that the MSE of an unbiased estimator using $N$ outcomes is bounded by
\begin{equation}
    \Delta^2 \hat{p}_H \geq \frac{p_H(1-p_H)}{N}.
\end{equation}
To remain somewhat general, the data collected is from $N$ coin tosses, $h$ of which resulted in heads and $N-h$ of which resulted in tails, which occurs with probability $\binom{N}{h}p_H^h(1-p_H)^{N-h}$. Using the locally optimized estimation strategy, Eq.~\eqref{eq:localestimator}, the estimator is
\begin{equation}
    \hat{p}_H^{\text{Local}} = p_{H,0} + \frac{1}{N \mathcal{I}_\text{coin}} \frac{\partial \ln p_H^{h}(1-p_H)^{N-h}}{\partial p_H} \bigg|_{p_H \rightarrow p_{H,0}}=\frac{h}{N},
\end{equation}
which is unbiased because
\begin{equation}
    \expval*{\hat{p}_H^{\text{Local}}} = \sum_{h=0}^N \binom{N}{h} p_H^h(1-p_H)^{N-h} \frac{h}{N}= p_H.
\end{equation}
Furthermore, the estimator is efficient because it saturates the CRB
\begin{equation}
    \Delta^2 \hat{p}_H^{\text{Local}} = \sum_{h=0}^N \binom{N}{h} p_H^h(1-p_H)^{N-h} \big( \frac{h}{N} - p_H)^2= \frac{p_H(1-p_H)}{N}=\frac{1}{N \mathcal{I}_\text{coin}}.
\end{equation}

Despite the fact that the estimator was initially constructed using a local approximation, the estimator is independent of $p_{H,0}$, and is thus globally optimized. In addition, the same estimator is realized using the maximum likelihood estimation strategy, see Fig.~(\ref{fig:likelihoodplots}). The biased coin exemplifies the underlying nature of the frequentist approach: as $N$ increases, the quantity $\frac{h}{N}$ converges to the quantity $p_H$, from which the (albeit simple) probability density function can be reverse engineered.

\subsection{The Bayesian Approach}

The \textit{Bayesian approach} is typically used to estimate unknown parameters which are stochastic. In other words, the latent parameters are themselves a random variable and have an intrinsic probability distribution $p(\theta )$ - which should to be confused with $p(\mathbf{x} | \theta)$. Therefore, the observed data $\mathbf{x}$ is dependent on specific realisations of $\theta$. Consequently, a well-constructed estimator within the Bayesian approach aims to minimize the MSE for all values (global) of $\theta$, and not subjected to local values like the frequentist approach. To achieve this, the Bayesian approach minimizes the average of a cost function $C(\hat{\theta},\theta)$ \cite{kay1993, trees2007}
\begin{equation}
    \label{eq:costfunction}
    \expval*{C(\hat{\theta},\theta)}=\int  p(\theta) \bigg( \int  p(\mathbf{x}|\theta) \big( C(\hat{\theta}(\mathbf{x}),\theta) \mathrm{d} \mathbf{x} \bigg) \mathrm{d} \theta.
\end{equation}
In principle, a cost function is a generalisation of the MSE for the frequentist approach. It is a function which decreases as $\hat{\theta}$ approaches $\theta$. The MSE is an example of a cost function, so too is the absolute error $C=|\hat{\theta}-\theta|$. Different cost functions are tailored to specific probability density functions to take advantage of specific symmetries or properties.

By merging the two probability distributions, the average cost can be interpreted as an average over the simultaneous realisations of $X$ and $\theta$. According to Bayes' theorem (hence the name of this approach), the joint probability distribution can be interpreted in two ways
\begin{equation}
    p(\mathbf{x},\theta)=p(\mathbf{x} |\theta)p(\theta)=p( \theta | \mathbf{x})p(\mathbf{x}),
\end{equation}
thus the average cost can be written as
\begin{equation}
    \expval*{C(\hat{\theta},\theta)}=\int  p(\mathbf{x}) \bigg( \int \mathrm{d} \theta p(\theta | \mathbf{x}) C(\hat{\theta}(\mathbf{x}),\theta) \mathrm{d} \theta \bigg) \mathrm{d} \mathbf{x}.
\end{equation}
The average cost can then be minimized through standard optimization techniques, i.e by solving the equation
\begin{equation}
    \label{eq:bayesianestimator}
    \frac{\partial}{\partial \hat{\theta}} \int  p(\theta | \mathbf{x}) C(\hat{\theta}(\mathbf{x}),\theta) \mathrm{d} \theta = 0,
\end{equation}
where the quantity $p(\theta | \mathbf{x})$ can be computed using Bayes' theorem
\begin{equation}
    p(\theta | \mathbf{x}) = \frac{p(\mathbf{x}|\theta)p(\theta)}{p(\mathbf{x})}=\frac{p(\mathbf{x}|\theta)p(\theta)}{\int  p(\mathbf{x}|\theta)p(\theta) \mathrm{d} \theta}.
\end{equation}
A priori knowledge of $p(\theta)$ is needed to evaluate Eq.~\eqref{eq:bayesianestimator}, which is why the Bayesian approach is often used in tandem with adaptive techniques. The estimator continually outputs a new probability density function $p(\theta)$ based on the previous density function and collected data, and as the number of repetitions increases it will converge towards the correct value. There are precision bounds similar to the CRB within the Bayesian framework, but they are dependent on the cost function \cite{bobrovsky1987}. More information about Bayesian inference can be found in \cite{trees2007}.

\section{Quantum Estimation Theory}

\begin{figure}[ht]
    \centering
    \begin{subfigure}{.49\textwidth}
        \centering
        \includegraphics[width=\textwidth]{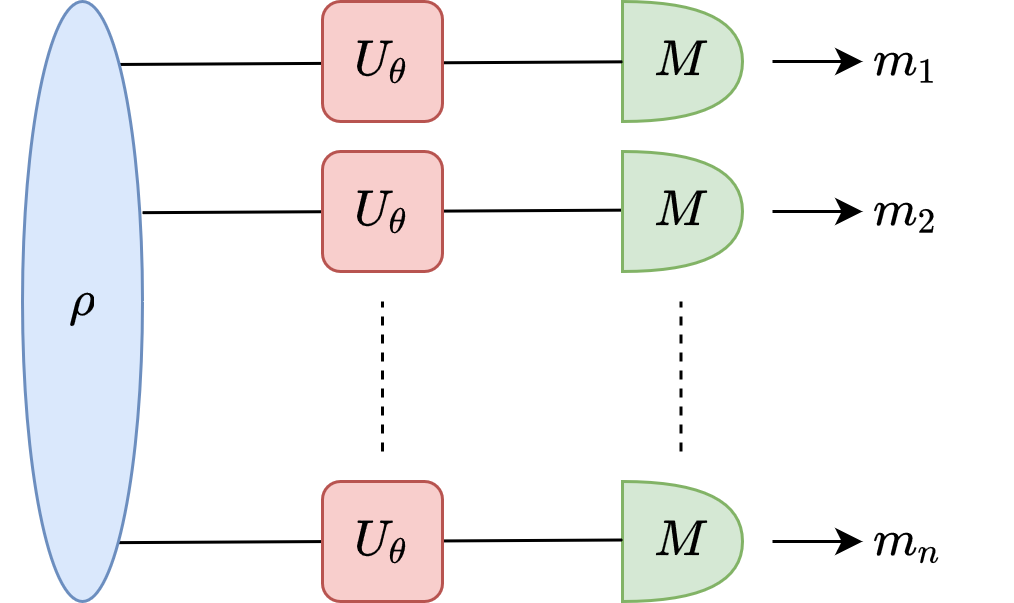}
        \caption{Phase estimation.}
    \end{subfigure}
        \begin{subfigure}{.49\textwidth}
        \centering
        \includegraphics[width=\textwidth]{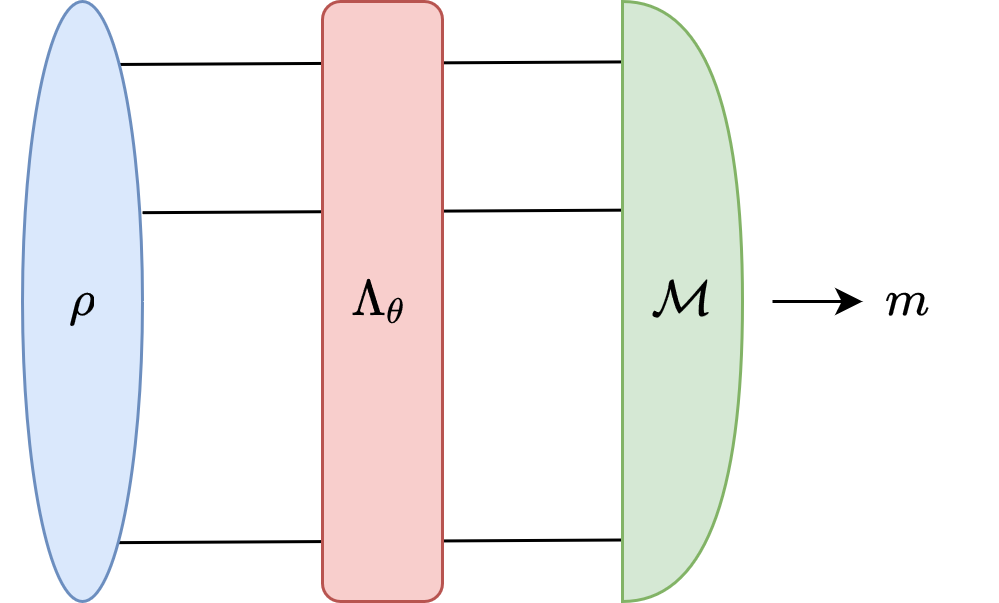}
        \caption{General framework.}
    \end{subfigure}
    \caption{Diagrams of the prepare, encode and measure segment of a quantum parameter estimation problem. (a) The canonical example of quantum metrology is phase estimation \cite{giovannetti2006}, in which a phase $\theta$ is independently and identically encoded into each of the $n$ qubits of $\rho$ through a unitary $U_\theta$. Each of the qubits are individually measured in accordance with a PVM $M$. (b) A more general framework involves $\theta$ being encoded through a general CPTP map $\Lambda_\theta$, which does not necessarily act identically on the $n$ qubits. Additionally, the PVM $M$ is replaced by a POVM $\mathcal{M}$. The generalized setting depicted in (b) is allowable in the realm of quantum mechanics, but unlike the problem of phase estimation, it is difficult to compare to classical setting. Further, highly entangling operations and measurements are not feasible for current quantum technologies \cite{calsamiglia2001}, so it is often more practical to consider the simplistic setting of phase estimation as a benchmark for quantum metrology.}
    \label{fig:MetrologySchematics}
\end{figure}

In the quantum setting, the foundations of the parameter estimation problem remains mostly unchanged from the classical setting \cite{helstrom1969,holevo1982}. An unknown parameter $\theta$ governs an $n$ qubit quantum state $\rho_\theta$, the individual qubits can be measured with respect to a PVM $M$, and the measurement outcomes $m_1,\ldots,m_n$ are used to construct an estimate $\hat{\theta}$\footnote{The assumptions that the qubits are acted on independently and identically (both the encoding and the measurement) are unnecessary and impose a limit on the most general framework of a quantum parameter estimation scheme, see Fig.~(\ref{fig:MetrologySchematics}). These assumptions are introduced to provide a natural extension from a classical framework to a quantum framework.}. The main difference from the classical setting is that the measurement outcomes (analogous to $\mathbf{x}$) are not necessarily independent from each other because of entanglement. As a result, estimates can be made with a super-classical precision known as the \textit{Heisenberg limit}.

A quantum parameter estimation problem can be viewed as a two step process. The first is the `prepare, encode and measure' step, which is inherently quantum by construction and depicted in Fig.~(\ref{fig:MetrologySchematics}). The second is the statistical inference step, which is uniquely classical, thus the techniques discussed in the the previous section can be applied. Therefore, using a frequentist approach with an unbiased estimator, if the quantum portion is repeated $\nu$ times, the MSE is bounded by a quantum version of the CRB, otherwise known as the quantum Cramér-Rao bound (QCRB)
\begin{equation}
    \label{eq:QCRB}
    \Delta^2 \hat{\theta} \geq \frac{1}{\nu \mathcal{I}(\rho_\theta,\mathcal{M})} \geq \frac{1}{\nu \mathcal{Q(\rho_\theta)}},
\end{equation}
where $\mathcal{Q}$ is the quantum Fisher information (QFI), which is the FI maximized over all POVM's $\mathcal{M}$ \cite{braunstein1994}. Evidently, the goal of finding an optimal estimator $\hat{\theta}$ naturally divides into a classical goal and a quantum goal. The classical goal is to devise an optimal estimation technique, e.g. a locally optimized estimator or the maximum likelihood estimator, whilst the quantum goal is to find an optimal combination of initialized states $\rho$ and POVM $\mathcal{M}$. For the task of phase estimation, this the QCRB can be saturated using highly entangled states, such as the GHZ state or NOON states, and a local measurement strategy \cite{giovannetti2004}. In general, the QFI is a highly non-linear equation, and there is no universal optimization strategy which is applicable to an arbitrary encoding $\Lambda_\theta$. There are different mathematical techniques to approximately solve this optimization problem \cite{gill2013,koczor2020,meyer2021}.

The quantum metrology schematics in Fig.~(\ref{fig:MetrologySchematics}) are idealized settings. In reality, it is much more complicated: environmental decoherence occurs in simultaneity with the parameter encoding, resulting in noisy measurement statistics and added uncertainty \cite{escher2011a, escher2011b, demkowicz2012}. More so, quantum technologies are not perfect, and an error may be introduced in either the quantum state preparation step or quantum measurement step. This more realistic noisy scenario is explored in greater detail in \textbf{Chapter~5}.

\subsection{Inferring an Estimate from an Observable}

A simple frequentist estimation strategy used in quantum metrology is to construct an estimator for the expectation value of an observable $O$ and infer the value of the latent parameter from this estimate \cite{toth2014}. Assuming that $O$ is chosen appropriately, the expectation value $\expval{O}=\Tr ( O \rho_\theta)$ will be a function of $\theta$, denoted by $f(\theta)=\Tr ( O \rho_\theta)$. An estimate of $f(\theta)$, $\hat{f}$, can be inverted to obtain $\hat{\theta}=f^{-1}(\hat{f})$. The estimator $\hat{f}$ is designed using the frequentist philosophy: with sufficient data $\nu \gg 1$, the frequency of the measurement results will mimic the true probability density function. Denote the eigenvalues of $O$ as $\{ \lambda_j \}$ with corresponding eigenvectors $\{ \ket{\phi_j} \}$
\begin{equation}
    O= \sum_j \lambda_j \dyad{\phi_j}.
\end{equation}
The state $\rho_\theta$ is measured with respect to the eigenbasis of $O$. The results are recorded as $m_1, \ldots, m_\nu$: if the $k$th measurement results in $\ket{\phi_j}$, then $m_k=\lambda_j$ and the maximum likelihood estimate can be written as
\begin{equation}
    \hat{f} = \frac{1}{\nu} \sum_{k=1}^\nu m_k.
\end{equation}
This is an unbiased estimate because
\begin{equation}
    \mathbb{E}(\hat{f})= \frac{1}{\nu} \sum_{k=1}^\nu \mathbb{E}(m_k) = \frac{1}{\nu} \sum_{k=1}^\nu \sum_{j} \lambda_j \Tr ( \rho_\theta \dyad{\phi_j} ) = \frac{1}{\nu} \sum_{k=1}^\nu \expval{O} = \expval{O},
\end{equation}
and the MSE is proportional to the variance of $O$
\begin{equation}
    \Delta^2 \hat{f} = \frac{\Delta^2 O}{\nu} = \frac{\Tr(O^2 \rho_\theta) - \Tr(O \rho_\theta)^2}{\nu}.
\end{equation}

An issue with this estimation technique is that $f(\theta)$ is not necessarily an invertible function, and thus $\hat{f}=f^{-1} (\hat{f})$ may be ambiguous. That is of course, unless one has a priori knowledge of $\theta \approx \theta_0$ such that one can properly define a local inverse in the region surrounding $f(\theta_0)$. Assuming this is true and that the MSE is small, $\Delta^2 \hat{f} \ll 1$, then by the central limit theorem $\hat{f}$ fluctuates close to $\expval{O}$, validating the first order Taylor approximation
\begin{equation}
    \label{eq:TaylorExpansion}
    \hat{\theta}=f^{-1}(\hat{f}) \approx f^{-1}(\expval{O}) + \frac{\partial f^{-1}(\hat{f})}{\partial \hat{f}} \Big|_{\hat{f} \rightarrow \expval{O}} (\hat{f} - \expval{O})=\theta + \frac{1}{\frac{\partial \expval{O}}{\partial \theta}}(\hat{f}-\expval{O}).
\end{equation}
It follows from the above approximation that the estimator $\hat{\theta}$ is unbiased and has MSE
\begin{equation}
    \label{eq:ErrorPropagation}
    \Delta^2 \hat{\theta} = \frac{\Delta^2 \hat{f}}{|\frac{\partial \expval{O}}{\partial \theta}|^2}= \frac{\Delta^2 O}{\nu |\frac{\partial \expval{O}}{\partial \theta}|^2}.
\end{equation}

\begin{figure}[!ht]
    \centering
    \includegraphics[width=0.85\textwidth]{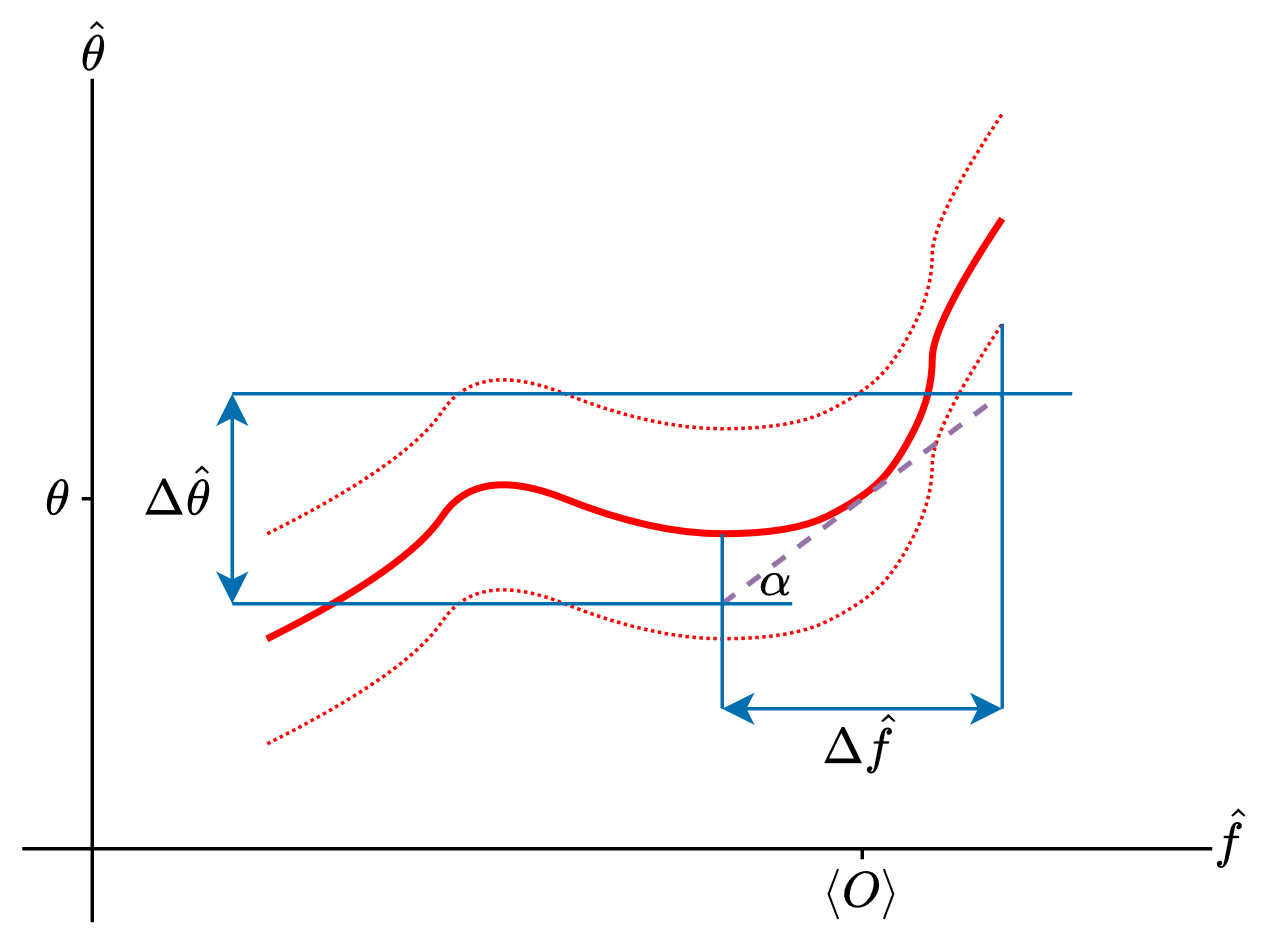}
    \caption{Graphical calculation of the MSE $\Delta^2 \hat{\theta}$ using the error propagation formula. The solid red curve depicts $\hat{\theta}=f^{-1}(\hat{f})$, which at the point $\hat{f}\rightarrow \expval{O}$ has a tangent with angle $\alpha$, therefore, $\frac{\Delta \hat{\theta}}{\Delta \hat{f}}=|\tan \alpha | = | \frac{\partial \expval{O}}{\partial \theta} | ^{-1}$.}
    \label{fig:ErrorPropagation}
\end{figure}

Eq.~\eqref{eq:ErrorPropagation} is the error propagation formula, which quantifies the amount that $\hat{\theta}$ fluctuates around $\theta$ in terms of the fluctuations of $\hat{f}$ around $\expval{O}$ \cite{ku1966}. A geometric intuition of the formula is depicted in Fig.~(\ref{fig:ErrorPropagation}). The term in the denominator $|\frac{\partial \expval{O}}{\partial \theta}|^2$ encapsulates the difficulty of inverting a function when there is uncertainty. The effects of uncertainty are amplified near a local maxima or minima, but diminish as $|\frac{\partial \expval{O}}{\partial \theta}| \rightarrow \infty$.

\subsection{Quantum Fisher Information}

Using the semantics of quantum information theory, the explicit expression for the FI with respect to a POVM $\mathcal{M}$ with outcomes $\{ E_m \}$ can be written as
\begin{equation}
     \mathcal{I}(\rho_\theta,\mathcal{M}) = \int \frac{\big( \Tr ( E_m \dot{\rho}_\theta ) \big)^2}{\Tr ( E_m \rho_\theta )} \mathrm{d}m,
\end{equation}
where the notation $\dot{\square}=\frac{\partial \square}{\partial \theta}$ is used for conciseness. Just as the FI is interpreted as an information measure, so too is the QFI \cite{barndorff2000}. Eq.~\eqref{eq:AngularDistance} suggests that the POVM which maximizes the distinguishability between the probability density functions associated to $\rho_\theta$ and $\rho_{\theta+\delta \theta}$ will similarly maximize the FI. This is a principle idea behind the derivation of the closed form expression of the QFI \cite{braunstein1994}.

The derivation begins by defining the superoperator
\begin{equation}
    \mathcal{R}_{\rho_\theta} (O) = \frac{1}{2}(\rho_\theta O + O \rho_\theta),
\end{equation}
whose inverse\footnote{The inverse $\mathcal{R}_{\rho_\theta}^{-1} (O)$ is not always well defined for all $O$, however the quantity used in the derivation of the QFI, $\mathcal{R}_{\rho_\theta}^{-1} (\dot{\rho}_\theta)$, always converges to a well-defined Hermitian operator.} is
\begin{equation}
    \label{eq:Rinverse}
    \mathcal{R}_{\rho_\theta}^{-1} (O) = \sum_{j,k} \frac{2}{\lambda_j+\lambda_k} O_{jk} \dyad{j}{k},
\end{equation}
where $\rho_\theta = \sum_{j} \lambda_j \dyad{j}$ is the orthonormal expansion of $\rho_\theta$ and $O_{jk}=\mel{j}{O}{k}$. A property of $\mathcal{R}$ is that for any Hermitian $A$ and $B$, $\Tr \big( AB \big)=\text{Re} \big[ \Tr \big( \rho_\theta A \mathcal{R}_{\rho_\theta}^{-1}(B) \big) \big]$, from which it follows that the FI can be written as
\begin{equation}
    \begin{split}
    \mathcal{I}(\rho_\theta,\mathcal{M}) &= \int \frac{ \text{Re} \big[ \Tr \big(  \rho_\theta E_m \mathcal{R}_{\rho_\theta}^{-1}( \dot{\rho}_\theta) \big) \big]^2}{\Tr ( E_m \rho_\theta )} \mathrm{d}m \\
    &\leq \int \frac{ \big| \Tr \big(  \rho_\theta E_m \mathcal{R}_{\rho_\theta}^{-1}( \dot{\rho}_\theta) \big) \big|^2}{\Tr ( E_m \rho_\theta )} \mathrm{d}m \\
    &= \int \frac{ \big| \Tr \big(  \sqrt{\rho_\theta} \sqrt{E_m} \sqrt{E_m} \mathcal{R}_{\rho_\theta}^{-1}( \dot{\rho}_\theta) \sqrt{\rho_\theta} \big) \big|^2}{\Tr ( E_m \rho_\theta )} \mathrm{d}m.
    \end{split}
\end{equation}
The final step in the derivation uses the Cauchy-Schwarz inequality $|\Tr(A^\dagger B)|^2 \leq \Tr(A A^\dagger) \Tr (B B^\dagger)$ with $A=\sqrt{E_m}\sqrt{\rho_\theta}$ and $B=\sqrt{E_m} \mathcal{R}_{\rho_\theta}^{-1}( \dot{\rho}_\theta) \sqrt{\rho_\theta}$,
\begin{equation}
    \begin{split}
    \mathcal{I}(\rho_\theta,\mathcal{M})  &\leq \int \Tr \big( E_m \mathcal{R}_{\rho_\theta}^{-1}( \dot{\rho}_\theta) \rho_\theta \mathcal{R}_{\rho_\theta}^{-1}( \dot{\rho}_\theta) \big) \mathrm{d} m \\
    &= \Tr \big( \mathcal{R}_{\rho_\theta}^{-1}( \dot{\rho}_\theta) \rho_\theta \mathcal{R}_{\rho_\theta}^{-1}( \dot{\rho}_\theta) \big) \\
    &= \mathcal{Q} ( \rho_\theta ).
    \end{split}
\end{equation}

The Hermitian operator $\mathcal{R}_{\rho_\theta}^{-1}( \dot{\rho}_\theta)$ is the symmetric logarithmic derivative. The QCRB can be saturated by setting $\mathcal{M}$ to be the measurement in the eigenbasis of  $\mathcal{R}_{\rho_\theta}^{-1}( \dot{\rho}_\theta)$ \cite{braunstein1994, luo2000, matsumoto2002}. Unfortunately, but not surprisingly, such a measurement is encumbered by the usual quandary of the frequentist approach: the measurement basis is dependent on $\theta$\footnote{Similar to how a locally optimized estimator, Eq.~\eqref{eq:localestimator}, approximately saturates the CRB, measuring in the eigenbasis of $\mathcal{R}_{\rho_\theta}^{-1}( \dot{\rho}_\theta)|_{\theta \rightarrow \theta_0}$ approximately saturates the QCRB.}. Furthermore, this measurement strategy is very sophisticated and out of reach for current technologies \cite{calsamiglia2001}. Fortunately, this is not the unique measurement strategy which saturates the QCRB \cite{giovannetti2004}. As mentioned, the quantum goal of parameter estimation problems is to determine feasible measurement schemes which best saturate the QCRB.

A closed form expression for the QFI can be derived using the definition of $\mathcal{R}_{\rho_\theta}^{-1}( \dot{\rho}_\theta)$, Eq.~\eqref{eq:Rinverse},
\begin{equation}
    \mathcal{Q}(\rho_\theta) = \sum_{j} \frac{\dot{\lambda}_j^2}{\lambda_j} + 2\sum_{j,k} \frac{(\lambda_j-\lambda_k)^2}{\lambda_j+\lambda_k} | \braket{\dot{j}}{k}|^2.
\end{equation}
The first sum is reminiscent of the classical FI and quantifies the amount of extractable information from the eigenvalues $\{ \lambda_j \}$.  Whilst the second sum accounts for quantum effects such as superposition and entanglement and quantifies the amount of extractable information from the quantum states $\{ \ket{j} \}$. To a certain extent, the classical term is limited to `amplitudes', while the quantum term has access to `amplitudes' and `phases'. As such, the quantum term is significantly more influential than the classical term, this is reinforced by the convexity property of the QFI \cite{alipour2015}
\begin{equation}
    \mathcal{Q}\big( p \rho_1 + (1-p) \rho_2 \big) \leq p \mathcal{Q} ( \rho_1 ) + (1-p) \mathcal{Q} ( \rho_2 ).
\end{equation}
For the special case of pure states $\rho_\theta = \dyad{\psi_\theta}$, the expression is much more aesthetically pleasing. It follows from $\rho_\theta^2=\rho_\theta$ that $\dot{\rho}_\theta= \rho_\theta \dot{\rho}_\theta + \dot{\rho}_\theta \rho_\theta$ and thus $\mathcal{R}_{\rho_\theta}^{-1} ( \dot{\rho}_\theta ) = 2 \dot{\rho}_\theta$. The QFI simplifies greatly to
\begin{equation}
    \label{eq:QFIpure}
    \mathcal{Q} ( \ket{\psi_\theta} ) = 4 \Tr ( \rho_\theta \dot{\rho}_\theta^2 ) = 4 \big( \braket*{\dot{\psi}_\theta} - | \braket*{\dot{\psi}_\theta}{\psi_\theta}|^2 \big).
\end{equation}
In fact, it was shown that a similar expression holds for arbitrary mixed states \cite{escher2011b}
\begin{equation}
    \mathcal{Q}( \rho_\theta) = \min_{\ket{\Psi_\theta}} 4 \big( \braket*{\dot{\Psi}_\theta} - | \braket*{\dot{\Psi}_\theta}{\Psi_\theta}|^2 \big),
\end{equation}
where the minimization is taken over all possible purifications, Eq.~\eqref{eq:purification}, of $\rho_\theta$.

\subsection{Geometric Perspectives of the QFI}

The representation introduced in \textbf{Chapter 2} is that a quantum state $\ket{\psi}$ can be thought of as a vector which is an element of a Hilbert space $\mathcal{H}$. An alternative to this is a geometric representation, where $n$ qubit quantum states are thought to be elements of the complex projective space $\mathbb{CP}^{n}$ \cite{wootters1981, petz1996, grabowski2005}. Pure states reside on the surface of this Riemannian manifold and mixed states in the interior, the $n=1$ case is the well-known Bloch sphere portrayed in Fig.~(\ref{fig:blochsphere}). $\mathbb{CP}^{n}$ is equipped with an infinitesimal metric called the Fubini-Study metric $ds^2$, which is called the Bures metric \cite{bures1969, sommers2003} when it is extended to include the interior. Such a metric allows one to compare neighbouring quantum states $\rho_\theta$ and $\rho_{\theta+\delta \theta}$, analogous to the FI metric for (classical) statistical manifolds, it can be shown that $ds^2=\frac{1}{4}\mathcal{Q(\rho_\theta)} \delta \theta^2$ \cite{facchi2010, sidhu2020}.

\begin{figure}
    \centering
    \includegraphics[width=0.6\textwidth]{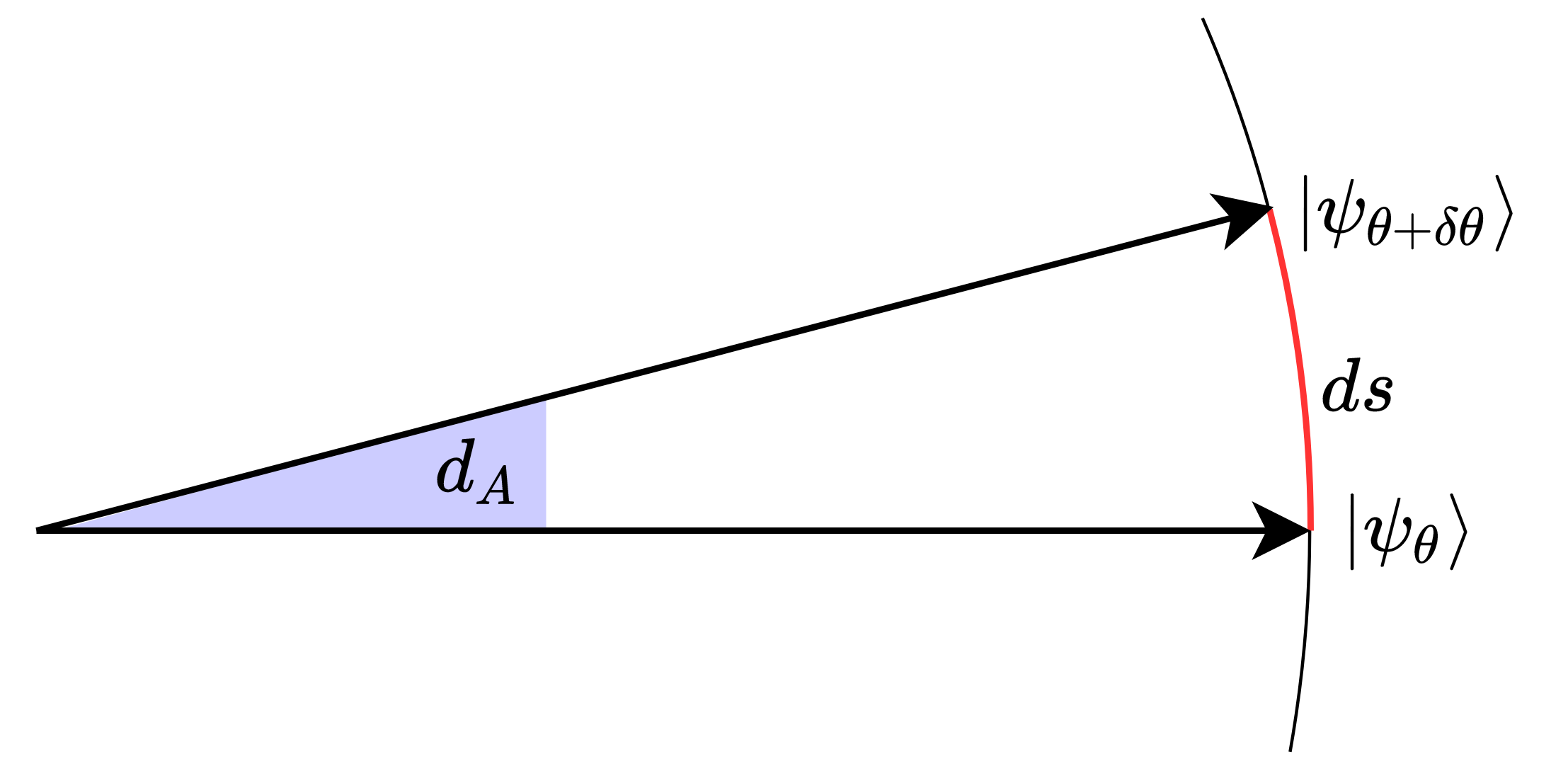}
    \caption{The length of the chord between $\ket{\psi_\theta}$ and $\ket{\psi_{\theta+ \delta \theta}}$ is $2 \sin \frac{d_A}{2}$. For small $d_A$, the length of the chord is approximately equal to the length of the geodesic, $ds \approx 2 \sin \frac{d_A}{2} \approx d_A $. This idea generalizes to higher dimensional abstract surfaces $\mathbb{CP}^n$ as well as their interior points (mixed states).}
    \label{fig:GeomQFI}
\end{figure}

The Bures angle $d_A$ is the angle between the rays of $\rho_1$ and $\rho_2$, explicitly \cite{amari2016, bengtsson2017}
\begin{equation}
    d_A(\rho_1, \rho_2 ) = \arccos \sqrt{\mathscr{F}(\rho_1,\rho_2)},
\end{equation}
where $\mathscr{F}$ is the fidelity, Eq.~\eqref{eq:fidelity}. For neighbouring quantum states, the Bures angle can be approximated two different ways. The first way is by using a first order Taylor expansion
\begin{equation}
    d_A(\rho_\theta, \rho_{\theta + \delta \theta} ) = \sqrt{ 2-2\sqrt{\mathscr{F}(\rho_\theta, \rho_{\theta + \delta \theta})}} + \mathcal{O}( \delta \theta^2).
\end{equation}
The second is a geometric approximation using the Bures metric (and by extension the QFI), the intuition of which is given in Fig.~(\ref{fig:GeomQFI})
\begin{equation}
    d_A ( \rho_{\theta}, \rho_{\theta+\delta \theta} ) = ds + \mathcal{O}(\delta \theta^2 ) =\frac{1}{2}\sqrt{\mathcal{Q}(\rho_{\theta}, \rho_{\theta+\delta \theta} )} \delta \theta + \mathcal{O}(\delta \theta^2 ).
\end{equation}
A new expression for the QFI is obtained by merging the two equations \cite{sidhu2020}
\begin{equation}
    \label{eq:QFIlimit}
    \mathcal{Q}(\rho_\theta) = \lim_{\delta \theta \rightarrow 0} 8\frac{1-\sqrt{\mathscr{F}(\rho_\theta, \rho_{\theta + \delta \theta})}}{\delta \theta^2},
\end{equation}
which can be useful to derive analytic bounds for the QFI and other information theoretic quantities \cite{suzuki2019, tsang2020}. A corollary of Eq.~\eqref{eq:QFIlimit} is the concavity of the QFI under CPTP maps $\mathcal{E}$
\begin{equation}
    \mathcal{Q} \big( \mathcal{E}(\rho_\theta) \big) \leq \mathcal{Q} \big( \rho_\theta \big),
\end{equation}
which follows from the monotonicity of the fidelity  $\mathscr{F} \big( \mathcal{E}(\rho_1),\mathcal{E}(\rho_2) \big) \geq \mathscr{F} \big( \rho_1, \rho_2 \big)$. If $\mathcal{E}$ is thought of as an interaction with an environment (\textbf{Chapter~5}) or a malicious adversary (\textbf{Chapter~6}), then the concavity of the QFI can be understood as information about $\theta$ being lost to these outside sources.

\subsection{Ultimate Precision: The Heisenberg Limit}

To recapitulate: the CRB is a bound on the MSE by optimizing over estimation strategies, and the QCRB extends the bound by optimizing over measurement strategies. The next natural extension is to optimize over initialized quantum states, to find the true limit of precision attainable through quantum mechanics. The upper bound for which is referred to as the \textit{Heisenberg Limit} (HL) \cite{yurke1986, holland1993}.

Originally, the HL was derived within the framework of phase estimation. In the phase estimation problem, a phase $\theta$ is encoded into each qubit of an $n$ qubit pure state $\ket{\psi}$ by a unitary $U_\theta =e^{-i \theta H}$, where the Hamiltonian $H$ acts independently and identically on all $n$ qubits. The QFI can be calculated to be
\begin{equation}
    \mathcal{Q} = 4 \big( \mel*{\psi}{H^2}{\psi} - |\mel*{\psi}{H}{\psi}|^2 \big) = 4 \Delta^2 H.
\end{equation}
The etymology of the term `Heisenberg limit' stems from the fact that the QCRB (with $\nu=1$) can be manipulated to mimic the the Heisenberg uncertainty principle
\begin{equation}
    \Delta^2 \hat{\theta} \Delta^2 H \geq \frac{1}{4}.
\end{equation}
The QFI for phase estimation can be maximized by setting $\ket{\psi}$ to be a highly entangled state, such as the GHZ state for qubit systems or the NOON state for photonic systems, which results in $\mathcal{Q} = n^2$. Hence, the ultimate allowable precision by quantum mechanics (the HL) is
\begin{equation}
    \label{eq:HL}
    \Delta^2 \hat{\theta}_\text{HL} = \frac{1}{\nu n^2}.
\end{equation}
The HL offers a quadratic improvement compared to the standard quantum limit (SQL), where the $\ket{\psi}$ is limited to separable states
\begin{equation}
    \label{eq:SQL}
    \Delta^2 \hat{\theta}_\text{SQL} = \frac{1}{\nu n}.
\end{equation}
The SQL is also referred to as the classical limit or the shot-noise limit \cite{xiao1987}.

For qubit (and qudit) systems\footnote{For CV systems a quantum advantage can be achieved with squeezing \cite{yurke1986,ono2010}.}, entanglement is a crucial resource for quantum metrology \cite{pezze2009, pezze2018}. In fact, the quadratic tendencies of the QFI of a quantum state for phase estimation can be bounded with respect to the geometric measure of entanglement $G$\footnote{The geometric measure of entanglement for a pure state $\ket{\psi}$ is $G(\ket{\psi})=1-\max_{\ket{\phi}} |\braket{\phi}{\psi}|^2$, where $\ket{\phi}$ is maximized over all fully separable states. The definition is extended to mixed states by finding the convex roof of the geometric measure of entanglement over all possible statistical ensembles \cite{wei2003}.} \cite{augusiak2016} 
\begin{equation}
    \mathcal{Q}(\rho_\theta) \leq n + 8n^2 \sqrt{G(\rho_\theta)}.
\end{equation}
It is worth stressing that entanglement may be a necessary condition to surpass the SQL but it is not a sufficient condition \cite{hyllus2010, oszmaniec2016}. Additionally, the bounds in Eq.~\eqref{eq:HL} and Eq.~\eqref{eq:SQL} are exclusive to the problem of phase estimation with an iid encoding. The QFI can surpass $n^2$ for non-linear $H$ \cite{luis2004, boixo2007, choi2008,braun2018}, and scenarios can be devised in which entanglement is not a necessary resource \cite{tilma2010}.

\subsection{Bayesian Approach to Quantum Metrology}

In the quantum version of the frequentist approach, the MSE is minimized by optimizing over all possible POVM's and input quantum states. The quantum version of the Bayesian approach \cite{holevo1982, jarzyna2015, rubio2018} is enhanced in an analogous fashion. As the estimator is updated adaptively, so too can the initialized quantum state as well as choice of POVM.

For parameter estimation problems which exhibit periodicity, such as phase estimation, the circular cost function
\begin{equation}
    C_\circ ( \hat{\theta}, \theta ) = 4\sin^2 \Big( \frac{\hat{\theta}-\theta}{2} \Big)
\end{equation}
is a natural choice as a figure of merit \cite{demkowicz2011, demkowicz2015}, and converges to the MSE as $\hat{\theta}$ approaches $\theta$. If the initial choice of input quantum state and POVM are $\rho_\theta=U_\theta \rho_0 U_\theta^\dagger$ and $\int E_m \mathrm{d}m$ respectively, then the average cost is
\begin{equation}
    \expval{C_\circ } = \int p(\theta) \bigg( \int \Tr ( \rho_\theta E_m ) C_\circ (\hat{\theta}(m),\theta)\mathrm{d}m \bigg) \mathrm{d}\theta,
\end{equation}
which is invariant when replacing the POVM $\{ E_m \}$ with a covariant POVM $\{ E_{\hat{\theta}} \}$ \cite{holevo1982, derka1998, chiribella2004, chiribella2005}
\begin{equation}
    E_{\hat{\theta}}=U_{\hat{\theta}} \Sigma U_{\hat{\theta}}^\dagger
\end{equation}
and $\Sigma$ is the positive-semi definite operator defined for a specific $\hat{\theta}$
\begin{equation}
    \Sigma = \int U_{\hat{\theta}(m)}^\dagger E_m U_{\hat{\theta}(m)} \mathrm{d} m.
\end{equation}
This re-parametrization allows the average cost to be expressed as
\begin{equation}
\begin{split}
    \expval{C_\circ } &= \int \int p(\theta) \Tr ( \rho_\theta E_{\hat{\theta}} ) C_\circ ( \hat{\theta}, \theta ) \mathrm{d} \hat{\theta} \mathrm{d} \theta \\
    &= \int p(\theta) \Tr ( \rho_\theta \Sigma ) 4 \sin^2 \frac{\theta}{2} \mathrm{d} \theta.
\end{split}
\end{equation}
By optimizing the above expression, the initialized quantum state $\rho_0$ and POVM characterized by $\Sigma$ can be updated adaptively \cite{derka1998, chiribella2005}.

As mentioned, the Bayesian statistical inference approach addresses the issues inherent to the frequentist approach: lack of a priori knowledge \cite{kolodynski2010, demkowicz2011} and limited resources \cite{rubio2020b}. The work presented in the subsequent chapters exclusively focus on the frequentist approach, as such, an interesting future perspective would be to generalize some of the findings to the Bayesian approach. Specifically in \textbf{Chapter~6}, where the estimation process is adapted in some capacity to account for the cryptographic framework.

\subsection{Multiple Parameters}

In the interest of simplicity, this chapter introduced the problem of parameter estimation with a single unknown parameter. The problem naturally generalizes to include multiple latent parameters $\theta \rightarrow \boldsymbol{\theta}=\{\theta_1,\ldots, \theta_m \} $, where the goal extends to devising estimators for each parameter $\hat{\theta}(\mathbf{x}) \rightarrow \hat{\boldsymbol{\theta}}(\mathbf{x})=\{ \hat{\theta}_1(\mathbf{x}), \ldots, \hat{\theta}_m(\mathbf{x}) \}$. There are two major quandaries which arise in the multiparameter setting. First, the parameters may be statistically dependent on one another, which adds ambiguity when trying to interpret the observed data $\mathbf{x}$. Second, it is not always possible to simultaneously construct an efficient estimator for each unknown parameter.

Within the frequentist inference framework, assuming that each estimator satisfies the unbiased estimator constraint, $\mathbb{E}( \hat{\boldsymbol{\theta}} ) =\boldsymbol{\theta} $, the generalization of the QCRB is the matrix equation \cite{yuen1973, helstrom1974, tsang2011}
\begin{equation}
    \label{eq:multiQCRB}
    \textbf{Cov}(\boldsymbol{\theta}) \geq \frac{1}{\nu} \boldsymbol{\mathcal{I}}^{-1} (\boldsymbol{\theta}) \geq \frac{1}{\nu} \boldsymbol{\mathcal{Q}}^{-1} (\boldsymbol{\theta}),
\end{equation}
where $\textbf{Cov}(\boldsymbol{\theta})$ is the covariance matrix with entries $\textbf{Cov}(\boldsymbol{\theta})_{i,j}=\expval*{(\hat{\theta}_i-\theta_i)(\hat{\theta}_j-\theta_j)}$, $\boldsymbol{\mathcal{I}} (\boldsymbol{\theta})$ is the FI matrix
\begin{equation}
    \boldsymbol{\mathcal{I}} (\boldsymbol{\theta})_{i,j} =  \int  p(\mathbf{x}|\boldsymbol{\theta}) \Big( \frac{\partial \ln p(\mathbf{x}|\boldsymbol{\theta}) }{\partial \theta_i} \Big) \Big( \frac{\partial \ln p(\mathbf{x}|\boldsymbol{\theta}) }{\partial \theta_j} \Big) \mathrm{d} \mathbf{x},
\end{equation}
and $\boldsymbol{\mathcal{Q}} (\boldsymbol{\theta})$ is the QFI matrix
\begin{equation}
    \boldsymbol{\mathcal{Q}} (\boldsymbol{\theta})_{i,j} = \Tr \Big( \mathcal{R}^{-1}_{\rho_{\boldsymbol{\theta}}} \big( \frac{ \partial \rho_{\boldsymbol{\theta}}}{\partial \theta_i} \big) \rho_{\boldsymbol{\theta}} \mathcal{R}^{-1}_{\rho_{\boldsymbol{\theta}}} \big( \frac{ \partial \rho_{\boldsymbol{\theta}}}{\partial \theta_j} \big) \Big).
\end{equation}
The diagonal elements of an invertible positive semi-definite matrix $M$ satisfy $M_{i,i} \geq 1/M_{i,i}$, and equality holds for each $i$ when $M$ is diagonal. Hence, when the $m$ parameters are statistically independent from each other,  Eq.~\eqref{eq:multiQCRB} reduces to the QCRB for each individual parameter $\theta_i$. Multiparameter estimation sustains additional complications in the quantum context with respect to optimizing quantum states and measurements. The superoperator $\mathcal{R}^{-1}_{\rho_{\boldsymbol{\theta}}} \big( \frac{ \partial \rho_{\boldsymbol{\theta}}}{\partial \theta_i} \big)$ is the symmetric logarithmic derivative with respect to $\theta_i$, and measuring in all of these bases will saturate the QCRB. However, these measurements may not be compatible and thus cannot be realized in simultaneity \cite{vidrighin2014, crowley2014, ragy2016}.

Despite the additional complexity of simultaneously estimating multiple parameters, multiparameter quantum metrology is an active research topic \cite{szczykulska2016, nichols2018, albarelli2019, rubio2020b, meyer2021}. With respect to phase estimation, when the phase encoding unitaries do not commute, it is more efficient to estimate them in simultaneity rather than independently \cite{ baumgratz2016}; when the encoding unitaries do commute, one can devise a simultaneous estimation strategy which is at least as efficient as estimating the phases independently \cite{humphreys2013}. Multiparameter quantum metrology is a natural framework for eigenvalue estimation of higher dimensional unitaries \cite{fujiwara2001, ballester2004, berry2015, baumgratz2016} and for spatially distributed estimation problems \cite{eldredge2018, ge2018, proctor2018, zhuang2018, rubio2020a, guo2020}. The subsequent research chapters focus on the single parameter setting. Nonetheless, the mathematical techniques and derivations can easily be adapted to the multiple parameter setting. Formally addressing these generalizations is a future perspective of the works presented.

\section{Example Applications of Quantum Metrology}

The final section of this chapter explores well-known applications of quantum metrology where the concepts and tools that were introduced are put into practise. The examples chosen, phase estimation and amplitude estimation, have a simple mathematical formalism and highlight the novelty of a quantum parameter estimation problem. Specifically, phase estimation, which is indisputably the canonical usage of quantum metrology \cite{caves1981, giovannetti2004}, clearly showcases the advantages a quantum system can provide. Additionally, phase estimation is the core problem of \textbf{Chapter~4} and \textbf{Chapter~5}. The example of amplitude estimation, although not present in the subsequent research chapters, is included to showcase a simple usage which is not phase estimation.

\subsection{Phase Estimation (Photonic Interferometry)}

Phase estimation is a benchmarking problem for quantum metrology \cite{holland1993}. It encapsulates a variety of applications for quantum sensors, notably magnetometry \cite{taylor2008, wasilewski2010, sewell2012, brask2015, razzoli2019}, frequency estimation for optomechanical sensors \cite{zheng2016, djorwe2019, tsang2013}, spectroscopy \cite{meyer2001, leibfried2004, kira2011, dorfman2016,shaniv2018} and unitary tomography \cite{svore2013, o2019}. The premise of phase estimation is simple, yet the results are an elegant display of a quantum advantage. A relative phase is encoded into a quantum system via a physical interaction, and using highly entangled states and a simple measurement strategy, the QCRB can be saturated to attain the HL \cite{giovannetti2006}.

\begin{figure}
    \centering
    \includegraphics[width=0.8\textwidth]{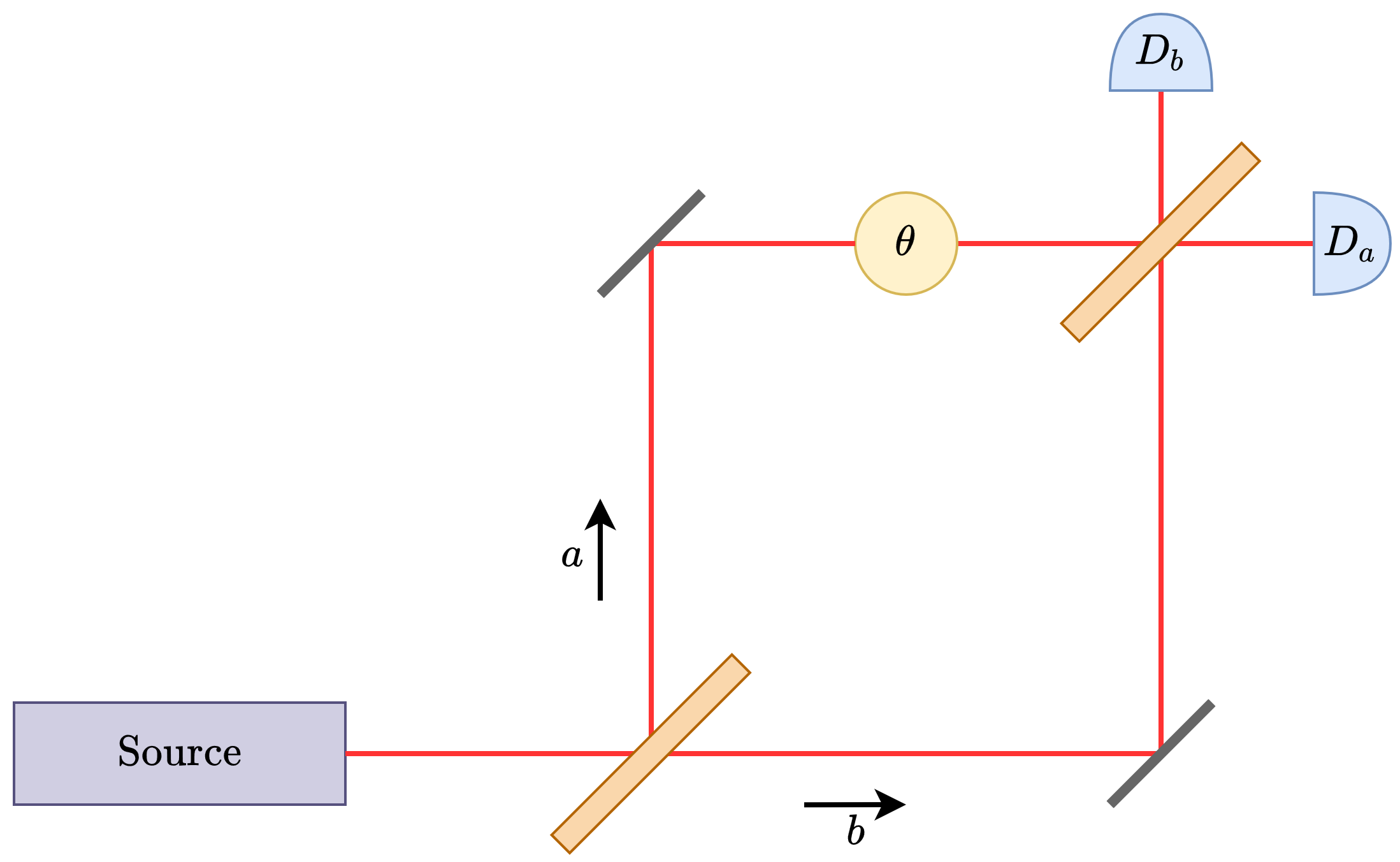}
    \caption{Schematic of a Mach-Zehnder interferometer. A photonic source is fired onto a 50:50 beam splitter, which reflects a photon with 50\% probability (path $a$), otherwise the photon is transmitted (path $b$). A phase shift of $\theta$ is introduced uniquely on path $a$, after the paths interfere via a second beam splitter. The quantum state is measured with photon-counters, whose outcomes $j$ are dependent on the relative phase $\theta$.}
    \label{fig:Interferometer}
\end{figure}

For photonic sources, a relative phase can be introduced using a Mach-Zehnder interferometer \cite{zetie2000}, depicted in Fig.~\ref{fig:Interferometer}. A photonic quantum state passes through a 50:50 beam splitter, in which photons are either transmitted or reflected, both with probability of one half, where reflection induces a phase shift of $\pi/2$ \cite{dowling2008}. After passing through the first beam splitter, a single photon will be in a superposition of the two possible modes, labelled by the respective paths $\ket{a}$ and $\ket{b}$. Since photons are indistinguishable particles, it is customary to write an $n$ photon quantum state as
\begin{equation}
    \ket{\psi}=\sum_{k=0}^n \alpha_k \ket{k,n-k},
\end{equation}
where the notation $\ket{k,n-k}$ denotes the quantum state with $k$ photons in mode $\ket{a}$ and $n-k$ photons in mode $\ket{b}$. The path-dependent phase shift is represented by the unitary
\begin{equation}
    U_\theta =  \exp (-i\frac{\theta}{2}\big(\dyad{a}-\dyad{b}\big))^{\otimes n},
\end{equation}
thus
\begin{equation}
    U_\theta \ket{k,n-k} = e^{-i\frac{2k-n}{2}\theta}\ket{k,n-k}.
\end{equation}
After the phase shifts, the two paths interfere by passing through a second beam splitter. This causes the detection probabilities to be dependent on $\theta$, from which the statistics can be used to generate an estimate $\hat{\theta}$. The precision of the estimate is ultimately bounded by the QFI
\begin{equation}
    \label{eq:qfiEx1}
    \mathcal{Q}(\ket{\psi}) = \sum_{k=0}^n |\alpha_k|^2 (2k-n)^2-\Big( \sum_{k=0}^n |\alpha_k|^2 (2k-n) \Big)^2 = 4 \sum_{k=0}^n |\alpha_k|^2 k^2 - 4 \Big( \sum_{k=0}^n |\alpha_k|^2 k \Big)^2 .
\end{equation}

Suppose that the source of photons is uncorrelated such that the quantum state after passing through the first beam splitter is
\begin{equation}
    \ket{\psi}_\text{sep}=\frac{1}{\sqrt{2^n}}\Big( \ket{a}+\ket{b} \Big)^{\otimes n}=\sum_{k=0}^n \sqrt{2^{-n}\binom{n}{k}} \ket{k,n-k}.
\end{equation}
Using Eq.~\eqref{eq:qfiEx1}, the QFI can be computed to be equal to the SQL, $\mathcal{Q}_\text{sep} = n$. In contrast, if the quantum state is initialized in the NOON state \cite{dowling2008, matthews2016, zhang2018}
\begin{equation}
    \label{eq:N00N}
    \ket{\psi}_\text{ent}=\frac{1}{\sqrt{2}}\Big( \ket{n,0}+\ket{0,n} \Big),
\end{equation}
then the HL is attained\footnote{This has been experimentally accomplished using $n=4$ photons \cite{nagata2007}}, $\mathcal{Q}_\text{ent}=n^2$. The QCRB can be saturated using the previously described method of inferring an estimate from an observable. The chosen observable is the parity of the detected photons in detector $a$, labelled $D_a$ in, Fig.~\ref{fig:Interferometer},
\begin{equation}
    O_\text{parity} = \sum_{k=0}^n (-1)^k \dyad{k,n-k}.
\end{equation}
Assuming no dark counts, if $D_a$ detects $j$ photons, $D_b$ is fixed to $n-j$ photons, hence the parity of $D_b$ could have equally been considered. After intercepting the second beam splitter, the quantum state (up to a global phase) is
\begin{equation}
\begin{split}
    \ket{\psi_\theta}_\text{ent} &= \frac{1}{\sqrt{2^{n}}} \sum_{k=0}^n \sqrt{\binom{n}{k}} \Big( e^{-in \theta/2} i^{n-k} + e^{in \theta/2} i^k \Big) \ket{k,n-k} \\
    &= \frac{1}{\sqrt{2^{n-1}}} \sum_{k=0}^n \sqrt{\binom{n}{k}} \cos \Big( \frac{n\theta}{2}+\frac{\pi}{4}(2k-n) \Big) \ket{k,n-k},
\end{split}
\end{equation}
from which it can be computed that
\begin{equation}
    \expval*{O}_\text{ent}=\cos \big( n \theta - \frac{\pi n}{2} \big).
\end{equation}
If the prepare and measure protocol is repeated $\nu$ times, the MSE of the estimator is
\begin{equation}
    \Delta^2 \hat{\theta}_\text{ent} = \frac{\Delta^2 O_\text{ent}}{\nu \big|\frac{\partial \expval*{O}_\text{ent}}{\partial \theta} \big|^2}=\frac{1}{\nu n^2},
\end{equation}
thus the QCRB is saturated with a simple measurement strategy. This example highlights the achievability of a quantum advantage, but simultaneously the locality of the frequentist approach. The periodicity of $\expval*{O}_\text{ent}$ implies that a priori knowledge of $\theta$ is required to an order of $2 \pi/n$. If $\theta$ is completely unknown, a frequentist approach can still be employed using varying $n=2^m$, where successive rounds of estimation are used to estimate the $m$th (binary) digit of $\theta$ \cite{kitaev1995}.

In this example, a relative phase is encoded using a photonic source by means of a Mach-Zehnder interferometer. Notably though, analogous results are obtained using spin systems \cite{toth2014}. In fact, the NOON state, Eq.~\eqref{eq:N00N}, can be interpreted as a GHZ state
\begin{equation}
    \ket{\text{GHZ}} =\frac{1}{\sqrt{2}}( \ket{0}^{\otimes n}+\ket{1}^{\otimes n}),
\end{equation}
which when used for phase estimation, similarly saturates the HL.

\subsection{Amplitude Estimation (Thermometry)}

After phase estimation, the next obvious example of a quantum metrology problem is amplitude estimation. A well-known example of amplitude estimation is quantum thermometry \cite{correa2015, de2018, mehboudi2019}, where the unknown parameter in question is temperature. Temperature is a seemingly intuitive notion ever present in our daily lives, and measuring this quantity may appear trivial. For every day objects, an infra-red thermometer converts infra-red radiation into a voltage which is converted into a temperature. This is done extremely quickly and accurately. Even so, in the `very cold' regime near zero Kelvin, measuring temperature is a complicated task, but very necessary for modern technologies such as superconductors. 

Experimental implementations of quantum thermometry differ greatly \cite{neumann2013, toyli2013, kucsko2013}, yet the underlying principle is straightforward \cite{mehboudi2019}. An $N$ level system interacts with an external source at temperature $T$. Eventually, the collective ensemble of the system and the bath will reach thermal equilibrium, and the state of the system is given by the Gibbs ensemble
\begin{equation}
    \rho_T = \frac{1}{Z} e^{-\frac{H}{k_B T}},
\end{equation}
where $k_B$ is the Boltzmann constant, $H$ is the system Hamiltonian
\begin{equation}
    H=\sum_{k=1}^N \epsilon_k \dyad{\epsilon_k}
\end{equation}
with energy eigenvalues $\epsilon_k$ and eigenstates $\ket{\epsilon_k}$, and $Z= \Tr e^{-\frac{H}{k_B T}}$ is the partition function. Within the standard convention of setting $k_B=1$, the derivative of the quantum system with respect to temperature is
\begin{equation}
    \label{eq:thermalequilibrium}
    \dot{\rho}_T= \frac{H}{T^2} \rho_T - \frac{\expval*{H}}{T^2} \rho_T=\frac{1}{2T^2}\big((H-\expval*{H})\rho_T+\rho_t (H-\expval*{H}) \big).
\end{equation}
From which it is clear that the symmetric logarithmic derivative is
\begin{equation}
    \mathcal{R}_{\rho_T}^{-1}( \dot{\rho}_T) = \frac{1}{T^2}(H-\expval*{H}),
\end{equation}
therefore
\begin{equation}
    \label{eq:QFIthermometry}
    \mathcal{Q}(\rho_T)=\Tr \big( \mathcal{R}_{\rho_T}^{-1}( \dot{\rho}_T) \rho_T \mathcal{R}_{\rho_T}^{-1}( \dot{\rho}_T) \big)=\frac{1}{T^4} \Tr \big( (H-\expval{H})^2 \rho_T\big) = \frac{\Delta^2 H}{T^4}.
\end{equation}
In fact, the heat capacity
\begin{equation}
    \frac{\partial \expval{H}}{\partial T} = \frac{\Delta^2 H}{T^2}
\end{equation}
is directly proportional to the QFI \cite{phillies1984}, hence the QCRB can be re-written as
\begin{equation}
    \Delta^2 \hat{T} \geq \frac{1}{\nu \mathcal{Q}} = \frac{\Delta^2 H}{\nu|\frac{\partial \expval{H}}{\partial T} |^2}.
\end{equation}
Which suggests that the QCRB can be saturated by using the previously described estimation strategy of inferring the temperature from an observable, Eq.~\eqref{eq:ErrorPropagation}. In this instance the observable is the energy of the system \cite{jahnke2011}, $H$, which may not be surprising because of how intertwined energy and temperature are as quantities within the realm of statistical mechanics\footnote{On a macroscopic scale, temperature is an average quantity of a system composed of many many particles. This definition is somewhat ambiguous on a microscopic scale. In Eq.~\eqref{eq:thermalequilibrium}, temperature can be interpreted as a variable which governs the probability of the quantum system occupying a specific energy eigenstate.}.

Analogous to a GHZ state or NOON state being the optimal probe for phase estimation, Eq.~\eqref{eq:QFIthermometry} can be maximized to find the optimal probe for thermometry. The solution is an effective two level system with a single eigenstate having an energy of $\epsilon_-$ and $N-1$ eigenstates having a degenerate energy of $\epsilon_+$, the relative error of such a probe is $\Delta^2 \hat{T}/T^2 = \mathcal{O}(1/\log N)$ \cite{correa2015, mehboudi2019}. Note that the information presented in this example holds only for fully thermalized systems, which may be a time consuming process. The analysis is significantly more complex for partially thermalized systems \cite{correa2015}.

%% file: Figures/Chapter3/FIplots.tex
\begin{tikzpicture}
\begin{axis}[every axis plot post/.append style={
  mark=none,domain=0:6,samples=100,smooth},
width=11cm,
height=6cm,
xmin=0,
xmax=6,
ymin=0,
ymax=3,
xlabel=$\sigma$,
ylabel=$\mathcal{I}$,
axis x line*=bottom, 
axis y line*=left, 
legend cell align={left},
legend style={row sep=5pt}
]
\addplot {1/x^2};
\addplot {1/(2*x^2)};

\legend{$\mathcal{I} \big( p(X | \mu ) \big) = \frac{1}{\sigma^2}$,$\mathcal{I} \big( p(X | \sigma ) \big) = \frac{1}{2\sigma^2}$}
\end{axis}
\end{tikzpicture}

%% file: Figures/Chapter3/Likelihood1.tex
\begin{tikzpicture}
\begin{axis}[every axis plot post/.append style={
  mark=none,domain=0:1,samples=100,smooth},
width=7.5cm,
height=7.5cm,
xmin=0,
xmax=1,
ymin=0,
ymax=0.17,
ytick=\empty,
xlabel=$p_H$,
ylabel=$L$,
axis x line*=bottom, 
axis y line*=left, 
legend cell align={left},
legend style={row sep=5pt}
]
\addplot {x^2*(1-x)^2};
\addplot {x^3*(1-x)};

\legend{${(\text{Heads, Tails}): (2, 2)}$, ${(\text{Heads, Tails}): (3, 1)}$}
\end{axis}
\end{tikzpicture}

%% file: Figures/Chapter3/Likelihood2.tex
\begin{tikzpicture}
\begin{axis}[every axis plot post/.append style={
  mark=none,domain=0:1,samples=100,smooth},
width=7.5cm,
height=7.5cm,
xmin=0,
xmax=1,
ymin=0,
ymax=3.6*10^(-29),
ytick=\empty,
xlabel=$p_H$,
ylabel=$L$,
axis x line*=bottom, 
axis y line*=left, 
legend cell align={left},
legend style={row sep=5pt}
]
\addplot {x^60*(1-x)^40};
\addplot {x^63*(1-x)^37};

\legend{${(\text{Heads, Tails}): (60, 40)}$, ${(\text{Heads, Tails}): (63, 37)}$}
\end{axis}
\end{tikzpicture}

%% file: Chapters/Chapter4-GraphStates.tex
\chapter{Graph States as a Resource for Quantum Metrology}

It is obviously very desirable to have an easy to implement quantum resource with a large span of applications. One class of quantum states which satisfies these criteria are graph states: a versatile resource for quantum computation and quantum communication. In this chapter, to help determine the full extent of the applicability of qubit graph states, we explore their practicality for the quantum metrology problem of phase estimation. Before beginning this work, it had been shown that cluster states (a subset of graph states) are an efficient resource for certain quantum metrology problems, namely with a non-local parameter encoding scheme \cite{rosenkranz2009} or after undergoing local rotations \cite{friis2017}. We consider the standard (local) phase estimation problem and are able to quantify the effectiveness of a general graph state based on the shape of the corresponding graph \cite{SM20}. Since our work has been published, others have explored the practicality of continuous variable graph states for quantum metrology \cite{wang2020}.

\section{Graph States}

Graph theory \cite{west2001} is a rich and diverse branch of mathematics. A \textit{graph} is a structure used to model pairwise relationships with respect to a set of elements. A graph is devised of two types of elements: i) a set of \textit{vertices} (or nodes) which are connected with ii) \textit{edges}\footnote{This is the simplest description of a graph. More general graphs can have edges with assigned weights and/or a direction. These extra parameters are unnecessary for the scope of our work.}. Graphs are a customary tool in mathematics - they are the standard representation of the popular travelling salesmen problem and minimum colouring problem, for example. Outside of mathematics, graph theory is a tool to model all sorts of relations. In computer science, it can model the flow of information, where vertices are websites and an edge a hyperlink from one website to another. In animal biology, a vertex could signify a geographical region and the edges denote migration patterns for a species. With the broad scope of utility and existing research surrounding graph theory, it is unsurprising that it is used in the field of quantum information \cite{hein2004}.

\textit{Graph states} are an incredible useful resource in quantum information \cite{hein2006}. In the language of graph theory, quantum systems (qubits, qudits, CV states) are the vertices and entangling operations are the edges. These quantum states have a wide range of applications, including, but not limited to, cryptography \cite{markham2008, qian2012}, verification \cite{markham2020}, quantum networks \cite{pirker2018, meignant2019, hahn2019}, t-designs \cite{mezher2018} and error correction \cite{schlingemann2001a}. Marginally more complex graphs, where the vertices are either a quantum system, a quantum operation or a quantum measurement, is the foundation of measurement based quantum computing \cite{raussendorf2001, raussendorf2003, van2006}. In terms of implementation, graph states have been experimentally constructed using trapped ions \cite{barreiro2011, lanyon2013}, superconducting circuits \cite{song2017, gong2019}, squeezed states of light \cite{yokoyama2013, chen2014} and photons \cite{lu2007, gu2019, russo2019}.

\subsection{Graphical Representation}

\begin{figure}[ht]
\begin{subfigure}{.24\textwidth}
  \centering
  \includegraphics[width=.9\linewidth]{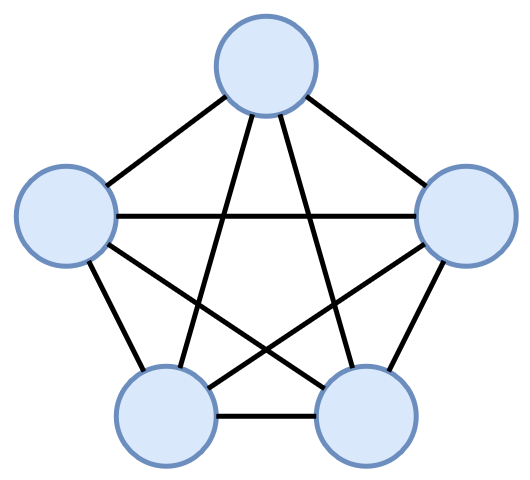}  
  \caption{Complete graph.}
  \label{fig:graph1}
\end{subfigure}
\begin{subfigure}{.24\textwidth}
  \centering
  \includegraphics[width=.9\linewidth]{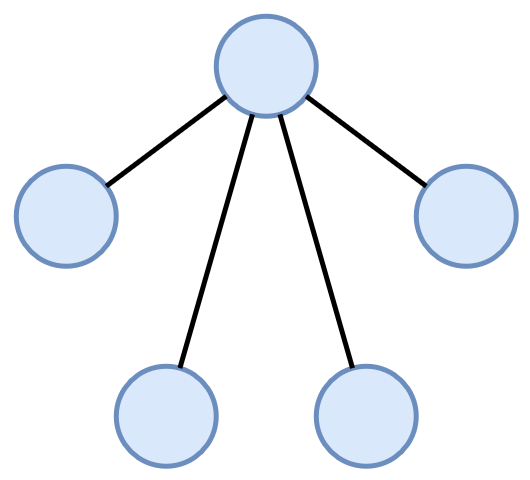}  
  \caption{Star graph.}
  \label{fig:graph2}
\end{subfigure}
\begin{subfigure}{.24\textwidth}
  \centering
  \includegraphics[width=.9\linewidth]{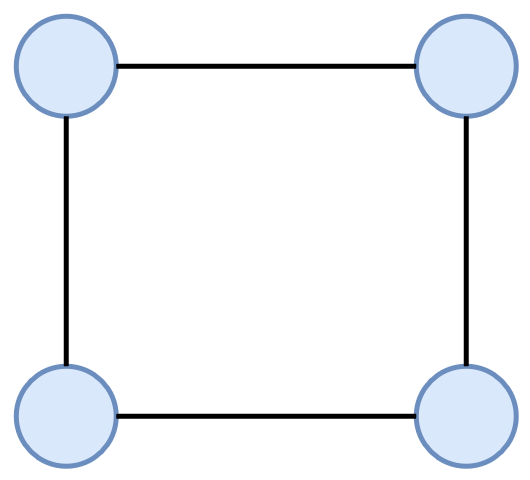}  
  \caption{Cyclic graph.}
  \label{fig:graph3}
\end{subfigure}
\begin{subfigure}{.24\textwidth}
  \centering
  \includegraphics[width=.9\linewidth]{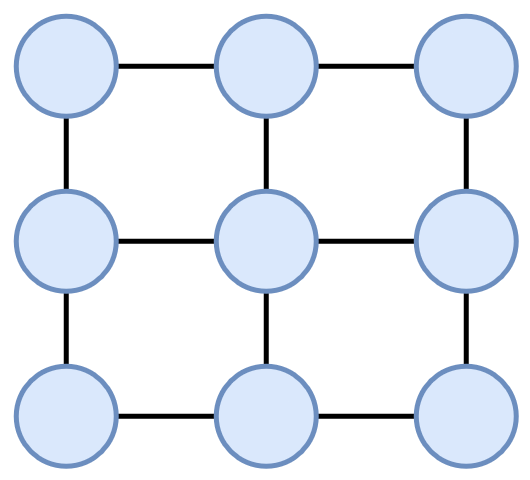}  
  \caption{Lattice graph.}
  \label{fig:graph4}
\end{subfigure}
\caption{Graphical representation of frequently used graphs in quantum information. (a) A complete graph (or fully connected graph) is where each vertex is connected to every other vertex; $E_a= \{(v_i,v_j) \; \forall v_i,v_j \in V_a\}$. (b) A star graph is where there is a central vertex which forms an edge with all of the other vertices; $E_b= \{(v_1,v_j) \; \forall j \geq 2 \}$. (c) A cyclic graph is where the vertices are connected by a ring like series of edges; $E_c= \{(v_j,v_{j+1})\}$. (d) A lattice graph is where the vertices are arranged in an array. By using only single qubit Clifford operations, the complete graph state (a) and star graph state (b) can be transformed into the (all important) GHZ state using local Clifford operations. A cyclic graph (c) is one of the simplest models for a chain of quantum repeaters \cite{azuma2017}. A lattice graph state (d) wrapped around itself to form a torus is the basic structure of topological quantum error correcting codes \cite{bombin2006, bombin2007}. All four graphs could be used to represent a quantum network with varying complexities and purposes.}
\label{fig:graph_examples}
\end{figure}

Formally, a graph is a set of vertices $V=\{ v_1, \ldots, v_n \}$ and edges $E= \{e_1, \ldots, e_m \}$, where each edge $e_j=(v_{j_1},v_{j_2})$ is a length-2 tuple of two vertices. The graph is denoted by $G=(V,E)$. In quantum information, the set of vertices correspond to quantum systems (for this work these are qubits) and the edges correspond to an entangling operation. Each qubit is prepared in the $\ket{+}$ state, and a controlled-$Z$ operation is performed on the $i$th and $j$th qubit if $(v_i, v_j) \in E$. As an example, consider a 3 qubit star graph state, Fig.~(\ref{fig:graph2}), where the first qubit is the central qubit. The bra-ket representation of this quantum state is
\begin{equation}
    \label{eq:examplegraphstate}
    \begin{split}
        \ket{G} &= CZ_{(1,2)} CZ_{(1,3)} \ket{+++} \\ &= \frac{1}{\sqrt{8}} \big( \ket{000}+\ket{001}+\ket{010}+\ket{011}+\ket{100}-\ket{101}-\ket{110}+\ket{111}\big).
    \end{split}
\end{equation}

Other graphical nomenclature used is this chapter is `neighbourhood' and `isolated vertex'. The neighbourhood of a vertex $v$, denoted as $N(v)$, is the set of vertices which are connected to $v$: $N(v)= \{ u \in V | \; (u,v) \in E \}$. A vertex $w$ is said to be isolated if it has an empty neighbourhood: $|N(w)|=0$.

\subsection{Stabilizer Representation}

Graph states belong to a larger family of quantum states called \textit{stabilizer states}. The (general) stabilizer formalism does not have a graphical and illustrative analogue, instead, it is built upon mathematical symmetries and elegance. Stabilizer formalism, originally developed for quantum error correction \cite{gottesman1997} and adapted for measurement based quantum computing \cite{raussendorf2001}, is efficiently simulated \cite{aaronson2004} and verifiable \cite{pallister2018, markham2020} due to the fact that the underlying structure is symmetries arising from the Pauli group. The set of stabilizer states is closed under local Clifford operations, and a large number of stabilizer states are highly entangled.

A quantum state $\ket{\psi}$ is said to be stabilized by an operator $S$ if $S\ket{\psi} = \ket{\psi}$. An $n$ qubit quantum state $\ket{\psi}$ is a stabilizer state if it is stabilized by $n$ non-identity stabilizing operators $g_1, \ldots, g_n$ which i) all commute, ii) are multiplicatively independent and iii) are elements of the Pauli group $\pm \mathcal{P}_n$ \cite{garcia2017}. The bra-ket representation of a stabilizer state is
\begin{equation}
    \label{eq:stabrepresentation}
    \dyad{\psi} = \frac{1}{2^n} \prod_{j=1}^n \big( \mathbb{I} + g_j \big) = \frac{1}{2^n} \sum_{S \in \mathcal{S}} S,
\end{equation}
where $\mathcal{S}$ is the stabilizer group of $\ket{\psi}$. Each $S \in \mathcal{S}$ stabilises $\ket{\psi}$ and is multiplicatively generated by $g_1, \ldots, g_n$ (hence the name generators). The generators are not necessarily unique, but the corresponding stabilizer group is unique. For example, the stabilizer group of the $3$ qubit GHZ state, Eq.~\eqref{eq:GHZstate}, can be written as $\langle X_1 X_2 X_3, Z_1 Z_2, Z_2 Z_3 \rangle$ or as $\langle X_1 X_2 X_3, -Y_1 Y_2 X_3, -X_1 Y_2 Y_3 \rangle$ (here the subscripts indicates which qubit a Pauli operator is acting on.

It is easy to verify that for all $1 \leq j \leq n$, operators of the form
\begin{equation}
    \label{eq:graphstabilizer}
    g_j=X_j \bigotimes_{k \in N(j)} Z_k,
\end{equation}
stabilize a graph state with neighbourhoods $N(1),\ldots,N(n)$. This follows from
\begin{equation}
    X_j CZ_{(l,m)} X_j = \begin{cases}
Z_l CZ_{(l,m)} & \text{if } j=m \\
Z_m CZ_{(l,m)} & \text{if } j=l \\
CZ_{(l,m)} & \text{otherwise}
\end{cases},
\end{equation}
thus
\begin{equation}
\begin{split}
    X_j \bigotimes_{k \in N(j)} Z_k \ket{G} &= \Big( \bigotimes_{k \in N(j)} Z_k \Big)  \Big( \prod_{(l,m) \in E}  X_j CZ_{(l,m)} X_j \Big) X_j \ket{+}^{\otimes n} \\
    &= \Big( \bigotimes_{k \in N(j)} Z_k \Big)^2  \Big( \prod_{(l,m) \in E}  CZ_{(l,m)} \Big) \ket{+}^{\otimes n} \\
    &= \ket{G}.
\end{split}
\end{equation}
The operators of the form Eq.~\eqref{eq:graphstabilizer} all commute and are multiplicatively independent, and hence correspond to the stabilizer group for a graph state. The stabilizer representation of the graph state from Eq.~\eqref{eq:examplegraphstate} is
\begin{equation}
    \dyad{\psi} = \frac{1}{8}\big(\mathbb{I} +X_1 Z_2 Z_3 \big)\big(\mathbb{I} +Z_1 X_2 \big)\big(\mathbb{I} +Z_1 X_3 \big).
\end{equation}
In fact, every stabilizer state can be transformed to a (not necessarily unique) graph state \cite{schlingemann2001b} by constructing a locally acting Clifford operator $C \in \mathcal{C}_1^{\otimes n}$ which maps the generators for a stabilizer state to generators of the form in Eq.~\eqref{eq:graphstabilizer}.

Because all non-identity Pauli operators have a trace of zero, it follows that for any Pauli operator $Q$ and stabilizer state $\ket{\psi}$ with stabilizer group $\mathcal{S}$
\begin{equation}
    \label{eq:traceprop}
    \expval{Q}{\psi} = \begin{cases}
1 & \text{if } Q \in \mathcal{S}\\
-1 & \text{if } -Q \in \mathcal{S} \\
0 & \text{otherwise}
\end{cases}.
\end{equation}

\section{Graph States for Phase Estimation}

In order to gauge the practicality of graph states for quantum metrology, we restrict the problem to phase estimation. As discussed in \textbf{Chapter~3}, phase estimation is versatile in its applications and the expression for the QFI is much more manageable than the general expression. Having said that, it is still not obvious which quantum states achieve a quantum advantage when it comes to phase estimation. Of course, for qubit systems, entanglement is a required resource to surpass the SQL. However, entanglement does not guarantee Heisenberg-like scaling; it was shown in  \cite{oszmaniec2016} that, on average, a randomly selected entangled quantum state would not attain a quantum advantage (even with the allowance of local unitary transformations). Notably though, it was shown in the same study that most symmetric states\footnote{Symmetric states, sometimes called permutation invariant states, are a class of quantum states which remain unchanged when any number of subsystems are swapped with one another.} are (up to local unitary transformations) an efficient resource for phase estimation. It is no surprise that the standard resources for phase estimation are highly symmetric, eg. the GHZ state \cite{giovannetti2004, giovannetti2006}, half-Dicke state \cite{toth2014} and spin squeezed states \cite{gross2012, zhang2014}. A sensible conclusion is that entanglement paired with symmetry makes for an efficient resource for phase estimation.

The canonical phase estimation problem encodes an unknown phase $\theta$ through a unitary of the form
\begin{equation}
    \label{eq:unitary}
    U_\theta = e^{-i \theta \sum_{j=1}^n H_i} = \big( e^{-i \theta H} \big)^{\otimes n},
\end{equation}
where $H_j=H \; \forall j$ are locally acting Hermitian operators. In \cite{SM20}, we set $H=\frac{1}{2} X$, as this choice leads to an easily described class of states which approximately saturate the HL. That being said, the solutions and results can be generalized to any Hermitian generator by rotating beneficial graph states appropriately. Using Eq.~\eqref{eq:QFIpure}, the QFI of an $n$ qubit graph state $\ket{G}$ is
\begin{equation}
    \mathcal{Q}(G) = \sum_{i,j=1}^n \Big( \expval{X_i X_j}{G} - \expval{X_i}{G} \hspace{-5pt} \expval{X_j}{G} \Big).
\end{equation}
This equation can be evaluated using the trace property of stabilizer states, Eq.~\eqref{eq:traceprop}, and the stabilizer group of graph states. For a graph without any isolated vertices $\expval{X_i}{G}=0$ for all $j$. The quantity $X_i X_j$ stabilizes $\ket{G}$ if and only if the neighbourhood of the $i$th qubit is equal to the neighbourhood of the $j$th qubit. By construction, the negation, $-X_i X_j$, never stabilizes $\ket{G}$. Therefore, the QFI of a graph state $\ket{G}$ with no isolated vertices is equal to the number of ordered pairs $(i,j)$ such that $N(i)=N(j)$. For the sake of a mathematical expression
\begin{equation}
    \label{eq:qfigraph}
    \mathcal{Q}(G) = \sum_{i,j=1}^n \delta_{N(i),N(j)},
\end{equation}
where $\delta_{x,y}$ is the Kronecker delta which evaluates to $1$ if $x=y$ and $0$ otherwise. One can conclude that the graph states, although not totally symmetric states, still require a form of internal symmetry (i.e pairs of qubits with equal neighbourhoods) to attain a quantum advantage.

As an example, all of the external vertices of a $n$ qubit star graph state, Fig.~(\ref{fig:graph2}), have the same neighbourhood (the central vertex). Thus, the QFI is $\mathcal{Q}(G_\text{star})=(n-1)^2+1$, which is approximately equal to the HL. This is unsurprising as it is a highly symmetric state. Conversely, an $n$ qubit cyclic graph state, Fig.~(\ref{fig:graph3}), may appear to be highly symmetric at a graphical level (rotational symmetry), it does not have any permutation symmetries. An $n$ qubit lattice graph state, Fig.~(\ref{fig:graph4}), similarly does not have any permutations and also is limited by the SQL. This is in accordance with \cite{friis2017}, where it is stated that unmodified cluster states are not good resources for quantum metrology. Note that this does not contradict the results of \cite{rosenkranz2009}, where an unconventional parameter encoding scheme is used.

Because of the choice of $H_i=\frac{1}{2} X_i$, many highly symmetric states do not achieve a quantum advantage. For example, the complete graph, Fig.~(\ref{fig:graph1}), is invariant under any permutation and achieves the SQL. However, using the alternative choice for the unitary encoding $H_i = \frac{1}{2} Y_i$, the the QFI of the complete graph is the HL and the QFI of the star graph is the SQL.  This alternative choice Hamiltonian also leads to an alternative, but more complicated, topological expression: the QFI is equal to the number of ordered pairs $(i,j)$ such that $N(i) \cup \{ i \} = N(j) \cup \{ j \}$. The problem concerning the choice of the encoding Hamiltonian vanishes by allowing for local transformations before the parameter is encoded. With this assumption, a graph state (or more generally, a stabilizer state) $\ket{\psi}$ is a practical resource for phase estimation if there exists a $C \in \mathcal{C}_1^{\otimes n}$ such that $C \ket{\psi}$ is a graph state whose corresponding graph has many pairs of vertices with identical neighbourhoods.  Logically, the final possibility to examine is when the encoding Hamiltonian is set to $H_i=\frac{1}{2} Z_i$. However a quick computation leads to the conclusion that this choise leads to a QFI equal to the SQL of $\mathcal{Q}=n$ for any graph state.

\subsection{Generalization to Stabilizer States}

The QFI of a graph state was computed by finding the overlap of the Pauli operators of $\pm X_i$ and $\pm X_i X_j$ with the stabilizer group. This argument is not unique to graph states and can be made for any stabilizer state, further it can be reverse engineered to determine the number of stabilizer states which achieve a desired level of QFI.

Begin by defining the sets
\begin{equation}
    A= \{X_1 X_2, X_1 X_3, \ldots, X_1 X_n \},
\end{equation}
and
\begin{equation}
        B_k = \{ Q_1 \ldots Q_k | Q_j \in \{Y,Z \} \; \forall 1\leq j \leq k \},
\end{equation}
where $k > 1$. The set $A$ is ordered such that if a group is generated with the first $k-1$ elements, it will contain all operators of the form $X_iX_j$ with $1 \leq i,j \leq k$ (and $i \neq j$). Each $b \in B_k$ will commute with all elements of said group. We do not allow $Q_j = \mathbb{I}$ as then there exists a $b \in B_k$ which anti-commutes with certain operators.

Next construct the stabilizer group
\begin{equation}
    \label{eq:sgroup}
    \mathcal{S} = \langle a_1, \ldots a_{k-1}, b, bg_1, \ldots, b g_{n-k} \rangle,
\end{equation}
where $a_1, \ldots, a_{k-1}$ are all unique operators from the set $A$, $b \in B_k$ and $g_1, \ldots, g_{n-k}$ act exclusively on the final $n-k$ qubits of the quantum state and are the generators for an $n-k$ qubit stabilizer state. By construction, $\mathcal{S}$ does not contain any stabilizer of the form $\pm X_i$ or $-X_i X_j$. Therefore, similar to a graph state with no isolated vertices, the QFI is equal to the number of stabilizers of the form $X_i X_j$, which is $k^2$ by construction. To determine a bound on the number of stabilizers states which achieve a QFI of $n^{2-\varepsilon}$, labelled via $\tilde{N}(n;\varepsilon)$, we count the total number of possible stabilizer groups which is of the form of Eq.~\eqref{eq:sgroup} with $k \geq n^{1-\varepsilon/2}$. Mathematically, one obtains
\begin{equation}
    \label{eq:numberofstab}
    \tilde{N}(n;\varepsilon) \geq \sum_{k \geq n^{1-\varepsilon/2}} \binom{n-1}{k-1} 2^k s_{n-k},
\end{equation}
where $s_m$ is the number of $m$ qubit stabilizer states \cite{aaronson2004}
\begin{equation}
    s_m = 2^m \prod_{j=0}^{m-1} (2^{n-j}+1).
\end{equation}

It is quite apparent that $\tilde{N}(n;\varepsilon) \ll s_n$ for small $\varepsilon$. This is because of a few different factors. The first is that most quantum states do not saturate the HL \cite{oszmaniec2016}. The second is that the bound in question is restricted to the problem of phase estimation via the specific unitary encoding with $H_i = \frac{1}{2}X_i$. Third, to simplify the mathematics, it was demanded that operators of the form $\pm X_j$ or $-X_j X_k$ were not in the stabilizer group; discrediting some stabilizer states which would still achieve the necessary QFI. In retrospect, a tighter bound could have been achieved by allowing for other encoding operations and a more concrete mathematical analysis.

\section{Bundled Graph States}

As it was formerly mentioned, graph states are a resource with many applications. It would be very desirable and convenient if a specific graph state with a specific application was also a practical resource for quantum metrology. Evidently from Eq.~\eqref{eq:qfigraph} and the previously mentioned examples, most graph states are not a good resource for quantum metrology, at least not before undergoing some sort of transformation \cite{friis2017}. In order to capitalize on graph states which are multipurpose, we provide a recipe to transform any graph into a (larger) graph which is practical for quantum metrology. The new graph maintains the underlying structure of the old graph and can still be used for the original purpose.

We name the new constructed graph a \textit{bundled graph state}, as the construction process involves replacing individual qubits by a bundle of qubits with identical neighbourhoods in order to maximize the QFI.

\subsection{Construction}

\definecolor{green1}{HTML}{33FF33}
\definecolor{blue1}{HTML}{7EA6E0}

\begin{figure}
    \centering
    \includegraphics[width=0.9\linewidth]{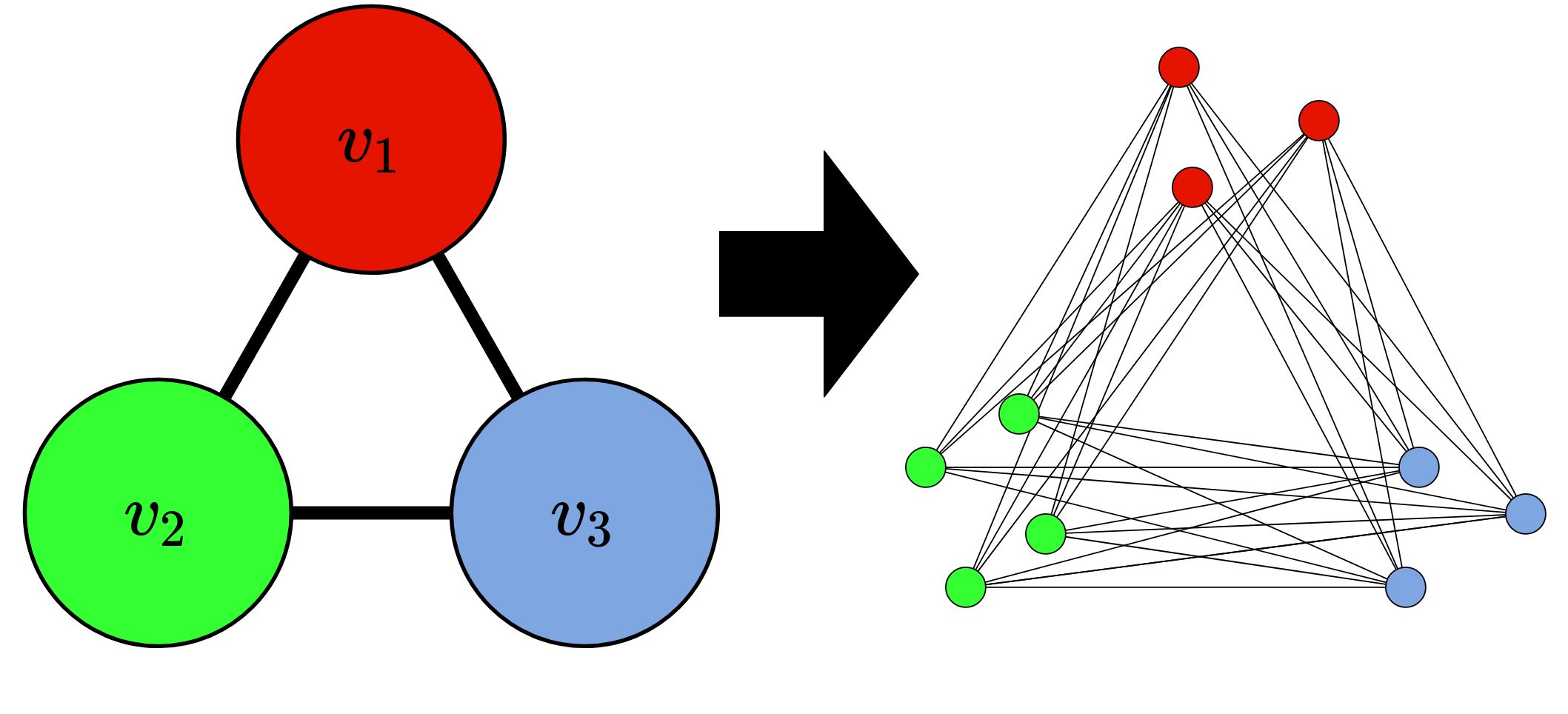}
    \begin{tabular}{ c | c | c | c }
        \textbf{Vertex} & \tikz[baseline=-0.75ex]\draw[black,fill=red] (0,0) circle (1ex); & \tikz[baseline=-0.75ex]\draw[black,fill=green1] (0,0) circle (1ex); & \tikz[baseline=-0.75ex]\draw[black,fill=blue1] (0,0) circle (1ex); \\
        \hline
        \textbf{Quantity} & $n_1=3$ & $n_2=4$ & $n_3=3$ \\  
    \end{tabular}
    \caption{Transforming a $k=3$ vertex graph into a $n=10$ vertex bundled graph. The QFI of the corresponding bundled graph state (in this specific example) is  $\mathcal{Q}=n_1^2+n_2^2+n_3^2 \approx n^{1.5}$.}
    \label{fig:bundlegraph}
\end{figure}

The recipe transforms a smaller graph $G=(V,E)$ with $k$ vertices (none of which are isolated) into a larger graph $G_\text{bundled} = (V^\prime, E^\prime)$ with $n \geq k$ vertices.

\begin{enumerate}
    \item Begin with any $k$ qubit graph state $G=(V,E)$ with no isolated vertices.
    \item Vertex $v_i$ is replaced with $n_i$ vertices, labelled $v_i^{(1)}, \ldots, v_i^{(n_i)}$, with $\sum_{i=1}^k n_i = n$.
    \item If $(v_i,v_j) \in E$ then $(v_i^{(a)},v_j^{(b)}) \in E^\prime \; \forall a,b$.
\end{enumerate}

The resulting graph $G_\text{bundled} = (V^\prime, E^\prime)$ has vertices
\begin{equation}
    V^\prime = \{ v_1^{(1)}, \ldots, v_1^{(n_1)}, \ldots, v_k^{(1)}, \ldots, v_k^{(n_k)} \}
\end{equation}
and edges
\begin{equation}
    E^\prime = \{ (v_i^{(a)},v_j^{(b)} ) \; \forall a,b \; | \; (v_i,v_j) \in E \}.
\end{equation}
The constructed bundled graph has many vertices with identical neighbourhoods: $N(v_i^{(a)}) = N(v_i^{(b)}) \; \forall i,a,b$. The above recipe is depicted in Fig.~(\ref{fig:bundlegraph}), in which an $n=10$ vertex bundled graph from a smaller $k=3$ vertex graph.

The QFI of a bundled graph state satisfies
\begin{equation}
    \label{eq:bundledqfi}
    \mathcal{Q}(G_\text{bundled}) \geq \sum_{i=1}^k n_i^2 \geq \frac{n^2}{k} = n^{2-\log_n k}.
\end{equation}
If $k \ll n$, the resulting bundled graph approximately saturates the HL and the underlying structure of the graph state is preserved. Needless to say, the QFI is still dependent on the shape of the original graph. For example, a bundled cyclic graph state, where each of the $k$ bundles contains an equal number of $n/k$ qubits has a QFI
\begin{equation}
    \mathcal{Q}(G_\text{cyclic,bundled})=\frac{n^2}{k}.
\end{equation}
A bundled star graph, built in the same manner, has a QFI
\begin{equation}
    \mathcal{Q}(G_\text{star,bundled})=n^2\big(1-\frac{1}{k}\big)^2+\frac{n^2}{k}.
\end{equation}
Unsurprisingly, $\mathcal{Q}(G_\text{star,bundled}) \geq \mathcal{Q}(G_\text{cyclic,bundled})$, this is due to the fact that the underlying structure of the star graph state contained symmetries, whereas the cyclic graph state did not. Nevertheless, both the bundled star graph state and the bundled cyclic graph state have a Heisenberg-like QFI.

\section{Robustness}

An important criteria for quantum states to possess to be a practical resource for quantum metrology is robustness against noise. Environmental noise is the primary obstacle for current quantum metrology technologies \cite{escher2011b, demkowicz2012, tsang2013}. This topic, along with error correction based noise mitigation strategies is explored in much more detail in \textbf{Chapter~5}. In this chapter, a different approach to noise is taken: which is pinpointing resources which have a naturally built-in robustness. Two noise models are explored: i) iid dephasing, and ii) a finite number of erasures. These two noise models are frequently used in other noisy phase estimation problems \cite{demkowicz2012, kolodynski2013}. In particular, the GHZ state is famously fragile against the effects of loss and becomes useless for quantum metrology in a lossy environment \cite{kolodynski2013}. We subject the graph states to the noise models to having the unknown parameter $\theta$ encoded. The QFI calculations for noisy graph states can be found in \textbf{Appendix~A}.

As expected, the shape of a graph greatly influences the severity of the noise on the corresponding graph state. Without loss of generality, we again only consider graphs which have no isolated vertices. To bound the QFI as elegantly as possible, we partition the vertices of a graph $G=(V,E)$ into disjoint subsets $U_1,U_2,\ldots,U_l,\ldots$ such that $\bigcup_{l} U_l = V$. The vertices are partitioned in accordance to commonly shared neighbourhoods, hence, if $v_i \in U_a$ and $v_j \in U_b$, then $N(v_i)=N(v_j)$ if $a=b$ and $N(v_i) \neq N(v_j)$ if $a \neq b$. We write that $|U_l|=u_l$ and the shared neighbourhood of $U_l$ is $M_l$ with $|M_l|=m_l$.

\definecolor{green2}{HTML}{008000}
\definecolor{HLmagenta}{HTML}{ff00ff}
\definecolor{SQLorange}{HTML}{ffa500}

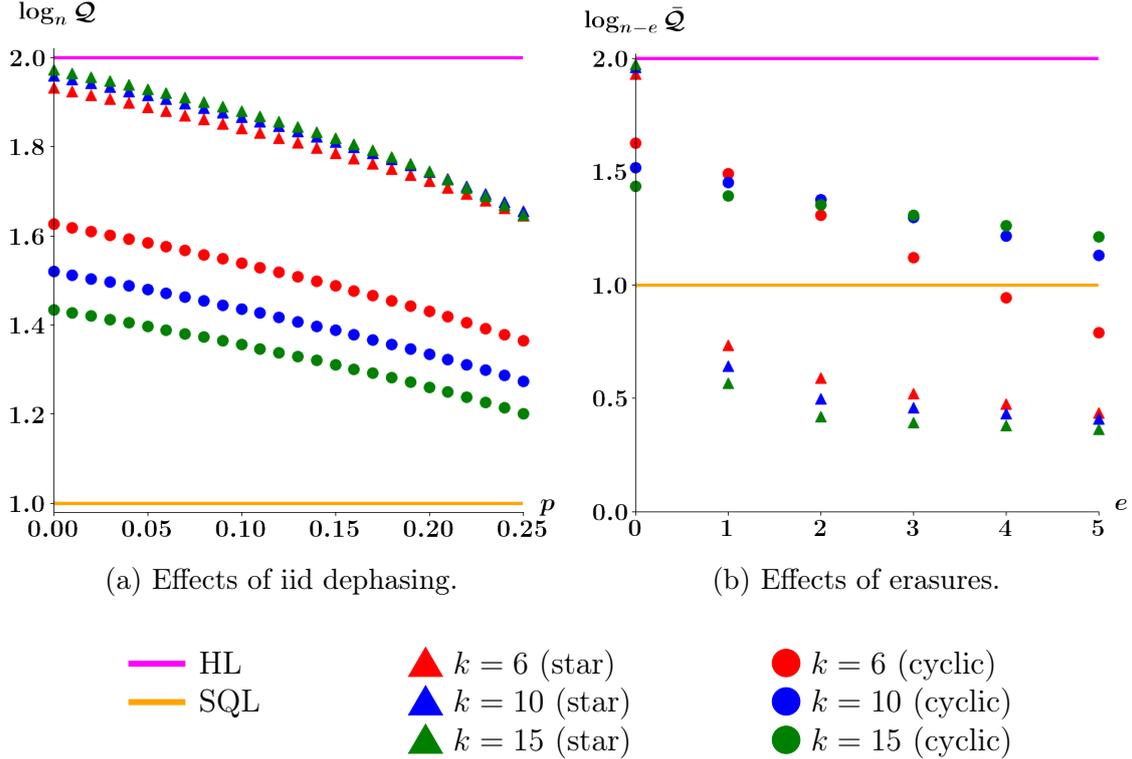
\begin{figure}[!ht]
\centering
\begin{subfigure}{.495\textwidth}
  \centering
  \includegraphics[width=.99\textwidth]{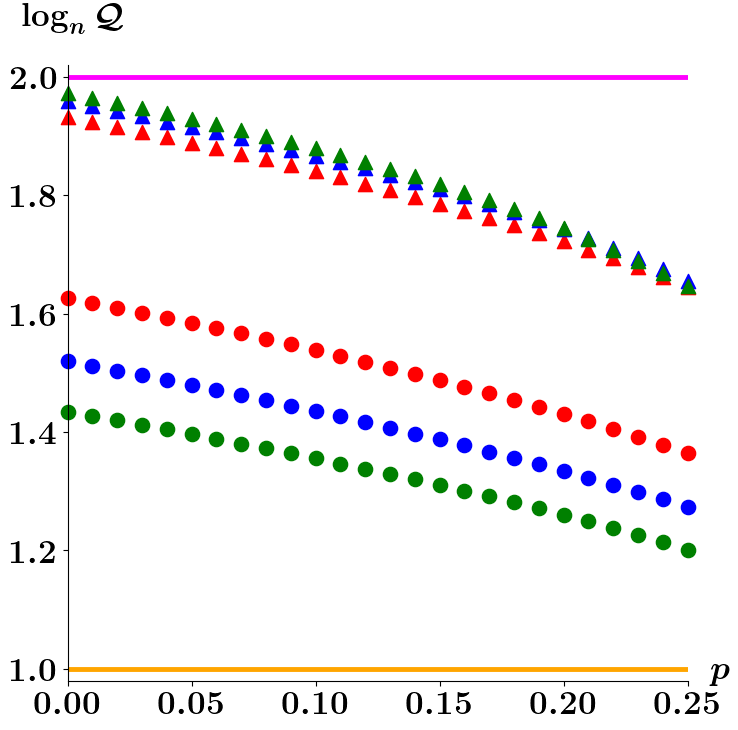}
  \caption{Effects of iid dephasing.}
  \label{fig:dephasingplot}
\end{subfigure}
\begin{subfigure}{.495\textwidth}
  \centering
  \includegraphics[width=.99\linewidth]{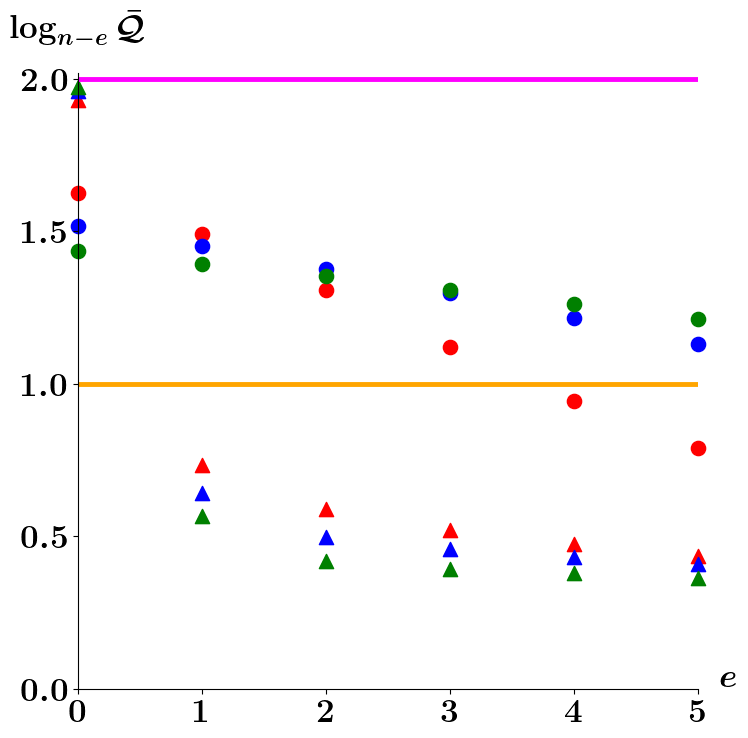}  
  \caption{Effects of erasures.}
  \label{fig:erasureplot}
\end{subfigure}

\vspace{3.2ex}

\begin{tabular}{ l m{1cm}  l m{1cm}  l }
        \begin{tikzpicture}[baseline=-0.7ex]
            \draw[line width=0.8 mm, color=HLmagenta] (0,0) -- (0.75,0);
        \end{tikzpicture} HL & &
        \scalebox{0.75}{
            \begin{tikzpicture}[baseline=-0.45ex]
            \node[tri,red,fill=red]{};
            \end{tikzpicture}}  $k=6$ (star) & &
        \tikz[baseline=-0.75ex]\draw[red,fill=red] (0,0) circle (1ex);  $k=6$ (cyclic) \\
        \begin{tikzpicture}[baseline=-0.7ex]
            \draw[line width=0.8 mm, color=SQLorange] (0,0) -- (0.75,0);
        \end{tikzpicture} SQL & &
        \scalebox{0.75}{
            \begin{tikzpicture}[baseline=-0.45ex]
            \node[tri,blue,fill=blue]{};
            \end{tikzpicture}}  $k=10$ (star) & &
        \tikz[baseline=-0.75ex]\draw[blue,fill=blue] (0,0) circle (1ex);  $k=10$ (cyclic) \\
          & &
        \scalebox{0.75}{
            \begin{tikzpicture}[baseline=-0.45ex]
            \node[tri,green2,fill=green2]{};
            \end{tikzpicture}}  $k=15$ (star) & &
        \tikz[baseline=-0.75ex]\draw[green2,fill=green2] (0,0) circle (1ex);  $k=15$ (cyclic) \\
    \end{tabular}
\caption{Robustness of an $n=120$ qubit bundled star graph states and bundled cyclic graph states subjected to (a) iid dephasing and (b) $e \leq 5 $ erasures. In both scenarios, the bundled graphs have $k$ equal size bundles of $n/k$ qubits. After being subjected to iid dephasing (a), $\log_n \mathcal{Q}$ decreases linearly for small $p$. This is expected from Eq.~\eqref{eq:qfidephasing}, and ultimately, a quantum advantage is maintained. To gauge the effects of erasures (b), the quantity $\log_{n-e} \mathcal{Q}$ is plotted, where $\bar{\mathcal{Q}}$ is the average QFI of the bundled graph state after $e$ erasures - it is necessary to take the average prior to the logarithm to avoid the problem of $\log_n 0$. Because bundled star graph states have an enormous amount of symmetry, a single erasure (regardless of where it occurs) will cause the QFI to fall below the SQL. In contrast, the bundled cyclic graph are more resilient to noise and can maintain an advantage after a small number of erasures; furthermore, the amount of qubits which are (on average) affected by the erasures decrease as $k$ increases.}
\label{fig:robustnessplots}
\end{figure}

\subsection{IID Dephasing}

After being subjected to iid dephasing, each qubit has probability $p$ of being dephased with respect to the $Z$ operator. Define $Z_{\vec{j}}$ to be the Pauli operator which applies $Z$ to all qubits indexed in $\vec{j}$ with $|\vec{j}|=j$. Post iid dephasing, an $n$ qubit graph state $\ket{G}$ is mapped to
\begin{equation}
    \ket{G} \rightarrow  \sum_{\vec{j}} p^j (1-p)^{n-j} Z_{\vec{j}} \dyad{G} Z_{\vec{j}},
\end{equation}
which has a QFI of
\begin{equation}
    \begin{split}
       \mathcal{Q}(G^\text{dephasing}) &\geq \sum_{l} \big( (1-2p)^2u_l^2+4p(1-p)u_l \big) \big( 1-(2p(1-p)+1/2)^{m_l} \big) \\ 
       &\geq (1-2p)^2 \big( 1-(2p(1-p)+1/2)^m \big) \mathcal{Q}(G),
    \end{split}
\end{equation}
where $m=\min_l m_l$. The quantity $(2p(1-p)+1/2)^m$ is approximately zero for large enough $m$ and small enough\footnote{$(2p(1-p)+1/2)^m < 0.006$ for $m \geq 10$ and $p \leq 0.05$.} $p$, using this approximation in tandem with the QFI of a bundled graph state Eq.~\eqref{eq:bundledqfi},
\begin{equation}
    \label{eq:qfidephasing}
    \mathcal{Q}(G^\text{dephasing}_\text{bundled} ) \geq (1-2p)^2 \frac{n^2}{k} = n^{2-\log_n k-\frac{4}{n}p + \mathcal{O}(p^2)}.
\end{equation}
Therefore, for small $p$, bundled graph states retain a quantum advantage for phase estimation. This is shown in Fig.~(\ref{fig:dephasingplot}), where the QFI of bundled star graph states and bundled cyclic graph states surpass the SQL for $p \leq 0.25$.

\subsection{Erasures}

A qubit becomes unusable after undergoing erasure, to model this the erased qubits are traced out
\begin{equation}
    \ket{G} \rightarrow \Tr_{\vec{e}} \dyad{G},
\end{equation}
where $\vec{e}$ indexes which qubits are erased. This maps the above state into an equally weighted mixed state\footnote{The mixed state in Eq.~\eqref{eq:erasedgraph} is left as an $n$ qubit state for clarity. The traced out systems are equivalent to maximally mixed states, $\mathbb{I}/2$, which are irrelevant with respect to the QFI.}
\begin{equation}
    \label{eq:erasedgraph}
    2^{-|L_{\vec{e}}|} \sum_{\vec{j} \subseteq L_{\vec{e}} } Z_{\vec{j}} \dyad{G} Z_{\vec{j}},
\end{equation}
where $L_{\vec{e}}$ is the set of vertices corresponding to the traced out qubits as well as their neighbourhoods and the sum is taken over all possible subsets of $L_{\vec{e}}$, denoted with $\vec{j} \subseteq L_{\vec{e}}$. As a consequence, the QFI is extremely dependent on the shape of the graph. In general
\begin{equation}
    \mathcal{Q}(G^{\text{erasures } \vec{e}}) = \sum_{l} h_l (\vec{e}),
\end{equation}
where
\begin{equation}
  h_l (\vec{e})  = \begin{cases}
    u_l^2 & \text{if } M_l \nsubseteq L_{\vec{e}} \text{ and } U_l \nsubseteq L_{\vec{e}}\\
    u_l & \text{if } M_l \nsubseteq L_{\vec{e}} \text{ and } U_l \subseteq L_{\vec{e}} \\
    0 & \text{otherwise}
    \end{cases}.
\end{equation}

An interpretation, is that the `noise' produced by an erasure effects all the similar qubits and propagates to the shared neighbourhood. Therefore, bundled graphs which were constructed from graphs that did not originally possess much symmetry are more robust against erasures then bundled graphs constructed from graphs with preexisting symmetries. This is witnessed in Fig.~(\ref{fig:erasureplot}), in which bundled cyclic graphs of varying size maintain a quantum advantage up to $e=3$ erasures, in contrast, the QFI of the analogous bundled star graphs is below the SQL after a single erasure.

A possible method to circumvent erasure errors is to construct graph states with two types of qubits. One type would be used for metrology but prone to noise (e.g. the spin of an electron), and the other type is more naturally robust to noise but not used for metrology (e.g. the spin of a neutron). By constructing a hybrid graph state one could reduce the propagation of noise caused by the erasure of a sensing qubit. Graphically, this transformation can be described as adding a `naturally robust' vertex in the center of each edge. If the naturally robust qubits are immune to erasures, the size of $L_{\vec{e}}$ would reduce drastically resulting in a higher QFI (on average).

\section{Saturating the QCRB}

Another important criteria for a quantum state to have in order to qualify as a practical resource for quantum metrology is the existence of a simple measurement scheme to saturate the QCRB. For a graph state with no isolated vertices, $\ket{G}$, this can be executed by measuring in the basis of a stabilizer, $S_M$, which consists entirely of $Y$ and $Z$ operators. Observe that the expected value of the observable (with respect to the phase encoded graph state) is
\begin{equation}
    \begin{split}
        \expval{S_M} &= \bra{G} U_\theta^\dagger S_M U_\theta \ket{G} \\
        &= \bra{G} (U_\theta^\dagger)^2 S_M \ket{G} \\
        &= \bra{G} (U_\theta^\dagger)^2 \ket{G} \\
        &= \sum_{j=0}^\infty \frac{(i\theta)^j}{j!}\bra{G} \Big( \sum_{i=1}^n X_i \Big)^j \ket{G}
    \end{split}
\end{equation}
For a graph state with no isolated vertices, the second order term is proportional to the QFI. Because the expectation value of an observable is real valued, the sum of all odd terms must be zero. Hence the above simplifies to
\begin{equation}
    \begin{split}
        \expval{S_M} = 1- \frac{\theta^2}{2}\mathcal{Q}(G) + \mathcal{O}(\theta^4).
    \end{split}
\end{equation}

Using the error propagation formula, the variance of the estimate scales as
\begin{equation}
    \frac{\Delta^2 S_M}{| \partial_\theta \expval{S_M} |^2 }= \frac{\theta^2 Q(G) + \mathcal{O}(\theta^4)}{\theta^2 Q(G)^2 + \mathcal{O}(\theta^4)} = \frac{1}{\mathcal{Q}(G)} + \mathcal{O}(\theta^2) \approx \frac{1}{\mathcal{Q}(G)}.
\end{equation}
The above approximation is only valid when the phase being estimated is very small, $\theta \approx 0$, fortunately this is naturally the regime explored for phase estimation. If the unknown phase is large, but it is known up to approximation because of a pre-existing model or another estimate, then the quantum state can be first transformed by local unitary operations such that the effective phase is small.

The condition that a graph state has a stabilizer $S_M$ which only consists of $Y$ and $Z$ operators is necessary for $S_M U_\theta = U_\theta^\dagger S_M$. Such a stabilizer is not guaranteed to exist and depends on the shape of the graph. For example, it always exists for bundled star graph states\footnote{Take the product of the generator of a central qubit and a generator of an external qubit.}, but for bundled star graph states, it exists only when $k=0 \mod 4$\footnote{Divide the bundles into sequences of four. Take the product of a generator from the two central bundles from each group of four.}. If the graph state does not have such a stabilizer, a solution can be remedied using an ancillary qubit. Let $S_M$ be the stabilizer with as many $Y$ or $Z$ operators, in any index which there is not $Y$ or $Z$ operator, entangle the corresponding qubit to the new ancillary qubit with a controlled-$Z$ operation. This will form an $n+1$ qubit graph state where the stabilizer $g_{n+1}S_M$ consists of entirely of $Y$ and $Z$ operators, thus the new graph, which would have a very similar structure to the original graph, can approximately saturate the QCRB with a simple single qubit measurement scheme.

\section{Quantum Sensing Networks}

An immediate application for graph states with respect to parameter estimation problems and metrology is quantum sensing networks \cite{komar2014, komar2016, eldredge2018, ge2018, proctor2018, zhuang2018, qian2019, rubio2020a, guo2020}. A quantum network is collection of nodes and edges, where the nodes have some quantum functionality and an edge represent some form of connection between a pair of nodes, this can be either entanglement of a quantum channel \cite{kimble2008,  van2012, schoute2016, wehner2018}. This is a much more general framework than the graph state framework, nonetheless the similarities between the two simplify the adaptation of a graph state into a quantum network \cite{meignant2019, hahn2019}. A quantum sensing network is a quantum network designed for quantum parameter estimation. Quantum sensing networks come in two flavours depending on the functionality of the nodes and edges.

\begin{figure}[!ht]
    \centering
    \begin{subfigure}{.495\textwidth}
        \centering
        \includegraphics[width=.99\textwidth]{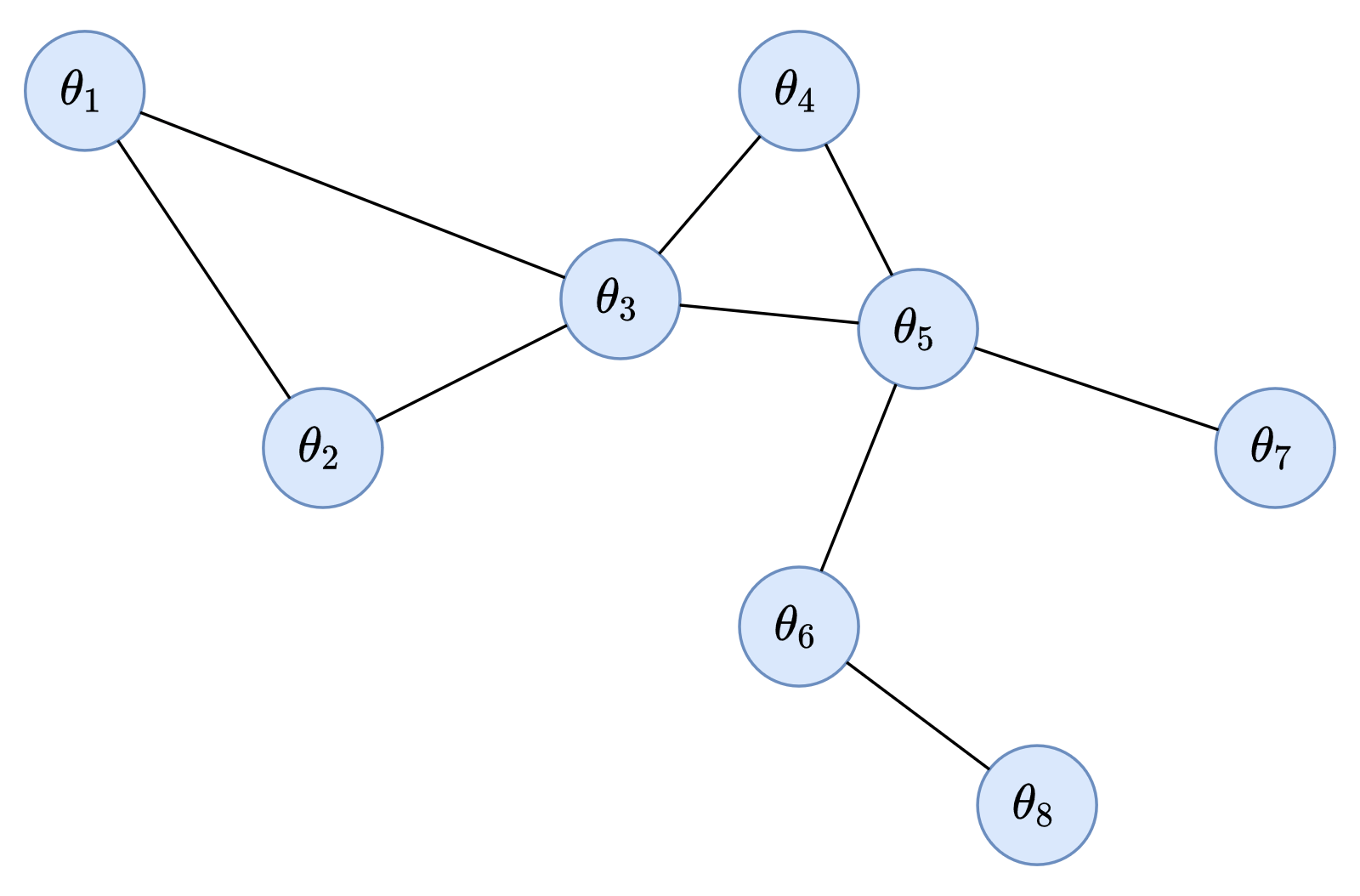}
        \caption{Type 1 Quantum Sensing Network.}
        \label{fig:SensingNetwork_Type1}
    \end{subfigure}
    \begin{subfigure}{.495\textwidth}
        \centering
        \includegraphics[width=.99\linewidth]{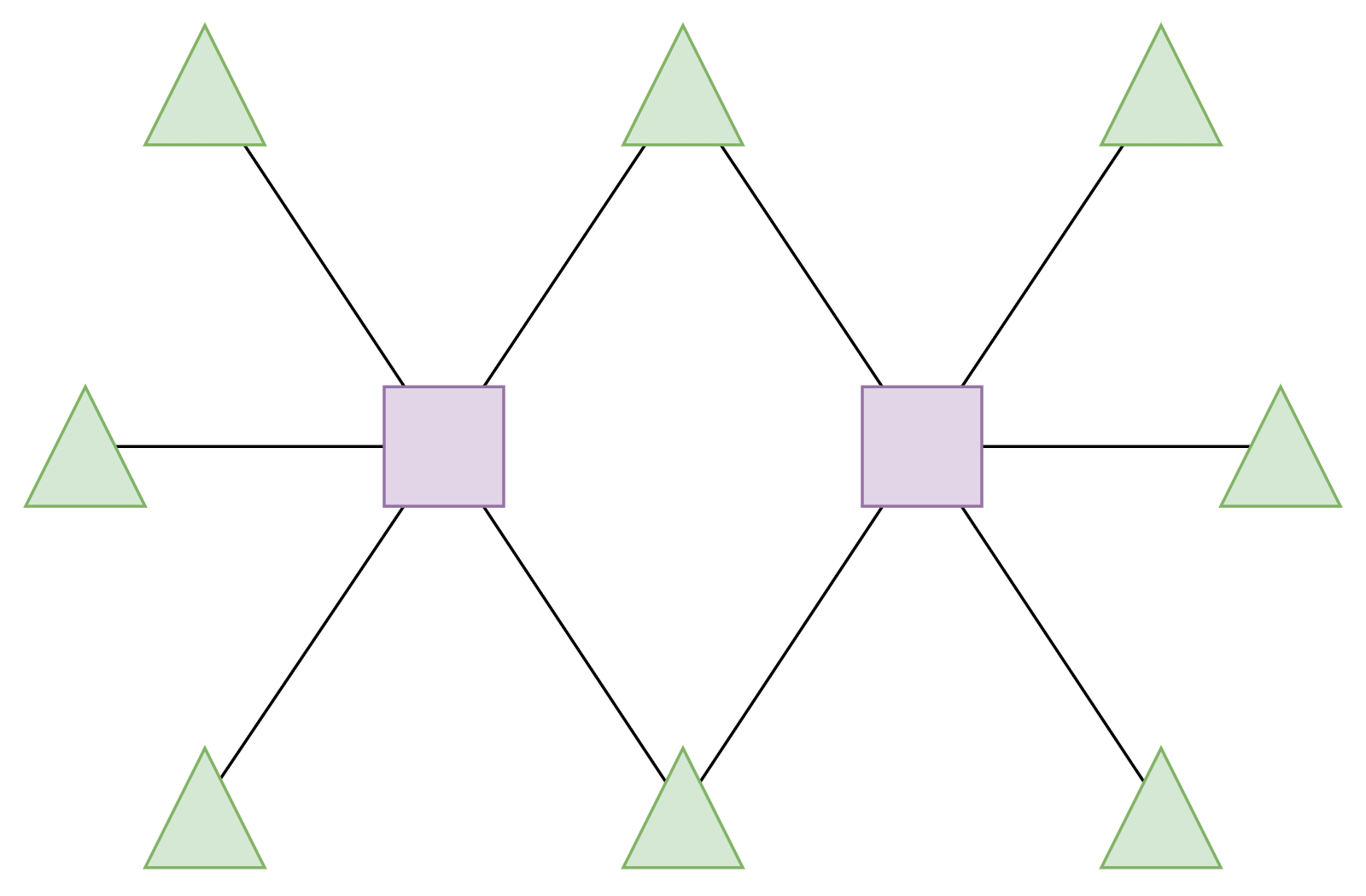}  
        \caption{Type 2 Quantum Sensing Network.}
        \label{fig:SensingNetwork_Type2}
    \end{subfigure}
    \caption{There are two main descriptions for quantum sensing networks. The first type (a) of quantum sensing networks aligns with the description of a graph state, where a node represents a qubit and an edge represents an entangling operation. This type of quantum network is a popular framework for multiparameter quantum problems \cite{proctor2018, rubio2020a}, as each node can encode a different unknown parameter $\theta_j$. The second type (b) of quantum sensing networks resembles the usual description of quantum networks \cite{van2012}, where an edge represents a quantum channel between two nodes. In particular, not all nodes are created equal and serve different purposes. In this schematic the central square nodes distribute quantum states to the exterior triangle nodes where an unknown parameter is encoded, the encoded quantum states are then returned to a central node to be measured.}
    \label{fig:SensingNetworks}
\end{figure}

The first type of quantum sensing networks is when the nodes represent a quantum state the edges represent a form of entanglement, depicted in Fig.~(\ref{fig:SensingNetwork_Type1}). Evidently, graph states are a subset of possible quantum sensing networks. Quantum sensing networks, with a suitable choice for entangling operations and initialized quantum states, have been shown to be an effective resource for multiparameter quantum metrology problems \cite{proctor2018, rubio2020a}. Graph states are no exception, and thus a future natural direction for this work is to formally classify utility of graph states for quantum metrology in the multiparameter setting. Likely, the most efficient graphs will resemble bundled graph states where a different parameter is encoded into a different bundle.

The second type of quantum sensing networks is when the nodes have different technological functionality and edges represent a quantum channel, depicted in Fig.~(\ref{fig:SensingNetwork_Type2}). For example, the authors  of \cite{komar2014, komar2016} construct a quantum network where a central node is much more powerful technologically than the exterior nodes. There, the central node prepares quantum states, which are then distributed to exterior nodes where a local phase is encoded, after which the encoded quantum state is returned to the central node to be measured. A current project of mine is combining the results from this chapter and the cryptographic protocols outlined in \textbf{Chapter 6} to devise a notion of a secure quantum sensing network for phase estimation problems.

\section{Discussion}

To recapitulate, graph states are applicable to many disciplines of quantum information and can be implemented with different technologies. In our work \cite{SM20}, we showed that quantum metrology problems can be added to the versatility of graph states. This was done by constructing a class of graph states, called bundled graph states, which have a Heisenberg-like QFI with respect to phase estimation. By design, bundled graph states can have any desired underlying structure, making them multi purposeful.

In addition to the Heisenberg-like QFI, graph states are robust against iid dephasing and (conditional on the shape of the graph) a small number of erasures. As a comparison, the GHZ state is similarly robust against dephasing but cannot tolerate a single erasure \cite{toth2014}. Even though we explored specific error models, we expect similar robustness results in other settings. For example errors during or after the parameter encoding, or a spatially correlated noise model \cite{jeske2014}, in which vertices (or bundle of vertices) of a graph is subjected to different error rates.

Lastly, a simple measurement scheme is presented to approximately saturate the QCRB. Even though this can always be done in theory by measuring in the basis of the symmetric logarithmic derivative \cite{braunstein1994}, doing so is unfeasible for real world quantum technologies. The measurement scheme we present uses local Pauli measurements, which is realizable for real world quantum technologies \cite{waldherr2012}.

There are a number of exciting future perspectives for graph states and quantum metrology. One direction is to explore more general scenarios, such as metrology problems other than phase estimation or phase estimation with non-local parameter encoding unitaries \cite{luis2004}. Another direction is to adapt the underlying structure of a graph state to that of a quantum sensing network \cite{komar2014, komar2016, eldredge2018, ge2018, proctor2018, zhuang2018, qian2019, rubio2020a, guo2020}. Likely, the most efficient graph states to adapt to a quantum sensing network problem is bundled graph states. This is because the inherent symmetries which boost their utility for quantum metrology remains unchanged in a multiparameter setting. Of course, this needs to be shown formally and may not be so straightforward to devise a measurement scheme with compatible measurements.

%% file: Chapters/Chapter5-QECC.tex
\chapter{Limits of Error Correction for Quantum Metrology}

Noise is the greatest obstacle for quantum metrology that limits the achievable precision and sensitivity \cite{escher2011a, escher2011b, demkowicz2012}. As a noisy system evolves in time, it becomes more and more difficult to distinguish the effects of the encoding Hamiltonian and the effects of noise \cite{haase2016}. A proposed solution to mitigate the effects of noise is to repeatedly perform quantum error correction \cite{kessler2014, dur2014, arrad2014, lu2015}. Recently, it has been shown that if the encoding Hamiltonian and the environmental noise satisfy an orthogonality condition, then the HL may be recovered indefinitely \cite{demkowicz2017, zhou2018}. This euphonic conclusion has the added caveat that the assumed frequency of which error correction is performed is infinite. Needless to say, this is an impractical assumption for current quantum technologies, where the rate of implementable error correction is on a similar time scale to the dephasing rates of spin qubits \cite{cramer2016, ofek2016} and superconducting qubits \cite{dutt2007, taminiau2014}.

In this chapter, we determine the limitations of error correction enhanced quantum metrology by accounting for imperfections of near term quantum technologies. These include a non-infinitesimal wait time between applications of error correction, noisy ancillary qubits and imperfect error correction operations. The work done in \cite{kessler2014} makes similar assumptions, however higher order error terms are ignored, which is equally presumptuous as infinitely frequent applications of error correction.

\section{Environmental Noise and Errors}

Quantum systems are extremely sensitive to small perturbations. These perturbations can arise from interactions with external degrees of freedom, e.g. an electron getting excited by an incident photon, or from the finite precision in which quantum operations and control can be performed. These interactions alter the evolution of a quantum system in an undesirable fashion, where the final quantum state is not the targeted quantum state in an idealistic scenario. This is perhaps the biggest hurdle in creating quantum technologies \cite{suter2016}, to such an extent that many have come to accept the current inevitability of errors and search for problems which may be solved with noisy intermediate-scale quantum (NISQ) technologies \cite{preskill2018, torlai2020, bharti2021}.

The standard nomenclature for `interactions with external degrees of freedom' is environmental noise. Models for open quantum systems subject to environmental noise comes in many flavours \cite{gardiner1991, breuer2002, clerk2010} and ultimately depend on the type of quantum technology. Photonic systems are prone to lossy effects \cite{wang2014}, whereas spin systems are prone to decoherence effects \cite{zurek2006}. Similarly, the consequences of noise is model dependent, but in principle entanglement in composite systems is lost, and the likely reason why quantum effects are not observed at a macroscopic scale \cite{schlosshauer2005, zurek2006}.

\subsection{Noisy Quantum Metrology}

In the past decade, the effects of noise on quantum metrology problems have been well established \cite{escher2011a, escher2011b, demkowicz2012, chaves2013, tsang2013, kolodynski2013, jeske2014, kolodynski2014, demkowicz2015, haase2016}. Optical systems are prone to loss and diffusion \cite{lee2009, demkowicz2009, knysh2011, zhang2013, demkowicz2015}, while atomic systems are prone to dephasing and decoherence \cite{shaji2007, borregaard2013, macieszczak2014, zheng2014}. In principle, as a noisy system evolves in time, it becomes more difficult to extract information about the encoded unknown parameter(s), and as a consequence of lost entanglement, the sensitivity is limited to that achievable by classical approaches \cite{chin2012}. 

The canonical example of a noisy quantum metrology scheme involves $n$ qubits governed by two interactions. The first is a signal $\omega$ which causes a detuning in each of the qubits, represented by $H=\frac{\hbar \omega}{2}\sum_{m=1}^n Z_m$. The second, an interaction with the environment which causes dephasing with rate $\gamma$ in the $X$ direction. Lastly, the qubit evolves in accordance to its natural resonance frequency, which is assumed to be known to a high degree of precision. In the rotating reference frame, where the natural frequency of the qubit is suppressed, the Lindbladian master equation can be written as \cite{rivas2012}
\begin{equation}
    \label{eq:MasterNoisy}
    \frac{d \rho}{d t} = -\frac{i}{\hbar} [H,\rho] + \gamma \sum_{m=1}^n (X_m \rho X_m - \rho ).
\end{equation}
After time $t$ the QFI of the system can be computed (see \textbf{Appendix B}) to be
\begin{equation}
    \label{eq:noisymetro_noecc}
    Q_\text{noisy} = n^2 t^2 \Big(1-\big(2-\frac{4}{3n} \big) \gamma t \Big) + \mathcal{O} ( t^4 ).
\end{equation}
In the short time limit, where the first two non-zero terms of the Taylor expansion
dominate the behaviour of the QFI, the HL is lost once the quantity $2 \gamma t$ becomes large. This is true regardless of the value of $\omega$, as depicted in Fig.~(\ref{fig:QFI_noECC}). This is not a practical time scale for quantum metrology \cite{huelga1997}, specifically in the interest of small values of $\omega$ where it is necessary for the system to evolve for a long enough time to distinguish between the effects of the signal and imperfections of real world measurement technologies.

\definecolor{green2}{HTML}{008000}

\begin{figure}
    \centering
    \begin{minipage}[c]{.65\textwidth}
    \includegraphics[width=\textwidth]{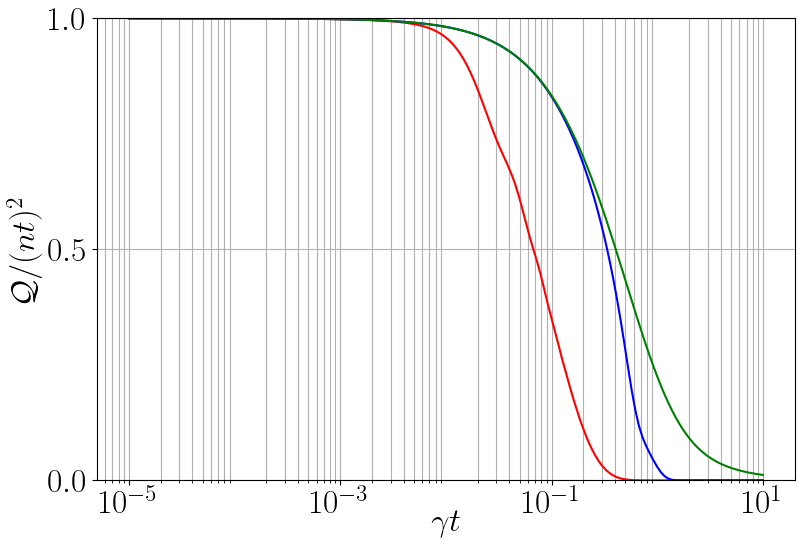}
    \end{minipage}
    \begin{minipage}[c]{.2\textwidth}
    \begin{tabular}{c l}
        \tikzcircle[red,fill=red]{0.7ex} & $\omega/\gamma = 100$ \\
         & \\
        \tikzcircle[blue,fill=blue]{0.7ex} & $\omega/\gamma = 1$ \\
         & \\
        \tikzcircle[green2,fill=green2]{0.7ex} & $\omega/\gamma = 1/100$ \\
    \end{tabular}
    \end{minipage}
    \caption{Normalized QFI $\mathcal{Q}/(nt)^2$ after an $n=10$ qubit GHZ state is used for phase estimation in the presence of environmental decoherence, Eq.~\eqref{eq:MasterNoisy}. Regardless of the signal-to-noise ratio, $\omega/\gamma$, the QFI tends to zero around $2\gamma t \approx 1$.}
    \label{fig:QFI_noECC}
\end{figure}

There are a number of proposed strategies to mitigate the effects of noise. A passive approach is to engineer the noise model so that is better suited for quantum metrology. For example, non-Markovian noise models\footnote{A Non-Markovian noise model is one which does not use the Born-Markov \cite{kolovsky2020} approximation to formulate the master equation.} can be tailored to outperform standard Markovian noise models \cite{chin2012, berrada2013}. Similarly, decoherence free subspaces (where qubit dephasing is not independent) outperform the standard uncorrelated depashing noise models \cite{dorner2012}. A more active approach is to monitor the effects of the environment using continuous measurements \cite{clerk2010, plenio2016, albarelli2018, rossi2020}.

Quantum control is a very promising technique to suppress the effects of noise on quantum metrology \cite{sekatski2017}. Broadly speaking, the quantum system is occasionally modified, and if done appropriately can reduce the impact of noise. \cite{zheng2015} proposes feedback to based off of a coupled to a cavity or reservoir to `reverse' the effects of noise. Dynamical decoupling protocols \cite{rong2011, souza2012, sekatski2016} apply a sequence of unitary operation in rapid succession can cancel out the effects of noise. Signal amplification in optical systems can be used to mitigate the effects of loss \cite{caves1981, ou2012, frascella2021}. This chapter focuses on incorporating quantum error correction as a means of control.

\section{Quantum Error Correction}

The fragility of quantum systems is a major obstacle for quantum computing \cite{unruh1995, raimond1996}. Suppose each quantum operation has a small probability of being done incorrectly: $\varepsilon \ll 1$; embedding an error in the quantum system. Then the probability that no errors occur after $N$ operations is $(1-\varepsilon)^N$, which will decrease to zero as $N$ increases. \textit{Quantum error correction} \cite{devitt2013} is a vital tool developed to combat the effects of noise and actualize fault tolerant quantum computing \cite{preskill1998}. Using clever encoding schemes, in which quantum states are encoded into larger systems (often called `logical' quantum states), the effects of environmental decoherence can be reduced substantially enough such that arbitrarily long protocols and computations can be fulfilled.

The \textit{no-cloning theorem}\footnote{It is impossible to create an independent and identical copy of an arbitrary unknown quantum state.} \cite{wootters1982} and the \textit{collapse of the wave-function} are the predominant reasons as to why classical error correction techniques cannot be seamlessly integrated into a quantum framework. Furthermore, qubits are susceptible to bit flips and phase flips, for which there is no classical analogue for the latter. Despite the challenges and constraints, primitive error correction models and protocols were established in the 1990's \cite{shor1995, steane1996a, steane1996b, calderbank1996, bennett1996, gottesman1996, gottesman1997, kitaev1997}. In the last two and half decades, the field of quantum error correction has flourished. Nowadays, a range of error correction protocols exist, such as topological codes \cite{kitaev1997, bombin2006, bombin2007}, permutation invariant codes \cite{pollatsek2004, ouyang2014, ouyang2019} and approximate codes \cite{leung1997, schumacher2002}; each with their own advantages and disadvantages. In addition, quantum error correction has been experimentally demonstrated using different resources, such as spin qubits \cite{dutt2007, taminiau2014, cramer2016}, continuous variable optical systems \cite{aoki2009} and superconducting circuits \cite{reed2012, ofek2016}.

\subsection{Example: Bit-Flip Code}

A bit-flip error, denoted by $\mathcal{E}$, maps the quantum state $\ket{0}$ to the quantum state $\ket{1}$ and vice versa. If $p$ is the probability of a bit-flip error, then
\begin{equation}
    \mathcal{E}(\rho) = (1-p) \rho + p X \rho X.
\end{equation}
The three qubit bit-flip code \cite{gottesman1997} is a rudimentary error correcting code designed to correct a single bit-flip error. The physical states $\ket{0}$ and $\ket{1}$ are encoded\footnote{The encoding can be implemented with two ancillary $\ket{0}$ states and controlled-$X$ operations.} into three qubit logical states $\ket{0_L}=\ket{000}$ and $\ket{1_L}=\ket{111}$ respectively, and in general,
\begin{equation}
    \ket{\psi} = \alpha \ket{0} + \beta \ket{1}  \rightarrow \ket{\psi_L} = \alpha \ket{0_L}+\beta \ket{1_L}. 
\end{equation}
Each of the three qubits physical qubits are independently susceptible to a bit-flip error
\begin{equation}
\begin{split}
    & \mathcal{E}( \dyad{\psi_L} ) = \\
    & (1-p)^3 \dyad{\psi_L} \\
    + & p(1-p)^2 \big(X_1 \dyad{\psi_L}X_1 + X_2 \dyad{\psi_L}X_2 + X_3 \dyad{\psi_L}X_3 \big) \\
    + & p^2(1-p) \big(X_1 X_2 \dyad{\psi_L}X_1 X_2 + X_1 X_3 \dyad{\psi_L}X_1 X_3 + X_2 X_3 \dyad{\psi_L} X_2 X_3 \big) \\
    + & p^3 X_1 X_2 X_3 \dyad{\psi_L} X_1 X_2 X_3,
\end{split}
\end{equation}
equivalently, there is a probability: $(1-p)^3$ of no errors occurring, $3p(1-p)^2$ of exactly one error occurring, $3p^2(1-p)$ of exactly two errors occurring, or $p^3$ of exactly three errors occurring. Assuming that $p$ is small, it is far more likely that $0$ or $1$ errors occur than $2$ or $3$ errors occur. Thus, by comparing the parity of the three qubits (which can be done using non-destructive and entangled measurements), one can apply a `majority-is-correct' correction rule, and (with high probability) recover the quantum state\footnote{The quantum state is recovered if zero or one errors occurred, but not if two or zero errors occurred. If $p \ll 1$, the former scenario is much more likely.}. Formally, this measurement is better known as a \textit{syndrome measurement} or \textit{syndrome diagnosis}, the measurement results are better known as \textit{error syndromes} and the correction rule is better known as a \textit{recovery operation}. The error syndromes and recovery operations of the bit-flip code are listed in Table~(\ref{tab:bitflip}).

\renewcommand*{\arraystretch}{1.5}
\begin{table}[ht]
    \centering
    \begin{tabular}{c|c}
        Error Syndrome & Recovery Operation \\
        \hline
        $\dyad{000}+\dyad{111}$ & $\mathbb{I}$ \\
        $\dyad{100}+\dyad{011}$ & $X_1$ \\
        $\dyad{010}+\dyad{101}$ & $X_2$ \\
        $\dyad{001}+\dyad{110}$ & $X_3$
    \end{tabular}
    \caption{The error syndromes and corresponding recovery operations for the bit-flip code.}
    \label{tab:bitflip}
\end{table}

\renewcommand*{\arraystretch}{1}

The bit-flip code, is not `technically' an error correction code, because, although it can correct bit-flip errors, it cannot correct any error, for example phase-flips. One can correct a single phase-flip using a similarly constructed phase-flip code \cite{gottesman1997}. Notably the 9-qubit code is constructed by superimposing the phase-flip and bit-flip code, which can correct any single qubit error \cite{shor1995}. It was later shown that any single qubit error can be corrected using a more compact code of five qubits \cite{laflamme1996}.

\section{Error Correction Enhanced Quantum Metrology}

It was shown in \cite{kessler2014} that repeated applications of error correction can be used to significantly increase the sensitivity of a quantum probe for quantum metrology. Since then, the extent of error correction enhanced quantum metrology has been well explored\footnote{As it happens, the converse setting of using mathematical techniques of quantum metrology for quantum error correction has also been explored in \cite{kubica2021}, where QFI bounds were used to provide a proof of the approximate Eastin-Knill Theorem.}: the general limitations have been established \cite{dur2014, arrad2014, lu2015, demkowicz2017, zhou2018, zhou2020}, and codes have been engineered for specific scenarios \cite{herrera2015, matsuzaki2017, layden2018, layden2019, zhuang2020, wang2021}. Error correction enhanced magnetometry has been experimentally realized in \cite{unden2016}, where the sensing time exceeded the natural dephasing times of the spin qubits.

For general Markovian noise, quantum error correction can be used to correct errors which can be distinguished from the Hamiltonian which encodes the signal (transverse noise). When the signal Hamiltonian and environmental noise commute (parallel noise), error correction cannot be used. Parallel noise can be corrected for non-Markovian noise models \cite{layden2018, layden2019} or using continuous measurements \cite{albarelli2018}. 

\subsection{Theoretical Limitations: Recovering the HL}

Recall from \textbf{Chapter 2} that the dynamics of a general Markovian noise model are governed by the master equation
\begin{equation}
    \dot{\rho}(t) = -\frac{i}{\hbar} [H, \rho(t)] + \sum_{j=1}^{d^2-1} \gamma_j \big[ L_k \rho(t) L_k^\dagger - \frac{1}{2} \big\{ \rho(t),L_k L_k^\dagger \big\} \big],
\end{equation}
$L_1,\ldots,L_{d^2-1}$ are Lindblad operators. It follows that, for a small time $\tau$, the evolution can be written as
\begin{equation}
    \label{eq:smallevo}
    \rho(t+\tau) = \rho(t)-\frac{i}{\hbar} [H, \rho(t)] \tau + \sum_{j=1}^{d^2-1} \gamma_j \big[ L_k \rho(t) L_k^\dagger - \frac{1}{2} \big\{ \rho(t),L_k L_k^\dagger \big\} \big] \tau + \mathcal{O} ( \tau^2 ).
\end{equation}

It was shown in \cite{demkowicz2017, zhou2018} that for a general transverse noise model, an error correction code can be constructed, which when applied, will not interrupt the encoding Hamiltonian, i.e
\begin{equation}
    \rho(t)-\frac{i}{\hbar} [H, \rho(t)] \tau
\end{equation}
and correct first order errors, i.e
\begin{equation}
    \sum_{j=1}^{d^2-1} \gamma_j \big[ L_k \rho(t) L_k^\dagger - \frac{1}{2} \big\{ \rho(t),L_k L_k^\dagger \big\} \big] \tau.
\end{equation}
The distinguishable criteria (transverse noise) is called Hamiltonian-not-in-Linblad span in \cite{zhou2018}, because the necessity condition is rephrased as
\begin{equation}
    H \notin \text{span} \{\mathbb{I},L_j,L_j^\dagger,L_j^\dagger L_k \},
\end{equation}
where the span is taken over all subscripts $j$ and $k$. It is demonstrated that, if the frequency at which error correction is performed is fast enough such that the higher order evolution terms are negligible, $\mathcal{O}(\tau^2) \rightarrow 0$, then the HL can be maintained indefinitely.

\subsection{Practical Limitations: Current Quantum Technologies}

Unfortunately, the mathematical assumption of arbitrarily fast error correction does not coincide with current quantum technologies. In fact, higher order error terms should not be ignored whatsoever, reason being that current error correction rates scale similarly to current dephasing rates \cite{dutt2007, schindler2011, taminiau2014, cramer2016, ofek2016}. The experimental realization of error correction enhanced quantum metrology \cite{unden2016} had a wait time between periods of error correction of 20$\mu$s (or 50kHz) - comparable to the reported decoherence rate of 30kHz. This experiment used a single NV center as a sensor and performed two applications of error correction.

Even supposing that the higher order terms $\mathcal{O}(\tau^2)$ in Eq.\eqref{eq:smallevo} are negligible compared to the first order approximation, the argument in itself falls short of expectations. If $t$ is the sensing time, and $\tau$ is the time between applications of error correction; the assumption of $\tau$ being arbitrarily small is equivalent with the number of rounds of error correction, $t/\tau$, being arbitrarily large. Although the higher order evolution term is negligible after a single round of error correction, which in turn adds a negligible amount of uncertainty to the final quantum state, this does not necessarily imply that the total uncertainty added to the quantum state after $t/\tau$ rounds is also negligible.

Furthermore, current quantum error correction technologies are not perfect. Ancillary qubits are also encumbered to the effects of noise. Syndrome diagnosis and recovery operations cannot be implemented with perfect fidelity. These imperfections will hinder the utility of the quantum state for quantum metrology.

\section{Our Model}

A more pragmatic approach for error correction enhanced quantum metrology is to make no assumptions regarding the time between applications of error correction and draw conclusions from an exact solution. It should be noted that the noise models in \cite{demkowicz2017, zhou2018} are completely general. Inevitably, obtaining an exact solution for an arbitrary noise model is infeasible, which is why we use a relevant noise model: dephasing in a direction orthogonal to the signal, see Eq.~\eqref{eq:MasterNoisy}. 

Similarly, we make use of a realizable error correction code: a parity check code \cite{hsieh2009,fujiwara2015, roffe2018}. A parity check code makes use of an ancillary qubit which is less sensitive to environmental interactions (and thus less noisy). For example, the experiment in \cite{unden2016} used an electron spin for sensing and a nuclear spin as the ancillary qubit. In each application of error correction, the syndrome diagnosis outputs the parity between individual sensing qubits and the ancillary qubit. The subsequent recovery operation will correct any qubits which demonstrated a difference in parity by applying an $X$ operation. 

\begin{figure}
    \centering
    \includegraphics[width=0.85\textwidth]{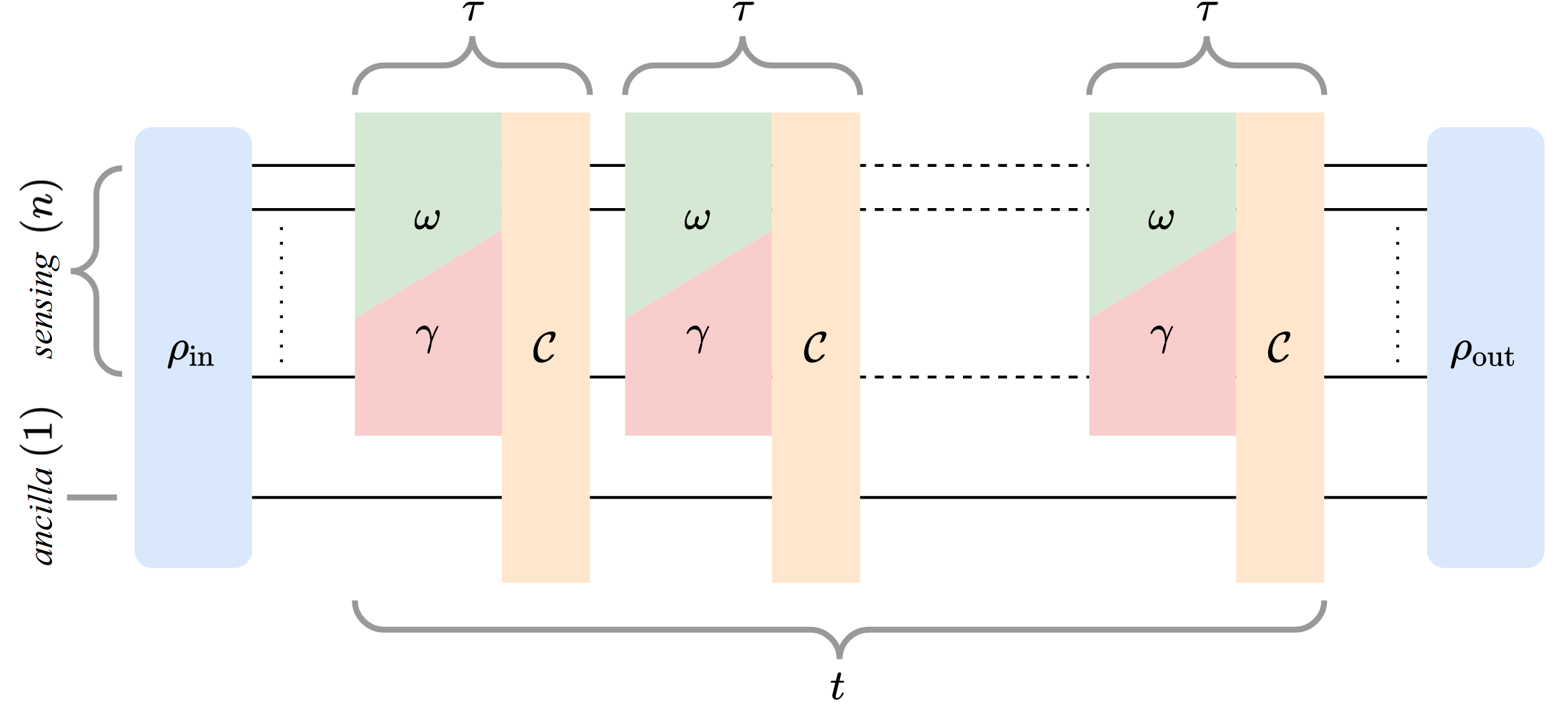}
    \caption{Schematic of our error correction enhanced quantum metrology model. The input state, $\rho_\text{in}$, is initialized as an $n+1$ qubit GHZ state composed of $n$ sensing qubits and one ancillary qubit. The sensing qubits are influenced by a signal $\omega$ and dephasing $\gamma$. The parity check code, denoted by $\mathcal{C}$, is repeatedly applied after a given time $\tau$ to mitigate the effects of dephasing. The final quantum state used for parameter estimation, $\rho_\text{out}$, undergoes $t/\tau$ rounds of error correction. The scheme can easily be generalized; allowing for arbitrary input states, error correction strategies and more ancillary qubits.}
    \label{fig:ECCQM}
\end{figure}

In our model, exhibited in Fig.~(\ref{fig:ECCQM}), the quantum state is initialized as an $n+1$ qubit GHZ state, where $n$ qubits are used for sensing and the remaining one qubit (which is more resistant to environmental noise) acts as an ancilla for error correction. The sensing qubits are influenced by a signal $\omega$ and dephasing with rate $\gamma$. The sensing qubits evolve per Eq.~\eqref{eq:MasterNoisy} for time $\tau$, after which the parity check code is applied; the procedure is then repeated $t/\tau$ times where $t$ is the total sensing. Without loss of generality, it is assumed that $t/\tau$ is an integer. The set-up is similar to that of \cite{kessler2014}, however the authors disregard higher order error terms, which is similarly presumptuous to assuming an arbitrarily small $\tau$.

To augment the reality of our model, we account for other hindrances current error correction technologies are burdened by: noisy ancilla and imperfect syndrome diagnosis. The noisy ancilla is subjected to a dephasing rate $\xi$, which changes the master equation to
\begin{equation}
    \label{eq:MasterNoisyWithAncilla}
    \frac{d \rho}{d t} = -\frac{i}{\hbar} [H,\rho] + \gamma \sum_{m=1}^n (X_m \rho X_m - \rho )+\xi(X_{n+1} \rho X_{n+1} - \rho ),
\end{equation}
where the ancillary qubit is indexed by the subscript $n+1$. Imperfect syndrome diagnosis is simulated by assuming that the syndrome diagnosis is incorrect with probability $p$, which results in an unnecessary recovery operation (or lack thereof).

\section{Results}

A completely general result for the final quantum state after $t/\tau$ rounds of error correction with a noisy ancilla and imperfect syndrome diagnosis is derived in \textbf{Appendix B}. The general solution is quite complicated and difficult to analyse. For clarity, each subcase is analysed individually: i) ideal error correction ($\xi=0$, $p=0$), ii) noisy ancilla ($\xi \neq 0$, $p=0$),  and iii) imperfect syndrome diagnosis ($\xi=0$, $p \neq 0$).

\begin{figure}[!ht]
    \centering
    \includegraphics[width=\textwidth]{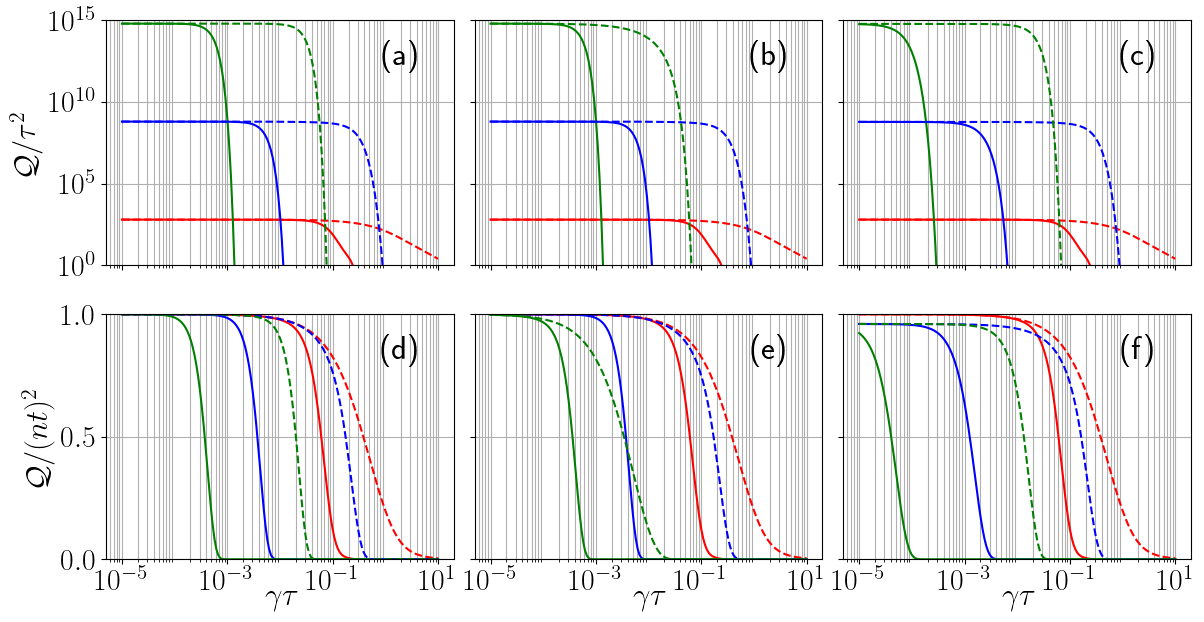}
    
    \vspace{0.7ex}

    \begin{tabular}{ l m{1cm}  l }
        \begin{tikzpicture}[baseline=-0.45ex]
            \draw[line width=0.3 mm, color=black] (0,0) -- (0.7,0);
        \end{tikzpicture} $\omega/\gamma=20$ & &
        \tikz[baseline=-0.75ex]\draw[red,fill=red] (0,0) circle (1ex);  No QECC \\
        \begin{tikzpicture}[baseline=-0.45ex]
            \draw[line width=0.3 mm, color=black, densely dashed] (0,0) -- (0.7,0);
        \end{tikzpicture} $\omega/\gamma=1/20$ & &
        \tikz[baseline=-0.75ex]\draw[blue,fill=blue] (0,0) circle (1ex);  $t=10^3 \tau$ \\
          & &
        \tikz[baseline=-0.75ex]\draw[green2,fill=green2] (0,0) circle (1ex);  $t=10^6 \tau$ \\
    \end{tabular}
    \caption{Plot of $\mathcal{Q}/\tau^2$ for an $n=25$ qubit GHZ state after undergoing repeated error correction with (a) ideal error correction, (b) a noisy ancilla ($\xi/\gamma=10^{-4}$), and (c) imperfect syndrome diagnosis ($p=0.01$), with total sensing times $t/\tau=10^3,10^6$. The characteristics of a noisy state without the inclusion of a quantum error correction code (QECC) after sensing time $t=\tau$ is also displayed.    As the total sensing time $t$ increases, the necessary rate at which error correction is needed to maintain the HL increases. Hence the reason why the curve with $t=10^6 \tau$ begins to decrease before the curve with $t=10^3 \tau$, which similarly begins to decrease before the curve without the application of the error correction code.    The curves are cutoff when $\mathcal{Q}/\tau^2=1$ for clarity purposes. Additionally, we illustrate the corresponding normalized QFI curves, $\mathcal{Q}/(nt)^2$, in plots (d), (e) and (f) respectively, to emphasize the deviation from the HL.}
    \label{fig:QFIPlotsECC1}
\end{figure}
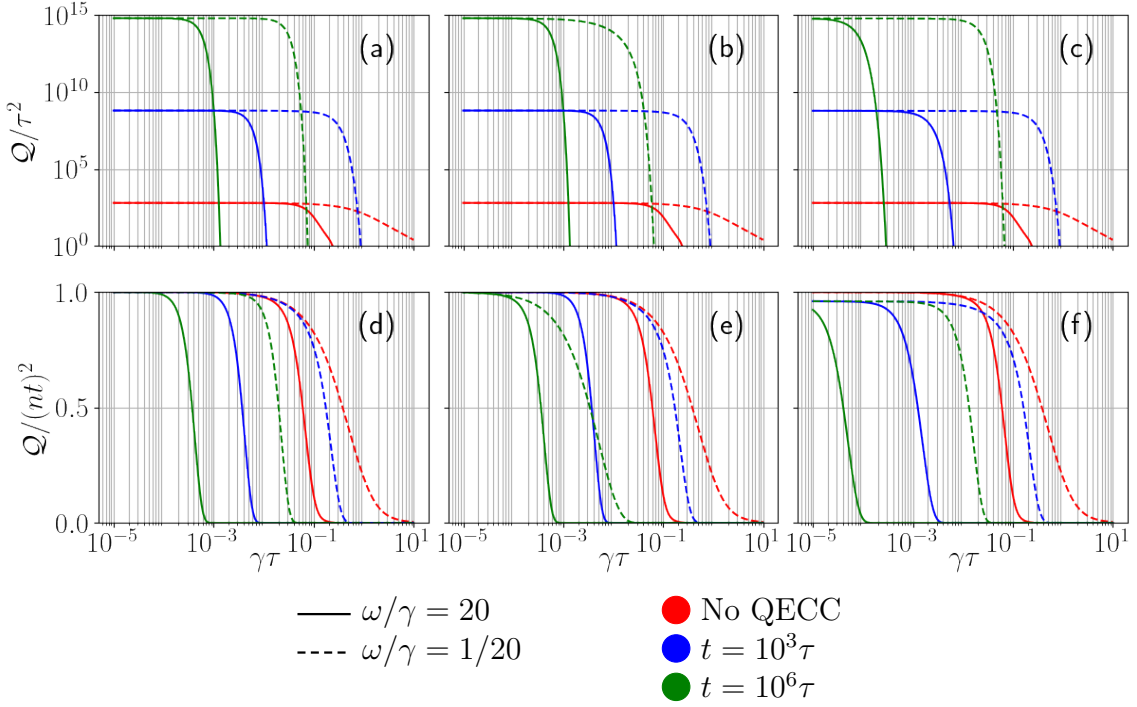

\begin{figure}[!ht]
    \centering
    \includegraphics[width=\textwidth]{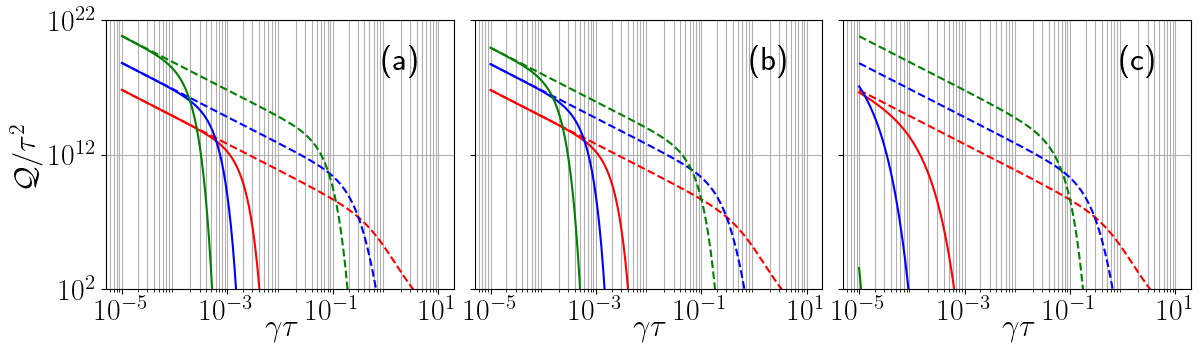}
    
    \vspace{0.7ex}

    \begin{tabular}{ l m{1cm}  l }
        \begin{tikzpicture}[baseline=-0.45ex]
            \draw[line width=0.3 mm, color=black] (0,0) -- (0.7,0);
        \end{tikzpicture} $\omega/\gamma=20$ & &
        \tikz[baseline=-0.75ex]\draw[red,fill=red] (0,0) circle (1ex);  $\gamma t = 100$ \\
        \begin{tikzpicture}[baseline=-0.45ex]
            \draw[line width=0.3 mm, color=black, densely dashed] (0,0) -- (0.7,0);
        \end{tikzpicture} $\omega/\gamma=1/20$ & &
        \tikz[baseline=-0.75ex]\draw[blue,fill=blue] (0,0) circle (1ex); $\gamma t = 1000$ \\
          & &
        \tikz[baseline=-0.75ex]\draw[green2,fill=green2] (0,0) circle (1ex);  $\gamma t = 10000$ \\
    \end{tabular}
    \caption{An alternative perspective on illustrating the tendencies of $\mathcal{Q}/\tau^2$ for an $n=25$ qubit GHZ state after repeated applications of error correction. Here, the total sensing time $\gamma t$ is held constant and deviations from the linear curve on the log-log plot represent the QFI tending away from the HL and towards a QFI of zero. The same scenarios are plotted: (a) ideal error correction, (b) a noisy ancilla ($\xi/\gamma=10^{-4}$), and (c) imperfect syndrome diagnosis ($p=0.01$). Without any error correction, the QFI after total sensing times $\gamma t=100,1000,10000$ is effectively zero. Note that the three values chosen for the total sensing time, $\gamma t$, deviate less than the values chosen in Fig.~(\ref{fig:QFIPlotsECC1}). This choice was intentional to properly illustrate the scenario of imperfect syndrome diagnosis. Regardless, the curves display here have an analogous curve with similar tendencies displayed in Fig.~(\ref{fig:QFIPlotsECC1}).}
    \label{fig:QFIPlotsECC2}
\end{figure}
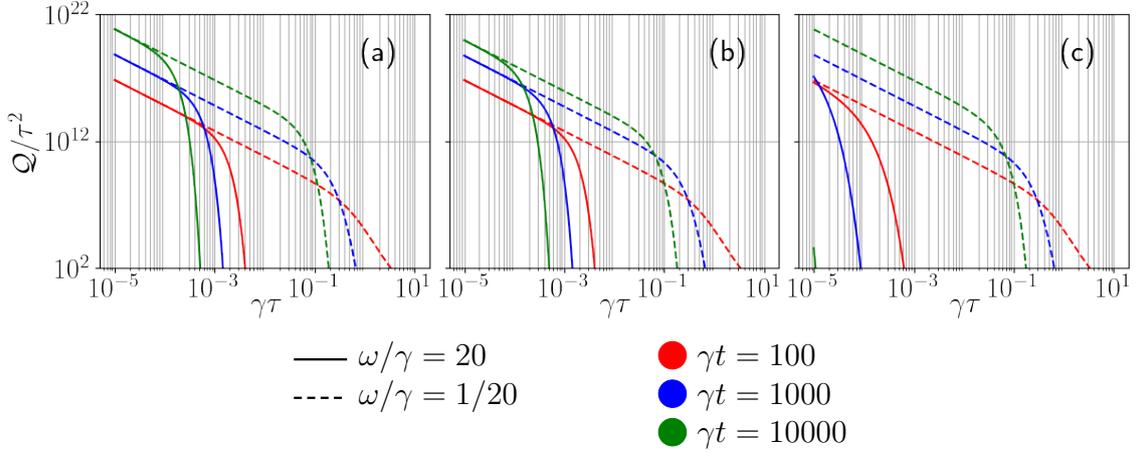

\subsection{Ideal Error Correction}

In the ideal error correction scenario (noiseless ancilla and perfect error correction), after $t/\tau$ rounds of error correction, the final quantum state can be expressed as a bipartite mixed state
\begin{equation}
\label{eq:finalstate}
\rho = \frac{1+r^{nt/\tau}}{2}\dyad*{\psi_+} + \frac{1-r^{nt/\tau}}{2} \dyad*{\psi_-},
\end{equation}
where
\begin{equation}
\ket*{\psi_\pm}= \frac{1}{\sqrt{2}} \big(\ket{0}^{\otimes n+1} \pm e^{i n \phi t/\tau} \ket{1}^{\otimes n+1} \big),
\end{equation}
and
\begin{equation}
\label{eq:small_r}
 re^{\pm i \phi}= e^{-\gamma \tau} \big(\cos (\Delta \tau )+ \frac{\gamma \pm i \omega}{\Delta}\sin ( \Delta \tau ) \big),
\end{equation}
with $\Delta=\sqrt{\omega^2-\gamma^2}$. There is no mathematical issue when $\omega^2 < \gamma^2$, in this regime the trigonometric functions are replaced by the corresponding hyperbolic functions (as per their definition). Because the quantum state is evaluated immediately after the $t/\tau$th application of error correction, a mixture of GHZ-like states is obtained. Assuming that $\gamma \tau > 0$ and $\omega \neq 0$, it follows that
\begin{equation}
    r^2 = e^{-2 \gamma \tau} \big( 1 + \frac{\gamma}{\Delta} \sin ( 2 \Delta \tau ) + \frac{2\gamma^2}{\Delta^2} \sin^2 ( \Delta \tau ) \big) < e^{-2 \gamma \tau} \big( 1 + \sinh (2 \gamma \tau) + \sinh^2 ( \gamma \tau) \big)=1,
\end{equation}
consequently, the quantum state becomes more mixed (and less useful for quantum metrology) once the quantity $n t/\tau$ becomes very large.

The QFI of the quantum state in Eq.~\eqref{eq:finalstate} can be written in the form
\begin{equation}
\label{eq:ECCQFI1}
\mathcal{Q}_1 = n^2 t^2 r^{2nt/\tau} f,
\end{equation}
where for small times $\tau$,
\begin{equation}
\label{eq:propconst1}
f = 1-2\gamma \tau + \mathcal{O} (\tau ^2 ).
\end{equation}
It is immediately clear that a Heisenberg level of precision is obtained if two conditions are met. The first being that $2\gamma \tau \ll 1$; it was derived in \cite{demkowicz2017, zhou2018} and is equivalent to the constraint for noisy metrology without quantum error correction, Eq.~\eqref{eq:noisymetro_noecc}. The second condition is $r^{2nt/\tau} \approx 1$, which suggests that the HL cannot be maintained indefinitely in a noisy environment (because $r^2 < 1$) and that the QFI will eventually tend to zero. For small $\tau$ we have
\begin{equation}
    r^{2nt/\tau} = 1-\frac{4}{3}n (\omega \tau)^2 \gamma t+ \mathcal{O} (\tau^3 ),
\end{equation}
meaning the second condition can be written as $\frac{4}{3} n \omega^2 \tau^2 \gamma t \ll 1$. This condition goes unnoticed in \cite{demkowicz2017, zhou2018} because it is of second order with respect to $\tau$.

Both of these conditions are illustrated in Fig.~(\ref{fig:QFIPlotsECC1}a), Fig.~(\ref{fig:QFIPlotsECC1}d) and Fig.~(\ref{fig:QFIPlotsECC2}a), where the QFI of an $n=25$ qubit GHZ state is plotted. In the regime $\omega^2 \gg \gamma^2$ ($\omega/\gamma = 20$), the HL of precision is lost once $r^{2nt/\tau}$ begins to tend to zero.

For each value of $r$ there is a critical value which the exponent will take such that the QFI will begin to rapidly converge to zero. Hence the difference in values of $\gamma \tau$ for when the curves with $\omega/\gamma = 20$ in Fig.~(\ref{fig:QFIPlotsECC1}a), Fig.~(\ref{fig:QFIPlotsECC1}d) and Fig.~(\ref{fig:QFIPlotsECC2}a) deviate from the HL.

In the regime $\omega^2 \ll \gamma^2$ ($\omega/\gamma = 1/20$), the HL level of precision is lost once $\gamma \tau \approx 10^{-2}$, regardless of if $t=10^3\tau$ or $t=10^6\tau$. The stark contrast in the families of curves ($\omega^2 \gg \gamma^2$ versus $\omega^2 \ll \gamma^2$) is due to larger deviations from the ideal case when $\omega^2 \gg \gamma^2$. Information about $\omega$ is stored in the relative phase, $n \phi t/\tau $, and if an error does occur between applications of error correction, the phase will deviate further from the ideal case. Thus, each round of error correction introduces a small amount of variance to the phase which scales with the magnitude of $\omega$.

In the noisy scenario without error correction, the optimal sensing time (which maximizes the QFI) is $t_\text{opt} \approx 1/(n \gamma)$ \cite{chaves2013}. The analogous quantity for the error correction enhanced setting can be computed by first realizing that $\frac{\partial f}{\partial t} = \mathcal{O} ( \tau )^2$, therefore the optimal sensing time is obtained by (approximately) maximizing the quantity $t^2r^{2nt/\tau}$. The resulting optimal sensing time is
\begin{equation}
    t_\text{opt} = \frac{1}{\frac{2}{3}n \gamma \omega^2 \tau^2+\mathcal{O} ( \tau^3 )}.
\end{equation}
As expected, $t_\text{opt}$ increases as $\tau$ decreases, and decreases as $n$ increases. The dependence on $\omega$ is linked to the effective variance in the phase of the quantum state.

\subsection{Noisy Ancilla}

The inclusion of the noisy ancilla alters the QFI to be
\begin{equation}
\label{eq:ECCQFI2}
\mathcal{Q}_2 =n^2 t^2 r^{2nt/\tau}(f-g\xi) + \mathcal{O} ( \xi^2 ),
\end{equation}
where $g$ is bounded by
\begin{equation}
\Big(\frac{2}{3}-7\gamma \tau \Big)t \leq g + \mathcal{O} (  \tau^2 ) \leq \frac{5}{2}(t+\tau),
\end{equation}
which can be interpreted as another necessary condition to obtain a Heisenberg-like scaling: $\xi t \ll 1$. This is not very surprising, since error correction will becomes less effective as time increases, and ultimately become ineffectual once the ancilla decoheres, $t \approx 1/\xi$. The new condition is displayed in Fig.~(\ref{fig:QFIPlotsECC1}b) and Fig.~(\ref{fig:QFIPlotsECC1}e), in which the ancillary qubit is set to have a dephasing rate 10000 times weaker than the sensing qubits. The noisy ancilla causes the HL to be lost sooner when compared to the case with a noiseless ancilla. The impact is more pronounced for the curve with $\omega/\gamma=1/20$ and $t=10^6 \tau$, where the loss of the HL is strictly due to $\xi t$ becoming too large instead of $\gamma \tau$. In Fig.~(\ref{fig:QFIPlotsECC2}b), where the total sensing time is static (and thus $\xi t$ is a constant), the QFI curve with $\gamma t =10000$ is noticeable shifted when compared to the same curve with ideal error correction in Fig.~(\ref{fig:QFIPlotsECC2}a). The problem of noisy ancillary qubits can be overcome by occasionally re-initializing the ancillary qubit (before it becomes too noisy) using an additional layer of error correction.

\subsection{Imperfect Syndrome Diagnosis}

The second hindrance explored is the inclusion of imperfect syndrome diagnosis due to flaws in the error correction hardware. To model this, for each instance of error correction, there is a probability $p$ that the parity measurement between a sensing qubit and the ancillary qubit is incorrect. Hence, if there is a difference parity, then no error correction is performed with probability $p$. Similarly, if there is no difference in parity (and no correction is needed), there is also a probability $p$ that an unnecessary correction is performed. An unnecessary correction (or lack thereof) will subject the quantum state to additional noise. Furthermore, each round of error correction introduces a small amount of variance to the quantum state due to the imperfect hardware, which will grow as the number of rounds of error correction increases. With the inclusion of imperfect syndrome diagnosis, the QFI is
\begin{equation}
\label{eq:ECCQFI3}
\mathcal{Q}_3=n^2 t^2 (rq)^{2nt/\tau}h,
\end{equation}
with
\begin{equation}
q^{2nt/\tau}=1-4p(1-p)\omega^2 t \tau  + \mathcal{O} ( \tau^2 ),
\end{equation}
and
\begin{equation}
h=(1-2p)^2f+4p\Big( \frac{1-p}{n}+1-2p \Big) \frac{\tau}{t} + \mathcal{O} ( \tau^2 ).
\end{equation}
The inclusion of imperfect syndrome diagnosis makes the true HL unattainable; $\mathcal{Q} \rightarrow n^2 t^2 (1-2p)^2$ as $\tau \rightarrow 0$. The multiplicative factor $(1-2p)^2$ is a result of the added uncertainty from each application of error correction. The exponential term in Eq.~\eqref{eq:ECCQFI3}, $(rq)^{2nt/\tau}$, must be approximately equal to $1$ to achieve Heisenberg-like precision. This is a more strict version of $r^{2nt/\tau} \approx 1$, and is again due to deviations in the relative phase, which are amplified by the imperfect syndrome diasgnosis. This stronger condition can be seen in Fig.~(\ref{fig:QFIPlotsECC1}c), Fig.~(\ref{fig:QFIPlotsECC1}f) and Fig.~(\ref{fig:QFIPlotsECC2}c), in which the probability of faulty syndrome diagnosis is $1\%$. The additional condition of $q^{2nt/\tau} \approx 1$ is more pronounced in the regime where $\omega^2 \gg \gamma ^2$. The upper bound of precision is displayed in Fig.~(\ref{fig:QFIPlotsECC1}f); as $\gamma \tau \rightarrow 0$, $\mathcal{Q}/(nt)^2 \rightarrow (1-2p)^2 \approx 0.96$.

\subsection{Fisher Information}

Given that the achievable precision of a metrology problem is also constrained by the estimation strategy, a more practical figure of merit is the Fisher information with respect to implementable estimation strategies. Consider measuring the output quantum state, Eq.~\eqref{eq:finalstate}, in the basis spanned by $\{ \ket{\alpha_+}, \ket{\alpha_-} \}^{\otimes (n+1)}$, in which
\begin{equation}
    \ket{\alpha_\pm} = \frac{1}{\sqrt{2}}(\ket{0} \pm e^{i \alpha} \ket{1}),
\end{equation}
and reverse engineering the measurement results to estimate $\omega$. Because of the symmetry of $\rho$, one only needs to consider the projectors of the form
\begin{equation}
    E_j = \dyad{\alpha_+}^{\otimes n+1-j} \otimes \dyad{\alpha_-}^{\otimes j},
\end{equation}
where
\begin{equation}
\Tr \big( E_j \rho \big) = \frac{1+(-1)^j R \cos \big( \theta - \alpha \big)}{2^{n+1}},
\end{equation}
with $R=r^{nt/\tau}$ and $\theta=n \phi t/\tau$. The Fisher information of this estimation strategy is
\begin{equation}
    \mathcal{I} = \sum_j \frac{\Tr \big( E_j \dot{\rho} \big)^2}{\Tr \big( E_j \rho \big)}=\frac{\Big( \dot{R} \cos \big( \theta - \alpha \big) - R \dot{\theta}\sin \big( \theta - \alpha \big) \Big)^2}{1-R^2 \cos \big( \theta - \alpha \big)},
\end{equation}
where the notation $\dot{\square}=\partial_\omega \square$ is used for conciseness. If $\alpha$ is chosen such that $\cos(\theta-\alpha) \approx 0$, then this estimation strategy approximately saturates the QFI
\begin{equation}
    \mathcal{I} = \mathcal{Q} + \mathcal{O} ( \tau ^2 ).
\end{equation}
Of course, this requires exact knowledge of $\omega$ to implement perfectly, which defeats the purpose of quantum metrology. However, this could be implemented with a high degree of precision using an adaptive estimation strategy \cite{gill2005, fujiwara2006, wiseman2009, pang2017}. On a similar note, saturating the QCRB requires that the value of $\gamma$ is precisely known. Any uncertainty in the noise model will naturally translate to uncertainty in the estimation of $\omega$. Alternatively, if $\gamma$ is unknown, one can consider estimating both $\omega$ and $\gamma$ in simultaneity, i.e. consider the setting as a multiparameter quantum metrology problem.

\subsection{QFI and Entanglement}

In \textbf{Chapter 3} a relationship between the geometric measure of entanglement $G$ and the QFI is given by \cite{augusiak2016}
\begin{equation}
    \mathcal{Q}(\rho_\theta) \leq n + 8n^2 \sqrt{G(\rho_\theta)}.
\end{equation}
This is an inequality and not an equality because entanglement is a necessary condition and not a sufficient condition \cite{oszmaniec2016}. However, the relationship between QFI and entanglement is much more pronounced for quantum states of the form
\begin{equation}
\rho = \frac{1+R}{2}\dyad*{\psi_+} + \frac{1-R}{2} \dyad*{\psi_-},
\end{equation}
with
\begin{equation}
\ket*{\psi_\pm}= \frac{1}{\sqrt{2}} \big(\ket{0}^{\otimes N} \pm e^{i \theta} \ket{1}^{\otimes N} \big).
\end{equation}
Hence, $\rho$ is a highly entangled pure state when $R=1$ and a mixture of two separable states when $R=0$. Using the recipe for rank-2 mixed symmetric mixed states in \cite{das2016}, the geometric measure of entanglement for the above quantum state is
\begin{equation}
    G(\rho) = \frac{1}{2}(1-\sqrt{1-R^2}).
\end{equation}
Therefore, the `quantum part' of the QFI can be written
\begin{equation}
    R^2 \dot{\theta}^2 = 4 G (1-G) \dot{\theta}^2.
\end{equation}
This result, albeit interesting, is mostly a bi-product of the fact that the initialized quantum state was a maximally entangled GHZ state. It is not surprising that the deterioration of the entanglement and the loss of the HL are dependent on the same quantity $R^2$. In fact, many quantities which measure some aspect of `quantum-ness' are similarly dependent, such as purity
\begin{equation}
    \Tr \rho^2 = \frac{1+R^2}{2},
\end{equation}
and Von Neumann entropy
\begin{equation}
    -\Tr \rho \log \rho = -\frac{1+R}{2} \log \Big( \frac{1+R}{2} \Big) -\frac{1-R}{2} \log \Big( \frac{1-R}{2} \Big).
\end{equation}

\section{Current Technologies}

\begin{figure}
    \centering
    \begin{minipage}[c]{.77\textwidth}
    \includegraphics[width=\textwidth]{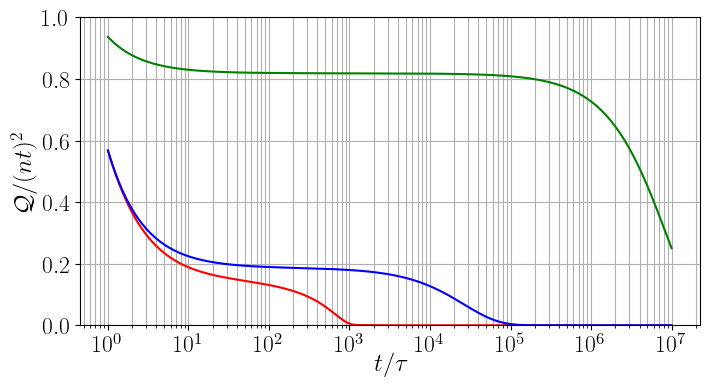}
    \end{minipage}
    \begin{minipage}[c]{.2\textwidth}
    \begin{tabular}{c l}
         & $\gamma \tau =1$ \\
        \tikzcircle[red,fill=red]{0.7ex} & $\xi \tau = 0.002$ \\
         & $p=0.06$ \\\vspace{1ex}
         & \\
         & $\gamma \tau =1$ \\
        \tikzcircle[blue,fill=blue]{0.7ex} & $\xi \tau = 0$ \\
         & $p=0.001$ \\\vspace{1ex}
         & \\
         & $\gamma \tau =0.1$ \\
        \tikzcircle[green2,fill=green2]{0.7ex} & $\xi \tau = 0$ \\
         & $p=0.001$ \\\vspace{1ex}
    \end{tabular}
    \end{minipage}
    \caption{Normalized QFI in the small signal regime $\omega/\gamma =0.01$. \tikzcircle[red,fill=red]{0.7ex} Today's quantum technologies ($\gamma^{-1} = \tau = 10^{-6}$s, $\xi^{-1}= 5\times10^{-4}$s, $p=0.06$) \cite{dutt2007, taminiau2014} suggest a QFI of $\sim 20\%$ of the HL can be attained for sensing times $t=10^1 \tau$. \tikzcircle[blue,fill=blue]{0.7ex} With improved error correction hardware and a noiseless ancilla, this can be sustained for a sensing time $t=10^3 \tau$. \tikzcircle[green2,fill=green2]{0.7ex} The QFI is significantly improved when the rate of error correction is increased by a factor of ten.}
    \label{fig:QFIBenchmarking}
\end{figure}

In \cite{dutt2007, taminiau2014}, the electron spin of a nitrogen-vacancy center is entangled to carbon-13 nuclear spins. The nuclear spins act as ancillary qubits, and error correction is performed on the electron spin using the parity check code. The reported dephasing rates are $\gamma^{-1} \sim 10^{-6}$s and $\xi^{-1} \sim 5\times10^{-4}$s. The error correction is being performed on a comparable timescale of $\tau \sim 10^{-6}$s, with infidelity reported at $p=0.06$ \cite{taminiau2014}. In Fig.~(\ref{fig:QFIBenchmarking}), this data is used to benchmark the abilities of current error correction technologies as a means of enhancing the precision of a noisy quantum metrology scheme in the regime $\omega^2 \ll \gamma^2$. Because the error correction rate is similar to the dephasing rate, the HL is unattainable. This is still the case with better hardware: $p=0.001$ and $\xi=0$ (the latter is justified by regularly re-initializing the ancilla). Notably, if the error correction rate increases by a factor of ten, the achievable QFI is $80\%$ of the HL for a total sensing time of $t=10^5 \tau$. This greatly outclasses the precision achieved in current experiments \cite{taylor2008, wasilewski2010, razzoli2019}. Although this result is promising, it is important to realize that experiments are hindered by more than what is considered in Fig.~(\ref{fig:QFIBenchmarking}), such as parallel noise, imperfect gate fidelity when applying the recovery operations and flaws in the quantum state initialization.

\section{Other Noise Mitigation Strategies}

In striving for an exact solution, it was necessary to consider a specific noise model and a specific error correction protocol. Whereas \cite{demkowicz2017, zhou2018} make no assumptions regarding the noise model. Although a completely general result is more satisfying, it is unfeasible with our methods. Nevertheless, our model and mathematical methodology are easy to adapt. One can substitute any combination of noise model and error correction strategy in place of iid dephasing and the parity check code respectively. In fact, repeated error correction can be forgone entirely and replaced with a suitable quantum control technique \cite{sekatski2017}, such as dynamical decoupling \cite{rong2011, sekatski2016} or reservoir engineering \cite{schirmer2010, zheng2015}.

We conjecture that, just as with discrete applications of the parity check code, for any noise mitigation strategy, the QFI will depend on term similar to $r^{2nt/\tau}$: one which suggests that there exist a critical time where the QFI begins to tend to zero. In fact, using the $n$ qubit bit flip code \cite{gottesman1997} yields the results
\begin{equation}
\mathcal{Q} = n^2 t^2 r^{2nt/\tau}f+\mathcal{O} (  \tau^{\frac{n-1}{2}} ).
\end{equation}
Hence, for large $n$ the QFI using the bit flip code is effectively the same as if one utilizes the parity check code. The reasoning supporting the aforementioned conjecture is that any errors which occur will cause the relative phase to deviate from the ideal value, and the deviation will remain even after performing a correction. Thus, after each round of error correction, the variance in the phase will increase, which propagates to an increase in variance in the eventual estimate of $\omega$. This conjecture is easily extended to any realistic noise model; as one expects the relative phase to deviate from the ideal value after performing error correction, regardless of the noise model.

We are not suggesting that the parity check code is the most efficient noise mitigation strategy (for transverse noise) at retaining the HL. For example, an adaptive parameter estimation \cite{gill2005, fujiwara2006, wiseman2009, pang2017} could be used to supplement the parity check code by incorporating an unitary operation which approximately corrects the deviations in the relative phase. This strategy is more difficult to implement, as the unitary rotations would be quite small and unlikely to be accurately realizable with current quantum hardware. Indisputably, as quantum technologies continue to improve, and the frequency at which these noise mitigation tools can be applied increases, so too does our ability to maintain the HL for increased sensing times.

\section{Discussion}

Our analysis is in agreement with previous results \cite{demkowicz2017, zhou2018}, which suggests that the inequality $2 \gamma \tau \ll 1$ is crucial for an error correction enhanced quantum metrology scheme to maintain a Heisenberg-like scaling. However, the findings in \cite{demkowicz2017, zhou2018} are based on the assumption that higher order terms are negligible, $\mathcal{O}( \tau^2 ) \rightarrow 0$, and as a result, Heisenberg-like scaling can be maintained permanently with repeated applications of (arbitrarily fast) error correction. This is not in accordance with today's quantum technologies, as the rate at which error correction can be performed is on a similar order of magnitude to the dephasing rate of physical qubits \cite{dutt2007, schindler2011, taminiau2014, cramer2016, ofek2016}. When the assumption $\mathcal{O}( \tau^2 ) \rightarrow 0$ is discarded, a second necessary condition to maintain the Heisenberg-like scaling emerges, $r^{2nt/\tau} \approx 1 \rightarrow \frac{4}{3} n \omega^2 \tau^2 \gamma t \ll 1$. 

Whenever an error occurs, it causes the phase to deviate from the ideal value of $n \omega t$, which is why the HL cannot be maintained indefinitely. That being said, in practise, no quantum metrology requires indefinite sensing time. For spin quibts, a more appropriate upper bound could be the relaxation time, which is typically a few orders of magnitude larger than the dephasing time \cite{wang2017}. With the limitations of current technologies, Fig.~(\ref{fig:QFIBenchmarking}), this may as well be indefinite.

We specifically analyse the effects of repeated applications of error correction for the specific case when the probe state is initialized in an $n$ qubit GHZ state. A logical generalization is to expand the results to a broader scope of initial states; such as squeezed states \cite{gross2012, zhang2014}, symmetric states \cite{toth2014}, or bundled graph states (\textbf{Chapter 4}). It is possible that these quantum states (which do not achieve the true HL, but do achieve a quantum advantage and Heisenberg-like scaling) are more robust to the effects of noise and can maintain a quantum advantage for a longer total sensing time when enhanced by error correction.

Further, we chose to analyse a dephasing noise model, something which is more applicable to atomic systems \cite{shaji2007, borregaard2013, macieszczak2014, zheng2014}. A future perspective is to consider noise models more relevant to optical systems, such as loss and phase diffusion \cite{lee2009, demkowicz2009, knysh2011, zhang2013, demkowicz2015}. Error correction codes for continuous variable systems are typically more complex \cite{paz1998, sarovar2005, zhuang2020}; it is not obvious how our results translate to these systems, if at all.

%% file: Chapters/Chapter6-Crypto.tex
\chapter{Quantum Cryptography for Quantum Metrology}

Quantum channels are likely to be the most vulnerable aspect of quantum communication protocols. Without proper cryptographic precautions, a malicious adversary can intercept the information being sent through a quantum channel while the honest parties remain none the wiser. As quantum network sensing and spatially distributed schemes become increasingly popular \cite{komar2014, proctor2018, rubio2020a}, it is important to verify which techniques from quantum cryptography are compatible with quantum metrology.

Until very recently, quantum metrology and quantum cryptography were non-overlapping disciplines. Gradually, the idea of security has been introduced to quantum metrology by considering scenarios involving unsecured quantum channels \cite{xie2018, huang2019}, delegated measurements to an untrusted party \cite{takeuchi2019, okane2020, yin2020}, or unwanted eavesdroppers \cite{kasai2021}. Although this direction is new and exciting, the aforementioned references fail to quantify the effects a malicious adversary poses to the quantum metrology problem, i.e the effects on the estimate and its precision. This chapter addresses this problem by linking the cryptographic notion of soundness to an overall uncertainty added to the quantum resource, which propagates to the quantum metrology problem. 

In addition to developing a toolbox for the merging of quantum cryptography and quantum metrology, several cryptographic protocols are devised for a variety of cryptographically motivated settings. Such as quantum metrology with an unsecured quantum channel \cite{SMK21} and quantum metrology with delegated tasks \cite{SM21}. The protocols devised are completely private, meaning that even if a malicious adversary intercepts the quantum data, they cannot interpret it, and maintain the integrity of the underlying metrology problem with no more than a quadratic increase in the number of resources. More so, (most of) the protocols devised take into account the limitations of real world quantum hardware and use nothing more complex than local Clifford operations.

\section{Quantum Cryptography}

\textit{Cryptography} is the practise and study of data security. For a long time, up until the advent of the computer, cryptography was synonymous with encryption - a method to cipher and decipher a message. Without knowledge of the cipher, an adversary could not intercept and learn the contents of the message. Nowadays, in the digital age, cryptography is much more than just encryption, yet the general philosophy of data security remains. Sophisticated techniques are manufactured for a range of tasks, such as sender/receiver authentication, secure data storage, secure computation, et cetera. Cryptography is undeniably essential for safeguarding confidential information and establishing trust between severs in the digital era.

\textit{Quantum cryptography} is the natural generalization of cryptography where quantum mechanical properties are allowed to be exploited. The quantum framework is accompanied by advantages and disadvantages alike. It is advantageous as quantum systems have built-in security aspects due to the no-cloning theorem and the collapse of the wave function. It is disadvantageous in the fact that an adversary with a quantum computer is much more powerful than an adversary with a classical computer. For example, the modern (classical) RSA encryption scheme is based on the difficulty of factoring large numbers efficiently \cite{rivest1978}; this encryption scheme can be broken with Shor's factoring algorithm\footnote{No need for panic; Shor's factoring algorithm is very much out of reach for modern quantum technologies.} \cite{shor1994}. As such, quantum cryptography differentiates from classical cryptography in the notion of security. A cryptographic protocol is said to be \textit{computationally secure} if it is immune to an adversary with `reasonable' computational power and time. Whereas quantum cryptography protocols opt for \textit{unconditionally security}, which is to say that no assumptions are made about the adversaries' computational power and time.

The premise of the first formulation of quantum cryptography \cite{wiesner1983} was simple but powerful: by randomly encoding the bit `0' (`1') in either $\ket{0}$ ($\ket{1}$) or $\ket{+}$ ($\ket{-}$), then the value of the bit is completely concealed from a malicious adversary if they are not aware of the preparation basis. This result stems from the uncertainty principle, Eq.~\eqref{eq:Uncertainty}, which has no classical analogue. This concept paved the way to the famous BB84 protocol for quantum key distribution \cite{bennett1984}. Since then the applicability of quantum cryptography has thrived \cite{broadbent2016, pirandola2020}; for example: quantum money \cite{aaronson2009, bozzio2018}, quantum coin flipping \cite{ambainis2004, pappa2014}, verification of quantum processes \cite{ying2013, gheorghiu2019, zhu2019a} and blind quantum computing \cite{broadbent2009, barz2012, fitzsimons2017}.

Just as (classical) cryptography is essential for confidentiality and trust in the digital era, so too is quantum cryptography in the quantum era. This is the core idea supporting the integration of quantum cryptography into a quantum metrology problem. If the problem involves multiple parties or communication through a quantum channel, then it is imperative to use quantum cryptography to certify the results and maintain a notion of privacy. Otherwise, a malicious adversary who intercepts an encoded quantum state can either bias the estimation result or estimate the latent parameter themselves. However, the problem is not as simple as using existing cryptographic protocols; in addition to adding security and privacy, the cryptographic protocol must not interfere with the mechanisms of the quantum metrology problem.

\section{Cryptographic Figures of Merit}

There is no unique cardinal figure of merit for cryptographic protocols due to the sheer vastness of quantum cryptography in both functionality and perspectives. Ergo, a suitable figure of merit for a cryptographic protocol should be relevant to the scope of the protocol and provide a method of comparison between similar protocols. The protocols we devise for quantum metrology take inspiration from quantum message authentication \cite{barnum2002}, so it is natural use the same figures of merit: \textit{privacy} and \textit{soundness}. These are both commonly used for most cryptographic protocols whose aim is to verify/authenticate/certify a process. Other than quantum messages \cite{barnum2002}, examples include quantum state preparation \cite{zhu2019a, zhu2019b} and quantum computation \cite{fitzsimons2017}. Providentially, the soundness of a protocol can be related to the additional bias and uncertainty of the quantum metrology problem.

\subsection{Privacy}

Privacy is a straightforward concept which quantifies the amount of information a malicious eavesdropper can extract from a message (quantum or otherwise). The protocols outlined in this chapter are all \textit{completely private}, this is to say that an eavesdropper can extract no information. If an eavesdropper can access the quantum state $\rho_E$, then this is achieved if
\begin{equation}
    \mathbb{E} ( \rho_E ) = \mathbb{I}/d,
\end{equation}
where $d$ is the dimension of $\rho_E$. Thus, a protocol is completely private when the expected quantum state accessible to an eavesdropper is indistinguishable from the maximally mixed state.

Having a completely private protocol is paramount for quantum metrology, as this prevents an eavesdropper from learning anything about the unknown parameter for themselves. This was overlooked in the first work which established the idea of quantum metrology integrated into a cryptographic framework \cite{huang2019}, in the appendix of \cite{SMK21} we show that their protocol is not completely private and that an eavesdropper can go completely undetected while learning performing parameter estimation for themselves.

\subsection{Soundness}

For authentication schemes, the soundness of a cryptographic  protocol is the standard figure of merit used to judge the security of a protocol \cite{barnum2002}. In essence, the soundness of a protocol quantifies the ability of a malicious adversary to alter the quantum state whilst remaining undetected. The formal mathematical definition of soundness varies depending on the formulation of the cryptographic protocol \cite{barnum2002, fitzsimons2017, zhu2019a, takeuchi2019b}, and is sometimes referred to as verifiability \cite{gheorghiu2019}. The version used in the work presented in this thesis uses a slightly modified version of the definition presented in \cite{barnum2002}.

Authentication schemes have two outputs: a binary accept or reject clause as well as a quantum output. The quantum output varies as per the protocol, in this chapter, it will either be a quantum state or a measurement result. The protocols are also equipped with ancillary qubits, which are constructed to have a deterministic measurement outcome in an ideal scenario in which there is no malicious activity. If the expected measurement result is observed, the outcome of `accept' is assigned. However, if an unexpected result is observed, then it must be the result of malicious activity, and the outcome of `reject' is assigned. For the sake of unconditional security, no assumptions are made with respect to the computational power of a malicious adversary. More so, it is assumed that a malicious adversary is completely familiar with the inner mechanisms of the protocol. In order to dissuade a malicious adversary, the protocols are supported by a set of classical keys $\mathcal{K}$, where each key alters the protocol differently. Before implementing a protocol, a key is chosen at random, and even if a malicious adversary may have access to the set of possible keys, it is assumed that they do not have access to the random choice. If there are multiple trusted parties who need access to the key, it is assumed that the key can be shared securely. This can be accomplished through a secure classical channel or quantum key distribution \cite{bennett1984}. For all intents and purposes we assume that a malicious adversary cannot access the (random) choice classical key.

The mathematical definition of soundness is a bound on the probability of witnessing `accept', while the quantum output, $\rho_\text{out}$ is simultaneously far from the ideal $\rho_\text{id}$. In \cite{barnum2002}, the protocol is constructed for $\rho_\text{id}$ being a pure state, and the `distance' is recorded in $1-\Tr(\rho_\text{id} \rho_\text{out})$. In \cite{SM21}, the outputs are not necessarily pure states, and we generalize the `distance' as $1-\mathscr{F}(\rho_\text{id},\rho_\text{out})$. Both expressions are equivalent in the event that $\rho_\text{id}$ is a pure state. Mathematically, a protocol has soundness $\delta$ if 
\begin{equation}
    \label{eq:soundness}
    \frac{1}{|\mathcal{K}|} \sum_{k \in \mathcal{K}} p_\text{acc} (k, \Gamma ) \cdot \Big( 1- \mathscr{F} \big(\rho_\text{id}, \rho_\text{out} ( k,\Gamma ) \big) \Big) \leq \delta.
\end{equation}
Here, $\Gamma$ represents any possible attack a malicious adversary may perform, and $k \in \mathcal{K}$ is the specific key chosen. The probability of the protocol outputting `accept', $p_\text{acc}(k,\Gamma)$, and the output $\rho_\text{out}(k,\Gamma)$ are dependent on both of these quantities. A well designed protocol should be provide a sense of security for all malicious attacks, thus Eq.~\eqref{eq:soundness} must hold for all $\Gamma$.

When it can be written that $p_\text{acc} (k , \Gamma ) \geq \alpha $, then Eq.~\eqref{eq:soundness} can be transformed into the inequality
\begin{equation}
    \label{eq:soundnessfidelity}
    1-\mathbb{E} \Big(  F \big(\rho_\text{id}, \rho_\text{out} \big) \Big) \leq \frac{\delta}{\alpha},
\end{equation}
where the dependence of $\rho_\text{out}$ on the key $k$ and the attack $\Gamma$ has been omitted for clarity. The quantity $\alpha$ is sometimes referred to as the statistical significance \cite{zhu2019a}. This alternative formulation permits more easily permits the use of other common figures of merit which are intertwined with the soundness and statistical significance \cite{zhu2019a, zhu2019b}. More so, it will be shown that Eq.~\eqref{eq:soundnessfidelity} can be manipulated to determine the utility of $\rho_\text{out}$ for quantum metrology. This is done by bounding the trace distance, which can be found using the the Fuchs-van de Graaf inequalities \cite{fuchs1999}, Eq.~\eqref{eq:fuchs}, and the arithmetic-quadratic mean inequality
\begin{equation}
    \label{eq:soundnesstracedistance}
    \mathbb{E} \Big(  \mathscr{D} \big(\rho_\text{id}, \rho_\text{out} \big) \Big) \leq  \sqrt{ \mathbb{E} \Big(  \mathscr{D} \big(\rho_\text{id}, \rho_\text{out} \big)^2 \Big) } \leq \sqrt{1-\mathbb{E} \Big(  F \big(\rho_\text{id}, \rho_\text{out} \big) \Big)}  \leq \sqrt{\frac{\delta}{\alpha}}.
\end{equation}

\section{Cryptographic Quantum Metrology}

The first adaptation of a quantum metrology problem in a cryptographic framework can be found in \cite{komar2014}. In the article, an entangled state is distributed from a central node to several exterior nodes, where a local phase is encoded and sent back to the central node for phase estimation. The authors propose occasionally distributing non-entangled decoy qubits throughout the sensing network. These decoy qubits have a deterministic measurement and are used to detect and thwart malicious activity. As this was not focal point of \cite{komar2014}, the `proof' of security is substandard, nonetheless the protocol was a good starting point for a cryptographic framework of quantum metrology.

The concept was later picked up in \cite{huang2019}, where two honest parties wish to perform phase estimation over an unsecured quantum channel. Alice sends a non-encoded quantum state to Bob, who encodes a phase using a unitary, and sends the quantum state back to Alice to be measured. The quantum states are sent back and forth through an unsecured quantum channel. The authors of \cite{huang2019} suggest a simple protocol to prevent a malicious adversary from intercepting the channel and tampering with the results. In each use of the quantum channel, Alice randomly prepares one of four quantum states: either a decoy quantum state $\ket{0}^{\otimes n}$ or $\ket{1}^{\otimes n}$, which will not serve any utility for phase estimation, or a GHZ state $\ket{\psi_+}= \frac{1}{\sqrt{2}}(\ket{0}^{\otimes n}+\ket{1}^{\otimes n})$ or $\ket{\psi_-}= \frac{1}{\sqrt{2}}(\ket{0}^{\otimes n}-\ket{1}^{\otimes n})$. Additionally, Bob will either randomly encode the unknown phase $\theta$, or a phase $\phi$ which maps $\ket{\psi_\pm}$ to $\ket{\psi_\mp}$. Even though this protocol is more sophisticated than what was presented in \cite{komar2014}, we show in \cite{SMK21} that it is vulnerable to a malicious attack which is undetectable by Alice and Bob. Additionally, \cite{huang2019} and many others who have since investigated `cryptographic quantum metrology' \cite{xie2018, takeuchi2019, okane2020, yin2020, kasai2021} fail to elaborate on the ramifications on the underlying metrology problem.

In a cryptographic framework, many of the concepts from estimation theory discussed in \textbf{Chapter 3} have to be altered in some capacity. This is because there is no guarantee that the resource used for the parameter estimation problem is the ideal resource. To fit the language of statistics, the cryptographic framework of quantum metrology injects uncertainty into the estimate. This additional uncertainty can be bounded by taking proper precautions and employing appropriate cryptographic protocols. However, this uncertainty in the resource leads to ambiguity with respect to the construction of an estimator; it is not immediately obvious how to select a measurement or how to process the measurement data. Assuming that the additional uncertainty is small, the most straightforward strategy is to process the data as if it was the ideal resource. Evidently, the unbiased condition, Eq.~\eqref{eq:unbiased}, is not necessarily satisfied. Since an unbiased estimator is integral to saturate the CRB, the QFI would be a naive choice of a figure of merit for quantum metrology within a cryptographic framework. Instead, we introduce the concept of \textit{integrity} in \cite{SMK21} as a figure of merit. Integrity refers to the ability of a cryptographic protocol to retain the quantum state and functionality in the presence of malicious adversaries. In this chapter, the notation $\square^\prime$ is used to signify the quantity $\square$ in a cryptographic framework. For example, $\hat{\theta}$ is an estimator with a MSE of $\Delta^2 \hat{\theta}$ in an ideal framework and $\hat{\theta}^\prime$ is an estimator with a MSE of $\Delta^2 \hat{\theta}^\prime$ in the cryptographic framework. The integrity of the cryptographic quantum metrology problem is measured in two ways, the first is the added bias
\begin{equation}
    \big| \mathbb{E}( \hat{\theta}^\prime ) - \mathbb{E}( \hat{\theta}) \big|,
\end{equation}
and the second is the increase in the MSE
\begin{equation}
    \big| \Delta^2 \hat{\theta}^\prime  - \Delta^2 \hat{\theta} \big|.
\end{equation}
For simplicity, we restrict estimation strategies, in which the value of the unknown parameter is inferred from expectation value of an observable $O$. The specific details of this strategy can be found in \textbf{Chapter 3}.

\subsection{Bounding the Integrity}

\begin{figure}[!ht]
    \centering
    \begin{subfigure}{.8\textwidth}
        \centering
        \includegraphics[width=.99\textwidth]{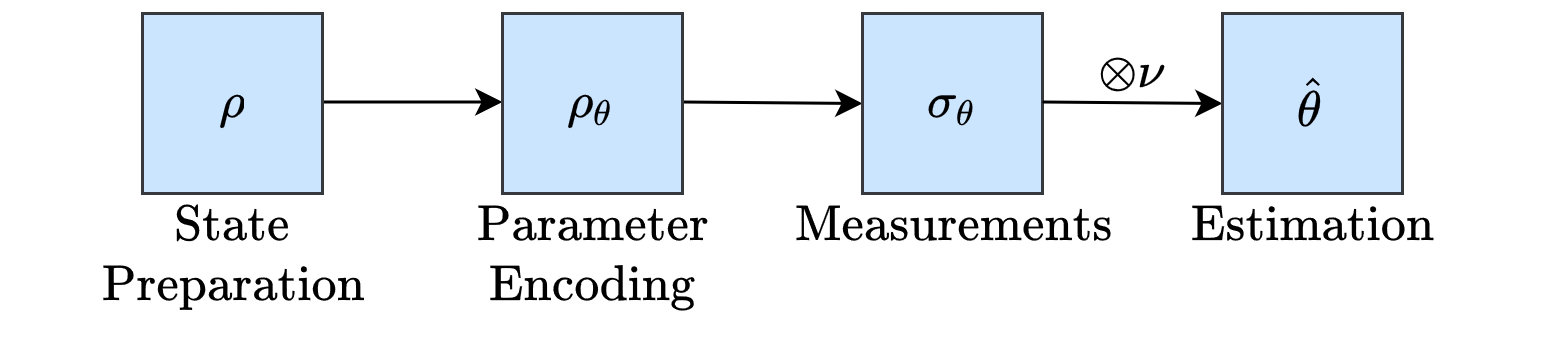}
        \caption{Quantum Metrology in an Ideal Framework.}
        \label{fig:MetroIdeal}
    \end{subfigure}
    
    \vspace{20pt}
    
    \begin{subfigure}{.8\textwidth}
        \centering
        \includegraphics[width=.99\textwidth]{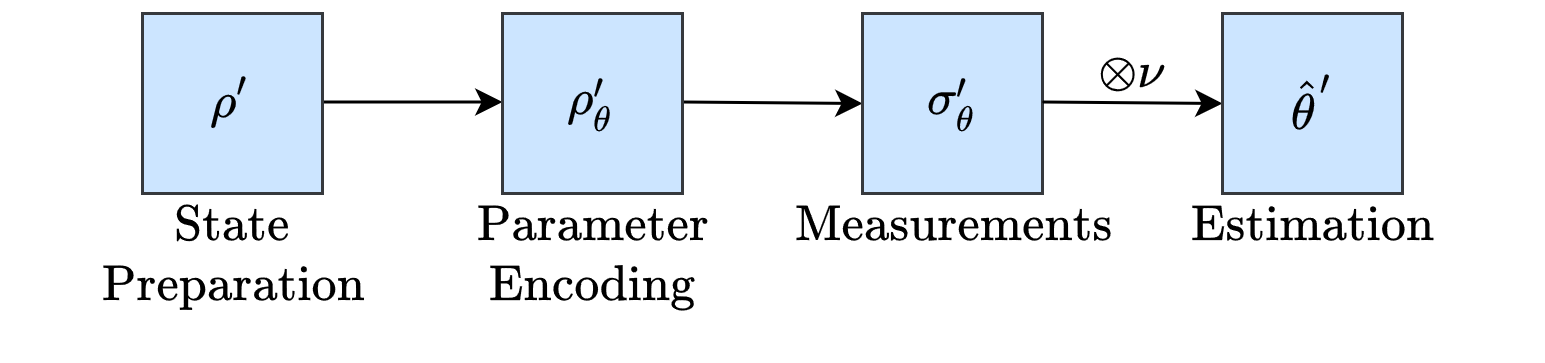}  
        \caption{Quantum Metrology in a Cryptographic Framework.}
        \label{fig:MetroCrypto}
    \end{subfigure}
    \caption{Comparison between a quantum metrology problem in an ideal framework (sans malicious adversary) and a cryptographic framework (potentially a malicious adversary). In the ideal framework (a), a quantum state $\rho$ is prepared, then an unknown parameter is encoded into the quantum states through a CPTP map $\rho_\theta = \Lambda_\theta(\rho)$, finally a measurement $\mathcal{M}$ is performed on the encoded quantum state. After $\nu$ repetitions, the measurement results are used to construct an estimate $\hat{\theta}$. In the cryptographic framework (b), a malicious adversary can intercept and alter the process at any step of the problem. For example, the state preparation can be done by an untrusted source, or an unsecured quantum channel may be intercepted. In fact, the subscript $\theta$ in the cryptographic framework is somewhat misleading as there is no guarantee that either $\rho_\theta^\prime$ or $\sigma_\theta^\prime$ is dependent on $\theta$. Additionally, the assumption of an iid process is discarded in the cryptographic framework: $\rho^\prime$ in the first round may be different from the $\rho^\prime$ in the second round (or any other round). Note that this figure depicts a completely general cryptographic setting, while the latter sections of this chapter explore specific cryptographic settings, in which it will be clear how and when a malicious adversary may alter the quantum metrology problem.}
    \label{fig:MetrologyFrameworkComparison}
\end{figure}

As Fig.~(\ref{fig:MetrologyFrameworkComparison}) suggests, a quantum metrology problem can be altered at many stages of the estimation process by a malicious adversary: state preparation, parameter encoding, or the measurements. Because the measurement is the final `quantum step' in the process before creating the estimate, the measurement statistics in the cryptographic framework must resemble the measurement statistics in the ideal framework. Otherwise, the estimate would not be practical.

Even though measurement results are a classical quantity, the measurement statistics can always be written as a mixed state with no coherence terms, where the amplitudes correspond to probabilities of witnessing a certain measurement outcome. If the observable the estimate is being inferred from has an eigenbasis with projectors $\{ E \}$, then the corresponding measurement statistics of an encoded quantum state $\rho_\theta$ is
\begin{equation}
    \mathcal{M}(\rho_\theta) = \sum_E E \rho_\theta E.
\end{equation}
For example if $\ket{\psi}=\alpha \ket{0}+ \beta \ket{1}$ is measured with respect to the computational basis, then the measurement statistics are $\mathcal{M}(\ket{\psi})=|\alpha|^2 \dyad{0} + |\beta|^2 \dyad{1}$. Similarly, the measurement statistics in a cryptographic framework can always be derived from an arbitrary (not necessarily encoded) quantum state $\rho_\theta^\prime$. Mathematically, we demand that
\begin{equation}
    \label{eq:cryptographicTDrequirement}
    \frac{1}{\nu} \sum_{j=1}^\nu \mathscr{D}\big( \mathcal{M}(\rho_\theta), \mathcal{M}(\rho_\theta^{\prime (j)}) \big) \leq \varepsilon \ll 1,
\end{equation}
where $\rho_\theta^{\prime (j)}$ is a quantum state which outputs the measurement statistics of the $j$th round of the prepare, encode and measure portion of the quantum metrology problem in the cryptographic framework. Eq.~\eqref{eq:soundnessfidelity} and Eq.~\eqref{eq:soundnesstracedistance} suggests that this can be achieved by implementing appropriate cryptographic protocols.

Suppose that the measurements are done in a secure fashion without malicious interference. If the encoded quantum states in the cryptographic framework obey the analogous restriction
\begin{equation}
    \label{eq:cryptographicTDrequirementOLD}
    \frac{1}{\nu} \sum_{j=1}^\nu \mathscr{D}\big(\rho_\theta, \rho^{\prime (j)}_\theta \big) \leq \varepsilon \ll 1,
\end{equation}
then Eq.~\eqref{eq:cryptographicTDrequirement} will still hold because of the monotonicity of the trace distance, Eq.~\eqref{eq:TDcontractive},
\begin{equation}
    \mathscr{D}\big( \mathcal{M}(\rho_\theta), \mathcal{M}(\rho_\theta^{\prime (j)}) \big) \leq \mathscr{D}\big(\rho_\theta, \rho^{\prime (j)}_\theta \big).
\end{equation}
The same argument holds if the malicious interference is localised to the state preparation step in Fig.~(\ref{fig:MetroCrypto}). In fact, Eq.~\eqref{eq:cryptographicTDrequirementOLD} was the imposed inequality in \cite{SMK21}, however, we needed to generalize to Eq.~\eqref{eq:cryptographicTDrequirement} in \cite{SM21} because we explore the possibility of delegating the measurement step to an untrusted party. In either case, the trace distance was chosen because of the relationship to distance of resulting classical probability distributions, Eq.~\eqref{eq:TraceDistanceAlt}: if $\varepsilon$ is small, then any measurement will give rise to similar probability distributions \cite{nielsen2002}.

To properly gauge the effects of a malicious adversary, we examine a specific estimation strategy. We revisit that which was established in \textbf{Chapter~3}: inferring an estimate from an observable. That is, the expectation value of the observable $O$ is estimated and then inverted. This strategy was chosen due to the mathematical simplicity and for the fact that it can be used to saturate the HL. In the ideal framework, this initial estimate is labelled $\hat{f}$. Specifically
\begin{equation}
    \hat{f} = \frac{1}{\nu} \sum_{j=1} m_j,
\end{equation}
where $m_j$ is the eigenvalue associated to the $j$th measurement result, where $\mathbb{E}(m_j)=\Tr(O \rho_\theta)$. This is equivalent to $\mathbb{E}(m_j)=\Tr \big(O \mathcal{M}(\rho_\theta) \big)$. In the cryptographic framework, the analogous estimate is constructed
\begin{equation}
    \hat{f}^\prime = \frac{1}{\nu} \sum_{j=1} m_j^\prime,
\end{equation}
where $\mathbb{E}(m_j^\prime) = \Tr \big( O \mathcal{M}(\rho_\theta^{\prime (j)}) \big)$, which is then inverted as if it were the ideal framework. Assuming that $\varepsilon$ is sufficiently small, the first order Taylor expansion of $f^{-1}(\hat{f}^\prime)$
\begin{equation}
    \label{eq:CryptoTaylorApprox}
    \hat{\theta}^\prime = \theta + \frac{1}{\frac{\partial \expval{O}}{\partial \theta}}(\hat{f}^\prime-\expval{O})
\end{equation}
provides a valid approximation even in the cryptographic framework. Here, $\expval{O}$ is the expectation value with respect to the ideal framework, thus $\expval{O}=\Tr(O\rho_\theta)$. Eq.~\eqref{eq:CryptoTaylorApprox} suggests that in the cryptographic framework, the added bias is bounded by
\begin{equation}
\label{eq:biasbound}
\begin{split}
    \big| \mathbb{E}(\hat{\theta}^\prime) - \mathbb{E}(\hat{\theta}) \big| &= \frac{1}{|\frac{\partial \expval{O}}{\partial \theta}|} \big| \mathbb{E}(\hat{f}^\prime) - \mathbb{E}(\hat{f}) \big| \\
    &= \frac{1}{\nu|\frac{\partial \expval{O}}{\partial \theta}|} \big| \sum_{j=1}^\nu \Tr \big( O \mathcal{M}(\rho_\theta^{\prime (j)})-O\mathcal{M}(\rho_\theta) \big) \big| \\
    &\leq \frac{2o}{\nu|\frac{\partial \expval{O}}{\partial \theta}|}  \sum_{j=1}^\nu \mathscr{D}\big( \mathcal{M}(\rho_\theta), \mathcal{M}(\rho_\theta^{\prime (j)}) \big) \\
    & \leq \frac{2o \varepsilon}{|\frac{\partial \expval{O}}{\partial \theta}|},
\end{split}
\end{equation}
where $o$ is the maximum magnitude of the eigenvalues of $O$. Recall from \textbf{Chapter~3} that in the ideal framework
\begin{equation}
    \Delta^2 \hat{f} = \frac{\Tr(O^2 \rho_\theta)-\Tr(O \rho_\theta)^2}{\nu} = \frac{ \Tr ( \mathbf{O} \rho_\theta \otimes \rho_\theta )}{\nu},
\end{equation}
where $\mathbf{O}=O^2 \otimes \mathbb{I} - O \otimes O$. Note that the maximum magnitude of the eigenvalues of $\mathbf{O}$ is bounded below $2o^2$. In the cryptographic framework, the MSE is the sum of the variance and the square of the bias
\begin{equation}
\begin{split}
    \Delta^2 \hat{f}^\prime &= \mathbb{E} \big( (\hat{f}^\prime -f)^2 \big) \\
    &= \mathbb{E} \big( \hat{f}^\prime - \mathbb{E}(\hat{f}^\prime) \big)^2 + \big(\mathbb{E}(\hat{f}^\prime)-f \big)^2 \\
    &\leq \frac{1}{\nu^2}\sum_{j=1}^\nu \Tr \big( \mathbf{O} \mathcal{M}(\rho_\theta^{\prime (j)}) \otimes \mathcal{M}(\rho_\theta^{\prime (j)}) \big) + 4o^2 \varepsilon^2.
\end{split}
\end{equation}
It follows that the increase of the MSE is bounded by
\begin{equation}
\label{eq:MSEintegrity}
\begin{split}
    \big| \Delta^2 \hat{\theta}^\prime - \Delta^2 \hat{\theta} \big| &= \frac{1}{|\frac{\partial \expval{O}}{\partial \theta}|^2} \big| \Delta^2 \hat{f}^\prime - \Delta^2 \hat{f} \big| \\
    &\leq \frac{4o^2}{\nu^2 |\frac{\partial \expval{O}}{\partial \theta}|^2} \sum_{j=1}^\nu \mathscr{D}\big( \mathcal{M}(\rho_\theta) \otimes \mathcal{M}(\rho_\theta), \mathcal{M}(\rho_\theta^{\prime (j)}) \otimes \mathcal{M}(\rho_\theta^{\prime (j)}) \big) + \frac{4o^2 \varepsilon^2}{ |\frac{\partial \expval{O}}{\partial \theta}|^2}  \\
    &\leq \frac{8o^2 \nu^{-1} \varepsilon+4 o^2 \varepsilon^2}{|\frac{\partial \expval{O}}{\partial \theta}|^2},
\end{split}
\end{equation}
where the triangle inequality
\begin{equation}
    \mathscr{D}(\rho_1 \otimes \rho_1, \rho_2 \otimes \rho_2) \leq \mathscr{D}(\rho_1 \otimes \rho_1, \rho_1 \otimes \rho_2) + \mathscr{D}(\rho_1 \otimes \rho_2, \rho_2 \otimes \rho_2) = 2 \mathscr{D}(\rho_1, \rho_2)
\end{equation}
is used in the derivation of Eq.~\eqref{eq:MSEintegrity}.

Notice that as $\nu \rightarrow \infty$, the added bias in Eq.~\eqref{eq:biasbound} does not vanish, and as a consequence, neither does the increase in the MSE, Eq.~\eqref{eq:MSEintegrity}. This is due to the construction of the cryptographic framework, where Eq.~\eqref{eq:cryptographicTDrequirement} can be interpreted as an average amount of uncertainty in the measurement statistics. If the uncertainty in each round is constant, $\varepsilon$ is of course independent of $\nu$, which ultimately limits the achievable precision of the quantum metrology problem. For the functionality of said quantum metrology problem to be the same in the cryptographic framework when compared to the ideal framework, $\Delta^2 \hat{\theta}^\prime$ must scale similarly to $\Delta^2 \hat{\theta}$. This is equivalent to the difference in the MSE scaling similarly to $\Delta^2 \hat{\theta}$, which occurs when
\begin{equation}
    \label{eq:varepsilonbound}
    \varepsilon^2 \leq \nu^{-1}.
\end{equation}
The factor of $4o^2$ is ignored as it is dependent on the metrology portion of the problem whereas $\varepsilon$ is dependent on the cryptographic portion of the problem. The term $8o^2 \nu^{-1} \varepsilon$ term is ignored, as it is appropriately small if $\varepsilon^2 \leq \nu^{-1}$. 

It follows from the equations for the added bias and difference in MSE, Eq.~\eqref{eq:biasbound} and Eq.~\eqref{eq:MSEintegrity} respectively, along with the relationship between trace distance and soundness, Eq.~\eqref{eq:soundnesstracedistance}, that if a cryptographic protocol has soundness $\delta$ and statistical significance $\alpha$, then the integrity of the quantum metrology problem is represented by the added bias
\begin{equation}
    \big| \mathbb{E}( \hat{\theta}^\prime ) - \mathbb{E}( \hat{\theta}) \big| \leq \frac{2o}{|\frac{\partial \expval{O}}{\partial \theta}|} \sqrt{\frac{\delta}{\alpha}},
\end{equation}
and the the increase in the MSE
\begin{equation}
    \big| \Delta^2 \hat{\theta}^\prime  - \Delta^2 \hat{\theta} \big| \leq \frac{4o^2}{|\frac{\partial \expval{O}}{\partial \theta}|^2} \big( 2 \nu^{-1} \sqrt{\frac{\delta}{\alpha}} + \frac{\delta}{\alpha} \big).
\end{equation}
More so, Eq.~\eqref{eq:varepsilonbound} suggests that the effective functionality is retained when
\begin{equation}
    \label{eq:metrology_interity}
    \frac{\delta}{\alpha} \leq \nu^{-1}.
\end{equation}
It should be noted that the trace distance and soundness relationship, Eq.~\eqref{eq:soundnesstracedistance}, and the demanded proximity of the average measurement statistics, Eq.~\eqref{eq:cryptographicTDrequirement}, are not a function of the same quantities. The former is a function of the expected trace distance while the latter is simply the trace distance. This is because a metrology problem is designed for specific states, while it is atypical for a cryptography protocol to have a precise output. Although these ideologies may seem to contrast with each other, we propose two solutions to remedy the difference. The first is that the measurement statistics of each round can be interpreted as the average measurement statistics after implementing the protocol, from which the integrity relationships still hold because of the strong convexity of trace distance \cite{nielsen2002}
\begin{equation}
    \mathscr{D} \big( \rho_\text{id}, \mathbb{E}(\rho_\text{out}) \big) \leq \mathbb{E} \big( \mathscr{D} ( \rho_\text{id}, \rho_\text{out} ) \big).
\end{equation}
The second, is that for sufficiently large $\nu$, the law of large numbers dictates that the proximity of the average measurement statistics will tend towards the expected value
\begin{equation}
    \frac{1}{\nu} \sum_{j=1}^\nu \mathscr{D}( \sigma_\theta, \sigma_\theta^{\prime (j)}) \approx \mathbb{E} \big( \mathscr{D}( \sigma_\theta, \sigma_\theta^{\prime}) \big).
\end{equation}

\section{Quantum Metrology over an Unsecured Quantum Channel}

The first cryptographic setting established in this chapter is when the quantum metrology problem uses an unsecured quantum channel \cite{SMK21}. In quantum sensing networks, the quantum channels will likely be the most vulnerable to malicious attacks \cite{komar2014}, so it important to include a cryptographic protocol to carry out the metrology problem in a secure fashion. This was the basis of the work presented in \cite{huang2019}, however, as described above, the authors fail to create a secure protocol. To achieve a notion of security, the protocols presented in this section take inspiration from quantum authentication schemes \cite{barnum2002, broadbent2016b}. Quantum authentication schemes are cryptographic protocols designed to send quantum states across an unsecured quantum channel in a private and secure fashion, which is precisely the nature of the task at hand.

\subsection{The Protocols}

Two protocols are presented for the task of quantum metrology over an unsecured quantum channel: i) a modified version of the trap code \cite{broadbent2013}, and ii) a modified version of the Clifford code \cite{aharonov2017}. From a functional stand point the two protocols are nearly identical, however the encryption and decryption methods vary drastically from a complexity standpoint and ease of implementation. The encryption scheme for the trap code is restricted to locally acting Clifford operations, $C \in \mathcal{C}_1^{\otimes m}$. In contrast, the encryption scheme for the Clifford code is an arbitrary $C \in \mathcal{C}_m$. As expected, the Clifford code leads to a much stronger soundness statement, due to the additional entanglement gained from the encryption.

\begin{figure}[!h]
    \centering
    \begin{subfigure}{.9\textwidth}
        \centering
        \includegraphics[width=.99\textwidth]{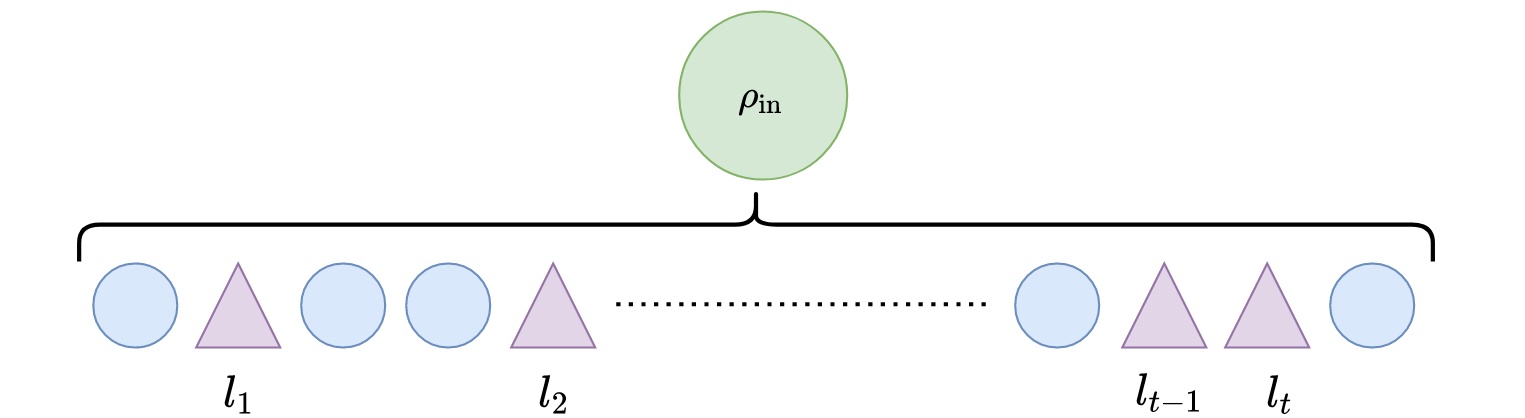}
        \caption{Initialized quantum state before using the unsecured quantum channel.}
        \label{fig:QM_UnsecuredChannel_source}
    \end{subfigure}
    
    \vspace{20pt}
    
    \begin{subfigure}{.9\textwidth}
        \centering
        \includegraphics[width=.99\textwidth]{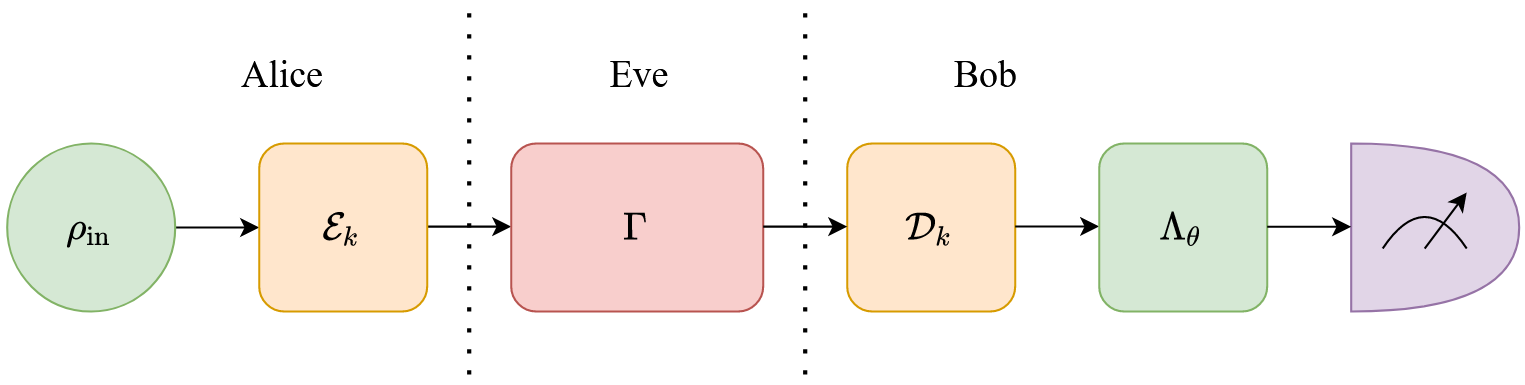}  
        \caption{Schematic of the protocols.}
        \label{fig:QM_UnsecuredChannel_protocol}
    \end{subfigure}
    \caption{(a) Alice prepares the quantum state $\rho_\text{in}$, which is a combination of $t$ ancillary flag qubits (randomly positioned) as well as the quantum state $\rho$ intended for quantum metrology. The flag qubits are indexed at positions $l_1,l_2,\ldots,l_t$. (b) Before utilizing the quantum channel, Alice and Bob randomly select a classical key $k$. This classical key corresponds to the encryption operation ($\mathcal{E}_k$) performed by Alice, and the decryption operation ($\mathcal{D}_k=\mathcal{E}_k^\dagger$) performed by Bob upon receipt. A malicious eavesdropper, labeled Eve, has complete access to the quantum quantum channel. Without loss of generality, Eve can perform any CPTP map $\Gamma$ when interacting with the channel. Bob encodes the unknown parameter into the portion of the quantum state intended for quantum metrology. Finally, Bob measures the qubits accordingly: the ancillary flag qubits in the computational basis, and the metrology qubits in the appropriate basis to construct an estimate. If the flag qubit measurement is an unexpected output, then a malicious adversary must have tampered with the quantum channel.}
    \label{fig:QM_UnsecuredChannel}
\end{figure}

In this setting, Alice and Bob are the trusted parties who wish to execute a quantum metrology problem. They are separated by an unsecured quantum channel, which may be intercepted by a malicious eavesdropper, labelled Eve. Note that Alice and Bob share a secure classical channel to communicate classical information, such as the choice of the random key. This is a standard assumption in quantum cryptography.

To have the ability to detect Eve, Alice prepares an input state $\rho_\text{in}$, which is a combination of the quantum state intended for the metrology problem $\rho_\text{id}$, as well as $t$ ancillary flag qubits. An example of an input state is depicted in Fig.~(\ref{fig:QM_UnsecuredChannel_source}). The flag qubits are all initialized in the state $\ket{0}$, and upon receipt Bob measured the flag qubits in the computational basis. In an ideal setting, the measurement will ubiquitously witness the result $\ket{0}^{\otimes t}$; any other result suggests that the quantum channel was compromised. This deterministic measurement result aids in certifying whether or not Eve tampered with the quantum channel.

After preparing the input state $\rho_\text{in}$, Alice encrypts it using a random Clifford operation. The set from which the Clifford operation is chosen from is dependent on the protocol. Upon receipt, Bob decrypts the quantum state by applying the inverse operation applied by Alice. Bob then measures the ancillary flag qubits in computational basis. If the expected measurement result $\ket{0}^{\otimes t}$ is witnessed, Bob will utilize the remaining qubits for the quantum metrology problem, otherwise they are discard the quantum state as Eve must have tampered with the quantum channel. This process is illustrated in Fig.~(\ref{fig:QM_UnsecuredChannel_protocol}).

\vspace{\baselineskip}
\textit{Implementation Instructions:}
\begin{enumerate}

\item Prior to using the quantum channel, Alice and Bob randomly select a key $k \in \mathcal{K}$, which is linked to an encryption operator $\mathcal{E}_{k}$. Specific to trap code, the key also contains information about a tuple $\ell=(l_1,\ldots,l_t)$ of length $t$, this tuple contains the index locations of the ancillary flag qubits.

\begin{enumerate}
    \item For the trap code, $\mathcal{E}_k \in \mathcal{C}_1^{\otimes m}$.
    \item For the Clifford code, $\mathcal{E}_k \in \mathcal{C}_m$.
  \end{enumerate}

\item Alice creates the $m=n+t$ qubit state $\rho_\text{in}$ by inserting $t$ ancillary flag qubits $\ket{0}$ at the positions indexed by $\ell$, and the remaining $n$ qubit state $\rho$ is the quantum state designated for quantum metrology.

\begin{enumerate}
    \item For the trap code, it is important that $\ell$ is randomly chosen because the encryption operation does not generate entanglement.
    \item For the Clifford code, $\ell$ can be static. This is because the encryption will generate entanglement between the ancillary qubits and the rest of the quantum state.
  \end{enumerate}

\item Alice encrypts the input state by applying the Clifford operator $\mathcal{E}_k$ and sends the quantum state to Bob.

\item Upon receipt, Bob decrypts the quantum state by applying the inverse operator $\mathcal{E}_k^\dagger$ upon receipt.

\item Bob measures the ancillary flag qubits in the computational basis. The result is accepted if $\dyad{0}^{\otimes t}$ is measured. The quantum state is discarded otherwise.

\item If the result is accepted, Bob continues with the quantum metrology problem using the remaining qubits.

\end{enumerate}

The random choice of encryption operation makes it impossible for Eve to extract any information about $\rho_\text{in}$, meaning that protocols are completely private. To see explicitly why, consider $\rho_\text{in}$ in the Pauli basis
\begin{equation}
    \rho_\text{in} = \frac{1}{2^m}\sum_{P \in \mathcal{P}_m} \Tr (P \rho_\text{in}) P.
\end{equation}
Using the Clifford code, the expected quantum state available to Eve is
\begin{equation}
    \mathbb{E}( \rho_E ) = \frac{1}{2^m |\mathcal{C}_m|} \sum_{C \in \mathcal{C}_m} \sum_{P \in \mathcal{P}_m} \Tr (P \rho_\text{in}) C P C^\dagger.
\end{equation}
For every $P \neq \mathbb{I}$, the Clifford group can be partitioned into pairs of operators $(C_a,C_b)$ such that $C_a P C_a^\dagger = -C_b P C_b^\dagger$, hence the only-non vanishing term is $P=\mathbb{I}$, and thus Eve cannot distinguish the quantum state from the maximally mixed state. For the trap code, the Pauli and Clifford operations can be decomposed into local operations, $C=\bigotimes_{j=1}^m C_j$ and $P=\bigotimes_{j=1}^m P_j$. Here, the expected quantum state available to Eve is
\begin{equation}
    \label{eq:privacyTrapCode}
    \mathbb{E}( \rho_E ) = \frac{1}{2^m |\mathcal{C}_1|^m} \sum_{C \in \mathcal{C}_1^{\otimes m}} \sum_{P \in \mathcal{P}_m} \Tr (P \rho_\text{in}) \bigotimes_{j=1}^m C_j P_j C_j^\dagger.
\end{equation}
By the same intuition, the only non-vanishing term is when $P$ is identically the identity, and thus using the trap Code, Eve cannot distinguish the quantum state from the maximally mixed state. Even though no information about the unknown parameter with respect to the metrology problem is passed through the quantum channel, as per Fig.~(\ref{fig:QM_UnsecuredChannel_protocol}), having complete privacy is still important. The same protocol can be used in the nearly identical setting where it is Alice who encodes the unknown parameter. More so, it will be shown that the protocol can be extended to a setting where Alice and Bob use the same quantum channel twice, similar to the setting of \cite{huang2019}. In either of these two settings, having a completely private protocol prevents Eve from extracting information about the unknown parameter in question.

Privacy is achieved as a consequence of randomly sampling $\mathcal{E}_k$ from a large set of Clifford operations. For example, the set $\mathcal{C}_1^{\otimes m}$ has $24^m$ elements. Although we do not focus on the logistics of the classical channel in our protocol, it is important to acknowledge that the size of the classical key required is quite large. As an alternative, one can consider sampling $\mathcal{E}_k$ from a smaller set of $\mathcal{O}(m2^m)$ unitary operators, which approximately guarantees privacy \cite{hayden2004}. Although in doing so, other assumptions are needed, namely that Eve not having access to a quantum memory \cite{lupo2015}.

A derivation for the soundness\footnote{We derive the soundness with the assumption that the quantum state intended for quantum metrology, $\rho_\text{id}$, is a pure state, because this greatly simplifies the derivation. This is a logical assumption since pure states are superior resources for quantum metrology.} of the two protocols can be found in \textbf{Appendix~C}, in which it is shown that the soundness of the trap code is $\delta_{\text{trap}}=\frac{3n}{2t}$, and the soundness of the Clifford code is $\delta_{\text{Cliff}}=\frac{1}{2^t}$. Eq.~\eqref{eq:metrology_interity} states that the integrity of the underlying metrology problem is maintained when $\frac{\delta}{\alpha} \leq \nu^{-1}$. Equivalently, the number of ancillary flag qubits required is $t_\text{trap} \geq \frac{3n \nu}{2\alpha}$ for the trap code, and $t_\text{Cliff} \geq \log_2 \frac{\nu}{\alpha}$ using the Clifford code. In the ideal framework, the total number of qubits is $\nu n$, in the cryptographic framework it is $\nu (n+t)$. This is a quadratic increase in resources if using the trap code, and a log-linear increase in resources if using the Clifford code.

\subsection{Generalizations}

\begin{figure}[!ht]
    \centering
    \includegraphics[width=0.8\textwidth]{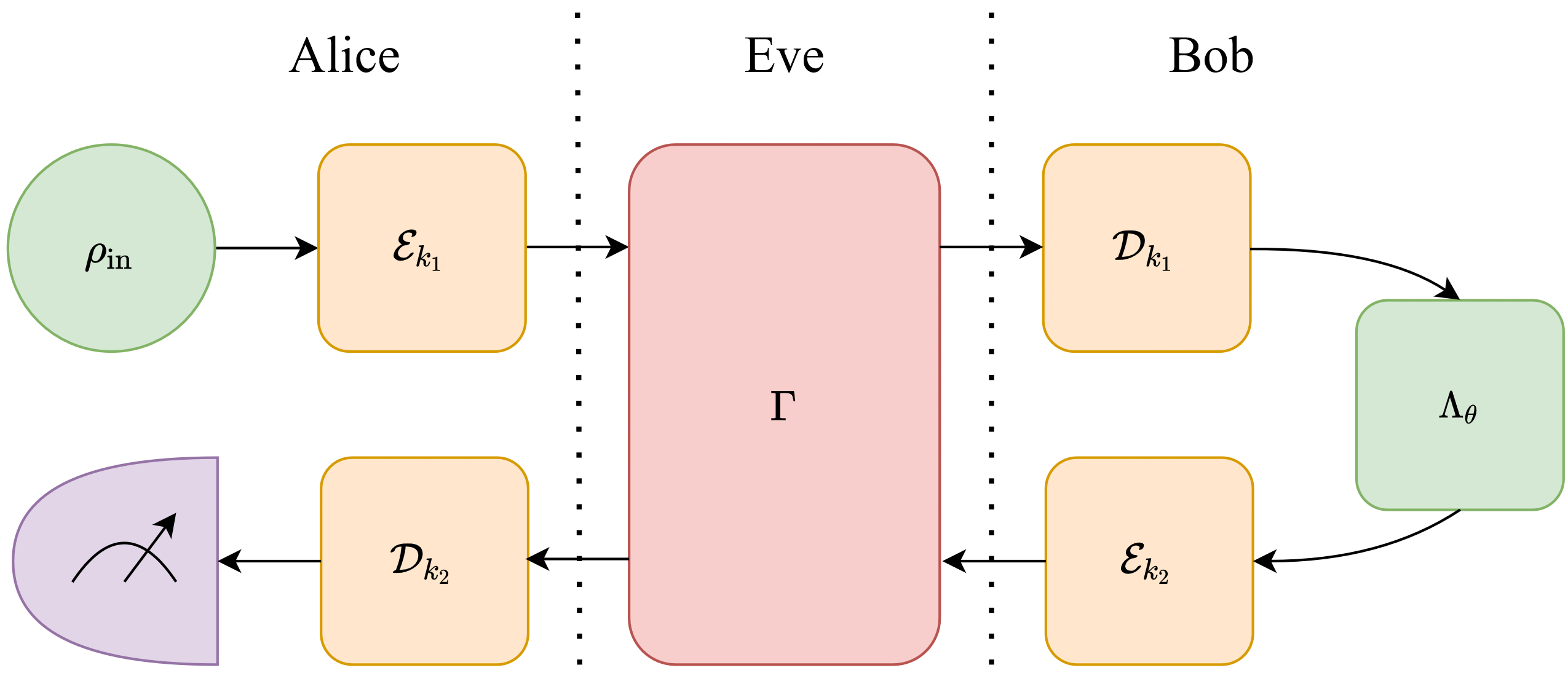}
    \caption{In the extended version of the protocol, the unsecured quantum channel is used twice. Alice sends the quantum state $\rho_\text{in}$ to Bob to be encoded, after which it is sent back to Alice. Because the quantum channel is used twice, the classical key shared by Alice and Bob describes the encryption and decryption operation for the first use of the quantum channel ($\mathcal{E}_{k_1}, \mathcal{D}_{k_1}$) and the second use of the quantum channel ($\mathcal{E}_{k_2}, \mathcal{D}_{k_2}$).}
    \label{fig:QM_channeltwice}
\end{figure}

The work in \cite{huang2019} addresses the distribution of entangled resources over quantum channels for quantum metrology, however, with a more restricted Bob, so that the measurement is also left to Alice, requiring the state be sent back to Alice once Bob has done the encoding. Both the trap code and Clifford code can be easily adapted to this setting. To do so, Alice and Bob perform a second encryption operation before the second usage of the quantum channel. The generalization of the protocol is illustrated in Fig.~(\ref{fig:QM_channeltwice}). Using two encryption operations is imperative for the success of the protocol; if it was just Alice who performed the encryption and the decryption, then Eve could simply apply a unitary on the use of the quantum channel, and its inverse on the second usage. This will not alter any of the ancillary flag qubits but can bias the qubits intended for quantum metrology. In \textbf{Appendix~C}, the soundness of the generalized protocols are computed to be $\delta_\text{trap}=\frac{9n}{4t}$ and $\delta_\text{Cliff}=\frac{1}{2^t}$.

\begin{figure}[!h]
    \centering        \includegraphics[width=.65\textwidth]{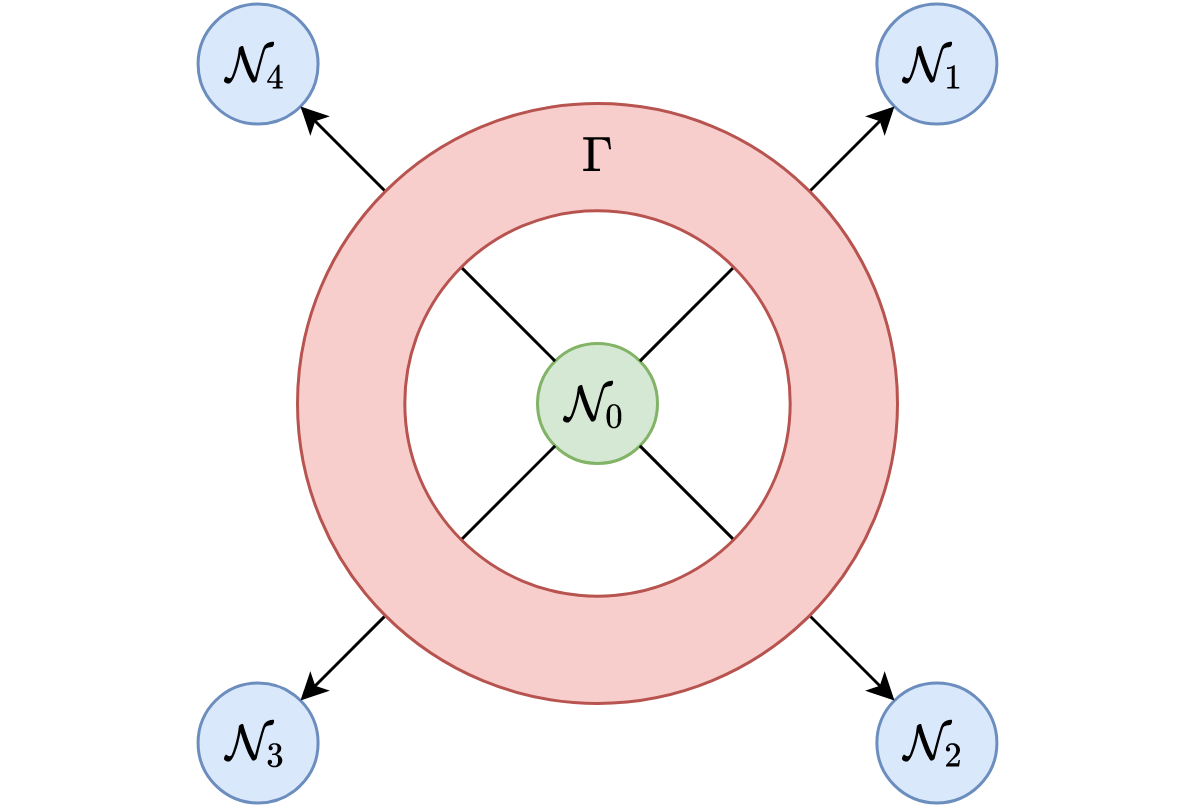}
    \caption{Generalization to a multipartite framework, where a central node $\mathcal{N}_0$ distributes a portion of a quantum state amongst external nodes $\mathcal{N}_1, \ldots, \mathcal{N}_k$ (in this illustration $k=4$). This distribution is done through quantum channels, and thus may be vulnerable to a malicious eavesdropper, whose (potential) interaction is depicted with a red ring. To ensure a sense of security, the trusted nodes can adopt the trap code since the decryption operations are all performed locally.}
    \label{fig:QM_multipartite}
\end{figure}

An alternative generalization is a multipartite setting, depicted in Fig.~(\ref{fig:QM_multipartite}). This would be a practical tool for any quantum sensing network problem \cite{komar2014, proctor2018, rubio2020a} with unsecured quantum channels. Here a central node $\mathcal{N}_0$ is connected to external nodes $\mathcal{N}_1,\ldots,\mathcal{N}_k$ via quantum channels, which may be simultaneously intercepted by a malicious adversary. The central node sends a portion of an entangled quantum state to each of the external nodes, after which the external nodes encode a local parameter on their portion of the quantum state for a spatially distributed quantum metrology scheme. The trap code can be adopted in this spatially distributed and multipartite framework since the decryption operations are local, and thus recover the same notions of privacy and soundness.

\section{Quantum Metrology with Delegated Tasks}

\begin{figure}[!t]
    \centering
    \includegraphics[width=0.9\textwidth]{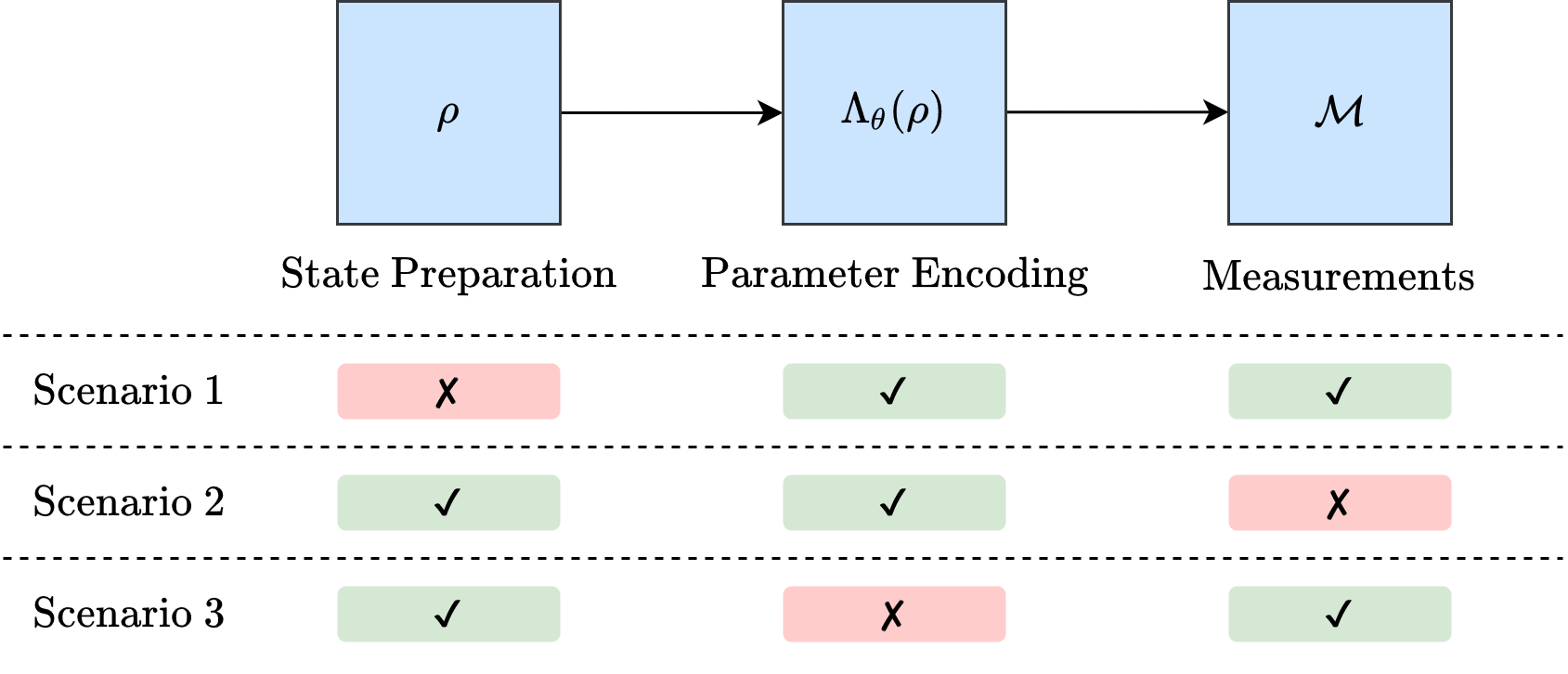}
    \caption{The different delegated quantum metrology scenarios addressed in the section. A quantum metrology problem can be decomposed into three (quantum) tasks: state preparation, parameter encoding and measurements. A red rectangle with a `\xmark'  indicates that the task is delegated to a third party, as opposed to a green rectangle with a `\checkmark' which indicates that the task is not delegated. In \textit{scenario 1}, state preparation is delegated and verification protocols \cite{zhu2019a, markham2020} are used to achieve a sense of security. In \textit{scenario 2}, the measurements are delegated and we devise an authentication based protocol to achieve a sense of security. Finally, in \textit{scenario 3}, the parameter encoding is delegated, and we discuss the impossibility of constructing a computationally secure protocol for such a scenario.}
    \label{fig:DelegatedScenarios}
\end{figure}

In the previous section, together the honest parties, Alice and Bob, had all of necessary quantum technologies to fully carry out a quantum metrology problem. In reality, fully implementing a quantum metrology problem is technologically demanding. Entangled quantum states must be generated and measured with high fidelity. The quantum internet \cite{wehner2018}, and other asymmetric quantum networks, is a possible solution where parties which lack the necessary hardware can delegate the desired task to another party in the network. Of course, when delegating tasks, it is important to be mindful of possible risks. Within the framework of quantum metrology, a malicious third party could bias the estimation results or conduct the estimate themselves. In this section, which is based off of work currently in preparation \cite{SM21}, we propose cryptographic protocols to allow for delegating a portion of the quantum metrology scheme to an untrusted third party. This is done by partitioning a quantum metrology problem into three tasks: state preparation, parameter encoding and measurements, and explore the repercussions when a specific task, or a combination, is delegated. The different scenarios are summarized in Fig.~(\ref{fig:DelegatedScenarios}). There is an additional task of processing the measurement results and creating the estimate, however we ignore this since it is inherently a classical computation.

\subsection{Delegated State Preparation}

The first scenario explored is when the task of quantum state preparation is delegated to an untrusted party. In the absence of a proper cryptographic protocol, the untrusted party could distribute any quantum state $\rho^\prime$, which could be preemptively biased to mask the true result of the parameter estimation. Because the metrology portion part of the problem has not yet come into effect, we can utilize one of the many existing quantum state verification protocols \cite{takeuchi2018, pallister2018, markham2020, liu2019, takeuchi2019b}, which ensure that the quantum state prepared is the desired quantum state.

Verification protocols are used to (as the name suggests) verify quantum states \cite{zhu2019a, zhu2019b}. Typically, this is done by requesting additional copies of the desired quantum state and by measuring the additional copies in specific bases. The measurement results are used to decide if the protocol is accepted or rejected. It should be noted that most verification protocols are tailored for specific classes of quantum states, such as graph states \cite{markham2020, takeuchi2019b} or Dicke states \cite{liu2019}. More general protocols tend to require significantly more resources to achieve the same level of soundness for arbitrary quantum states \cite{takeuchi2018, pallister2018}.

Of course, the soundness is dependent on the protocol chosen to be integrated into the cryptographic quantum metrology framework. For the sake of an example, consider the protocol outlined in \cite{markham2020}. The protocol is a verification protocol for graph states (and can thus be used for the bundled graph states introduced in \textbf{Chapter~4}), but naturally extends to all stabilizer states, including the GHZ state. The protocol takes advantage of the deterministic measurement results when a stabilizer state is measured in a basis of any of its stabilizers, Eq.~\eqref{eq:stabrepresentation}. In summary, the protocol requests $N$ copies of the desired stabilizer state, all but one (randomly selected) is measured with respect to an arbitrary stabilizer. The result is accepted if the $N-1$ measurements results all witness a $+1$ eigenvalue of their respective stabilizer. The protocol achieves a soundness of $\delta=1/N$. Therefore, the integrity of the underlying quantum metrology problem is maintained if
\begin{equation}
    N \geq \frac{\nu}{\alpha}.
\end{equation}
After $\nu$ repetitions of the protocol, this translates to a quadratic increase in resources compared to the ideal framework.

Quantum state verification uses several figures of merit (besides just soundness) which are intertwined \cite{zhu2019a,zhu2019b}. Specifically, the soundness is bounded for a fixed $N$, however the characterisation introduced in \cite{zhu2019a} permits the optimization of $N$ for a fixed $\delta$ and $\alpha$. For qubit stabilizer states the answer is $N=2(\ln2)^{-1} \delta^{-1} \ln \alpha^{-1}$. The bounds are different because the `worst case' attack which saturates the soundness for a fixed $N$ is different than the `worst case' attack for a fixed $\delta$.

\subsection{Delegated Measurements}

The second scenario we explore is the when the measurements are delegated to an untrusted third party. A simplistic version of this scenario with an honest-but-curious adversary has been explored \cite{takeuchi2019, okane2020, yin2020}, where the authors propose using a blind quantum computing protocol \cite{broadbent2009} to achieve privacy by masking the measurement results from an eavesdropper. However, blind quantum computing is not sufficient to achieve unconditional security, where no assumptions are made with respect to the adversary. For all intents and purposes, the untrusted party may return arbitrary measurement results, and without proper cryptographic precautions, the untrusted party can bias the estimate to their own accord. To combat this, the protocol we propose takes inspiration from verified blind quantum computing \cite{morimae2014, fitzsimons2017} and the protocol proposed for quantum metrology over an unsecured quantum channel.

\begin{figure}[!h]
    \centering
    \includegraphics[width=0.65\textwidth]{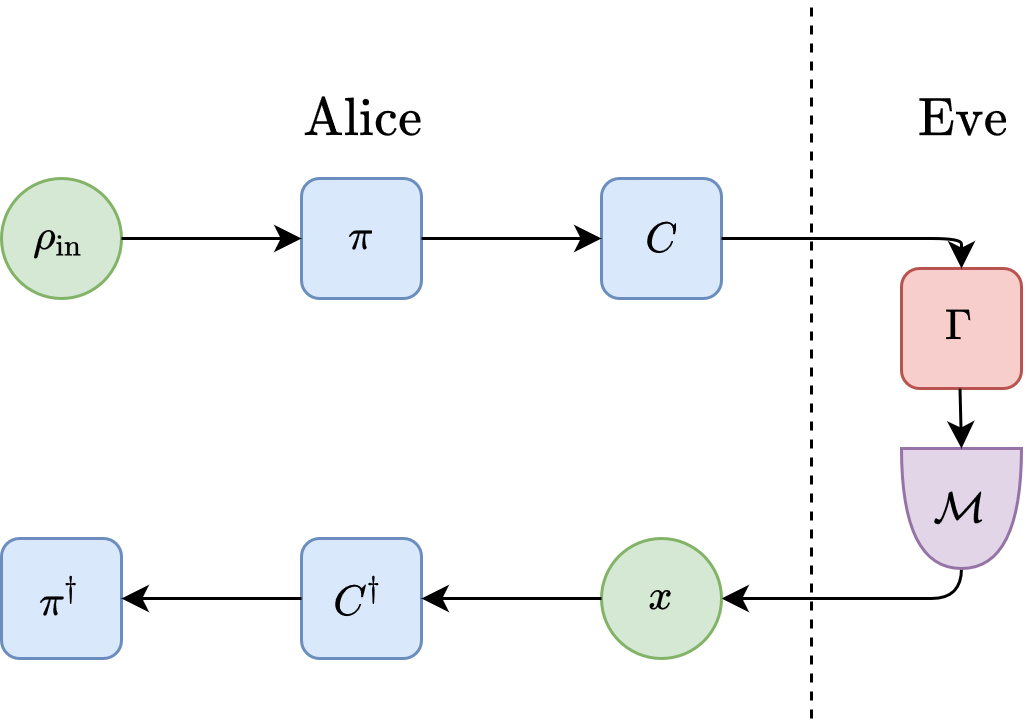}
    \caption{Alice prepares the quantum state $\rho_\text{in}$, which is a combination of an $n$ for quantum metrology and $t$ flag qubits. Before sending the quantum state to Eve, Alice randomly permutes the flag qubits amongst the encoded qubits and subsequently encrypts the quantum state by applying a random Pauli $C$. Without loss of generality, the measurement result, $x$, returned by Eve will coincide with the measurement statistics of $\mathcal{M}\big(\Gamma(C\pi\rho_\text{in}\pi^\dagger C^\dagger)\big)$, where $\Gamma$ is any CPTP map. Upon receipt, Alice performs classical post-processing on $x$ such that it can be properly interpreted. This is represented as applying $C^\dagger$ and $\pi^\dagger$ on $\mathcal{M}\big(\Gamma(C\pi\rho_\text{in}\pi^\dagger C^\dagger)\big)$.}
    \label{fig:DelegatedMeasProtocol}
\end{figure}

Using the same nomenclature as the protocol for quantum metrology over an unsecured quantum network: Alice is the trusted party who lacks the necessary quantum hardware to execute quantum measurements, and Eve is the untrusted party who is delegated the measurements task. In a trusted setting, Alice sends Eve an encoded quantum state $\rho_\theta$, and the probability of Eve returning a specific measurement result corresponds to the amplitudes of $\mathcal{M}(\rho_\theta)$. In the untrusted setting, Eve can return arbitrary measurement results, but without loss of generality, they will correspond to the amplitudes of $\mathcal{M}(\rho_\theta^\prime)$, where $\rho_\theta^\prime$ is an arbitrary and not necessarily encoded quantum state. In addition, Eve can perform the correct measurement, such that they can construct an estimate of $\theta$ for themselves and send a biased or nonsensical results back to Alice. To attain a notion of security and privacy, Alice employs the protocol illustrated in Fig.~(\ref{fig:DelegatedMeasProtocol}) and described below.

The protocol we outline is specific to the case when, in the ideal framework, Alice would request Eve to measure each qubit respect to the basis of a (non-identity) Pauli operator $P$. It can be adapted to other non-entangled measurements by appropriately rotating the encryption operations. We focus on simple measurements as the protocol is more tangible: it only requires local Clifford operations to encrypt the quantum state. It is also practical as feasible measurement strategies in quantum metrology are typically with respect to the eigenbasis of a Pauli operator because they are the simplest to implement.

\vspace{\baselineskip}
\textit{Implementation Instructions:}
\begin{enumerate}
    \item Alice prepares the $m=n+t$ qubit state $\rho_\text{in}=\rho_\theta \otimes \dyad{0}^{\otimes t}$. Here, $\rho_\theta$ is the $n$ qubit encoded quantum state and $\dyad{0}^{\otimes t}$ is used ancillary flag qubits as ancillary flag qubits because of their deterministic measurement outcome.
    
    \item Alice encrypts $\rho_\text{in}$ by first performing a permutation $\pi$ and then applies a random Clifford $C \in \mathcal{C}_1^{\otimes m}$. The permutation will insert the flag qubits at random positions so that Eve cannot distinguish between the encoded qubits and flag.
    
    \item Alice requests Eve to measure the quantum state in the basis of $C \pi P^{\otimes n} \otimes Z^{\otimes t} \pi^\dagger C^\dagger$, the measurement is represented by the map $\mathcal{M}$.
    
    \item Eve returns a measurement result $x$, which are derived from the measurement statistics $\mathcal{M}\big(\Gamma(C\pi\rho_\text{in}\pi^\dagger C^\dagger)\big)$, where $\Gamma$ is any CPTP map.
    
    \item Alice performs classical post-processing on $m$ to obtain the measurement result as if it had not been encrypted. With respect to the measurement statistics, this is represented by applying $\pi^\dagger C^\dagger$ to $\mathcal{M}\big(\Gamma(C\pi\rho_\text{in}\pi^\dagger C^\dagger)\big)$.
    
    \item Alice accepts the measurement result if, after post-processing, the measurement results of the $t$ flag qubits coincide with the expected result of $\dyad{0}^{\otimes t}$. Otherwise, Alice rejects the measurement results as Eve must have acted maliciously.
\end{enumerate}

The reason the protocol is designed\footnote{To adapt the protocol to more complex measurements, the encryption on the requested measurement basis would have to mimic the actions of an arbitrary Clifford operation on a Pauli operator.} for measuring with respect to an eigenbasis of a Pauli operator $P^{\otimes n}$, is because regardless of the encryption $C$, the requested measurement is an random string of Pauli operators, and Eve cannot decipher which measurements coincide with qubits for quantum metrology and which measurement results coincide with ancillary flag qubits. As a result, the protocol is completely private
\begin{equation}
    \mathbb{E}(\rho_E) =\mathbb{I}/2^m,
\end{equation}
where the above privacy statement can be shown using the same logic as the privacy of the trap code for quantum metrology over an unsecured quantum channel, Eq.~\eqref{eq:privacyTrapCode}. 

In \textbf{Appendix~C}, we show that the soundness of this protocol is bounded below the soundness of the trap code for quantum metrology over an unsecured quantum channel, and thus $\delta = \frac{3n}{2t}$. Therefore, the integrity of the underlying quantum metrology problem is maintained if $\frac{3n}{2\alpha t} \leq \nu^{-1} $. Equivalently, the number of ancillary flag qubits required is $t \geq \frac{3n \nu}{2\alpha}$. Thus, the cryptographic framework requires a quadratic increase in the number of resources to maintain the same level of precision as the ideal framework.

\subsection{Delegated Parameter Encoding}

The final scenario considered is when the task of parameter encoding is delegated to an untrusted third party. From a verification perspective, the goal is to assure that some output state $\rho_\text{out}$ is close to the ideal encoded state $\rho_\theta$ with high probability. Unsurprisingly, this is an impossible task from an information theoretic standpoint without having perfect knowledge of $\theta$, which would entirely defeat the purpose of quantum metrology. The impossibility of this task stems from the fact that an adversary can manipulate the lack of information about $\theta$ to their advantage. For example, an adversary can introduce a slight bias $\Lambda_{\theta+\delta \theta}$, encode a different parameter altogether $\Lambda_{\varphi}$, encode $\theta$ into a different quantum state $\tilde{\rho}$, or do nothing at all $\mathbb{I}$. Furthermore, there is no way of guaranteeing that an adversary acts identically each round.

Suppose that the information theoretic standpoint is abandoned and the abilities of the adversary are greatly limited to either applying $\Lambda_\theta$ or the identity $\mathbb{I}$. If one has a priori knowledge that $\theta \approx \theta_0$, a loose `accept' criteria is for the estimate to be within some range of $\theta_0$. This `protocol' can still be manipulated by an adversary if they learn the range of acceptance: $\mathbb{I}$ is applied a small number of times such that the expected estimate falls within the acceptance range despite the added bias.

Finally, if the adversary is further hindered by assuming that they cannot access any sort of classical information - such as an a priori approximation $\theta \approx \theta_0$, or the acceptance range of the aforementioned protocol - then one can continue on with the quantum metrology scheme. This is because in this specific setting, the effective encoding map is now the CPTP map
\begin{equation}
    \rho \rightarrow (1-p)\Lambda_\theta(\rho)+p\rho,
\end{equation}
where $p$ is the effective probability that the adversary does nothing, and hence applies $\Lambda_\theta$ with effective probability $1-p$. Here, the metrology problem of estimating $\theta$ has evolved into the multiparameter problem \cite{ragy2016} of estimating $\theta$ and $p$. However, in making these assumptions, we have ventured out of the realm of cryptographic quantum metrology and into a fusion of quantum channel tomography \cite{bendersky2008} and quantum metrology.

\section{Discussion}

The work presented in this chapter is a novel approach to immerse a general quantum metrology problem in a cryptographic framework. By demanding the final measurement statistics used to construct an estimate are close to that of the ideal framework (sans malicious adversary), the cryptographic notion of soundness can be related to the integrity of the quantum metrology problem. Within the frequentist approach, in an ideal framework, the estimate converges to the true value as $\nu \rightarrow \infty$, so any added uncertainty as a result of the cryptographic framework, $\varepsilon$, will be the factor which limits the precision in the cryptographic framework. The `cryptographic uncertainty' was presented as a result of the interference of a (potentially) malicious adversary, but in reality, the integrity statements hold for any resource satisfying Eq,~\eqref{eq:cryptographicTDrequirement}. For example, the uncertainty caused by faulty quantum hardware or environmental noise.

The soundness of the protocols are derived for unconditional security, i.e no assumptions about the adversary are made. Of course, by discarding this assumption and limiting the abilities of an adversary (for example, only local Clifford operations, etc), the soundness bounds can be greatly improved, thus reducing the number of ancillary flag qubits to maintain the functionality of the underlying quantum metrology. Additionally, the protocols are designed for qubit systems, which naturally generalize to qudit systems, however, the protocols do not easily translate to a continuous variable quantum system; properly deriving the analogous results is a future perspective of this work.

From a cryptographic standpoint, there are numerous ways to broaden the perspective of quantum metrology in a cryptographic framework. For example, the untrusted parties in the delegated task framework can be replaced by untrusted devices to attain a notion of device independent \cite{mayers1998, xu2014} quantum metrology. Alternatively, the notions of cryptography introduced can be further abstracted \cite{maurer2011} to attain a notion of quantum metrology in an abstract cryptographic framework.

At first glance of the integrity statements throughout this chapter, the statistical significance $\alpha$ may seem like an undesirable quantity and counter-intuitive to the unconditional security assumptions. However, a bound on the trace distance cannot be made in any other way. Consider the problem of performing quantum metrology over an unsecured quantum channel, if a malicious party replaces the quantum state by the maximally mixed state, then (regardless of the protocol) the measurement results of the `flag qubits' will result in accept with a very small but non-zero probability. In this example, the quantum state then used for quantum metrology would be useless. Formally, the statistical significance parameter $\alpha$ used throughout quantum authentication and verification \cite{zhu2019a, zhu2019b} is identical to the notion of confidence level $1-\alpha$ used in traditional statistics. In which, $\alpha$ is a pre-decided upon value related to the probability of rejecting the null-hypothesis, or in this case the outcome of the protocol.

The two protocols presented for quantum metrology over an unsecured quantum channel differ in practicability and efficiency. Although the Clifford code is more efficient, the required entanglement is highly impractical. In contrast the trap code is only slightly more demanding than non-secure versions, requiring only local Clifford operations for encryption. For the task of delegated measurements, we designed a protocol analogous to the trap code. We could have additionally made an analogous protocol to the Clifford code for the same task, however this would require Alice requesting highly entangled measurements to be performed, which seems more out of reach than a highly entangled unitary operation. These protocols can also be made somewhat robust to noise by tweaking acceptance parameters  \cite{unnikrishnan2020}.

In Fig.~(\ref{fig:DelegatedScenarios}), three scenarios for quantum metrology with delegated tasks are presented. Separately, we show that the task of state preparation and measurements can be delegated to an untrusted third party if reinforced with a proper cryptographic protocol. The natural question to ponder is if both tasks can be delegated to a third party, where the trusted party, Alice, can only perform the encoding map $\Lambda_\theta$ and a set of encryption operations. At first glance, this seems possible by fusing the verification protocol of \cite{markham2020} and the protocol presented for delegated measurements. A local Clifford encryption guarantees absolute privacy and soundness is easily derived for the case when $\Lambda_\theta=\mathbb{I}$. As the nature of $\Lambda_\theta$ should have little to no impact on the soundness, it ought to follow that a similar derivation can be performed for any CPTP encoding $\Lambda_\theta$. A future perspective is to prove and verify this claim.

%% file: Chapters/Chapter7-Remarks.tex
\chapter{Remarks}

\begin{quote}
    \textit{La volonté trouve, la liberté choisit. Trouver et choisir, c'est penser.}
    \begin{flushright} -Victor Hugo \end{flushright}
\end{quote}

Quantum metrology is a promising discipline of quantum information; it has a broad scope of applications in a variety of scientific fields and is currently witnessing an abundance experimental and theoretical developments. The objective of quantum metrology is to use quantum probes to estimate unknown parameters as accurately as possible. By capitalizing on quantum properties, it is possible to achieve a precision which is unobtainable using the best classical strategies. This thesis explored how other quantum techniques can be appropriately incorporated within the realm of quantum metrology. Specifically, the utility of graph states, the limitations of quantum error correction, and the consequences of a cryptographic framework. Within each scenario, the idealized `Heisenberg limit precision' is used as a figure of merit.

The work in this thesis is uniquely theoretical, even so, the general philosophy was to be relevant and applicable to the first generations of quantum hardware. Graph states can be constructed using only control-$Z$ operations and the QFI of graph states can be approximately saturated using single qubit measurements (\textbf{Chapter~4}). The error correction protocol in \textbf{Chapter~5} is currently realizable \cite{dutt2007, taminiau2014, waldherr2014}. Most of the cryptographic protocols presented in \textbf{Chapter~6} use local encryption/decryption operations and use local measurements on ancillary qubits. One concern may be that the noise models are `too idealized' and the adversarial tools are `too abstract', and in general these will be dependent on the quantum hardware. In reply, the models presented in this thesis are a baseline and can be straightforwardly adapted to better describe the desired setting - as any experimental setup will be extremely dependent on the implementation and the available technologies.

Another general philosophy we strove to maintain was a sense of generality from the perspective of quantum metrology. However, an arbitrary parameter encoding map $\Lambda_\theta$ is quite vague and the equations pertaining to parameter estimation are highly non-linear, making it difficult to draw conclusions from the most general situation. Instead, we often used the frequentist approach to phase estimation as a baseline example. This example, is canonical with quantum metrology and has several applications \cite{holland1993, paris2009, giovannetti2011}. Nonetheless, many of the mathematical tools and derivations can be adapted to specific $\Lambda_\theta$. In particular with respect to \textbf{Chapter~4}, if $\Lambda_\theta =e^{-i\theta G}$, then the QFI can be defined as a relationship between the stabilizer group of the initialized quantum state and the expansion of $G$ and $G^2$. As hypothesized in \textbf{Chapter~5}, similar results for the limitations of error correction are likely obtained regardless of $\Lambda_\theta$, note that it is necessary to implement an error correction protocol that does not interfere with $\Lambda_\theta$. 

Similarly, most of the mathematical tools and results can be adapted to the multiparameter quantum metrology. The main difficulty, incompatibility of simultaneous measurements \cite{ragy2016}, is a standard across the multiparameter framework and not inherent to any of the settings we explored. Although not explicitly proven, if there does exists a set of compatible measurements, then the integrity of the estimate for each parameter in a cryptographic framework will be of the form Eq.~\eqref{eq:biasbound} and Eq.~\eqref{eq:MSEintegrity}. The reason being that an equivalent derivation would follow from a secondary (compatible) measurement (and all subsequent measurements) because of the definition of the trace distance. A future perspective is to formally address this question.

Aside from the brief summary of Bayesian statistical interference in \textbf{Chapter~3}, this thesis is void of the Bayesian approach to quantum metrology \cite{holevo1982, jarzyna2015, rubio2018}. Because the work of \textbf{Chapter~4} is heavily influenced by the QFI, which is not used in the Bayesian approach, it is unclear if the shape of a graph can be related to the practicality of the corresponding graph state for phase estimation using a Bayesian approach. With respect to \textbf{Chapter~6}, it is impossible to gauge the integrity of a Bayesian quantum metrology problem within a cryptographic framework without first specifying a cost function, Eq.~\eqref{eq:costfunction}, and estimation strategy.

\section{Summary of Results and Future Perspectives}

\subsection{Chapter 4}

In \textbf{Chapter 4}, we demonstrate that graph states - in conjunction with their existing versatility - are a useful resource for quantum metrology. This is done by constructing a class of graph states, named bundled graph states, which possess a large amount of internal symmetry, and in consequence can approximately saturate the Heisenberg limit. More so, graph states are robust against dephasing noise and a small number of erasures, and the QFI can be approximately saturated with a simple measurement scheme. The robustness against a small number of erasures is compelling as the standard resource for phase estimation, GHZ states, lose all functionality after a single erasure \cite{toth2014}.

By construction, bundled graph states are a natural resource for multiparameter metrology, specifically in the context of quantum sensing networks \cite{komar2014, eldredge2018, proctor2018, rubio2020a}. Each bundle can be subjected to independent parameter encoding schemes. The robustness derivations can be generalized for noise models which are bundle dependent, and a compatible measurement scheme arises from the fact that not all parameters are encoded into each qubit.

\subsection{Chapter 5}

The limitations of error correction enhanced quantum metrology is outlined in \textbf{Chapter~5}. In contrast to previous results, which state that the Heisenberg limit can be permanently achieved if the signal and noise are orthogonal \cite{demkowicz2017, zhou2018}, we show that when hardware limitations are accounted for, the Heisenberg limit is eventually lost. As expected, if the frequency of error correction is high enough, the Heisenberg limit is achieved for a serviceable duration of time. Even though the focus is a single error correction protocol, we conjecture that the results translate to any error correction protocol or noise mitigation strategy, as small deviations of the phase caused by noise cannot be perfectly corrected. Eventually these small deviations accumulate enough such that the quantum state is useless for quantum metrology.

\subsection{Chapter 6}

In the presence of a (potential) malicious adversary, many notions of estimation theory have to be altered in some capacity. This is simply because there is no guarantee of having access to the ideal resource, leading to ambiguity in the construction of an estimator. In \textbf{Chapter~6}, we formalize the consequences of quantum metrology within a cryptographic framework. The idea is to use the same estimation strategy as if there was no malicious adversary, and if an appropriate protocol is used to detect any malicious alterations, then the soundness of said protocol can be linked to the integrity of the quantum metrology problem. Integrity is a concept used in quantum cryptography, which quantifies the ability to retain the functionality of the underlying process, in this case the underlying process is quantum metrology, and so we decided that the integrity will encapsulate any added bias and the difference in precision.

Additionally, in \textbf{Chapter 6}, we constructed several cryptographic protocols for cryptographic quantum metrology. The cryptographic protocols are each motivated by the absence of the necessary quantum hardware to fully execute the quantum metrology task, forcing interactions with a third party. For example, protocols to transmit quantum information across an unsecured quantum channel, and protocols to guarantee security when a task, either quantum state verification or quantum measurements, are performed by an untrusted party. It goes without saying that it is impossible to delegate the task of parameter encoding to an untrusted party, as this defeats the purpose of quantum metrology.

The immersion of quantum cryptography into quantum metrology is a novel area of research, and as such there are several future perspectives. Just as the first generations of quantum technologies will be limited in abilities, so too will be the abilities of a malicious adversary. In our work, in order to fulfill a notion of computational security, no assumptions about the malicious adversaries are made. By limiting the possible attacks a malicious adversary can perform (as one expects), an improved notion of cryptographic soundness is achieved. One concern of the quantum cryptography protocols presented in \textbf{Chapter~6} is the dependence on noiseless quantum operations, this restriction can be loosened by tweaking acceptance parameters \cite{unnikrishnan2020}. Lastly, a possible future research direction is to consider continuous variable resources, because it is not obvious if the analogous protocols will satisfy the same soundness inequalities.

\section{Secure Sensing Networks}

An ongoing project of mine is to adopt quantum sensing networks \cite{komar2014, eldredge2018, ge2018, proctor2018, zhuang2018, rubio2020a, guo2020} into the realm of quantum metrology in a cryptographic framework. This is a logical next step in the direction of cryptography and quantum metrology, as there exists a plethora of secure multipartite protocols across quantum networks \cite{pappa2012, huang2019b} and the foundation introduced in \textbf{Chapter~6} is easily adapted to the network setting. Additionally, the authors of \cite{komar2014}, who proposed a clock synchronization scheme across a quantum network, are the first to consider the security aspect of a quantum metrology problem.

The idea is straightforward: there is a central node who has the ability to prepare highly entangled quantum states, qubits are then distributed throughout the network for a multiparameter quantum metrology problem. We introduce two types of malicious adversaries: the first are eavesdroppers who can interact with the quantum channels, and the second are some of the exterior nodes of the network who may behave maliciously. By adapting the protocols introduced in \cite{SMK21,SM21} to the network setting, we establish the concept of a secure quantum sensing network.

In addition to the cryptographic notions of soundness, privacy and integrity, the notion of anonymity is introduced to quantum sensing networks \cite{unnikrishnan2019}. We define anonymity within quantum sensing networks to mean that the local parameters cannot be estimated from the measurement results, only a global parameter, for example, an average of the parameters. For example, the total power consumption of all appliances may be transmitted to a power supplier, but not the consumption of individual appliances. In some sense, anonymity is a form of privacy.

\clearpage

\thispagestyle{plain}

\vspace*{\fill}
\begin{quote}
    \textit{Somewhere, something incredible is waiting to be known.}
    \begin{flushright}-Carl Sagan\end{flushright}
\end{quote}
\vspace*{\fill}

%% file: Appendices/AppendixA.tex
\chapter{Robustness of Graph States Subjected to Noise}


Recall that a graph $G=(V,E)$ is divided into disjoint subsets $U_1,U_2,\ldots,U_l,\ldots$ such that $\bigcup_{l} U_l = V$. The vertices are partitioned in accordance to commonly shared neighbourhoods, hence, if $v_i \in U_a$ and $v_j \in U_b$, then $N(v_i)=N(v_j)$ if $a=b$ and $N(v_i) \neq N(v_j)$ if $a \neq b$. We write that $|U_l|=u_l$ and the shared neighbourhood of $U_l$ is $M_l$ with $|M_l|=m_l$.

In both proofs, sums are taken over all possible combinations of qubits, indexed by vectors. When these vectors are summed, it is taken modulo $2$. For example if $\vec{j}= \{ 1, 1 ,0 \}$ and $\vec{k} = \{ 1, 0 ,1 \}$, then $\vec{j}+\vec{k}=\{ 0, 1 , 1\}$. 

\section{Robustness Against IID Dephasing}

After a graph state undergoes iid dephasing, it can be expressed as
\begin{equation}
    \sum_{\vec{j}} p^j (1-p)^{n-j} Z_{\vec{j}} \dyad{G} Z_{\vec{j}}.
\end{equation}
Conveniently, the above quantum state is already expressed as a sum of orthogonal pure states, computing the QFI is then a straightforward use of the general expression
\begin{equation}
    \mathcal{Q}(G^\text{dephasing}) = \frac{1}{2} \sum_{\vec{j},\vec{k}} \frac{(\lambda_{\vec{j}}-\lambda_{\vec{k}})^2}{\lambda_{\vec{j}}+\lambda_{\vec{k}}} \big| \bra{G} Z_{\vec{j}} \sum_i X_i Z_{\vec{k}} \ket{G} \big|^2,
\end{equation}
where $\lambda_{\vec{j}}=\lambda_j=p^j(1-p)^{n-j}$. The only non-vanishing terms in the sum occurs when $\vec{j}+\vec{k}=M_l$ for some $l$. We divide $\vec{k}$ into three disjoint parts, $a$ qubits with a flipped phase from the set $U_l$, $b$ qubits with a flipped phase from $M_l$, and $c$ qubits from the remaining qubits
\begin{equation}
    \begin{split}
        \mathcal{Q}(G^\text{dephasing}) &= \frac{1}{2} \sum_{l} \sum_{\vec{k}} \frac{(\lambda_{\vec{k}+M_l}-\lambda_{\vec{k}})^2}{\lambda_{\vec{k}+M_l}+\lambda_{\vec{k}}} \big| \bra{G} Z_{\vec{k}+M_l} \sum_i X_i Z_{\vec{k}} \ket{G} \big|^2 \\
        &= \frac{1}{2} \sum_{l} \sum_{a=0}^{u_l} \sum_{b=0}^{m_l} \sum_{c=0}^{n-u_l-m_l} \frac{(\lambda_{a-b+c+m_l}-\lambda_{a+b+c})^2}{\lambda_{a-b+c+m_l}+\lambda_{a+b+c}} (u_l-2a)^2 \\
        &= \sum_{l} f_l g_l,
    \end{split}
\end{equation}
where
\begin{equation}
    f_l=u_l^2 (1-2p)^2 +4u_l p(1-p) \geq u_l^2 (1-2p)^2,
\end{equation}
and
\begin{equation}
    \begin{split}
        g_h &= \frac{1}{2} \sum_{j=0}^{m_l} \binom{m_l}{j} \frac{\big(p^{m_l-j}(1-p)^j - p^{j}(1-p)^{m_l-j} \big)^2}{p^{m_l-j}(1-p)^j + p^{j}(1-p)^{m_l-j}} \\
        &\geq 1-\big(2p(1-p)+1/2 \big)^{m_l} \\
        &\geq 1-\big(2p(1-p)+1/2 \big)^{m},
    \end{split}
\end{equation}
where $m=\min_l m_l$. Combining the bounds of $f_l$ and $g_l$ with the fact that $\sum_{l} u_l^2 = \mathcal{Q}(G)$, one obtains
\begin{equation}
    \mathcal{Q}(G^\text{dephasing}) \geq (1-2p)^2 \Big( 1-\big(2p(1-p)+1/2 \big)^{m} \Big) \mathcal{Q}(G).
\end{equation}

\section{Robustness Against Finite Erasures}

We return to the stabilizer representation to obtain a useful closed form expression for a graph state $\vec{G}$ subjected to erasures indexed by $\vec{e}$
\begin{equation}
    \ket{G} \rightarrow \Tr_{\vec{e}} \dyad{G} = \frac{1}{2^n} \sum_{S \in \mathcal{S}} \Tr_{\vec{e}} S.
\end{equation}
Recall that the stabilizer group $\mathcal{S}$ can be generated by generators $g_i = X_i \bigotimes_{j \in N(i)} Z_j$. Therefore each stabilizer $S$ can be written in the form
\begin{equation}
    S=g_1^{a_1}g_2^{a_2} \ldots g_n^{a_n},
\end{equation}
where $a_j \in \{ 0, 1 \}$. Thus, $\Tr_{\vec{e}} S$ vanishes under two conditions. The first is if $a_x=1$ for any $x$ indexed by $\vec{e}$. The second is if $\sum_{j \in N(x)} a_j \equiv 1 \mod 2$ for any $x$ indexed $\vec{e}$. Define the set $L_{\vec{e}}$ to be set of erased qubits and their neighbourhoods
\begin{equation}
    L_{\vec{e}} = \bigcup_{x \in \vec{e}} \{ x \} \cup N(x).
\end{equation}
Define $\tilde{Z}$ to be the set of all possible combination of $Z$ operators indexed by a subset of $L_{\vec{e}}$
\begin{equation}
    \tilde{Z} = \{ Z_{\vec{j}} \; | \; \vec{j} \subseteq L_{\vec{e}} \}.
\end{equation}
Any stabilizer $S$ which is traced out, i.e $\Tr_{\vec{e}} S = 0$, will commute with half of Pauli operators in $\tilde{Z}$ and anti-commute with the other half. Any stabilizer which is not traced out will commute with all of the operators. From which it follows that, the quantum state after going erasures indexed by $\vec{e}$ can be expressed as
\begin{equation}
    \label{eq:erasedstate}
    2^{-|L_{\vec{e}}|}\sum_{\vec{j} \in L_{\vec{e}}} Z_{\vec{j}} \dyad{G} Z_{\vec{j}}.
\end{equation}
As it was noted in the main text, the above mixed state is left as $n$ qubit state for clarity. The traced out systems are equivalent to maximally mixed states, $\mathbb{I}/2$, which are irrelevant with respect to the QFI.

The quantum state in Eq.~\eqref{eq:erasedstate} is written as a sum of orthonormal pure states. The QFI is thus
\begin{equation}
    \begin{split}
        \mathcal{Q}(G^{\text{erasures } \vec{e}}) &= \frac{1}{2} \sum_{\vec{j},\vec{k}} \frac{(\lambda_{\vec{j}}-\lambda_{\vec{k}})^2}{\lambda_{\vec{j}}+\lambda_{\vec{k}}} \big| \bra{G} Z_{\vec{j}} \sum_i X_i Z_{\vec{k}} \ket{G} \big|^2 \\
        &= \frac{1}{2} \sum_{l} \sum_{\vec{k}} \frac{(\lambda_{\vec{k}+M_l}-\lambda_{\vec{k}})^2}{\lambda_{\vec{k}+M_l}+\lambda_{\vec{k}}} \big| \bra{G} Z_{\vec{k}+M_l} \sum_i X_i Z_{\vec{k}} \ket{G} \big|^2,
    \end{split}
\end{equation}
where $2^{-|L_{\vec{e}}|}$ if $\vec{j} \subseteq L_{\vec{e}}$ and $0$ otherwise. It follows then that $\lambda_{\vec{k}+M_l}-\lambda_{\vec{k}}=0$ if $\vec{k},\vec{k}+M_l \subseteq L_{\vec{e}}$. Regardless of $\vec{k}$, this only occurs if $M_l \subseteq L_{\vec{e}}$. If $M_l \nsubseteq L_{\vec{e}}$, the sum over $\vec{k}$ depends on if $U_l \subseteq L_{\vec{e}}$ or $U_l \nsubseteq L_{\vec{e}}$
\begin{equation}
    \mathcal{Q}(G^{\text{erasures } \vec{e}}) = \sum_{l} h_l (\vec{e}),
\end{equation}
where
\begin{equation}
  h_l (\vec{e})  = \begin{cases}
    u_l^2 & \text{if } M_l \nsubseteq L_{\vec{e}} \text{ and } U_l \nsubseteq L_{\vec{e}}\\
    u_l & \text{if } M_l \nsubseteq L_{\vec{e}} \text{ and } U_l \subseteq L_{\vec{e}} \\
    0 & \text{otherwise}
    \end{cases}.
\end{equation}

%% file: Appendices/AppendixB.tex
\chapter{QFI of a Noisy GHZ State}


\section{Solving the Master Equation}

The dynamics of \textbf{Chapter 5} are governed by the master equation
\begin{equation}
\label{eq:MasterEqAppedix}
\frac{d \rho}{dt}=-\frac{i}{\hbar} [H, \rho ] + \gamma \sum_{m=1}^n (X_m \rho X_m - \rho),
\end{equation}
with $H = \frac{\hbar \omega}{2} \sum_{m=1}^n Z_m$. In this appendix, we derive the solutions to the dynamics, as well as the modified version in which error correction is incorporated. 
Without loss of generality, it can be assumed that the solution is of the form
\begin{equation}
\rho = \sum_{j,k} \alpha_{j,k} \dyad{j}{k},
\end{equation}
where $j,k \in \{0,1 \}^{\otimes n}$ are bit strings of length $n$. As such, one approach to solving the master equation, Eq.~\eqref{eq:MasterEqAppedix}, is to view it a system of linear differential equations with respect to the amplitudes $\alpha_{j,k}$. Because, the quantum state is initialized in a GHZ state, the only non-zero amplitudes are those of the form $\alpha_{j,j}$ or $\alpha_{j,\bar{j}}$, where $\ket{\bar{j}}=X^{\otimes n}\ket{j}$. Furthermore, the system of differential equation can be divided into two independent equations
\begin{align}
    \frac{d \vec{a}}{dt} &=A \vec{a}, \\
    \frac{d \vec{b}}{dt} &=B \vec{b},
\end{align}
where $\vec{a}$ ($\vec{b}$) is a vector of size $2^n$ containing all of the amplitudes of the form $\alpha_{j,j}$ ($\alpha_{j,\bar{j}}$), and
\begin{align}
A &= \sum_{m=0}^{n-1} \begin{pmatrix}
1 & 0 \\
0 & 1
\end{pmatrix}^{\otimes m} \otimes \begin{pmatrix}
-\gamma & \gamma \\
\gamma & -\gamma
\end{pmatrix} \otimes \begin{pmatrix}
1 & 0 \\
0 & 1
\end{pmatrix}^{\otimes n-m-1}, \\
B &= \sum_{m=0}^{n-1} \begin{pmatrix}
1 & 0 \\
0 & 1
\end{pmatrix}^{\otimes m} \otimes \begin{pmatrix}
-i \omega -\gamma & \gamma \\
\gamma & i \omega -\gamma
\end{pmatrix} \otimes \begin{pmatrix}
1 & 0 \\
0 & 1
\end{pmatrix}^{\otimes n-m-1}.
\end{align}
Both and $A$ and $B$ are time-independent, therefore the solutions are given by the corresponding matrix exponential: $\vec{a}=e^{At}\vec{a}_0$ and $\vec{b}=e^{Bt}\vec{b}_0$ (here $\vec{a}_0$ and $\vec{b}_0$ are the initial amplitude vectors), where
\begin{equation}
\label{eq:diffeqA}
e^{A t} = e^{-n \gamma t} \begin{pmatrix}
\cosh ( \gamma t ) & \sinh ( \gamma t ) \\
\sinh ( \gamma t ) & \cosh ( \gamma t )
\end{pmatrix}^{\otimes n} = e^{-n \gamma t} \begin{pmatrix}
c_\gamma & s_\gamma \\
s_\gamma & c_\gamma
\end{pmatrix}^{\otimes n},
\end{equation}
and
\begin{equation}
\label{eq:diffeqB}
e^{B t} = e^{-n \gamma t} \begin{pmatrix}
\cos ( \Delta t ) - i \frac{\omega}{\Delta} \sin ( \Delta t ) & \frac{\gamma}{\Delta} \sin ( \Delta t ) \\[5pt]
\frac{\gamma}{\Delta} \sin (\Delta t ) & \cos ( \Delta t ) + i \frac{\omega}{\Delta} \sin ( \Delta t )
\end{pmatrix}^{\otimes n} = e^{-n \gamma t} \begin{pmatrix}
x_- & y \\[5pt]
y & x_+
\end{pmatrix}^{\otimes n},
\end{equation}
with $\Delta=\sqrt{\omega^2-\gamma^2}$. Because this is a solution with complex solutions, there are no issues when $\gamma^2 > \omega^2$, this maps the usual trigonometric functions ($\cos$ and $\sin$) to their hyperbolic counterparts ($\cosh$ and $\sinh$). The notation - $c_\gamma = \cosh ( \gamma t )$, $s_\gamma = \sinh ( \gamma t )$, $y=\frac{\gamma}{\Delta} \sin (\Delta t )$ and $x_\pm = \cos ( \Delta t ) \pm i \frac{\omega}{\Delta} \sin ( \Delta t )$ - is used for conciseness.

\section{QFI without Error Correction}

In the case without error correction, one can simply use the solutions of the differential equations, Eq.~\eqref{eq:diffeqA} and Eq.~\eqref{eq:diffeqB}. The quantum state at time $t$ is given by
\begin{equation}
\rho = \frac{1}{2} \sum_j \lambda_{j,+} \dyad*{\psi_{j,+}}+\lambda_{j,-} \dyad*{\psi_{j,-}},
\end{equation}
with
\begin{equation}
\lambda_{j,\pm}=e^{-n \gamma t} \frac{s_j \pm r_j}{2},
\end{equation}
and
\begin{equation}
\ket{\psi_{j,\pm}}=\frac{1}{\sqrt{2}}\big(e^{-i \theta_j /2}\ket{j} \pm e^{+i \theta_j /2} \ket{\bar{j}} \big). 
\end{equation}
The factor of $1/2$ in front of the sum is to avoid double counting, because $\lambda_{j,\pm}=\lambda_{\bar{j},\pm}$ and $\ket{\psi_{j,\pm}}=\ket{\psi_{\bar{j},\pm}}$. The eigenvalues and eigenvectors are parameterized by
\begin{align}
s_j &= c_\gamma^{n-h_j} s_\gamma^{h_j}+c_\gamma^{h_j} s_\gamma^{n-h_j}, \\
r_j e^{\pm i \theta_j} &= x_\pm^{h_j} y^{n-h_j}+x_\mp^{n-h_j} y^{h_j},
\end{align}
where $h_j$ is the Hamming weight (number of $1$'s) of $j$. Using the general formula for the QFI, one obtains
\begin{equation}
\begin{split}
\mathcal{Q}_\text{noisy} &= \frac{1}{2}\sum_j \Bigg( \frac{\dot{\lambda}_{j,+}^2}{\lambda_{j,+}}  +  \frac{\dot{\lambda}_{j,-}^2}{\lambda_{j,-}} + 2\frac{(\lambda_{j,+}-\lambda_{j,-})^2}{\lambda_{j,+}+\lambda_{j,-}} \Big( \big| \braket*{\psi_{j,+}}{\dot{\psi}_{j,-}} \big|^2+\big| \braket*{\psi_{j,-}}{\dot{\psi}_{j,+}} \big|^2 \Big) \Bigg)\\
&= \frac{e^{-n \gamma t}}{2}\sum_j \frac{s_j \dot{r}_j^2}{s_j^2-r_j^2} +\frac{r_j^2}{s_j} \dot{\theta}_j^2 \\
&= n^2t^2\Big(1-\big(2-\frac{4}{3n}\big)\gamma t\Big)+ \mathcal{O}\big( t^4 \big),
\end{split}
\end{equation}
where the notation $\dot{\square}=\partial_\omega \square$ for clarity and the factor of $1/2$ in front of the sum is again used to avoid double counting.

\section{QFI using the Parity Check Code}

The overall dynamics are modified upon inclusion of error correction. The system evolves in accordance to the master equation, Eq.~\eqref{eq:MasterEqAppedix}, for time $\tau$, after which an error correction operation is performed. This process is repeated until the total time $t$ has passed (it is assumed that $t/\tau$ is an integer).

To incorporate the parity check code into the dynamics, the evolution of the ancillary qubit (indexed by $m=n+1$), which is subjected to dephasing with a rate of $\xi$, must be tracked. The matrix solutions of this altered system are
\begin{equation}
e^{A \tau} = e^{-(n \gamma+\xi) \tau} \begin{pmatrix}
c_\gamma & s_\gamma \\
s_\gamma & c_\gamma
\end{pmatrix}^{\otimes n} \otimes \begin{pmatrix}
c_\xi & s_\xi \\
s_\xi & c_\xi
\end{pmatrix},
\end{equation}
and
\begin{equation}
e^{B \tau} = e^{-(n \gamma+\xi) \tau} \begin{pmatrix}
x_- & y \\
y & x_+
\end{pmatrix}^{\otimes n} \otimes \begin{pmatrix}
c_\xi & s_\xi \\
s_\xi & c_\xi
\end{pmatrix},
\end{equation}
where $c_\xi = \cosh (\xi \tau)$ and $s_\xi = \sinh ( \xi \tau)$. Note that the other variables - $c_\gamma$, $s_\gamma$, $y$ and $x_\pm$ - are in terms of $\tau$ here (not $t$).

When using the parity check code, a correction is made on a sensing qubit if it has a different parity than the ancillary qubit. Imperfect syndrome diagnosis is simulated by adding a probability that the syndrome diagnosis outputs an incorrect result with probability $p$. The overall dynamics of the error correction can be translated into the matrix language
\begin{equation}
E =\begin{pmatrix}
1-p & 1-p \\
p & p \\
\end{pmatrix}^{\otimes n} \otimes \begin{pmatrix}
1 & 0 \\
0 & 0 \\
\end{pmatrix} + \begin{pmatrix}
p & p \\
1-p & 1-p \\
\end{pmatrix}^{\otimes n} \otimes \begin{pmatrix}
0 & 0 \\
0 & 1 \\
\end{pmatrix}.
\end{equation}

Combing everything, the amplitudes after time $t$, and therefore $t/\tau$ applications of the parity check code, is
\begin{equation}
\label{eq:AEvo}
\begin{aligned}
\vec{a} = \big( E e^{A\tau} \big)^{t/\tau} \vec{a}_0
=e^{-\xi t} \Bigg( & \begin{pmatrix}
1-p & 1-p \\
p & p
\end{pmatrix}^{\otimes n} \otimes \begin{pmatrix}
c_\xi & s_\xi \\
0 & 0
\end{pmatrix} \\
+& \begin{pmatrix}
p & p \\
1-p & 1-p
\end{pmatrix}^{\otimes n} \otimes \begin{pmatrix}
0 & 0 \\
s_\xi & c_\xi
\end{pmatrix} \Bigg)^{t/\tau} \vec{a}_0,
\end{aligned}
\end{equation}
and
\begin{equation}
\label{eq:BEvo}
\begin{aligned}
\vec{b} = \big( E e^{B\tau} \big)^{t/\tau} \vec{b}_0
=r^{nt/\tau}e^{-\xi t} \Bigg( & \begin{pmatrix}
(1-p)e^{-i\phi} & (1-p)e^{i\phi} \\
pe^{-i\phi} & pe^{i\phi}
\end{pmatrix}^{\otimes n} \otimes \begin{pmatrix}
c_\xi & s_\xi \\
0 & 0
\end{pmatrix} \\
+& \begin{pmatrix}
pe^{-i\phi} & pe^{i\phi} \\
(1-p)e^{-i\phi} & (1-p)e^{i\phi}
\end{pmatrix}^{\otimes n} \otimes \begin{pmatrix}
0 & 0 \\
s_\xi & c_\xi
\end{pmatrix} \Bigg)^{t/\tau} \vec{b}_0,
\end{aligned}
\end{equation}
with
\begin{equation}
    re^{\pm i \phi} = e^{-\gamma \tau} ( x_\pm + y ) = e^{-\gamma \tau} \Big( \cos ( \Delta \tau ) + \frac{\gamma \pm i \omega}{\Delta} \sin ( \Delta \tau) \Big).
\end{equation}

It easy to show using Eq.~(\ref{eq:AEvo}) that the final amplitude corresponding to the outer product $\dyad{j0}$ is equal to $\frac{(1-p)^{n-h_j}p^{h_j}}{2}$, and similarly the amplitude corresponding to the outer product $\dyad{\bar{j}1}$ is also equal to $\frac{(1-p)^{n-h_j}p^{h_j}}{2}$ - here $h_j$ is the Hamming weight of the bit string of the sensing qubits, and does not include the ancillary qubit. The solution to Eq.~(\ref{eq:BEvo}) is more complex. After the first round of error correction (and each subsequent round), the amplitude corresponding to the outer product $\dyad{j0}{\bar{j}1}$ is of the form $\frac{(1-p)^{n-h_j}p^{h_j} Re^{-i \theta}}{2}$, and the amplitude corresponding to the outer product $\dyad{\bar{j}1}{j0}$ is of the form $\frac{(1-p)^{n-h_j}p^{h_j} Re^{i \theta}}{2}$. By translating the problem to a recurrence relation between $Re^{-i \theta}$ and $Re^{i \theta}$, the problem becomes
\begin{equation}
\begin{aligned}
\begin{pmatrix}
Re^{-i \theta} \\
Re^{+i \theta}
\end{pmatrix}  &= r^{nt/\tau} e^{-\xi t}  \begin{pmatrix}
c_\xi q_- & s_\xi q_+ \\
s_\xi q_- & c_\xi q_+ \\
\end{pmatrix}^N \begin{pmatrix}
\upsilon_- \\
\upsilon_+
\end{pmatrix} \\
&=  r^{n t/\tau} e^{-\xi t}  \left(  \frac{\mu_+ \mu_-^N-\mu_-\mu_+^N}{\mu_+-\mu_-} \begin{pmatrix}
1 & 0 \\
0 & 1
\end{pmatrix} +  \frac{ \mu_+^N-\mu_-^N}{\mu_+-\mu_-} \begin{pmatrix}
c_\xi q_- & s_\xi q_+ \\
s_\xi q_- & c_\xi q_+ \\
\end{pmatrix}  \right)  \begin{pmatrix}
\upsilon_- \\
\upsilon_+
\end{pmatrix},
\end{aligned}
\end{equation}
with $N(nt/\tau-1)$, and
\begin{equation}
    q_\pm=(1-p)e^{\pm i \phi}+p e^{\mp i \phi},
\end{equation}
\begin{equation}
\mu_\pm = c_\xi \cos \phi \pm \sqrt{q_+ q_- s_\xi^2-(1-2p)^2 c_\xi^2 \sin^2 \phi},
\end{equation}
\begin{equation}
\upsilon_\pm = c_\xi e^{\pm i n \phi} + s_\xi e^{\mp i n \phi}. 
\end{equation}

In the general setting, with a noisy ancilla and imperfect error correction, the quantum state after time $t$ can be written as
\begin{equation}
\rho = \sum_j \lambda_{j,+} \dyad{\psi_{j,+}}+\lambda_{j,-} \dyad{\psi_{j,-}},
\end{equation}
where
\begin{equation}
\lambda_{j,\pm}=(1-p)^{n-h_j}p^{h_j}\frac{1\pm R}{2},
\end{equation}
and
\begin{equation}
\ket{\psi_{j,\pm}}=\frac{1}{\sqrt{2}} \big( e^{-i\theta/2} \ket{j0}\pm e^{+i\theta/2} \ket{\bar{j}1} \big).
\end{equation}
From which, the QFI can be computed
\begin{equation}
\begin{aligned}
\mathcal{Q} &= \sum_j  \Bigg( \frac{\dot{\lambda}_{j,+}^2}{\lambda_{j,+}} +  \frac{\dot{\lambda}_{j,-}^2}{\lambda_{j,-}}+2\frac{(\lambda_{j,+}-\lambda_{j,-})^2}{\lambda_{j,+}+\lambda_{j,-}} \Big( \big| \braket*{\psi_{j,+}}{\dot{\psi}_{j,-}} \big|^2+\big| \braket*{\psi_{j,-}}{\dot{\psi}_{j,+}} \big|^2 \Big) \Bigg)\\
&= \frac{\dot{R}^2}{1-R^2}+R^2 \dot{\theta}^2 .
\end{aligned}
\end{equation}

The various sub-cases are explored in the following subsections.

\subsection{Ideal Error Correction}

The simplest case is when $\xi=0$ and $p=0$. In this scenario $Re^{\pm i \theta}=\big( re^{\pm i \phi}\big)^{nt/\tau}$. The QFI simplifies greatly, it can be written in the form
\begin{equation}
    \mathcal{Q}_1 = n^2 t^2 r^{2nt/\tau} f,
\end{equation}
where
\begin{equation}
\label{eq:propconstant1}
f=\frac{1}{\tau^2}\Big( \frac{1}{1-r^{2nt/\tau}}\frac{\dot{r}^2}{r^2}+\dot{\phi}^2 \Big) = 1-2 \gamma \tau + \frac{7 \gamma^2 \tau^2}{3} + \frac{4\gamma \tau^2}{3nt} + \mathcal{O} \big(\tau^3 \big),
\end{equation}
and
\begin{equation}
r^{2nt/\tau}=1-\frac{4}{3} n t \gamma \omega^2 \tau^2 + \mathcal{O} \big( \tau^3 \big).
\end{equation}

\subsection{Noisy Ancilla}

The second case has a noisy ancillary qubit ($\xi \neq 0$). The analytic expression for $Re^{\pm i \theta}$ is quite complicated; to gauge the effects of the noise a Taylor expansion is performed
\begin{equation}
Re^{\pm i \theta} = (1-\xi t)\big( r e^{\pm i \phi} \big)^{nt/\tau} + \xi \tau \frac{\sin ( n t \phi/\tau )}{\sin ( n \phi )} \big( r e^{\mp i \phi} \big)^{n} + \mathcal{O} \big( \xi^2 \big).
\end{equation}
Which leads to a QFI of
\begin{equation}
\mathcal{Q}_2 =n^2 t^2 r^{2n t /\tau}(f-g \xi)+\mathcal{O} \big( \xi^2 \big),
\end{equation}
where
\begin{equation}
\begin{aligned}
g  = &\Big(\frac{n \omega t \big(1+3 \cos (2n \omega t)\big)+(n^2 \omega^2 t^2 -2) \sin (2 n \omega t)}{n^3 \omega^3 t^3}+2 \Big)t \\
& +   \frac{2\big(n \omega t \cos ( n \omega t )- \sin (n \omega t) \big)^2}{n^2 \omega^2 t^2} \tau  \\
& + \Big( \frac{(4n \omega t-2n^3 \omega^3 t^3)\cos (2n \omega t)  - (2-5n^2 \omega^2 t^2) \sin (2n \omega t)}{n^3 \omega^3 t^3} - 4 \Big) \gamma t \tau  + \mathcal{O} \big( \tau^2 \big).
\end{aligned}
\end{equation}
To simplify analysis, the following inequalities are used:
\begin{gather}
\frac{2}{3} \leq \frac{n \omega t \big(1+3 \cos (2n \omega t)\big)+(n^2 \omega^2 t^2 -2) \sin (2 n \omega t)}{n^3 \omega^3 t^3}+2  \leq \frac{5}{2}, \\
0 \leq \frac{2\big(n \omega t \cos ( n \omega t )- \sin (n \omega t) \big)^2}{n^2 \omega^2 t^2} \leq \frac{5}{2}, \\
-7 \leq \frac{(4n \omega t-2n^3 \omega^3 t^3)\cos (2n \omega t)-(2-5n^2 \omega^2 t^2) \sin (2n \omega t)}{n^3 \omega^3 t^3} -4 \leq 0,
\end{gather}
from which it follows that
\begin{equation}
\Big(\frac{2}{3}-7\gamma \tau \Big)t \leq g + \mathcal{O} \big( \tau^2 \big) \leq \frac{5}{2}(t+\tau).
\end{equation}

\subsection{Imperfect Error Correction}

The third case has imperfect syndrome diagnosis ($p \neq 0$). It is straightforward to show
\begin{equation}
    Re^{\pm i \theta} = \big( r e ^{\pm i \phi} \big)^{nt/\tau} \big( q_\pm e^{\mp i \phi} \big)^{n(t/\tau-1)}.
\end{equation}
Which leads to a QFI of
\begin{equation}
\mathcal{Q}_3 = n^2 t^2 (rq)^{2nt/\tau} h,
\end{equation}
with  $q^2=q_+ q_-$, which also satisfies
\begin{equation}
q^{2nt/\tau}=1-4p(1-p) \omega^2 t \tau+ \mathcal{O} \big( \tau^2 \big), 
\end{equation}
and
\begin{equation}
h =(1-2p)^2 f + 4p\Big( \frac{1-p}{n}+1-2p\Big) \frac{\tau}{t} + \mathcal{O} \big( \tau^2 \big).
\end{equation}

\section{QFI using the Generalized Bit Flip Code}

The generalized bit flip code \cite{gottesman1997} does not use an ancillary qubit, instead a global stabilizer measurement is made. The correction made maps the outer product $\dyad{j}{j}$ to $\dyad{0}^{\otimes n}$ if $h_j < n/2$ and $\dyad{1}^{\otimes n}$ if $h_j > n/2$ (it is assumed that $n$ is odd to avoid complications when $h_j = n/2$), it similarly maps $\dyad{j}{\bar{j}}$ to $\dyad{0}{1}^{\otimes n}$ if $h_j < n/2$ and $\dyad{1}{0}^{\otimes n}$ if $h_j > n/2$. This transformation can (just as with the parity check code) be represented as a matrix $E$, where $E^{(j,k)}$ is defined to be the entry of $E$ in the $j$th row and $k$th column
\begin{equation}
E^{(j,k)} = \begin{cases}
1, \text{ if } h_j=0 \text{ and } h_k < \frac{n}{2}\\
1, \text{ if } h_j=n \text{ and } h_k > \frac{n}{2}\\
0, \text{ otherwise}
\end{cases}.
\end{equation}

Using the same methodology as the parity check code (the main difference being that there is no ancillary qubit), the amplitudes of the final quantum state are given by
\begin{equation}
\vec{a} = \Big( E e^{A \tau} \Big)^{t/\tau} \vec{a}_0,
\end{equation}
and,
\begin{equation}
\vec{b} = \Big( E e^{B \tau} \Big)^{t/\tau} \vec{b}_0.
\end{equation}
The solution of $\vec{a}$ is trivial; the only non-zero entries are the first and last, both of which are equal to $1/2$. The solution for $\vec{b}$ is more complicated, but the only significant terms of the matrix are the four corner entries. By discarding the other entries, a reduced version of the problem is obtained
\begin{equation}
\vec{b}^\prime = \begin{pmatrix}
\eta_- & \zeta_+ \\
\zeta_- & \eta_+
\end{pmatrix}^{t/\tau} \begin{pmatrix}
1/2 \\
1/2
\end{pmatrix}
\end{equation}
where the first and second entry of $\vec{b}^\prime$ corresponds to the amplitudes of $\dyad{0}{1}^{\otimes n}$ and $\dyad{1}{0}^{\otimes n}$ respectively, and
\begin{equation}
\eta_\pm = e^{-n \gamma \tau }\sum_{m=0}^{\lfloor n/2 \rfloor} \binom{n}{m} x_\pm ^{n-m} y^m,
\end{equation}
\begin{equation}
\zeta_\pm = e^{-n \gamma \tau} \sum_{m=0}^{\lfloor n/2 \rfloor} \binom{n}{m} x_\pm ^{m} y^{n-m},
\end{equation}
and additionally, $\eta_\pm + \zeta_\pm = \big(r e^{\pm i \phi} \big)^n$. Because $\zeta_\pm \in \mathcal{O} \big( \tau^\frac{n+1}{2} \big)$, it follows that
\begin{equation}
\begin{pmatrix}
\eta_- & \zeta_+ \\
\zeta_- & \eta_+
\end{pmatrix}^{t/\tau} = \begin{pmatrix}
r^n e^{-ni \phi} & 0 \\
0 & r^n e^{+ni \phi}
\end{pmatrix}^{t/\tau} + \mathcal{O}\big((\tau ^\frac{n-1}{2} \big).
\end{equation}
Therefore, for large $n$ the final quantum state in this scenario is very similar to the final quantum state using the parity check code (with a noiseless ancilla and perfect error correction). Mathematically, it is equivalent up to $\mathcal{O}\big( \tau ^\frac{n-1}{2} \big)$. Thus, the QFI is similarly equivalent up to the same order.

%% file: Appendices/AppendixC.tex
\chapter{Soundness Proofs}


\section{Recurring Mathematical Tools}

Before introducing the derivations of the soundness proofs, some frequently recurring tools present in (most of) the proofs are explained here for the sake of organization.

\subsection{Twirling Lemmas}

The first type recurring tool used in the soundness proofs in this Appendix are Clifford twirling lemmas. The Pauli twirling lemma states \cite{dankert2009} that for any $m$ qubit quantum state $\rho$ and Pauli operators $Q, Q^\prime \in \mathcal{P}_m$, with $Q \neq Q^\prime$
\begin{equation}
    \label{eqn:PauliTwirl}
    \sum_{P \in \mathcal{P}_m} P Q P \rho P Q^\prime P = 0.
\end{equation}
The reason is that because $Q \neq Q^\prime$, $\mathcal{P}_m$ can be divided into four equal sets. One set of operators which commutes with both $Q$ and $Q^\prime$, one set which commutes $Q$ and anti-commutes with $Q^\prime$, one set which anti-commutes with $Q$ and commutes with $Q^\prime$, and one set which anti-commutes with both $Q$ and $Q^\prime$. All four of these are equal in size, from which it follows that the above sum is zero.

The proof of the Clifford twirling lemma \cite{dankert2009} is slightly more involved, but the statement is similar: for any $m$ qubit quantum state $\rho$ and Pauli operators $Q, Q^\prime \in \mathcal{P}_m$, with $Q \neq Q^\prime$
\begin{equation}
    \label{eq:CliffordTwirl}
    \sum_{C \in \mathcal{C}_m} C Q C^\dagger \rho C Q^\prime C^\dagger = 0.
\end{equation}
The basis of the proof is similar: for any $Q$ and $Q^\prime$ which are not equal and neither of which is the identity, the operators of the Clifford group can be partitioned into sets of four $C_a,C_b,C_c,C_d$ where
\begin{equation}
    \label{eq:miscgarb1}
    C_a Q C_a^\dagger = C_b Q C_b^\dagger = - C_c Q C_c^\dagger = - C_d Q C_d^\dagger,
\end{equation}
and
\begin{equation}
    \label{eq:miscgarb2}
    C_a Q^\prime C_a^\dagger = -C_b Q^\prime C_b^\dagger = C_c Q^\prime C_c^\dagger = - C_d Q^\prime C_d^\dagger.
\end{equation}
When either one of $Q$ or $Q^\prime$ is the identity, the idea is still true except the corresponding relationship, Eq.~\eqref{eq:miscgarb1} or Eq.~\eqref{eq:miscgarb2}, is no longer true by definition, $C \mathbb{I} C^\dagger=\mathbb{I}$.

A corollary of the Clifford twirling lemma is that the results still holds when the sum is restricted to locally acting Clifford operators
\begin{equation}
    \label{eq:LocalCliffordTwirl}
    \sum_{C \in \mathcal{C}_1^{\otimes m}} C Q C^\dagger \rho C Q^\prime C^\dagger = 0.
\end{equation}
This is apparent when $\rho$ is written as a sum over the Pauli group $P$, and all operators are decomposed into locally acting operators
\begin{equation}
    \sum_{C \in \mathcal{C}_1^{\otimes m}} C Q C^\dagger \rho C Q^\prime C^\dagger = \frac{1}{2^m} \Tr(P \rho) \sum_{P \in \mathcal{P}_m} \bigotimes_{j=1}^m \Big( \sum_{C_j \in \mathcal{C}_1} C_j Q_j C_j^\dagger P C_j Q_j^\prime C_j^\dagger \Big).
\end{equation}
Because $Q \neq Q^\prime$ there exists a $j$ such that $Q_j\neq Q_j^\prime$, the Clifford twirling lemma dictates
\begin{equation}
    \sum_{C_j \in \mathcal{C}_1} C_j Q_j C_j^\dagger P_j C_j Q_j^\prime C_j^\dagger =0,
\end{equation}
and thus the whole sum is zero, proving the locally acting Clifford twirling lemma. Note that $P_j$ not being a quantum state is irrelevant to the proof of the Clifford twirl.

\subsection{CPTP Representation of a Malicious Attack}

Another recurring mathematical tool used in the soundness proofs is to represent an arbitrary attack as a CPTP map $\Gamma$, which can be expanded in terms of a Kraus decomposition $\{A_\alpha\}$
\begin{equation}
    \rho \rightarrow \Gamma ( \rho) = \sum_{\alpha} A_\alpha \rho A_\alpha^\dagger,
\end{equation}
where $\sum_\alpha A_\alpha A_\alpha^\dagger = \mathbb{I}$. Next, a $m$ dimensional Kraus operator can be written with respect to the Pauli basis
\begin{equation}
    A_\alpha= \sum_{P \in \mathcal{P}_m} a_{\alpha,P} P,
\end{equation}
where $a_{\alpha,P}=2^{-m}\Tr(P A_\alpha)$. Therefore, in the Pauli representation, the action of $\Gamma$ is
\begin{equation}
    \Gamma(\rho) = \sum_{\alpha} \sum_{P,Q} a_{\alpha,P} a_{\alpha,Q}^* P \rho Q,
\end{equation}
where the asterisk denotes the complex conjugate. The completeness relationship reads
\begin{equation}
    \label{eq:KrausCompleteness}
    \sum_\alpha \sum_{P \in \mathcal{P}_m} |a_{\alpha,P} |^2=1.
\end{equation}

\section{Unsecured Quantum Channel}

For this protocol, we assume that the ideal output state $\rho_\text{id}$\footnote{We make this assumption for both the single use of the quantum channel, and the double use of the quantum channel.} is a pure state. This is logical assumption since pure states are superior candidates as a resource for quantum metrology. In doing so, the fidelity component of the soundness is equal to trace, greatly simplifying the expression
\begin{equation}
    \label{eq:soundnessexpressions}
    \begin{split}
    &\frac{1}{|\mathcal{K}|} \sum_{k \in \mathcal{K}} p_\text{acc} (k, \Gamma ) \cdot \Big( 1- \mathscr{F} \big(\rho_\text{id}, \rho_\text{out} ( k,\Gamma ) \big) \Big) \\
    =& \frac{1}{|\mathcal{K}|} \sum_{k \in \mathcal{K}} p_\text{acc} (k, \Gamma ) \cdot \Big( 1- \Tr \big(\rho_\text{id} \rho_\text{out} ( k,\Gamma ) \big) \Big) \\
    =& \frac{1}{|\mathcal{K}|} \sum_{k \in \mathcal{K}} \Tr \Big( \Pi_{\text{acc}}(k) \rho_f (k,\Gamma) \Big),
    \end{split}
\end{equation}
where $\rho_f(k,\Gamma)$ is understood as the final ensemble of both ancillary flag qubits and the qubits intended for quantum metrology, and $\Pi_\text{acc}(k)$ projects the ancillary flag qubits onto the `accept state', and the qubits intended for quantum metrology onto $\mathbb{I}-\rho_{\text{id}}$.

\subsection{Trap Code (Single Use)}

When using the trap code, there are a total of $\big|\mathcal{K}\big|=\big| \mathcal{C}_1^{\otimes m} \big| \binom{m}{t}$ possible classical keys; describing both the choices $k$ and $\ell$. The projector $\Pi_\text{acc}(k)$ depends only on the choice of $\ell$ and is independent of the choice of  encryption operation $C$. But for all intents and purposes, it can be expressed as $\Pi_\text{acc}(\ell) = \pi(\ell) \Pi \pi(\ell)^\dagger$ where $\pi(\ell)$ is the permutation corresponding to the random placement of the flag qubits and 
\begin{equation}
    \Pi = (\mathbb{I}-\rho) \otimes \dyad{0}^{\otimes t}.
\end{equation}

Upon receipt and decryption by Bob, the quantum state for a specific key $k$ and attack $\Gamma$ is
\begin{equation}
    \rho_f(k,\Gamma)=C^\dagger \Gamma \big( C \rho_{\text{in},\ell} C^\dagger \big) C,
\end{equation}
where $\rho_{\text{in},\ell}= \pi(\ell) \rho \otimes \dyad{0}^{\otimes t} \pi(\ell)^\dagger$. Thus the soundness is a bound on the quantity
\begin{equation}
    \frac{1}{\binom{m}{t}|\mathcal{C}_1|^m} \sum_{\ell} \sum_{C \in \mathcal{C}_1^{\otimes m}} \Tr \Big( \Pi \pi(\ell)^\dagger \rho_f (k,\Gamma) \pi(\ell) \Big).
\end{equation}
Because of the linearity of the trace, the sum over $C$ can be brought into the trace, which is simplified by expanding $\Gamma$ into a Kraus decomposition and the locally acting Clifford twirling lemma
\begin{equation}
\begin{split}
    \sum_{C \in \mathcal{C}_1^{\otimes m}} \rho_f (k,\Gamma) = & \sum_{C \in \mathcal{C}_1^{\otimes m}} \sum_{\alpha} C^\dagger A_\alpha C \rho_{\text{in},\ell} C^\dagger A_\alpha^\dagger C \\
    =& \sum_{C \in \mathcal{C}_1^{\otimes m}} \sum_{\alpha} \sum_{P_1,P_2 \in \mathcal{P}_m} a_{\alpha,P_1} a_{\alpha,P_2}^* C^\dagger P_1 C \rho_{\text{in},\ell} C^\dagger P_2 C \\
    =& \sum_{C \in \mathcal{C}_1^{\otimes m}} \sum_{\alpha} \sum_{P \in \mathcal{P}_m} |a_{\alpha,P}|^2 C^\dagger P C \rho_{\text{in},\ell} C^\dagger P C. \\
\end{split}
\end{equation}
The next simplification uses the fact that the single qubit Clifford group $\mathcal{C}_1$ maps any non-identity Pauli into an equal distribution over the set $\{ \pm X, \pm Y, \pm Z \}$. It then follows that if $P$ has $d(P)$ non-identity terms, then $CPC^\dagger$ is a similar Pauli operator $\tilde{P}$ (with a phase of $\pm 1$). Specifically, $\tilde{P}$ has the same number of non-identity terms which are indexed by the same positions as the non-identity terms of $P$. The notion of `similarity' is denoted using $
\sim$, for example $\mathbb{I} \otimes X \sim \mathbb{I} \otimes Y \sim \mathbb{I} \otimes Z$. Thus,
\begin{equation}
    \frac{1}{|\mathcal{C}_1|^m} \sum_{C \in \mathcal{C}_1^{\otimes m}} \rho_f (k,\Gamma) = \sum_{\alpha} \sum_{P \in \mathcal{P}_m}   \frac{|a_{\alpha,P}|^2 }{3^{d(P)}} \sum_{\tilde{P} \sim P} \tilde{P} \rho_{\text{in},\ell} \tilde{P}.
\end{equation}

Combining everything thus far
\begin{equation}
\begin{split}
    &\frac{1}{|\mathcal{K}|} \sum_{k \in \mathcal{K}} \Tr \Big( \Pi_{\text{acc}}(k) \rho_f (k,\Gamma) \Big)\\
    =& \frac{1}{\binom{m}{t}} \sum_\ell \sum_\alpha \sum_{P \in \mathcal{P}_m}  \frac{|a_{\alpha,P}|^2}{3^{d(P)}} \sum_{\tilde{P} \sim P}  \Tr \Big( \Pi \pi(\ell)^\dagger \tilde{P} \pi(\ell) \rho \otimes \dyad{0}^{\otimes t} \pi(\ell)^\dagger \tilde{P} \pi(\ell) \Big).
\end{split}
\end{equation}
If $d(P)=0$, then $P$ is identically the identity and the trace is zero. For any $d(P) >0$ and $s \leq d(P)$, there are $\binom{m-d(P)}{t-s}$ permutations $\ell$ where the non-identity terms of $P$ interact with $s$ trap qubits. The only $\tilde{P}$ which results in a non-zero trace is the unique possibility of $Z$ acting on all the $s$ trap qubits. Additionally, when $d(P) \leq t$ and $s=d(P)$ the trace is identically zero, since the Pauli is uniquely acting on the qubits intended for quantum metrology. Otherwise, the trace onto the first $n$ qubits can be bounded by $1$. Hence,
\begin{equation}
\begin{split}
    \frac{1}{|\mathcal{K}|} \sum_{k \in \mathcal{K}} \Tr \Big( \Pi_{\text{acc}}(k) \rho_f (k,\Gamma) \Big) &\leq \sum_{\alpha} \sum_{P \in \mathcal{P}_m} |a_{\alpha,P}|^2 \sum_{s=0}^{s_\text{max}} \frac{1}{3^s} \frac{\binom{m-d(P)}{t-s}}{\binom{m}{t}} \\
    &= \frac{1}{\binom{m}{t}} \sum_{r=1}^m c_r \sum_{s=0}^{s_\text{max}} \frac{1}{3^s} \frac{\binom{m-r}{t-s}}{\binom{m}{t}}.
\end{split}
\end{equation}
where $d(P)$ has been replaced by $r$ as it is no longer dependent on a specific Pauli $P$, $s_\text{max}=r-1$ if $r \leq t$ and $s_\text{max}=t$ otherwise, and $c_r$ is the sum of all $|a_{\alpha,P}|^2$ with $r$ total non-identity indices spanned by $P$. The completeness relationship, Eq.~\eqref{eq:KrausCompleteness}, guarantees that $c_r \leq 1$. Using the upper bound for $c_r$ and swapping the sums of $s$ and $r$, the inequality becomes
\begin{equation}
\begin{split}
    \frac{1}{|\mathcal{K}|} \sum_{k \in \mathcal{K}} \Tr \Big( \Pi_{\text{acc}}(k) \rho_f (k,\Gamma) \Big) &\leq \frac{1}{\binom{m}{t}} \sum_{s=0}^t \Big( \frac{1}{3} \Big)^s \sum_{r=s+1}^{m} \binom{m-r}{t-s} \\
    &=\frac{1}{\binom{m}{t}} \sum_{s=0}^t \Big( \frac{1}{3} \Big)^s \binom{m-s}{t-s+1} \\
    &= \frac{m-t}{t+1}\sum_{s=0}^t \Big( \frac{1}{3} \Big)^s \frac{(t+1)!(m-s)!}{(t-s+1)!m!} \\
    &= \frac{m-t}{t+1}+\frac{m-t}{t+1}\sum_{s=1}^t \Big( \frac{1}{3} \Big)^s \prod_{j=0}^{s-1} \frac{t+1-j}{m-j} \\
    &\leq \frac{m-t}{t+1}+\frac{m-t}{t+1}\sum_{s=1}^t \Big( \frac{1}{3} \Big)^s \big(\frac{t+1}{m}\big)^s \\
    &\leq \frac{3}{2} \frac{m-t}{t}. \\
\end{split}
\end{equation}

\subsection{Clifford Code (Single Use)}

Using the Clifford code, the key is solely dependent on the choice of $C \in \mathcal{C}_m$. The projector $\Pi_{\text{acc}}(k)=(\mathbb{I}-\rho) \otimes \dyad{0}^{\otimes t}$ is independent from this choice.

The derivation begins in a similar fashion to that of the trap code, where the output quantum state is simplified using the Clifford twirling lemma
\begin{equation}
    \sum_{C \in \mathcal{C}_1^{\otimes m}} \rho_f (C,\Gamma) = \sum_{C \in \mathcal{C}_m} \sum_{\alpha} \sum_{P \in \mathcal{P}_m} |a_{\alpha,P}|^2 C^\dagger P C \rho_{\text{in}} C^\dagger P C.
\end{equation}
However, in this case the above can be further simplified as the Clifford group maps any non-identity Pauli uniformly to all other non-identity Pauli operators (up to a phase of $\pm 1$)
\begin{equation}
    \frac{1}{|\mathcal{C}_m|} \sum_{C \in \mathcal{C}_m} C^\dagger P C \rho C^\dagger P C = \frac{1}{|\mathcal{P}_m|-1}\sum_{P^\prime \neq \mathbb{I} \in \mathcal{P}_m} P^\prime \rho P^\prime =\frac{1}{4^m-1} (2^m\mathbb{I}-\rho).
\end{equation}
Denoting $a=\sum_\alpha |a_{\alpha,\mathbb{I}}|^2$, the expected final state is
\begin{equation}
    \frac{1}{|\mathcal{C}_m|} \sum_{C \in \mathcal{C}_m} \rho_f(C,\Gamma) = a \rho_\text{in} +\frac{1-a}{4^m-1}(2^m\mathbb{I}-\rho_\text{in}),
\end{equation}
from which it follows that
\begin{equation}
    \frac{1}{|\mathcal{K}|} \sum_{k \in \mathcal{K}} \Tr \Big( \Pi_{\text{acc}}(k) \rho_f (k,\Gamma) \Big) = \Big(a-\frac{1-a}{4^m-1}\Big) \Tr \big( \Pi \rho_\text{in} \big)+2^m\frac{1-a}{4^m-1} \Tr \big( \Pi \big).
\end{equation}
The first trace is null because $\Tr\big(( \mathbb{I}-\rho)\rho \big)=0$ and the second trace is equal to $2^{m-t}-1$, hence
\begin{equation}
    \frac{1}{|\mathcal{K}|} \sum_{k \in \mathcal{K}} \Tr \Big( \Pi_{\text{acc}}(k) \rho_f (k,\Gamma) \Big) \leq \frac{2^m(2^{m-t}-1)}{4^m-1}  \leq \frac{2^m \cdot 2^{m-t}}{4^m} \leq 2^{-t}.
\end{equation}

\subsection{Trap Code (Double Use)}

In the double use of the quantum channel, depicted in Fig.~(\ref{fig:QM_channeltwice}), a malicious eavesdropper can interact with the quantum channel twice. These interactions can both be represented as CPTP maps, $\Gamma_1$ and $\Gamma_2$. For the trap code, the set of keys is now comprised of the random placement of the flags $\ell$, and two Clifford operations $C_1,C_2 \in \mathcal{C}_1^{\otimes m}$. The projector $\Pi_\text{acc}(k)$ is once again of the form $\pi(\ell) \Pi \pi(\ell)^\dagger$, with the change that $\Pi$ projects onto the encoded quantum state (which is still assumed to be a pure state)
\begin{equation}
    \Pi = (\mathbb{I}-\Lambda_\theta(\rho)) \otimes \dyad{0}^{\otimes t}
\end{equation}

After undergoing the final decryption by Alice, the final quantum state for a specific key $k$ and attacks $\Gamma_1$ and $\Gamma_2$ is given by
\begin{equation}
    \rho_f(k,\Gamma)=C_2^\dagger \Gamma_2 \big( C_2 \Lambda_\theta^{(\ell)} \big( C_1^\dagger \Gamma_2 \big( C_1 \rho_{\text{in},\ell} C_1^\dagger \big) C_1 \big) C_2^\dagger \big) C_2,
\end{equation}
where $\Lambda_\theta^{(\ell)}$ is the parameter encoding operation which only acts on the qubits intended for quantum metrology, if $\sigma$ is an $m$ dimensional quantum state then
\begin{equation}
    \Lambda_\theta^{(\ell)}(\sigma) = \pi(\ell) \big( \Lambda_\theta \otimes \mathbb{I} \big) \big( \pi(\ell)^\dagger \sigma \pi(\ell) \big) \pi(\ell)^\dagger,
\end{equation}
where the identity term in the above equation represents the identity channel acting on the final $t$ qubits. Using the same techniques used in the trap code (single use) proof, the sum over $C_1$ and $C_2$ can be brought into the trace. By representing $\Gamma_1$ and $\Gamma_2$ using Kraus decomposition $\{A_\alpha \}$ and $\{ B_\beta \}$ respectively, and further reducing these operators in the Pauli basis, the sum over $C_1$ and $C_2$ is greatly simplified thanks to the local Clifford twirling lemmas
\begin{equation}
\begin{split}
    &\sum_{C_1,C_2 \in \mathcal{C}_1^{\otimes m}} \rho_f(k,\Gamma) \\
    =& \sum_{C_1,C_2 \in \mathcal{C}_1^{\otimes m}} \sum_{\alpha,\beta} \sum_{P,Q \in \mathcal{P}_m}  |a_{\alpha,P}|^2  |b_{\beta,Q}|^2 C_2^\dagger Q C_2 \Lambda_\theta^{(\ell)} \big( C_1^\dagger P C_1 \rho_{\text{in},\ell} C_1^\dagger P C_1 \big) C_2^\dagger Q C_2,
\end{split}
\end{equation}
where $a_{\alpha,P} = \Tr(A_\alpha P)$ and $b_{\beta,Q} = \Tr(B_\beta Q)$, which satisfy the completeness relationships $\sum_{\alpha,P}|a_{\alpha,P}|^2=\sum_{\beta,Q}|b_{\beta,Q}|^2=1$. This is once again simplified using the notation of similar Pauli operators
\begin{equation}
    \frac{1}{|\mathcal{C}_1|^{2m}} \sum_{C_1,C_2 \in \mathcal{C}_1^{\otimes m}} \rho_f (\Gamma) = \sum_{\alpha,\beta} \sum_{P,Q \in \mathcal{P}_m}   \frac{|a_{\alpha,P}|^2 |b_{\beta,Q}|^2 }{3^{d(P)+d(Q)}} \sum_{\substack{\tilde{P} \sim P \\ \tilde{Q} \sim Q}} \tilde{Q} \Lambda_\theta^{(\ell)} \big( \tilde{P} \rho_{\text{in},\ell} \tilde{P} \big) \tilde{Q}.
\end{equation}

Combining everything thus far
\begin{equation}
\begin{split}
    &\frac{1}{|\mathcal{K}|} \sum_{k \in \mathcal{K}} \Tr \Big( \Pi_{\text{acc}}(k) \rho_f (k,\Gamma) \Big)\\
    =& \frac{1}{\binom{m}{t}} \sum_\ell \sum_{\alpha,\beta} \sum_{P,Q \in \mathcal{P}_m}   \frac{|a_{\alpha,P}|^2 |b_{\beta,Q}|^2 }{3^{d(P)+d(Q)}} \sum_{\substack{\tilde{P} \sim P \\ \tilde{Q} \sim Q}} \Tr \Big( \Pi \pi(\ell) \tilde{Q} \Lambda_\theta^{(\ell)} \big( \tilde{P} \rho_{\text{in},\ell} \tilde{P} \big) \tilde{Q} \pi(\ell)^\dagger \Big).
\end{split}
\end{equation}
To simplify the above expression, we use a slightly different argument to the protocol for the single use case. This time we define $r$ to be the number of non-identity indices spanned by $P$ or $Q$. For example the total number of non-identity indices spanned by $P=\mathbb{I} \otimes X \otimes Z$ and $Q=\mathbb{I} \otimes X \otimes \mathbb{I}$ is $r=2$. Again, for any $s \leq r$, there are $\binom{m-r}{t-s}$ permutations of the trap qubits where the non-identity indices spanned by $P$ or $Q$ interact with $s$ trap qubits. The number of $\tilde{P}$ and $\tilde{Q}$ which results in an accepted outcome is less than $(\frac{5}{9})^s\cdot{3^{d(P)+d(Q)}}$. This is because when the non-identity terms of $\tilde{P}$ and $\tilde{Q}$ do not overlap, the only accepted Pauli is $Z$ (which is $1/3<5/9$ of the possibilities), however, when there is overlap at said index, the Pauli's which are accepted are $ZZ$, $XX$, $YY$, $XY$ and $YX$ (which is $5/9$ of the possibilities). The orthogonal compliment portion of the projector $\Pi$ is again equal to zero when all of the $r$ non-identity terms interact with the trap qubits. Mathematically, in this version of the protocol, we obtain
\begin{equation}
\begin{split}
    \frac{1}{|\mathcal{K}|}\sum_{k \in \mathcal{K}} \Tr \big(  \Pi_\text{acc}(k)  \rho_f (k, \Gamma) \big) &\leq \sum_{\alpha,\beta} \sum_{P,Q}  |a_{\alpha,P}|^2  |b_{\beta,Q}|^2 \sum_{s=0}^{s_\text{max}} \Big( \frac{5}{9} \Big)^s \frac{\binom{m-r}{t-s}}{\binom{m}{t}} \\
    &= \sum_{r=0}^m c_r \sum_{s=0}^{s_\text{max}} \Big( \frac{5}{9} \Big)^s \frac{\binom{m-r}{t-s}}{\binom{m}{t}},
\end{split}
\end{equation}
once again $s_\text{max}=r-1$ for $r \leq t$ and $s_\text{max}=t$ otherwise, and here $c_r \leq 1$ is the sum of all $|a_{\alpha,P}|^2 |b_{\beta,Q}|^2$ with $r$ total non-identity indices spanned by $P$ and $Q$. Using the upper bound for $c_r$ and swapping the sums of $s$ and $r$ we obtain
\begin{equation}
\begin{split}
\frac{1}{|\mathcal{K}|}\sum_{k \in \mathcal{K}} \Tr \big( \Pi_\text{acc}(k) \rho_f (k, \Gamma) \big) &\leq \frac{1}{\binom{m}{t}} \sum_{s=0}^t \Big( \frac{5}{9} \Big)^s \sum_{r=s+1}^{m} \binom{m-r}{t-s} \\
&=\frac{1}{\binom{m}{t}} \sum_{s=0}^t \Big( \frac{5}{9} \Big)^s \binom{m-s}{t-s+1} \\
&= \frac{m-t}{t+1}\sum_{s=0}^t \Big( \frac{5}{9} \Big)^s \frac{(t+1)!(m-s)!}{(t-s+1)!m!} \\
&= \frac{m-t}{t+1}+\frac{m-t}{t+1}\sum_{s=1}^t \Big( \frac{5}{9} \Big)^s \prod_{j=0}^{s-1} \frac{t+1-j}{m-j} \\
&\leq \frac{m-t}{t+1}+\frac{m-t}{t+1}\sum_{s=1}^t \Big( \frac{5}{9} \Big)^s \big(\frac{t+1}{m}\big)^s \\
&\leq \frac{9}{4} \frac{m-t}{t+1} \\
&\leq \frac{9}{4} \frac{m-t}{t}. \\
\end{split}
\end{equation}

\subsection{Clifford Code (Double Use)}

In the double use of the quantum channel, the soundness derivation for the Clifford code is very similar to the proof in the single use case. Simplification using the Clifford twirling lemma leads to the expression
\begin{equation}
\begin{split}
    &\sum_{C_1,C_2 \in \mathcal{C}_m} \rho_f(k,\Gamma) \\
    =& \sum_{C_1,C_2 \in \mathcal{C}_m} \sum_{\alpha,\beta} \sum_{P,Q \in \mathcal{P}_m}  |a_{\alpha,P}|^2  |b_{\beta,Q}|^2 C_2^\dagger Q C_2 \Lambda_\theta \big( C_1^\dagger P C_1 \rho_{\text{in}} C_1^\dagger P C_1 \big) C_2^\dagger Q C_2,
\end{split}
\end{equation}
where it is understood that $\Lambda_\theta$ acts exclusively on the first $n$ qubits. Define $a=\sum_\alpha |a_{\alpha,\mathbb{I}}|^2$ and $b=\sum_\beta |a_{\beta,\mathbb{I}}|^2$. Using the same logic introduced in the single use case, the expected final state is
\begin{equation}
\begin{split}
    & \frac{1}{|\mathcal{K}|}\sum_{k \in \mathcal{K}} \rho_f(k,\Gamma) \\
    =& \Big(ab-\frac{a(1-b)+b(1-a)}{4^m-1}+\frac{(1-a)(1-b)}{(4^m-1)^2} \Big) \Lambda_\theta \big(\rho_\text{in} \big) \\
    &+\frac{(1-a)b+a(1-b)}{4^m-1} 2^m \mathbb{I}.
\end{split}
\end{equation}
Because
\begin{equation}
    \max_{0 \leq a,b \leq 1} \big( (1-a)b+a(1-b) \big)=1,
\end{equation}
it follows that
\begin{equation}
    \frac{1}{|\mathcal{K}|}\sum_{k \in \mathcal{K}} \Tr \big( \Pi_\text{acc}(k) \rho_f(k,\Gamma) \big) = \frac{(1-a)b+a(1-b)}{4^m-1} 2^{m}(2^{m-t}-1) \leq \frac{1}{2^t}.
\end{equation}

\section{Delegated Measurements}

After post-processing, the measurement result Alice receives stems from the measurement statistics of
\begin{equation}
    \pi^\dagger C^\dagger \mathcal{M} \big( \tilde{\rho}=\Gamma(C \pi \rho_\text{in} \pi^\dagger C^\dagger ) \big) C \pi,
\end{equation}
where $\mathcal{M}$ corresponds to measuring the quantum states in the basis of $C (P^{\otimes n} \otimes Z^{\otimes t}) C^\dagger$. If $\mathcal{M}_{\text{id}}$ is the measurement with respect to the basis of $P^{\otimes n} \otimes Z^{\otimes t}$, then
\begin{equation}
    \mathcal{M}(\sigma) = C \pi \mathcal{M}_\text{id}(\pi^\dagger C^\dagger \sigma C \pi ) \pi^\dagger C^\dagger,
\end{equation}
and thus, after post-processing, the measurement result Alice receives stems from the measurement statistics of
\begin{equation}
    \mathcal{M}_\text{id}\big( \pi^\dagger C^\dagger \Gamma(C \pi \rho_\text{in} \pi^\dagger C^\dagger ) C \pi \big).
\end{equation}

Suppose that Alice accepts measurement results, then the remaining measurement statistics will be of the form $\bar{\mathcal{M}}_\text{id}(\rho_\text{out}(k,\Gamma)$, where $\bar{\mathcal{M}}_\text{id}$ is restricted to the measurement results of the $n$ qubits for quantum metrology. The fidelity term in the soundness is bounded
\begin{equation}
    \mathscr{F} \Big( \bar{\mathcal{M}}_\text{id} \big( \rho_\text{id} \big), \bar{\mathcal{M}}_\text{id} \big( \rho_\text{out}(k,\Gamma) \big) \Big) \geq \mathscr{F} \Big(  \rho_\text{id} ,  \rho_\text{out}(k,\Gamma) \Big) = \Tr \Big(  \rho_\text{id}  \rho_\text{out}(k,\Gamma) \Big) ,
\end{equation}
due to the monotonicity of the fidelity. Thus, the soundness is again bounded by the quantity
\begin{equation}
    \label{eq:soundnessexpressions2}
    \begin{split}
    &\frac{1}{|\mathcal{K}|} \sum_{k \in \mathcal{K}} p_\text{acc} (k, \Gamma ) \cdot \Big( 1- \mathscr{F} \big(\mathcal{M}_1(\rho_\text{id}), \mathcal{M}_1(\rho_\text{out} ( k,\Gamma )) \big) \Big) \\
    \leq & \frac{1}{|\mathcal{K}|} \sum_{k \in \mathcal{K}} p_\text{acc} (k, \Gamma ) \cdot \Big( 1- \Tr \big(\rho_\text{id} \rho_\text{out} ( k,\Gamma ) \big) \Big) \\
    =& \frac{1}{|\mathcal{K}|} \sum_{k \in \mathcal{K}} \Tr \Big( \Pi_{\text{acc}}(k) \rho_f (k,\Gamma) \Big).
    \end{split}
\end{equation}
The same notation is used here as in Eq.~\eqref{eq:soundnessexpressions}, where
\begin{equation}
    \rho_f(k,\Gamma) = \pi^\dagger C^\dagger \Gamma(C \pi \rho_\text{in} \pi^\dagger C^\dagger ) C \pi 
\end{equation}
is understood as the ensemble of both ancillary flag qubits and the qubits intended for quantum metrology from which the measurement statistics are derived from, and
\begin{equation}
    \Pi_\text{acc}(k)=\Pi=(\mathbb{I}-\rho_\text{id}) \otimes \dyad{0}^{\otimes t}
\end{equation}
projects the ancillary flag qubits onto the `accept state', and the qubits intended for quantum metrology onto $\mathbb{I}-\rho_{\text{id}}$. More specifically, this combination of $\rho_f(k,\Gamma)$ and $\Pi_\text{acc}(k)$ is equivalent to that of the soundness derivaiton for the single use of the trap code over a quantum channel, and thus the same techniques mathematical techniques can be applied to find that the soundness is bounded by $\frac{3n}{2t}$.